\documentclass{article}
\usepackage{mathtext}
\usepackage{amssymb}
\usepackage{graphics,graphicx}
\usepackage{color}
\usepackage{amsbsy}
\usepackage[left=2.5cm,right=2.5cm,top=2.5cm,bottom=2.5cm,bindingoffset=0cm]{geometry}
\usepackage{amsmath}
\usepackage{ccaption}
\usepackage{epstopdf}
\usepackage{ dsfont }
\usepackage{enumerate}
\usepackage{soul}
\usepackage{verbatim}
\usepackage{ dsfont }
\usepackage[]{natbib}
\title{Dynamic phenomena and crack propagation in dissimilar elastic lattices}

\graphicspath{{./}{./Figures/}}

\author{A. Piccolroaz$^{a}$, N. Gorbushin$^{b}$, G. S. Mishuris$^{c}$ and  M. J. Nieves$^{d, e}$
\\{\small $^{a}$\emph{Dipartimento di Ingegneria Civile, Ambientale e Meccanica} }\\
{\small{\emph{Universit\`a di Trento, 77 I-38123 Trento, Italy}}} \\
$^{b}$ {\small \it Laboratoire de Physique et M\'ecanique des Milieux H\'et\'erog\'enes (PMMH UMR 7636) CNRS, }\\
{\small  \it ESPCI Paris, PSL Research University, 10 rue Vauquelin, 75005 Paris, France}\\
{\it $^{c}${\small Department of Mathematics, Aberystwyth University,} }
\\ {\small {\it Ceredigion SY23 3BZ, Wales, UK}}\\
$^{d}${\small{{\emph{School of Computing and Mathematics}},}} \\ {\small{\emph{Keele University, Keele ST5 5BG, UK}}}
\\
$^{e}${\small{\emph{Department of Mechanical, Chemical and}}}\\
{\small{\emph{Material Engineering, University of Cagliari,}}}{\small{\emph{ 09123, Cagliari, Italy}}}
}
\date{}

\begin{document}
\maketitle

\begin{abstract}
Dynamic Mode III interfacial fracture in a dissimilar square-cell lattice, composed of  two contrasting mass-spring lattice half-planes joined at an interface, is considered. The fracture, driven by a remotely applied load,
is assumed to propagate at a constant speed   along the interface. The choice of the load allows the solution of the problem to be matched with the crack tip field for a Mode III interfacial crack propagating between two  dissimilar continuous elastic materials. The lattice problem is reduced to a system of functional equations of the Wiener-Hopf type through the Fourier transform.
The derived solution of the system fully characterises the  process. 
{We demonstrate the existence of trapped vibration modes that propagate ahead of the crack along the interface during the failure process. In addition, we show as the crack propagates several preferential directions for wave radiation can emerge in the structured medium that are determined by the lattice dissimilarity.} The energy attributed to the wave radiation as a result of the fracture process is studied and admissible fracture regimes supported by the structure are identified. The results are illustrated by numerical examples that demonstrate the influence of the dissimilarity of the lattice on the existence of the steady failure modes and the lattice dynamics. 
\end{abstract}
\section{Introduction}

It is well known that fracture is a process accompanied by several phenomena occurring at
both 
macro- and micro-level scales {including wave radiation, crack surface roughening and crack branching}; for discussion, see, for example 
\citet{Bouchbinder} and \citet{Cox}.  Numerical treatment of such a problem can be usually performed with molecular dynamics simulations embedding complex lattice interactions. Often this can be highly demanding, with the accurate determination of the response of the medium being extremely challenging and difficult to verify.  This also presents additional difficulties in understanding the role of certain physical parameters, especially if they are numerous.  
Here, we develop a model that incorporates 
{phenomena} occurring at both the macro- and micro-scale in the fracture of an anisotropic elastic lattice,  whose far-field behaviour reflects the dynamic Mode III debonding (interfacial crack propagation) of two dissimilar elastic solids and whose microstructure allows us
 to characterise the micro-level features involved in the propagation of a crack for wide range of speeds. 
Further, the lattice studied here provides an efficient way to capture the response of the dissimilar medium undergoing failure and the role of the dissimilarity on global and local processes.

Analytical modelling of failure processes in lattice systems utilises the Wiener-Hopf technique, whereby the governing equations are reduced to a scalar Wiener-Hopf problem and solved as shown by \citet{LSbook}, who developed this approach for homogeneous lattices. Only few attempts have been made to fully analyse the failure of certain dissimilar structured systems using this approach, with the most relevant works being carried out for dissimilar chains by \citet{Gorbushinetaldissimilar} and \citet{Berinskiietal}. 
In general, when the structure undergoing failure is heterogeneous, matrix Wiener-Hopf problems can arise, for which the general procedure to their solution is not known. Heterogeneous lattice systems also play a prominent role in the design of structured metamaterials, capable of performing unconventional responses for a variety of practical purposes. Hence, analysing their response during failure and how they fail is important.



The bonding of dissimilar materials can be found in a wide variety of applications in industry including in joint assemblies, 
laminates, and the reinforcement of frame-like structures found in, for instance,  civil engineering. 
Understanding 
the interfacial crack mechanics is therefore of obvious importance.

Models of the debonding of dissimilar media also have  applications in describing intersonic rupture events associated with the deformations of the earth's crust as discussed by \citet{Rosakis} and \citet{EWangetal}.  These models describe the separation of dissimilar rocks, {caused by the slippage of faults within the earth's crust (see the work of \citet{Brietzkeetal}).  {They  may also be useful in the  analysis of failure in nanoscale multilayered materials, used in micromechanical and electronic devices, studied by \citet{Guoetal}.}

The propagation of interfacial cracks in dissimilar media is a complex phenomenon partly due to the differences in wavespeeds
 in the constituent materials. 
 As a consequence, the limiting speed of the crack is difficult to define. Additionally, different phenomena may occur in both components  of the dissimilar medium as the interfacial crack propagates as well as unexpected behaviour in the crack motion.
 {Molecular dynamics simulations presented by
 \citet{Beuhler} and \citet{Buehleretal}
  for Mode I and II cracks indicate the existence of
  several shock-like wave fields or Mach cones in their wake {within} the weaker material.} Large scale atomistic simulations of failure in dissimilar materials performed by 
\citet{BuehlerGao2} has shown transition mechanisms between  different crack speed regimes can occur discontinuously through crack nucleation mechanisms.  Atomistic simulations  conducted by \citet{Galletal} to analyse the tensile separation of two materials  has shown the interface may become rippled during the separation process.

In addition, the near tip field of a static Mode I and II crack  oscillates in a bimaterial  and this results from complex stress intensity factors as identified by \citet{Rice, Riceetal, England} and \cite{Williams}, although this phenomenon does not exist for Mode III cracks.
  Dynamic stress intensity factors for Modes I-III interfacial cracks in a dissimilar medium formed from two different anisotropic elastic solids have been identified by \citet{Yangetal}. 
  

Crack propagation in homogeneous elastic materials can occur in several regimes depending on 
the shear and pressure wave speeds of the medium. In a similar way, the crack propagation regimes for a dissimilar material depend on 
 the elastic properties of the constituent materials. As an example, for Mode III cracks, these regimes are defined by the minimum and maximum shear wave speeds of the materials forming the dissimilar solid.  This implies that the crack may travel with a speed that is 
 between 
 the two shear wave speeds;
  this is referred to as a transonic crack by \citet{Liuetal}. \citet{Chenetal} have presented an analytical investigation of  Mode III interfacial crack propagation in a continuous bimaterial subjected to gradually increasing uniformly distributed shear stress. \citet{YiShyongetal} presented the transient analysis of  a Mode III subsonic interfacial crack propagating in a bimaterial driven by concentrated dynamic loads.
In the subsonic regime, the crack propagates below the Rayleigh speed of the stiffer material. In this {scenario}, the problem is governed by equations of the mixed type, i.e. one elliptical equation for the stiffer material and a hyperbolic equation for the softer material, as described by  \citet{Yuetal}. This changes
the singularity of the near tip 
field as investigated in \citet{Huangetal2}. The notion of forbidden crack propagation speeds in models of  continuous bimaterials  has been addressed by \citet{Yuetal2} 
using a cohesive zone approach. Local fracture criteria for the dynamic  debonding of  dissimilar materials composed of anisotropic half-planes  has been developed by \citet{Willis}. Asymptotic analysis has been employed by \citet{Chengetal} to analyse significant  transient effects encountered near the tip of an interfacial crack propagating in a bimaterial. Experiments used to investigate subsonic and intersonic debonding regimes in bimaterials are reported by \citet{Chengetal, Lambrosetal} and \citet{Rosakisetal}. 


Several further studies should be mentioned. Mode I interfacial cracks propagating in particulate bimaterials have been investigated by \citet{Kileyetal}.  The separation of two different anisotropic media 
was addressed by \citet{Clements, Suo2} and \cite{Djokovicetal}. Mode III interfacial crack propagation in Cosserat materials has been studied in {\citet{Piccolroazetal2}}. For the study of interfacial crack-defect interaction in bimaterials see the work of \citet{Huangetal} and \citet{Piccolroaz}. A method for controlling the fracture at the interface of a dissimilar medium has been proposed by \citet{Birman} through the use of randomly distributed inclusions. \citet{BeomJang} have derived the solution for problem of a bimaterial containing an interfacial wedge crack subjected to  Mode III loading. Subsonic and transonic  crack propagation in magneto-electro-elastic bimaterials has been treated by \citet{MaSuFeng}.

However, as shown by \citet{Fineberg}, continuum mechanics models
are not capable of resolving many important questions that arise from experiments on fracture concerning the radiation of sound, the roughening of crack surfaces and the branching of a crack when it propagates. As an alternative,  the analytical and numerical  treatment  of failure propagation in a lattice can be used to provide useful insights in this regard. 
In contrast to a continuum, 
lattice modelling provides various flexibilities in capturing special responses attributed to the lattice geometry or composition.
In addition, several wave radiation processes that can  influence the direction or nature of the fracture propagation may also be examined as in the studies by \citet{Marder2} and  \citet{Marder}. Moreover, the lattice can be  physically constructed to 
improve the overall behaviour of the material for specific purposes, allowing it to surpass the performance of conventional materials (see for example the work of \citet{Berger}). Here, we focus on the dynamics 
of {a} crack propagating in a dissimilar lattice, whose microstructure embeds several classes of defects. Our aim is to understand how 
these factors affect
 the  dynamics of the system and the crack itself.


The first model of dynamic fracture  in a discrete medium that embeds  micro-level phenomena can be found in the works of \citet{LS83} and \citet{LS1}, where the analysis of dynamic Mode III fracture of a square-cell lattice driven by a remote load was considered. There, it was shown that by introducing a moving coordinate that follows the crack tip and employing the Fourier transform, the entire problem may be reduced to a scalar Wiener-Hopf equation. This functional equation contains information concerning possible micro-structural  lattice vibrations and when solved the lattice behaviour attained during the fracture propagation can be traced through the inverse Fourier transform.
This method has also been developed in detail in the monograph by \citet{LSbook}, where an 
in depth review of relevant works of
the last century is 
also 
given.

Following  the approach of 
 \citet{LS83} and \citet{LS1}, several studies of 
 various structured media undergoing dynamic failure caused by different loads  have appeared.
\citet{LS2} considered Mode I and II dynamic failure of a triangular-cell lattice representing the micro-structural model of an elastic continuum. Wave induced phase transitions in mass-spring chains were also studied by \citet{LS3}. The 
influence of periodically distributed lattice inhomogeneities in Mode III crack propagation and the associated wave radiation processes in square-cell lattices were 
investigated by \citet{MMS}. \citet{NMJM} have analysed  the influence of material's anisotropy  on Mode I fault propagation in elastic lattices.
Crack propagation driven by exponentially localised wave forms produced by oscillatory loads 
were studied in the works of  \citet{MMS1} and \citet{SMM}. The 
failure induced by moving loads 
in  a one-dimensional elastic chain  
was considered by \citet{GM2}. Phase transitions of  two-dimensional  lattices with local nonlinearities  
were treated by \citet{SlepyanAyzenbergStepanenko}.

{In the studies presented by \citet{LSbook},
 a crack  propagates in the lattice as a result of the sequential rupture or phase transition of links contained in two adjacent lattice rows. Alternate failure mechanisms characterising how a defect propagates in a lattice can also be considered, such as in the work of \citet{MMS2}, where a removal of a mass-spring chain 
  from a two-dimensional square-cell lattice is considered. } 
 {In addition, failure mechanisms leading to the propagation of bridge cracks  in lattice systems, representing simplified models of void nucleation and crack growth, 
} 
 were investigated by \citet{MMS3}. Different fracture criteria can also influence how steady state failure regimes are achieved; 
 this issue was examined,
 for example, by \citet{GGM}, where a time dependent fracture criterion simulating the discrete analogue of the incubation time failure criterion
 was considered. The role of different fracture criteria on the dynamic propagation of cracks in a micropolar medium have been examined by \citet{Morinietal} using a continuum model embedding intrinsic length scales describing the microstructure of the material. Approaches for treating the dynamics of damage in lattice systems composed of rods have been proposed by \citet{Cherkaevetal}. 

{A defect in a structured medium may also induce localised deformation modes,
with
large strains 
in the vicinity of the defect as shown by \citet{Osharovichetal} and \cite{Colquittetal}. 
Controlling interfacial waves in structured media also has  applications  in topological protection and the design of  seismic shields. Passive methods for controlling interfacial waves in structured plates have been proposed by \citet{MakwanaCraster} and this has been extended by \citet{MakwanaCraster2} to develop an algorithm for designing  interfacial wave networks. 
This has also been applied  by \citet{Makwanaetal}  to design interfacial wave systems  for photonic crystals.
Granular systems with rotational inertia have been shown to admit interfacial waveforms by
 \citet{Zheng}. \citet{Garau}  have demonstrated gyroscopic spinners attached to discrete elastic media can generate interfacial waves with preferential directionality depending on the frequency of the applied load.} {
 Interfacial dynamic regimes may also be crucial in the initiation and support of interfacial fracture, leading to possible non-steady failure regimes.}

The method developed in \citet{LSbook}  for analysing dynamic lattice fracture of structured systems  leads to a solution that may violate some of the model assumptions, such as the uniqueness of the crack front, as demonstrated by \citet{MG}. Thus, 
solutions 
must be checked against the modelling assumptions. Admissibility of {failure  modes} in square-cell lattices 
was investigated  by \citet{GM}.

Different intermolecular interactions  in a discrete system also 
affect phase transitions.
We refer to the work of  \citet{GM3}, where admissible failure regimes in chains with non-local  interactions 
were
studied. \citet{LSmaterial} considered the influence of distributed mass along the lattice connections 
on lattice fracture.
Further, one can achieve different failure modes when the spring connections in the lattice are replaced by
flexible
elements. Such media have a larger range of applicability as they can be used as simplified models of  civil engineering systems.
The collapse of bridges was modelled by \citet{BMS} and \citet{BGMS} using mass-beam systems subjected to gravitational or thermal loads, respectively. Extension of the analytical methods developed in the work of  \citet{LSbook} to mass-beam systems subjected to vibrating loads
was carried out by \citet{NMS}  and the transient analysis of such failure processes in flexural systems 
was given in the work of \citet{NMS2}.

The problems mentioned above, concerning the failure of two-dimensional structured media, have an inherent symmetry 
linked to type of lattice considered and the action of the applied load. This allows certain physical symmetry conditions to be applied in the study of a defect advancing inside the structure.
 In this case, they can be reduced to the scalar Wiener-Hopf problem.


\begin{figure}[htbp]
\center{\includegraphics[scale=1.3]{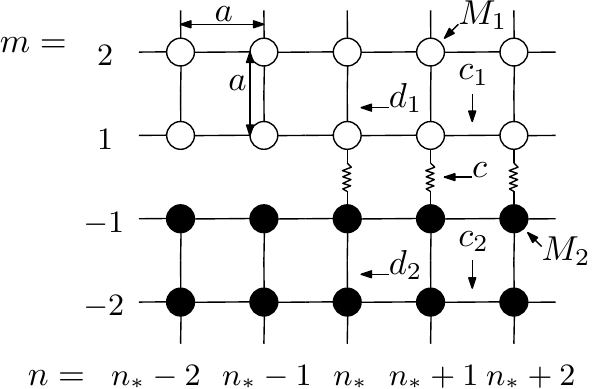}}
\caption[ ]{Dissimilar lattice composed of two different structured media that are 
joined by vertically aligned springs between the rows $m=-1$ and $m=1$; 
the defect 
propagates with a 
constant speed ${V}$ {in the direction of increasing $n$}.}
\label{fig:Infinite Lattice}
\end{figure}

 However, as mentioned, few solutions have been produced for the dynamic fracture propagation in a dissimilar structured lattice.
 Moreover, 
 such problems pose a considerable challenge, as they
 lead to matrix Wiener-Hopf problems. Unfortunately, the general factorisation procedure for this situation is not available, 
 although it 
  exists for certain specific matrix Wiener-Hopf problems, with some described by \citet{Rogosinetal}.
 For the analysis of problems  concerning dissimilar structured media, we mention
the static analysis of bi-material lattices or bi-crystals with interfacial cracks;
 a general method for determining the static Green's functions for such configurations
  was also developed by \citet{Tewary}.
Numerical approaches to modelling failure of dissimilar systems via the molecular dynamics approach 
are given
by \citet{Beuhler}. 

Here,
we treat
the analytical model of failure in a two-dimensional dissimilar lattice along an interface. 
 We focus on the simplest, Mode III, conditions.
 Two different semi-infinite lattices are joined by spring connections that form an interface between the lattices, as in Figure \ref{fig:Infinite Lattice}. There, if the interfacial crack bond stiffness $c$ is equal to the transverse bond stiffness  of either the upper or the  lower lattice, the interface may be considered as perfect, otherwise it is imperfect. {In this sense, a perfect interface can correspond to two scenarios and these are further explored here.}
 A  remote 
 load is 
 applied driving the crack (modeled by breakage of the spring connections) that is assumed to propagate at constant speed.  The choice of the load leads to a  solution that matches the behaviour observed near the tip of an Mode III interfacial crack propagating between two contrasting elastic solids.
The model may be interpreted as 
a discrete analogue of the dynamic failure of a continuous bi-material. 
We are able to reduce the governing equations to a triangular matrix Wiener-Hopf system for which the factorisation method is known (see  \citet{Rogosinetal}).

{
{Similar to  problems considered by \citet{LSbook} for homogenous lattices, here we show that vibrations occurring at both low and high frequencies can appear within the lattice as the crack propagates. However, for the dissimilar medium we identify novel effects such as trapped vibration modes that can propagate along the interface of the elastic lattices that is a feature not encountered in the corresponding continuum model.   Additionally, we illustrate how the dissimilarity affects the dynamics of the bulk medium undergoing failure.
 }
In particular, we:
\begin{enumerate}[$\bullet$]
\item show that the dissimilar lattice admits the discrete analogue of Stoneley waves, previously observed for continuous bi-materials by \citet{Stoneley} and  \citet{Barnett}. We also 
obtain
the conditions for the  lattice 
parameters (Figure \ref{fig:Infinite Lattice}) required to control the appearance of
interfacial wave regimes. Further, we demonstrate that the interfacial wave modes are linked to 
specific dispersive features of the medium, represented by the appearance of continuous or discontinuous optical curves.

\item determine important information related to the wave radiation processes that result from the 
propagation of a crack in the dissimilar structure. Two wave radiation regimes resulting from the failure process  are identified. They include (i) a regime  where waves are released within Mach cones emanating from the crack tip as this advances and (ii) a regime where waves radiated behind the crack tip are accompanied by interfacial waves that can be excited ahead of the crack tip as it advances. 

\item investigate the behaviour of the energy released during the lattice fracture in comparison to the energy released in the analogous continuum bi-material and we determine the validity of the constructed solution against the imposed failure conditions.
 
\end{enumerate}
The influence of the material parameters governing the lattice on the above effects are also identified.

 
%

{

}

{

The structure of the article is as follows. In Section \ref{sec2}, we 
formulate the problem and
 the governing equations for the  dissimilar  lattice undergoing steady state fracture propagation driven by a
  remote  load.
 We also  analyse the solution of the problem in the bulk lattice and reduce the 
 problem 
 to a Wiener-Hopf equation 
 by using the Fourier transform with respect to the moving coordinate connected 
 to the crack. 
 In Section \ref{Disp_prop}, we determine the  properties of the kernel function of the Wiener-Hopf equation and the dispersive nature of the dissimilar lattice. There, we also investigate the eigenmodes of two separate unbounded structures connected with the considered problem and show how their dispersive properties influence  dynamic interfacial failure  in the dissimilar lattice.
The Wiener-Hopf equation is then solved in Section \ref{solWH}, 
and the dynamic properties of the dissimilar lattice during the steady fracture process are discussed.
 Results concerning the energy released by the crack when propagating in the lattice are also obtained. The local-to-global energy release rate ratio linking characteristics of the interfacial crack propagation in the lattice and the analogous continuum is discussed in Section \ref{secERR}.
In Section \ref{sec3}, 
numerical examples are 
given that show how the behaviour of the system and wave radiation processes 
are affected by the properties of the dissimilar lattice. 
We also analyse the admissibility of  the associated failure modes.
In Section \ref{conclusions}, we 
summarize and discuss the findings. In the Appendix, we present the conditions on the lattice material parameters that provide this medium with the ability to support vibration modes that are exponentially localised along the interface. Finally, in the Supplementary Material we provide 
details of some derivations.
}

\section{Crack in a square-cell dissimilar  lattice}\label{sec2}
\subsection{The configuration of the dissimilar lattice with a moving crack}\label{sec2.1}
We consider a crack propagating in a dissimilar square-cell lattice, shown in Fig.~\ref{fig:Infinite Lattice}, composed of massless springs connecting a periodic square arrangement of point masses having positions $(m, n)\in  {\mathbb{Z}^2}$, $m\ne 0$. At the junctions with the index $m\ge 1$ ($m\le -1$) the nodes have mass $M_1$ $(M_2)$. All elastic connections in the lattice have length $a$. Additionally, horizontal and vertical links in  the regions characterised by $m\ge 1$ ($m\le -1$) have the stiffness $c_1$ and $d_1$ ($c_2$ and $d_2$), respectively. We define by
\begin{equation}
v_j=\sqrt{\frac{c_j a^2}{M_j}}, \quad {j=1,2}
\label{eq:ModelParameters}
\end{equation}
the shear wave speeds for upper and lower lattice half-planes, respectively. 

Between the rows $m=-1$ and $m=1$, the vertical links connecting the upper and lower parts of the lattice have a stiffness $c$. This part of the lattice forms an interface linking the upper and lower lattice half-planes.
In this interface,  a semi-infinite defect is assumed to propagate with a uniform speed ${V}$, whose front has the position defined by the index $n_*=n_*(t)$, $t\ge 0$. 
{The speed of
crack growth   is limited by a critical value, i.e.:
\begin{equation}
V<v_c,\quad v_c=\min\{v_1,v_2\}\;.
\label{eq:CriticalSpeeda}
\end{equation}}
The crack is assumed to propagate as a result of the action of an out-of-plane remote load prescribed at infinity (see the monograph by \citet{LSbook}) and this crack follows  a straight path as a result of the subsequent removal of vertical links connecting the {dissimilar regions}. 

\subsection{Model of crack propagation in a dissimilar lattice}
We study Mode III fracture of the dissimilar lattice and assume that crack propagates along the interface following a straight path. The  propagation of the fault is caused by an energy flux coming from infinity towards a crack tip.  {The energy flux imposed allows us to asymptotically match the field in the vicinity of the crack tip in the analogous continuum   with the far-field behaviour in the structured medium (see Figure \ref{fig:Infinite Lattice}).}
In the formulation that follows, this condition is not explicitly presented in the equations but will be taken into account in subsequent sections. {The approach adopted also allows us to perform multi-scale numerical studies. }

In this case, masses with $m\ge 1$, $n\in \mathbb{Z}$,  have the out-of-plane displacements $u_{m, n}$,  whereas  {$w_{-m, n}$} is used to represent the displacement of the masses in the lower part of the lattice defined by $m\le -1$,  $n\in \mathbb{Z}$. The crack will advance a distance $a$ when the fracture criterion
\begin{equation}
|u_{1,n_*}(t_*)-w_{1,n_*}(t_*)|=\epsilon_c,
\label{eq:FractureCondition}
\end{equation}
is achieved.  Here, $\epsilon_c$  represents the critical elongation of a spring connecting the upper and lower half-planes of the dissimilar lattice and  is considered to be a material property and  $t_*$ is the  fracture time. In addition, we set the requirement that
\begin{equation}
|u_{1,n_*}(t)-w_{1,n_*}(t)|<\epsilon_c\;,\quad n>n_*\;,
\label{eq:FractureCondition_1}
\end{equation}
at any moment during the steady state crack propagation. 
 We assume that the remote load applied to the lattice ensures
that the failure of a link along the interface is always achieved with a positive crack opening.
Therefore, we will neglect the absolute value applied in conditions (\ref{eq:FractureCondition}) and (\ref{eq:FractureCondition_1}). 
Fracture propagating as a result of an alternating critical strain can be achieved under specific loading configurations that lead to micro-structural oscillations within the medium, as shown by  \citet{MMS1} and \citet{Garaufracture}.

We call a solution to the problem \emph{admissible} if  criteria (\ref{eq:FractureCondition}) and (\ref{eq:FractureCondition_1}) are satisfied. Condition \eqref{eq:FractureCondition_1} ensures  the uniqueness of the crack tip. Indeed, when (\ref{eq:FractureCondition_1}) is violated one can detect a failure of a spring somewhere ahead of the main crack where $n>n_*$. In this scenario, we say the solution is \emph{not admissible}. In such cases a special analysis should be performed, whereas in this study we focus on steady state regimes, where the crack tip is uniquely defined and propagates with the constant speed.\footnote{{Since we consider  {a} dissimilar anistropic lattice with {an} interface the following five scenarios are possible. First is the failure of the interfacial links as described above. The other four concern  the failure of horizontal or vertical links in both of the lattices either above or below the interface. If one has a specific set of elastic and strength properties of the anisotropic dissimilar lattice with an interface, then it is possible to analyse all scenarios for the failure of the lattice. However, the analysis of all these possible failure conditions for the general dissimilar lattice composed of two different anisotropic bodies with an interface is rather ``mission impossible". As a result, we will restrict our attention to the analysis of the interface failure described {by} the conditions (\ref{eq:FractureCondition}) and (\ref{eq:FractureCondition_1}).}}
 
The governing equations for each mass in the lattice are as follows.
 The linear momentum balance  of the masses in the upper and lower half-spaces is given by
\begin{eqnarray}
&& M_1\ddot{u}_{m,n}=c_1(u_{m,n+1}+u_{m,n-1}-2u_{m,n})+d_1(u_{m+1,n}
+u_{m-1,n}-2u_{m,n})\;,\quad \text{ for } m> 1\;, \label{eqLattice1}\\[3mm]
&& M_2\ddot{w}_{-m,n}=c_2(w_{-m,n+1}+w_{-m,n-1}-2w_{-m,n})+d_2(w_{-m+1,n}
+w_{-m-1,n}-2w_{-m,n})\;,\quad \text{ for } m<-1\;, \nonumber \\
\end{eqnarray}
respectively. Along the rows of the lattice containing the defect, the governing equations are
\begin{eqnarray}
M_1\ddot{u}_{1,n}=c_1(u_{1,n+1}+u_{1,n-1}-2u_{1,n})-c(u_{1,n}-w_{1,n})H(n-n_*)+d_1(u_{2,n}-u_{1,n}),\quad \text{ for } m=1\;, 
\end{eqnarray}
and 
\begin{eqnarray}\label{eqLattice4}
M_2\ddot{w}_{1,n}=c_2(w_{1,n+1}+w_{1,n-1}-2w_{1,n})+c(u_{1,n}-w_{1,n})H(n-n_*)+d_2(w_{2,n}-w_{1,n}), \quad \text{ for }m=-1\;.
\end{eqnarray}
Here, the dots denote differentiation with respect to time $t$ and $H$ is the Heaviside function:
\begin{equation*}
H(x)=\left\{\begin{array}{ll}
1\;, & \quad \text{ for } x\ge 0\;,\\
0\;, & \quad \text{ otherwise\;.}
\end{array}\right.
\end{equation*}

\subsection{Normalisation of the governing equations}
Next, we obtain a dimensionless form of (\ref{eqLattice1})--(\ref{eqLattice4}) by setting 
\[\hat{t}=\sqrt{\frac{c_1}{M_1}} t\;, \qquad \hat{v}=\sqrt{\frac{M_1}{c_1 a^2}} V{=\frac{V}{v_1}}\;, \qquad \hat{u}_{m,n}(\hat{t})=\frac{u_{m, n}(t)}{a}\;,\qquad  \hat{w}_{m,n}(\hat{t})=\frac{w_{m, n}(t)}{a}\;, \]
and introducing the normalised  parameters corresponding to the shear wave speeds and the critical elongation
\[\hat{v}_j={\sqrt{\frac{M_1}{c_1 a^2}} v_j=\frac{v_j}{v_1}}\;, \qquad  \hat{\epsilon}_c=\frac{\epsilon_c}{a}\;.\]
Finally, the material contrast  parameters 
\begin{equation}\label{eqParam}
\hat{\beta}=\frac{M_1}{M_2}\;\qquad \hat{\gamma}=\frac{c_2}{c_1}, \qquad \hat{\mu}=\frac{c}{c_1}\qquad \hat{\alpha}_j=\frac{d_j}{c_1}\;,  \quad j=1,2\;,
\end{equation}
are used below.

{ We note that the subsequent results should hold in the case when the upper and lower lattices of the system are interchanged. This will be not reflected in the dimensionless equations due to the normalisation taken in time. However, to realise this symmetry  inherited from the physical consideration of the lattice one should rescale dimensionless time $t$ as follows:
\[
\hat{t}^{\rm new}=\sqrt{\hat{\beta} \hat{\gamma}}\, \hat{t}=\sqrt{\frac{c_2}{M_2}} t\;,
\]
where $\hat{t}^{\rm new}$ will be the new dimensionless time. Then, in order to represent the interchange of the upper and lower lattices, the normalised parameters in (\ref{eqParam}) should be linked to new material parameters as follows:
\[
\hat{\beta}^{\rm new}=\frac{1}{\hat{\beta}}\;,\quad \hat{\gamma}^{\rm new}=\frac{1}{\hat{\gamma}}\;,\quad \hat{\alpha}^{\rm new}_1=\frac{\hat{\alpha}_2}{\hat{\gamma}}\;, \quad \hat{\alpha}^{\rm new}_2=\frac{\hat{\alpha}_1}{\hat{\gamma}}\; \quad \text{ and } \quad \hat{\mu}^{\rm new}=\frac{\hat{\mu}}{\hat{\gamma}}\;. 
\]
If it is necessary one can always reduce to the case 
$\hat{\beta} \hat{\gamma} \le 1$ by interchanging the upper with the lower lattice and applying the above alternate normalisation.
}

The quantities appearing above with the symbol ``\,\,$\hat{}$\,\," are all dimensionless. We omit this symbol in going forward for ease of notation. Then the dimensionless equations of motion are
 \begin{eqnarray}
&& \ddot{u}_{m,n}= u_{m,n+1}+u_{m,n-1}-2u_{m,n}+\alpha_1(u_{m+1,n}
+u_{m-1,n}-2u_{m,n})\;,\quad \text{ for } m> 1\;, \label{eqLattice1a}\\[3mm]
&&\ddot{u}_{1,n}=u_{1,n+1}+u_{1,n-1}-2u_{1,n}-\mu (u_{1,n}-w_{1,n})H(n-n_*)+\alpha_1(u_{2,n}-u_{1,n}),\quad \text{ for } m=1\label{eqLattice2a}\;, \\ [3mm]
&&  \ddot{w}_{1,n}=\beta \big[\gamma (w_{1,n+1}+w_{1,n-1}-2w_{1,n})+\mu(u_{1,n}-w_{1,n})H(n-n_*)+ \alpha_2(w_{2,n}-w_{1,n})\big], \quad \text{ for }m=-1\label{eqLattice3a}\;,\nonumber\\
\end{eqnarray}
and 
\begin{eqnarray}\label{eqLattice4a}
 \ddot{w}_{-m,n}=\beta \big[\gamma (w_{-m,n+1}+w_{-m,n-1}-2w_{-m,n})+\alpha_2(w_{-m+1,n}
+w_{-m-1,n}-2w_{-m,n})\big]\;,\quad \text{ for } m<-1\;.
\end{eqnarray}
In addition, the constraints for the establishing the steady-state  crack growth remain as (\ref{eq:FractureCondition}) and (\ref{eq:FractureCondition_1}) in the dimensionless form. 




\subsection{Solution of the problem in the lattice half-planes}
\subsubsection*{The moving coordinate and additional assumptions}
We switch to a moving reference frame with the origin at the moving crack tip by the change of variable
\begin{equation}
\eta=n-vt,\quad v={\rm const}\;.
\label{eq:eta}
\end{equation}
In the  subsonic speed regime the normalised speed $v$, according to (\ref{eq:CriticalSpeeda}), satisfies the inequality
\begin{equation}\label{speedbound}
{ v<\min\{v_1, v_2\}}
\end{equation}
with the normalised shear wave speeds now being $v_1=1$, $v_2=(\beta \gamma)^{1/2}$. 

In the development below, we treat the coordinate $\eta$ in \eqref{eq:eta} as continuous (see the monograph by \citet{LSbook}). 
The displacements are also assumed to depend  on  $\eta$ only, so that
\begin{equation}\label{form1}
u_{m,n}(t)=u_m(\eta),\quad w_{-m,n}(t)=w_{-m}(\eta),
\end{equation}
and in this way we eliminate the explicit dependence on time $t$. An example of an analysis investigating the transition to the steady state regime can be found in the work of  \citet{GM2}.   In this setting, the  conditions in \eqref{eq:FractureCondition} and \eqref{eq:FractureCondition_1} are modified, having the form
\begin{equation}
u_{1}(0)-w_{1}(0)=\epsilon_c
\label{eq:FractureCondition_eta}
\end{equation}
and
\begin{equation}
u_{1}(\eta)-w_{1}(\eta)<\epsilon_c,\quad \eta>0.
\label{eq:FractureCondition_eta_1}
\end{equation}
In what follows, we quote some results 
that arise when attempting to reduce the problem to a Wiener-Hopf equation along the crack. Their derivation can found in the Supplementary Material \ref{SM1}. 

\subsubsection*{Transformed equations for the lattice half-spaces}
After introducing the dependencies in (\ref{form1}), and following the Fourier transform with respect to the moving coordinate $\eta$, one can show (\ref{eqLattice1a}) and (\ref{eqLattice4a}) reduce to 
\begin{equation}
\begin{gathered}
\left[(0+{\rm i}kv)^2+\omega_1^2(k)+2\alpha_1\right]U_m(k)=\alpha_1(U_{m+1}(k)+U_{m-1}(k)),\quad m>1,\\
\left[(0+{\rm i}kv)^2+\omega_2^2(k)+2\beta \alpha_2\right]W_{-m}(k)=\beta \alpha_2(W_{-m+1}(k)+W_{-m-1}(k)),\quad m<-1,
\end{gathered}
\label{eq:FourierInsideLattice}
\end{equation}
where the functions $U_m$ and $W_{-m}$ are the Fourier transforms of displacements  given by
\begin{equation}
U_m=\mathcal{F}[u_m]=\int_{-\infty}^{\infty}u_{m}e^{{\rm i}k\eta}\,d\eta\quad \text{ and } \quad  W_{-m}=\mathcal{F}[w_{-m}]=\int_{-\infty}^{\infty}w_{-m}e^{{\rm i}k\eta}\,d\eta\; \;, \label{FTeta}
\end{equation}
and
\begin{equation}
\omega_1^2(k)=4\sin^2{\left(\frac{k}{2}\right)}\quad \text{ and } \quad \omega_2^2(k)={4}{\beta}\gamma \sin^2{\left(\frac{k}{2}\right)},
\label{eq:DispersionRelations}
\end{equation}
which are the dispersion relations for the upper and lower half-planes of the lattice. Note that $(\beta\gamma)^{1/2}$ is  the ratio of the shear wave speed of the lower lattice half-plane to the wave speed in the upper lattice half-plane. In addition, in (\ref{eq:DispersionRelations}),  $k$ is the dimensionless wavenumber, with $\hat{k}={k}/a$ being the wavenumber with dimension. {As we show later}, the above relations enable us to partially characterise the dynamics of the system during the steady-state failure process.

Below we also use the half range Fourier transforms:
\[\{U^\pm_m, W^\pm_{-m}\}=\mathcal{F}[\{u_m, w_{-m}\}H(\pm \eta)]=\int_{-\infty}^\infty\{u_m, w_{-m}\}H(\pm \eta) e^{{\rm i}k\eta}\, d\eta\;,\]
where the functions labelled with a ``+" (``$-$") are analytic for ${\text{Im}}(k)>0$ (${\text{Im}}(k)<0$) and allow for the additive splits:
\begin{equation}\label{addsplitUW}
U_m(k)=U_m^+(k)+U_m^-(k)\;, \quad W_{-m}(k)=W_{-m}^+(k)+W_{-m}^-(k)\;. 
\end{equation}
Additionally, here the notation $``0+ {\rm i}k v"$, appearing  in (\ref{eq:FourierInsideLattice}) and throughout the analysis, represents the limit
\begin{equation}
0+ {\rm i}kv=\lim_{s\to0+}(s+ {\rm i}kv),
\label{eq:Limit_s}
\end{equation}
which is also connected with the causality principle {(see the book of \citet{LSbook}, Chapter 3)} and reflects the transition from the  transient  to the steady-state behaviour of the system.
\subsubsection*{Solution of the transformed equations for $m\ne \pm 1$}
 The coefficients in the system of linear equations (\ref{eq:FourierInsideLattice})  do not depend on the value of $m$ and $n$. Hence, it can be shown that their solution can be written as
\begin{equation}
\begin{gathered}
U_m(k)=[\lambda_1(k, 0+{\rm i}kv)]^{m-1}U_1(k),\quad m \ge 1,\\
W_{-m}(k)=[\lambda_2(k, 0+{\rm i}kv)]^{-m-1}W_1(k),\quad m\le -1,
\end{gathered}
\label{eq:LambdaIntroduction}
\end{equation}
where
\begin{eqnarray}
\lambda_j(k, s)=\frac{\sqrt{\Omega_j(k, s)+1}-\sqrt{\Omega_j(k, s)-1}}{\sqrt{\Omega_j(k, s)+1}+\sqrt{\Omega_j(k,s)-1}}\;, \quad j=1,2,
\label{eq:Lambda_final}
\end{eqnarray}
and 
\begin{equation}\label{eqOM}
\Omega_j(k, s)=\frac{1}{2\beta^{j-1}\alpha_j}\left(s^2+\omega_j^2(k)+2\beta^{j-1}\alpha_j\right) \quad \text{with}\quad  \Omega_j(k, s)=\frac{1}{2}\left(\lambda_j(k, s)+\frac{1}{\lambda_j(k, s)}\right)
\end{equation}
for $j=1,2$. Here, (\ref{eq:Lambda_final}) represents the characteristic root of the recursive equations (\ref{eq:FourierInsideLattice}) and we note 
\begin{equation}
|\lambda_j(k, 0+{\rm i}kv)|\le1,\quad j=1,2.
\label{eq:Lambda_condition}
\end{equation}

\subsection{Derivation of the scalar Wiener-Hopf problem}
We now turn to employing assumption (\ref{form1}) in  \eqref{eqLattice2a} and \eqref{eqLattice3a}, which describe the motion of the masses along the interface, to identify a Wiener-Hopf equation in terms of the Fourier transformed variables.  
\subsubsection*{Auxiliary functions}
In the subsequent analysis of these equations  it is useful to introduce linear combinations of displacements:
\begin{equation}\label{psiphi}
\psi(\eta)=u_1(\eta)-w_1(\eta)\quad \text{ and }\quad \phi(\eta)=u_1(\eta)+w_1(\eta)\;.
\end{equation}
The function $\psi(\eta)$ represents the crack opening for $\eta<0$ and the  elongation of the springs  for $\eta\ge 0$. As a result, conditions \eqref{eq:FractureCondition_eta} and \eqref{eq:FractureCondition_eta_1} can be reformulated in terms of function $\psi(\eta)$ as
\begin{equation}
\psi(0)=\epsilon_c,\quad  \text{ and } \quad  \quad \psi(\eta)<\epsilon_c,\,\eta>0\;.
\label{eq:FractureCondition_psi}
\end{equation}
The introduction of the function $\phi(\eta)$ provides a presentation of  subsequent  equations in a compact form. We define the Fourier transforms of $\psi$ and $\phi$ by
\begin{equation}
\Psi(k)=\int_{-\infty}^{\infty}\psi(\eta)e^{{\rm i}k\eta}\,d\eta \quad \text{ and }\quad  \Phi(k)=\int_{-\infty}^{\infty}\phi(\eta)e^{{\rm i}k\eta}\,d\eta\;.\label{FTPsiPhi}
\end{equation}
Later, we will use the following representation of the function $\Psi$ 
\begin{equation}
\Psi(k)=\Psi^+(k)+\Psi^-(k),\quad 
\Psi^{\pm}(k)=\int_{-\infty}^{\infty}\psi(\eta)H(\pm\eta)e^{{\rm i}k\eta}\,d\eta\;.
\label{psieq2}
\end{equation}
\subsubsection*{The matrix Wiener-Hopf problem}
Applying the Fourier transform in   $\eta$ to the combination of {(\ref{eqLattice2a}) and  (\ref{eqLattice3a}), }
 with (\ref{form1}) and using {(\ref{eq:LambdaIntroduction}) and (\ref{eq:Lambda_final})} yields
\begin{equation}
\begin{gathered}
\left[(0+{\rm i}kv)^2+\omega_1^2(k)\right]U_1(k)=-\mu\Psi^+(k)+\alpha_1(\lambda_1(k, 0+{\rm i}kv)-1)U_1\;,\\
\left[(0+{\rm i}kv)^2+\omega_2^2(k)\right]W_1(k)=\beta {\mu}\Psi^+(k)+\beta \alpha_2(\lambda_2(k, 0+{\rm i}kv)-1)W_1\;,
\end{gathered}
\label{eq:LatticeFourierTransform}
\end{equation}
where use was made of (\ref{eq:LambdaIntroduction})
given earlier (see Supplementary Material \ref{SM1} for the derivation of (\ref{eq:LatticeFourierTransform})). Owing to  (\ref{psiphi}) and (\ref{psieq2}), this system of equations can take the matrix form:
\begin{eqnarray}
&&\begin{bmatrix}
{\alpha_1}\mathcal{S}_1(k, 0+{\rm i}kv)+{\mu}& -{\mu}\\[3mm]
-\beta {\mu}&\beta \big[{\alpha_2}\mathcal{S}_2(k, 0+{\rm i}kv)+{\mu}\big]
\end{bmatrix}\begin{bmatrix}
U^+_1(k)\\[3mm]
W^+_1(k)
\end{bmatrix} \nonumber\\[3mm]
&&\qquad\qquad\qquad\qquad\qquad\qquad\qquad +\begin{bmatrix}
{\alpha_1}\mathcal{S}_1(k, 0+{\rm i}kv)& 0\\[3mm]
0&{\beta \alpha_2}\mathcal{S}_2(k, 0+{\rm i}kv)
\end{bmatrix}
\begin{bmatrix}
U^-_1(k)\\[3mm]
W^-_1(k)
\end{bmatrix}=\begin{bmatrix}
0\\[3mm]
0
\end{bmatrix}\label{matform1}
\end{eqnarray}
valid for $k \in \mathbb{R}$, with
\begin{equation}\label{eqSj}
{\mathcal{S}_j}(k, 0+{\rm i}kv)
=\frac{1-\lambda_j(k, 0+{\rm i}kv)}{\lambda_j(k, 0+{\rm i}kv)}\;.
\end{equation}
The left-hand side of the matrix equation (\ref{matform1}) shows the clear connection 
between two problems. The first is the vibration problem  for two separated lattice half-spaces, associated with the coefficient of the ``$-$" terms.
The second is the problem of wave propagation in a medium formed from the same two lattices joined by the interface, governed by the coefficient of the ``+" terms. These problems and the associated dynamics of the lattice  are investigated in Section \ref{Disp_prop}.

\subsubsection*{Reduction to a scalar Wiener-Hopf problem}

Fortunately, the matrix Wiener-Hopf problem in (\ref{matform1}) can be reduced to a scalar Wiener-Hopf problem. Note that \eqref{eqOM} allows us to  further  modify (\ref{eq:LatticeFourierTransform}) to
\begin{equation}
\begin{gathered}
U_1(k)=-\frac{\mu}{\alpha_1}\frac{\lambda_1(k, 0+{\rm i}kv)}{1-\lambda_1(k, 0+{\rm i}kv)}\Psi^+(k)=-\frac{\mu}{\alpha_1 {\mathcal{S}_1}(k, 0+{\rm i}kv)}\Psi^+(k)\;,\\
W_1(k)=\frac{\mu}{\alpha_2}\frac{\lambda_2(k, 0+{\rm i}kv)}{1-\lambda_2(k, 0+{\rm i}kv)}\Psi^+(k)=\frac{\mu}{\alpha_2 {\mathcal{S}_2}(k, 0+{\rm i}kv)}\Psi^+(k)\;.
\end{gathered}
\end{equation}
By combining this with (\ref{eq:Lambda_final}) and the Fourier transforms of (\ref{psiphi}), we arrive at an alternative form of the system (\ref{matform1}) described by  two decoupled equations
\begin{equation}
\Psi^-(k)+L(k, 0+{\rm i}kv)\Psi^+(k)=0\; ,
\label{eq:WienerHopf_1}
\end{equation}
and 
\begin{equation}
\Phi(k)=\Phi^+(k)+\Phi^-(k)=-M(k, 0+{\rm i}kv)\Psi^+(k)\; ,
\label{eq:WienerHopf_relation}
\end{equation}
where
\begin{equation}
L(k, s)=\frac{L_1(k, s)+L_2(k, s)}{2}\; ,\quad M(k)=\frac{L_1(k, s)-L_2(k, s)}{2}\;,
\label{eq:FunctionL1}
\end{equation}
and 
\begin{equation}
L_j(k, s)=1+\frac{2\mu}{\alpha_j{\mathcal{S}_j}(k, s)}=1+\frac{\mu}{\alpha_j}\Bigg[\sqrt{\frac{\Omega_j(k, s)+1}{\Omega_j(k, s)-1}}-1\Bigg]\;,
\quad j=1,2\;.
\label{eq:FunctionL2}
\end{equation}
{Later, we consider the solution of  \eqref{eq:WienerHopf_1} with a specific inhomogeneous right-hand side that corresponds to the action of a remote load at infinity. We note that when (\ref{eq:WienerHopf_1}) is solved, the function $\Phi(k)$ (and  $\Phi^{\pm}$) can be readily determined from  \eqref{eq:WienerHopf_relation}. 
{In order to solve the inhomogeneous form of the  Wiener-Hopf equation, the factorisation of function $L(k, 0+{\rm i}kv)$ is required into two functions $L_+$ and $L_-$, one analytic for $\text{Im}(k)\ge 0$ and the other for $\text{Im}(k)\le 0$, respectively.
The asymptotes  of $L_\pm$ and $L(k, 0+{\rm i}kv)$ are also required in the neighborhood of the origin, all their  other singular points and at infinity 
and this information  is closely related to the dispersive nature of the system.
 }
}

\section{Analysis of the kernel function $L(k, 0+{\rm i}kv)$}\label{Disp_prop}

\subsection{The dispersive properties of the lattice}\label{secdisp_description}
The dispersion relations for the dissimilar lattice are found by directly replacing $kv-{\rm i} 0$ by $\omega$ in the kernel function of (\ref{eq:WienerHopf_1}) and identifying its zero and singular points due to a standard procedure described by \citet{LSbook}. Before discussing the dynamics of the problem concerning the fault propagation in the dissimilar lattice, we first analyse the vibration modes arising from the two separate problems that are coupled  in (\ref{matform1}) as discussed above.

\subsubsection{Acoustic  branches corresponding to half-plane lattices with free boundaries} 
The real singular points of $L(k, {\rm i} \omega)$ correspond to the acoustic branches $\omega_j$, $j=1,2$, given in (\ref{eq:DispersionRelations}). Here, $\omega_1$ and $\omega_2$ determine waves propagating along the upper and lower crack faces, respectively. 
A representative example of these curves is presented in Figure \ref{disp1}(a).  Note that these dispersion relations are also associated with the singular points of $M(k, {\rm i}\omega)$ defined in (\ref{eq:FunctionL1}) and this fact is used later when identifying the form of the waves in the lattice. 
Along the branch $\omega_1$, ($\omega_2$) the exponent  $\lambda_1$  ($\lambda_2$) is constant and equal to unity.

\begin{figure}[htbp]
\begin{center}
{\includegraphics[width=1\textwidth]{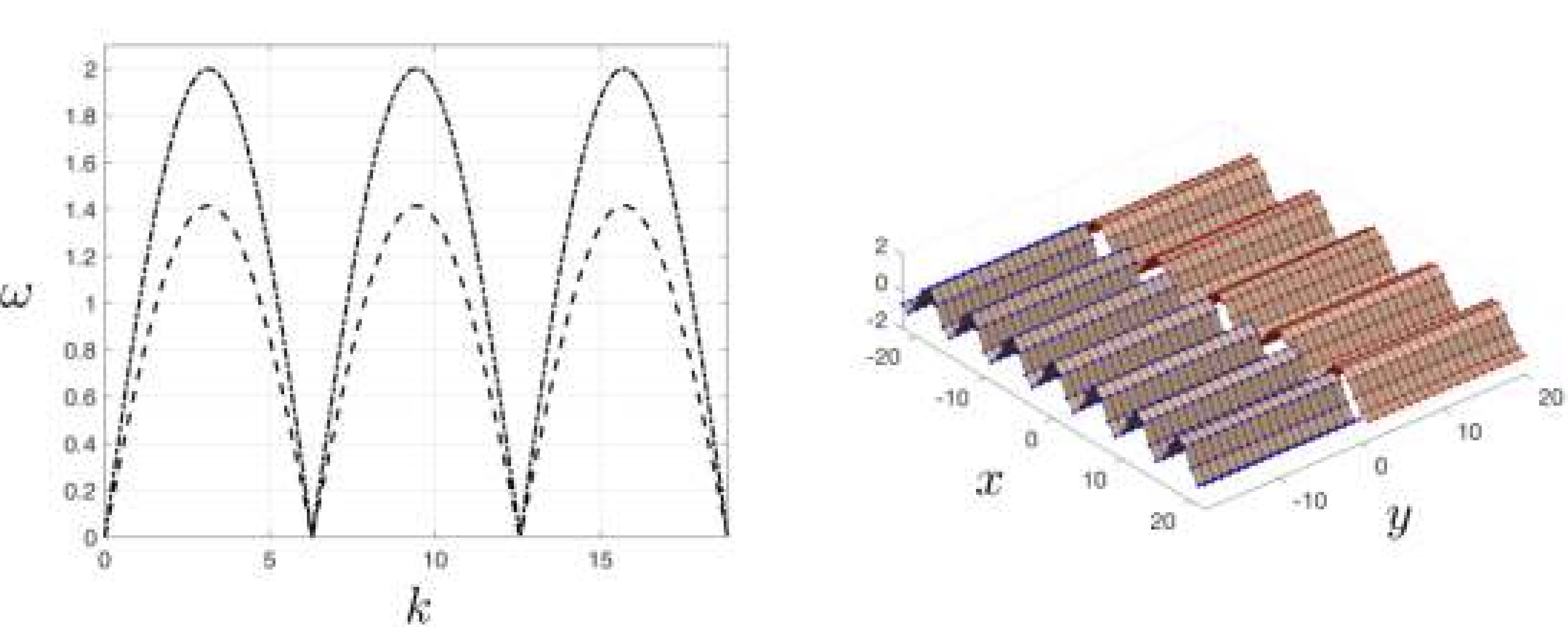}
 }
 (a)~~~~~~~~~~~~~~~~~~~~~~~~~~~~~~~~~~~~~~~~~~~~~~~~~~~~~~~~~~~~~~~~~~(b)
\caption{(a) Acoustic branches for the dissimilar lattice as functions of the wavenumber 
for  $\gamma \beta=1/2$. The dash-dot (dashed) curve corresponds to the relation $\omega_1$ $(\omega_2)$. (b) Eigenmodes with frequency  $\omega=0.7$ for the lattice half-planes  having  $\alpha_1=\alpha_2=1$, based on the results of the Supplementary Material \ref{SM2}. The upper half-plane lattice is represented by the red dots, and the lower half-plane lattice is associated with the blue dots. The wave number for the eigenmode in the upper and lower lattice is $k=0.72$ and $k=1.03$, respectively.}
\label{disp1}
\end{center}
\end{figure}
 We mention that the acoustic branches can be linked to the propagation of vibrations in the lattice half-planes having free boundaries. The corresponding eigenmodes can be computed using the algorithm outlined in the Supplementary Material \ref{SM2}. This has been used to compute the modes shown in Figure \ref{disp1}(b)
for a particular frequency inside the passband for both media, under the assumption the amplitude of vibration is unity in both cases. Note the amplitude of vibration is independent of $m$ for fixed $x$.
In all illustrations given here, increasing  $k$ within $0<k<\pi$, produces modes with
an increasing number of oscillations in the horizontal direction of the lattice. This can also be observed in Figure \ref{disp1}, where for $\gamma \beta<1$ the mode in the upper half-plane has a shorter wavelength in the horizontal direction than the mode in the lower half-plane. {There we define $x=n$, $n\in \mathbb{Z}$ and for $m \in \mathbb{Z}$, $y=m$ if $m\ge 1$ and $y=m+1$ if $m\le -1$.}

\subsubsection{Optical branches characterising waves in the intact material} 
The real zero points of $L(k, {\rm i}\omega)$ are found by solving the equation 
\begin{equation}\label{eqzeros}
F(k, \text{i}\omega)=0\;,
\end{equation}
for $\omega$ separated from zero, where 
\begin{equation}\label{eqzerosa}
F(k, {\rm i}\omega):=\left(1-\mu\Big[\frac{1}{\alpha_1}+\frac{1}{\alpha_2}\Big]\right)(1-\lambda_1(k, {\rm i}\omega))(1-\lambda_2(k, {\rm i}\omega))+\frac{\mu}{\alpha_1}(1-\lambda_2(k, {\rm i}\omega))+\frac{\mu}{\alpha_2}(1-\lambda_1(k, {\rm i}\omega))\;.
\end{equation}

Equation (\ref{eqzeros})  may have one, two or no real solutions and this number is dependent upon the material parameters of the dissimilar  lattice. When a solution exists it may not necessarily be defined for all $k$, but on a periodic disconnected set. When two solutions exist, we denote the corresponding branches on the dispersion diagram by $\omega^{\text{op}}_j$, $j=1,2,$ and in an interval $k$  where they co-exist we have $\omega_1^{\text{op}}>\omega_2^{\text{op}}$.


We now describe the nature of solutions for (\ref{eqzeros}).  First we note that solutions of (\ref{eqzeros}) can only appear when
\begin{equation}\label{con1}
\Lambda:=\mu\Big[\frac{1}{\alpha_1}+\frac{1}{\alpha_2}\Big]-2>0\;.
\end{equation}
This is a necessary condition. Unfortunately, due to the behaviour of $F(k, {\rm i}\omega)$ it is impossible to have a simple 
necessary and sufficient condition. For $\omega$ separated from zero, we have a sufficient condition dependent on the wavenumber $k$ in the form 
\begin{equation}\label{con1a}
\Lambda^2\ge {\frac{\mu^2}{\alpha_1^2}}\frac{\Omega_1(k, {\rm i }\omega)+1}{\Omega_1(k, {\rm i }\omega)-1}+{\frac{\mu^2}{\alpha_2^2}}\frac{\Omega_2(k, {\rm i }\omega)+1}{\Omega_2(k, {\rm i }\omega)-1}\;.
\end{equation}
  Here, this inequality provides a tool to estimate the domain in the $(k, \omega)$-plane where the optical curves can appear. To  accurately identify the domain one should consult (\ref{eqzeros}).
For reasons that will become apparent later, we present on the dispersion diagrams the optical branches $\omega_j^{\text{op}}$, $j=1,2,$ together with the acoustic curves $\omega_j$, $j=1,2$. In addition, we will see that $\omega_1^{\text{op}}$ is always situated above both acoustic branches
hence we will refer to this as the high frequency optical branch. In contrast, $\omega_2^{\text{op}}$ is always located between the curves $\omega_1$ and $\omega_2$, thus we will call this the low frequency optical curve.

\vspace{0.1in}{\bf The high frequency optical branch }$\omega_1^{\text{op}}$.
The high frequency optical curve may exist as a discontinuous curve, a continuous curve or may not exist at all. The appearance of this curve and its nature is dependent upon the lattice material parameters and the ranges of these parameters for $\omega_1^{\text{op}}$ to appear 
are specified in the tables of the Appendix.
 The left panels of Figure \ref{disp2_cont} show several examples of dispersion diagrams where the curve for $\omega^{\text{op}}_1$ is present  and thus (\ref{con1}) is satisfied.
 The presence of the latter is connected with
 waves that can propagate in the dissimilar lattice.

 In Figure \ref{disp2_cont}, the optical curve $\omega_1^{\text{op}}$ appears to be  continuous. As we will see later, this curve can be also piecewise defined for a given combination of material parameters (see Figure \ref{disp2}).
In the central panels of Figures \ref{disp2_cont}(a)--(c), the behaviour of the  exponents $\lambda_j$, $j=1,2,$ along the optical branch $\omega^{\text{op}}_1$  in the corresponding left panels is given.
Along the optical  branch $\omega^{\text{op}}_1$, the exponents $\lambda_j$, $j=1,2$, in (\ref{eq:Lambda_final}),  are not constant and are $2\pi$-periodic functions. Figure \ref{disp2_cont} shows that in the considered cases, the exponents are always negative and  their moduli can also be less than unity. In addition, it holds in all cases that $|\lambda_1|<|\lambda_2|$ for all $k$.  Eigenmodes for the dissimilar lattice  can be computed using the algorithm outlined in the Supplementary Material \ref{SM2}. There, the solution is modulated by positive powers of $\lambda_j$ in the direction perpendicular to the interface. Hence, in such situations the corresponding eigenmode in the intact lattice is an interfacial wave that is  exponentially localised near  interface  between $m=\pm 1$. When $|\lambda_1|<|\lambda_2|$ the mode decays faster into the bulk of the upper lattice than in the lower lattice.
Moreover, for a fixed $n$ in the lattice, the sign of the lattice displacements always alternate in the direction perpendicular to the interface.

\begin{figure}[htbp]
\begin{center}
{\includegraphics[width=1\textwidth]{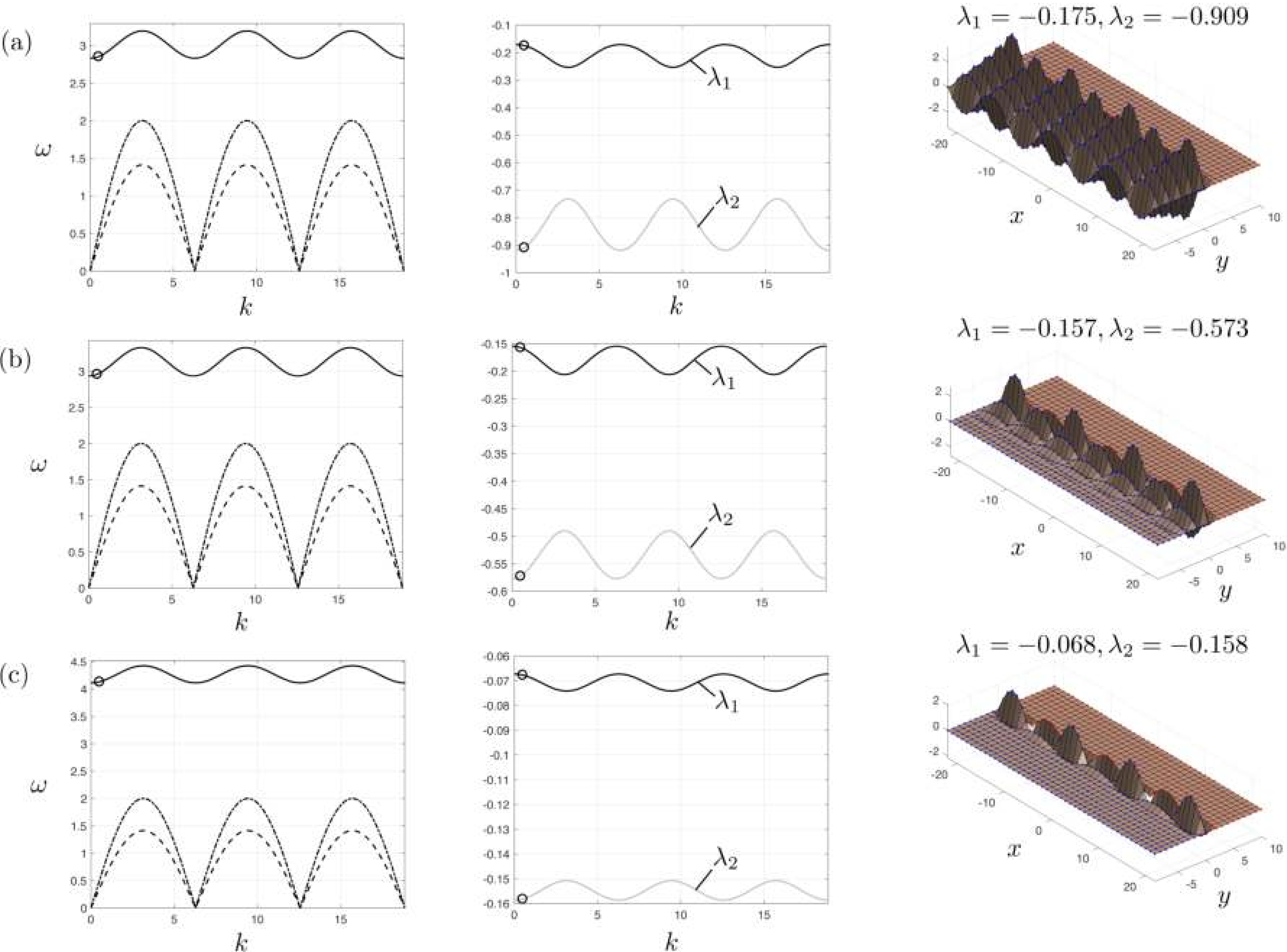}
 }
\caption{\emph{The dynamics of the  dissimilar lattice associated with the optical curve $\omega_1^{\text{op}}$ and its dependency on the interfacial bond stiffness}. In (a)--(c),  the left panels  shows the dispersion curve for the dissimilar lattice represented by  the optical branch $\omega_1^{\text{op}}$ given by solution of (\ref{eqzeros}) (solid curve). For comparison the acoustic branches $\omega_1$ (dash-dot curve) and  $\omega_2$ (dashed curve) in (\ref{eq:DispersionRelations}) are also shown. In the central panels, the trace of the exponents $\lambda_j$, $j=1,2,$ (in (\ref{eq:Lambda_final})) along the optical branch is provided as a function of the wave number.  In the right panels, we show the eigenmodes of the system that correspond to the frequency and wave number defined by the circle in dispersion diagram and the plot of the exponents $\lambda_j$, $j=1,2$ (highlighted by the circles in the central panel).   The computations are performed for $\beta=2$, $\gamma=1/4$, $\alpha_1=\alpha_2=1$ and (a) $\mu=8/5$, (b) $\mu=2$ and (c) $\mu=5$. }
\label{disp2_cont}
\end{center}
\end{figure}

\begin{figure}[htbp]
\begin{center}
{\includegraphics[width=1\textwidth]{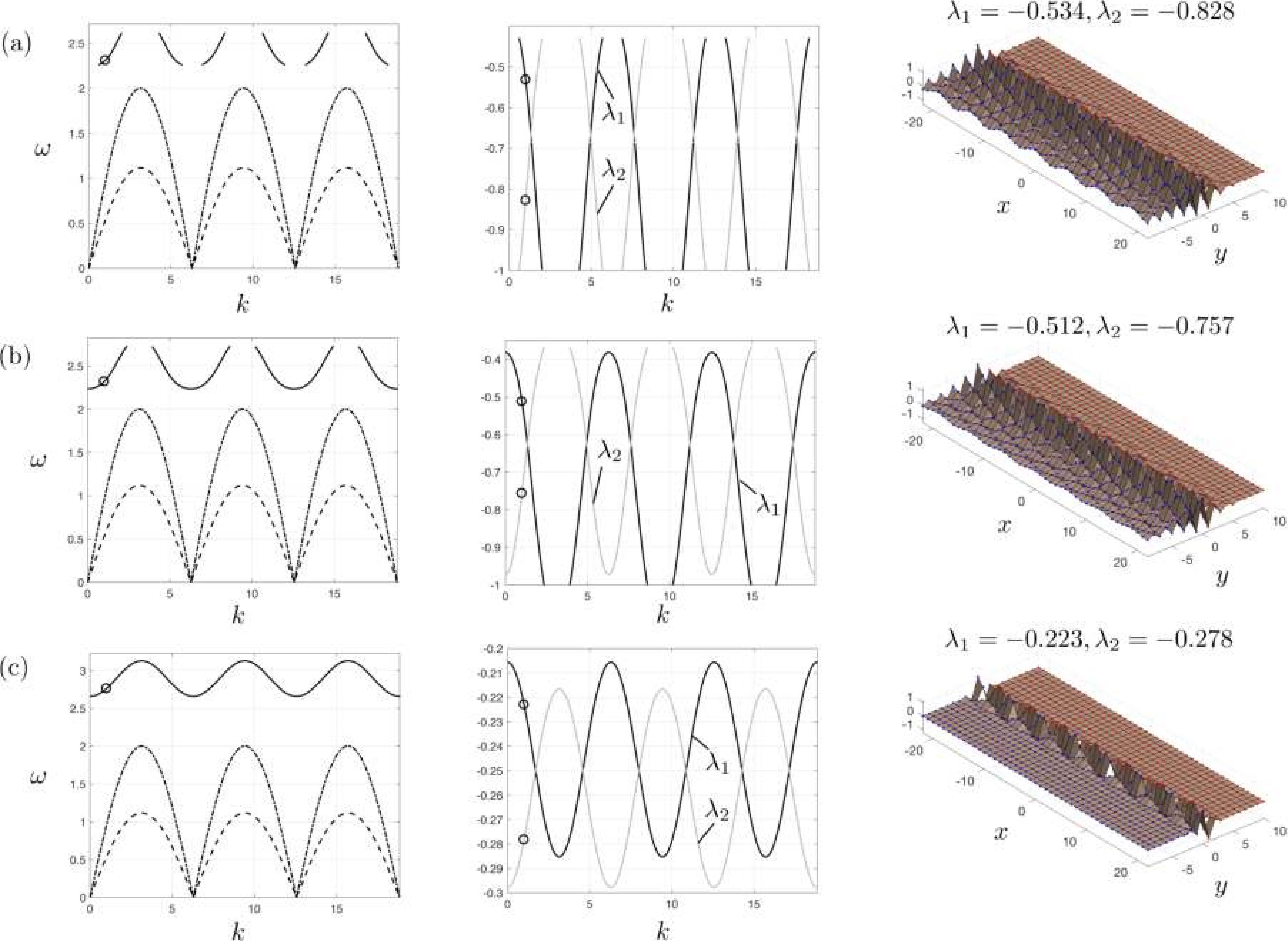}
 }
\caption{
\emph{Influence of the interfacial bond stiffness on  the discontinuous nature of the the optical curve $\omega^{\text{{\rm op}}}_1$ and the associated vibration modes.}
 In (a)--(d),  the left panel  shows the dispersion curve for the dissimilar lattice represented by the optical branch $\omega_1^{\text{op}}$ given by solution of (\ref{eqzeros}) (solid curve). We supply the curves of the acoustic branches $\omega_1$ (dash-dot curve) and  $\omega_2$ (dashed curve) in (\ref{eq:DispersionRelations}) for comparison. In the central panels, the trace of the exponents $\lambda_j$, $j=1,2,$ (in (\ref{eq:Lambda_final})) along the optical branch are shown as a function of the wavenumber.  In the right panels, we show the eigenmodes of the system that correspond to the frequency and wave number defined by the circle in dispersion diagram and the plot of the exponents $\lambda_j$, $j=1,2,$ highlighted by the circles in the central panel.   The computations are performed for $\beta=5/4$, $\gamma=1/4$, $\alpha_1=\alpha_2=1$ and (a) $\mu=5/4$, (b) $\mu=13/10$ and  (c) $\mu=5/2$. }
\label{disp2}
\end{center}
\end{figure}

Several eigenmodes of this type, computed according to the Supplementary Material \ref{SM2}
 are presented in the right panels of  Figure \ref{disp2_cont} at frequencies and wave numbers associated with the circles in the dispersion diagrams. In these computations, the wave amplitude for the upper lattice half-plane is chosen equal to unity (see (\ref{eigmode4})). 
  These computations demonstrate the lack of symmetry in the interfacial wave  modes  about the interface. The larger the magnitude of $\lambda_j$, $j=1,2$, the slower the decay of the mode into the lattice bulk from the interface, as evidenced in Figures \ref{disp2_cont}(a) and (b) in the lower lattice half-plane.  The modes presented in Figure \ref{disp2_cont} show that increasing the interfacial bond stiffness relative to the stiffness of the horizontal bonds in the upper lattice causes the localisation effect to be stronger.
Additionally, they indicate that the lower lattice half-plane can support interfacial wave modes with a weak localisation, whereas the upper lattice half-plane is almost undisturbed (see Figure \ref{disp2_cont}(a)). 
Figures \ref{disp2_cont}(c) illustrate examples of interfacial wave  modes highly localised at the interface, where the ambient lattice is almost undisturbed.

Finally, concerning the acoustic branch $\omega_1^{\text{op}}$, we mention that this can be a piecewise defined curve as shown in the dispersion diagrams in the left panels of Figure \ref{disp2}(a) and (b). The nature of the continuity of $\omega_1^{\text{op}}$ depends on the combinations of lattice material parameters. In the Appendix, we present tables indicating the combinations of these parameters that lead to continuous or discontinuous curves for $\omega_1^{\text{op}}$.
If $\mu$ is {sufficiently large}, a continuous optical branch  can exist, as shown in the left panel of Figure \ref{disp2}(c). In contrast to Figure \ref{disp2_cont}, the central panels of Figure \ref{disp2} show that the exponents $\lambda_j$, $j=1,2$, along the optical branch can be  comparable for some values of $k$ and physically the strength of mode decay into the bulk lattice can be approximately the same. Again, the modes in Figure \ref{disp2} demonstrate that the exponential localisation  of the interfacial waves increases with increase of $\mu$. There, the strength of the localisation is smaller in the lower lattice half-plane than in the upper lattice half-plane. On the other hand, the central panels in Figure \ref{disp2} show there exist wave numbers where the opposite scenario exists.


\vspace{0.1in}{\bf The low frequency optical branch }$\omega_2^{\text{op}}$. Next, we discuss the case when the optical branch $\omega_2^{\text{op}}$ can  appear. 
In fact, this may happen when 
\begin{equation}\label{con2}
\beta<\min\Big\{ \frac{1}{\gamma+\alpha_2},  \frac{\alpha_1+1-(\alpha_2+1)^2}{(\alpha_1+2\alpha_2)(\gamma+\alpha_2)}\Big\}\qquad \text{ or }\qquad \beta>\max\Big\{ \frac{1+\alpha_1}{\gamma},  \frac{(2\alpha_1+\alpha_2)(1+\alpha_1)}{(2\alpha_1+\alpha_2)\gamma-\alpha^2_1}\Big\}\;,
\end{equation}
and the interfacial bond stiffness is sufficiently large, as indicated in the Appendix.

Figure \ref{disp2opt} shows the dispersion curves in this case for several choices of the lattice parameters. We note that unlike $\omega^{\text{op}}_1$, which can be either a piecewise defined curve or a continuous curve depending on the choice of the lattice parameters, the optical branch $\omega^{\text{op}}_2$ is always a piecewise defined curve.  

In the left panels of Figure \ref{disp2opt}, the  optical branch   $\omega_2^{\text{op}}$ is always accompanied by the  continuous branch $\omega_1^{\text{op}}$. In the sense described above, the curve for  $\omega_2^{\text{op}}$ occupies a lower frequency range than $\omega_1^{\text{op}}$.  In the computations presented in Figure \ref{disp2opt}, $\omega^{\text{op}}_2$ is always contained in the regions defined by the acoustic branches. Figure \ref{disp2opt} shows that the width of $k$ occupied by $\omega^{\text{op}}_2$ increases with increase of $\mu$. Hence, interfacial bonds with a sufficiently high stiffness can lead to a scenario where the inhomogeneity has more potential to support interfacial vibrations.

We note that  Figure \ref{disp2opt} shows that $\omega_2^{\text{op}}$ is always concave. As we will see in the next section, curvature of the dispersion relations is important in the description of waves that appear during the steady  fracture propagation. In Figure \ref{disp2opt}, the location of the curve $\omega_2^{\text{op}}$ implies the lattice will support waves  with frequencies intermediate to those waves encountered for the free-boundary problem of the upper and lower lattices.

In the central panels of Figure \ref{disp2opt},  the exponents $\lambda_j$, $j=1,2,$ along the optical curve $\omega_2^{\text{op}}$ are presented. While the exponents are always negative along the branch $\omega_1^{\text{op}}$ (see Figures \ref{disp2_cont} and \ref{disp2}), for the  branch $\omega_2^{\text{op}}$ we have $\text{sgn}(\lambda_j)=(-1)^{j-1}$, $j=1,2$. As before, along the low frequency optical curve, $|\lambda_1|<|\lambda_2|$. In addition, in Figure \ref{disp2opt}, the magnitude of the exponents along $\omega_2^{\text{op}}$ are always larger than those associated with $\omega_1^{\text{op}}$. Hence, waves for the low-frequency optical branch are less localised than those  for the high-frequency optical branch.

\begin{figure}[htbp]
\begin{center}
{\includegraphics[width=1\textwidth]{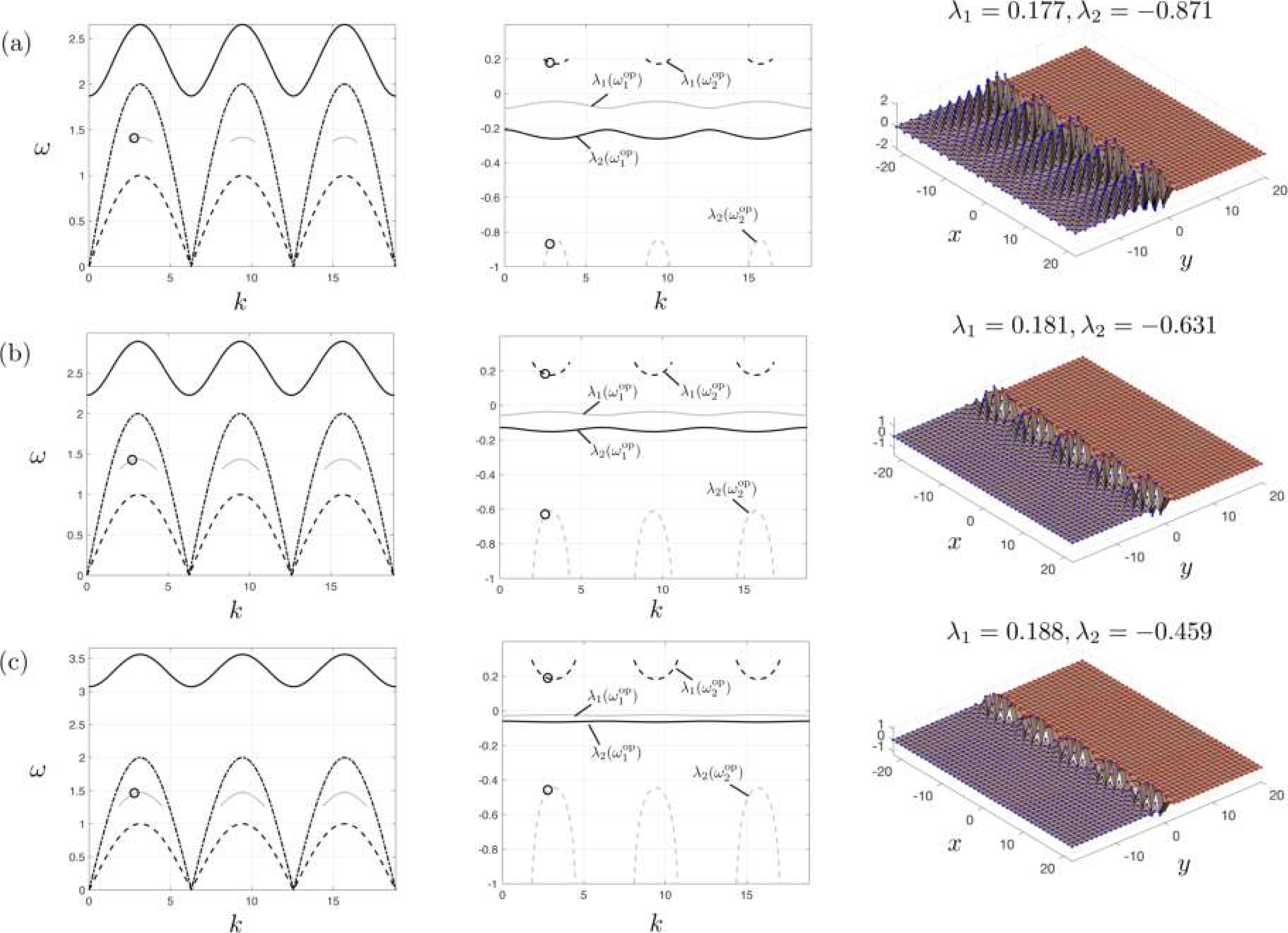}
 }
\caption{\emph{Behaviour of the second discontinuous optical branch $\omega_2^{\text{{\rm op}}}$ and the associated vibration modes as a function of the interfacial bond stiffness.} Here, $\omega_2^{\text{{\rm op}}}$ always appears together with a continuous optical branch $\omega_1^{\text{{\rm op}}}$.
In (a)--(d),  the left panels  show the dispersion curves for the dissimilar lattice represented by  the optical branches $\omega^{\text{op}}_1$ (black solid curve) and $\omega^{\text{op}}_2$ (grey solid curve)  given as solutions of (\ref{eqzeros}), presented alongside the acoustic branches $\omega_1$ (dash-dot curve) and  $\omega_2$ (dashed curve)
in (\ref{eq:DispersionRelations}). In the central panels, the trace of the exponents $\lambda_j$, $j=1,2,$ (in (\ref{eq:Lambda_final})) along the optical branch $\omega_2^{\text{op}}$ is shown as a function of the wave number.  In the right panels, we show the eigenmodes of the system that correspond to the frequency and wave number defined by the circle in dispersion diagram and the plot of the exponents $\lambda_j$, $j=1,2,$ highlighted by the circles in the central panel. 
The computations are performed for $\beta=\gamma =\alpha_1=\alpha_2=1/2$ and (a) $\mu=2$, (b) $\mu=3$ and (c) $\mu=6$. }
\label{disp2opt}
\end{center}
\end{figure}

\vspace{0.1in}{\emph{Eigenmodes   along the low frequency optical curve.}  The eigenmodes along $\omega_2^{\text{op}}$ in the dissimilar lattice are computed using the results of the Supplementary Material \ref{SM2}.
  Based on the analysis of the exponents $\lambda_j$, $j=1,2,$ in this scenario, the corresponding physical behaviour of the lattice again involves  interfacial wave modes that are exponentially localised about the interface. However, in this case, in the lower lattice half-plane these modes oscillate in the direction perpendicular to the interface. The upper lattice half-plane behaves in a similar way to  that encountered along the acoustic branches. The eigenmodes  for a single point along the low frequency optical branch are presented in Figure \ref{disp2opt}. Here, once more, increasing $\mu$ creates a highly localised interfacial wave mode, as in Figures \ref{disp2_cont} and \ref{disp2}.

Figures \ref{disp2_cont}--\ref{disp2opt} show there exist vibration modes associated with the optical branches where 
there can exist a disparity between the displacements along the rows containing the interface.  In particular, as will be shown later, these modes  may lead to regimes for steady crack propagation that violate the assumption (\ref{eq:FractureCondition_eta_1}) (see {Section \ref{sec3}}), due to the large elongations of the interfacial bonds that can arise ahead of the crack.

We also mention some examples where  only the dispersion curve $\omega_2^{\text{op}}$ can exist or it can exist alongside a piecewise defined $\omega_1^{\text{op}}$, as shown in Figure \ref{special_cases} and  computed according to the tables of the Appendix.

\begin{figure}[htbp]
\begin{center}
{\includegraphics[width=1\textwidth]{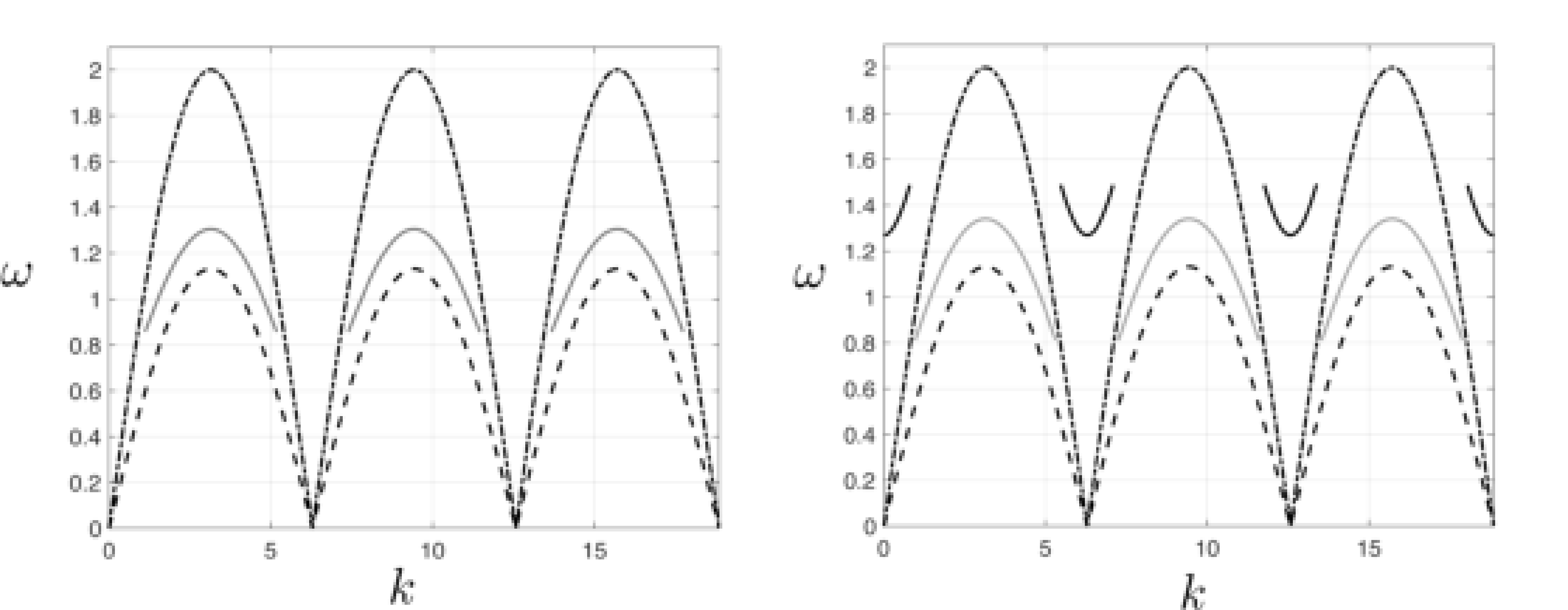}
 }~~~~(a)~~~~~~~~~~~~~~~~~~~~~~~~~~~~~~~~~~~~~~~~~~~~~~~~~~~~~~~~~~~~~~~~~~~~(b)
\caption{Dispersion curves for the cases when $\beta=4/5$, $\gamma=2/5$, $\alpha_1=2/5$, $\alpha_2=3/25$. In (a) $\mu=2/5$ and (b) $\mu=3/5$. The acoustic branches $\omega_1$ (dash-dot curve) and $\omega_2$ (dashed curve) are shown in addition to  (a) the low frequency optical branch $\omega_2^{\text{op}}$ (grey curves) and (b) the piecewise defined curve for the high frequency optical branch $\omega^{\text{op}}_1$ (solid black lines)  and $\omega^{\text{op}}_2$. }
\label{special_cases} 
\end{center}
\end{figure}


We also report that in the case of a lattice with $\mu=\alpha_1$ and $\mu=\alpha_2$, which corresponds to two media perfectly joined at  either $m=-1$ or $m=0$, the dispersive properties of these systems are different. 


{\bf Possible limiting positions of the optical curves \rm}$\omega^{\text{op}}_j$, $j=1,2.$ As mentioned before, it is difficult to  judge whether the optical branch exists and its location on the dispersion diagram. However, in some limiting cases, we can deliver more informative statements. For example, it is  easy to directly check that in the case of,  $\alpha_1\ll 1$, $\mu\gg 1$, with  finite $\alpha_2$ and $\mu \alpha_1>\alpha_2^2$,   the lower optical branch $\omega_2^{\text{op}}$ tends to the lowest acoustic curve covering all of the real axis with the exception of  the neighborhoods of the points $k=2\pi l$, $l\in \mathbb{Z}$, and thus this curve never reaches  the zero frequency.
On the other hand, for  $\alpha_2\ll 1$, $\mu\gg 1$, $\beta \alpha_2 \ll 1$, with $\alpha_1$ finite and  $\mu \alpha_2>\alpha_1^2$, the high frequency optical curve uniformly tends  to the highest acoustic curve and never reaches the zero frequency.
Moreover, if $\alpha_j$, $j=1,2,$ are finite and $\mu \gg 1$ then the high frequency optical curve approaches infinity.
 

{In summary, the above analysis of the dispersive properties of the medium  has provided information  concerning the eigenmodes of the dissimilar lattice and the half-plane lattices with free boundaries. As we will see later, this analysis will aid us in understanding how a dissimilar lattice with a moving interfacial crack behaves. 
There, the modes discussed here for two separate lattice systems are coupled through the equation (\ref{matform1}). The analysis presented here allows to characterise the complex physical phenomena in the dissimilar lattice undergoing fracture.
In particular, this phenomena is linked to the zero and singular points of the kernel function $L(k,{\rm i}kv)$, as shown in the Supplementary Material \ref{sec_crack_disp}. }

\subsection{Asymptotic behaviour of the function $L(k, 0+{\rm i}kv)$ and its factorisation}\label{asympL}
{Based on the analysis of the zero and singular points of the kernel function $L(k, {\rm i}kv)$ in the Supplementary Material \ref{sec_crack_disp}, we now state some results concerning the asymptotic behaviour of the function  $L(k, 0+{\rm i}kv)$ in the vicinity of these points.}

Firstly, one can write down the behaviour of functions $L_{1,2}(k, 0+{\rm i}kv )$ 
appearing in (\ref{eq:FunctionL1}) and (\ref{eq:FunctionL2}) at infinity as
\begin{equation}\label{Ljinf}
L_{j}(k, 0+{\rm i}kv)=1-\frac{2{\beta^{j-1}\mu_{j}}}{v^2k^2}+O\left(\frac{1}{k^4}\right)\;,\quad k\to\infty\;,\\
\end{equation}
and near zero as
\begin{equation}\label{Ljzero}
L_{j}(k, 0+{\rm i}kv)=2\mu \sqrt{\frac{{\beta^{j-1}}}{{{\alpha_j}(v_j^2-v^2)}}}\frac{1}{\sqrt{(0+{\rm i}k)(0-{\rm i}k)}}+1-\frac{\mu}{\alpha_j}+O(k^2)\;,\quad k\to0\;,
\end{equation}
for $j=1,2.$
The zero point corresponds to the loading of the structure at infinity (see also the Supplementary Material \ref{sec_crack_disp}), and is linked to the square root growth of the solution in the far-field as discussed below. 
From (\ref{Ljinf}) one can show that the asymptotes of $L(k, 0+{\rm i}kv)$ and $M(k, 0+{\rm i}kv)$ at infinity admit the form
\begin{equation}
L(k, 0+{\rm i}kv)=1-\frac{\mu(1+\beta)}{v^2k^2}+O\left(\frac{1}{k^4}\right),\quad k\to\infty,
\label{eq:Asymptotics_L_inf}
\end{equation}
and
\begin{equation}
M(k, 0+{\rm i}kv)=-\frac{\mu(1-\beta)}{v^2k^2}+O\left(\frac{1}{k^4}\right),\quad k\to\infty\;.
\label{eq:Asymptotics_M_inf}
\end{equation}
The asymptotes in (\ref{Ljzero}), show the behaviour of $L(k, 0+{\rm i}kv)$ and $M(k, 0+{\rm i}kv)$ near zero is given by:
\begin{equation}
L(k, 0+{\rm i}kv)=\frac{\Theta^2}{\sqrt{(0+{\rm i}k)(0-{\rm i}k)}}+1-\frac{\mu}{2}\left(\frac{1}{\alpha_1}+\frac{1}{\alpha_2}\right)+O(k^2),\quad k\to0,\label{eq:Asymptotics_L_zero}
\end{equation}
and
\begin{equation}
M(k, 0+{\rm i}kv)=\frac{ \Upsilon}{\sqrt{(0+{\rm i}k)(0-{\rm i}k)}}-\frac{\mu}{2}\left(\frac{1}{\alpha_1}-\frac{1}{\alpha_2}\right)+O(k^2),\quad k\to0,
\label{eq:Asymptotics_M_zero}
\end{equation}
where 
\begin{equation}
\Theta^2=\mu\left[\frac{1}{\sqrt{\alpha_1(v_1^2-v^2)}}+ \sqrt{\frac{\beta}{{\alpha_2(v_2^2-v^2)}}}\right] \quad \text{ and }\quad \Upsilon=\mu\left[\frac{1}{\sqrt{\alpha_1(v_1^2-v^2)}}- \sqrt{\frac{\beta}{{\alpha_2(v_2^2-v^2)}}}\right] \;.
\end{equation}
The function $L(k, 0+{\rm i}kv)$ also possesses the following properties
\begin{equation}\label{prop_L}
|L(k, 0+{\rm i}kv)|=|L(-k, 0-{\rm i}kv)|,\quad  \text{Arg}(L(k, 0+{\rm i}kv))=-\text{Arg}(L(-k, 0-{\rm i}kv)),
\end{equation}
where $\text{Arg}(L(k, 0+{\rm i}kv))$ is the continuous argument of $L(k, 0+{\rm i}kv)$,  and also the index (or winding number) of $L(k, 0+{\rm i}kv)$ is zero. Hence, we can factorise the kernel function in the form  $L(k, 0+{\rm i}kv)=L^+(k)L^-(k)$ with use of the Cauchy-type integral defined by:
\begin{equation}
L^\pm(k)=\exp{\left(\pm\frac{1}{2\pi {\rm i}}\int\limits_{-\infty}^{\infty}\frac{\text{log}\,{L(\xi, 0+{\rm i}\xi v)}}{\xi-k}\,d\xi\right)},\quad \pm {\text{Im }}k>0\;.
\label{eq:Factorisation}
\end{equation}
Here, in the sense of (\ref{eq:Limit_s}), the integral in the preceding expression is interpreted as
\[\int\limits_{-\infty}^{\infty}\frac{\text{log}\,{L(\xi, 0+{\rm i}\xi v)}}{\xi-k}\,d\xi=\lim_{s\to0^+}\int\limits_{-\infty}^{\infty}\frac{\text{log}\,{L(\xi, s+{\rm i}\xi v)}}{\xi-k}\,d\xi\;.\]
The factors  $L^{\pm}(k)$ are analytic in the half-planes defined by $\pm{\text{Im }k>0}$ and satisfy 
the following asymptotic relations:
\begin{equation}
L^{\pm}(k)-1\sim \pm {\rm i}\frac{Q}{k},\quad k\to\infty,\quad
Q=\frac{1}{\pi}\int\limits_{0}^{\infty}\text{log}\,{|L(\xi, 0+{\rm i}\xi v)|}\,d\xi.
\label{eq:Asymptotics_L+-_inf}
\end{equation}
One can also find an asymptotic behaviour of the factors $L^{\pm}$ in the neighbourhood of   zero  with the use of \eqref{eq:Asymptotics_L_inf} and the Sokhotski-Plemelj theorem, to obtain:
\begin{equation}
L^{\pm}(k)\sim R^{\pm1}\frac{\Theta}{\sqrt{0\mp {\rm i}k}},\quad k\to0,\quad
R=\exp{\left(\frac{1}{\pi}\int\limits_0^\infty\frac{\text{Arg}(L(\xi, 0+{\rm i}\xi v))}{\xi}\,d\xi\right)}\;,
\label{eq:Asymptototics_L+-_zero}
\end{equation}
where the argument $\text{Arg}(L(k, 0+{\rm i}kv))=\lim_{s\to 0^+}\text{Arg}(L(k, s+{\rm i}kv))$. Here, $Q$ and $R$ appear due to the properties $(\ref{prop_L})$. 

Finally, we mention the behaviour of  $L^{\pm}(k)$ near some of their zero and singular points identified in the Supplementary Material \ref{sec_crack_disp}. {In what follows, we consider those points that provide a singular behaviour in the solution of (\ref{eq:WienerHopf_1}) and (\ref{eq:WienerHopf_relation}) and that 
  reveal information about the behaviour of the lattice undergoing failure through its inverse Fourier transform. 
}
Note that $L(k, {\rm i}kv)$ is an even function of $k\in \mathbb{R}$, hence for every positive zero  or singular point of this function there is a negative point of the same type with equal magnitude. 
The function $L(k, 0+{\rm i}kv)$ has singular branch points at the  $k = \pm p_{2j+1}$, $0\le j \le f$, and these points can be identified from the dispersion diagram as outlined in the Supplementary Material \ref{sec_crack_disp}. {These wave numbers} are connected with intersection points of the ray $\omega=kv$ with the acoustic branch for the free boundary problem in the upper half of the dissimilar lattice system. Near these wave numbers we have
\begin{equation}\label{rLm1}
L^{-}(k)\sim \frac{\text{Const}}{\sqrt{0+{\text{i}(k\mp p_{2j+1})}}}\;, \quad \text{ for } 0\le j \le f\;.
\end{equation}
Additionally, in a similar way, owing to the dispersive properties of the lower lattice half-plane with free boundary, the points $k = \pm q_{2j+1}$, $0\le j \le g$, define another collection of branch points in $L^-(k)$ where
\begin{equation}\label{rLm2}
L^{-}(k)\sim \frac{\text{Const}}{\sqrt{0+{\text{i}(k\mp q_{2j+1})}}}\;, \quad \text{ for } 0\le j \le g\;.
\end{equation}
Here in (\ref{rLm1}) and (\ref{rLm2}) the ${\text{Const}}$ is used to denote different constants. Further, the integers $f$ and $g$ depend on the crack speed. The branch points described above determine waves in the lattice with a group velocity $v_g<v$ that propagate behind and away from the crack tip during the advancement of the crack.

Next, we consider the wave numbers that give rise to transmitted modes that are connected with the optical curves identified in Section \ref{secdisp_description}.  In particular, we are interested in the modes which generate the highly localised waves propagating  along  the interface.
For this phenomenon to occur in the lattice, referring to \citet[Chapter 3]{LSbook},  we require that 
\[\frac{\partial F}{\partial k}(k, \text{i}kv)\Big|_{k=\xi}\ne 0\;,\]
where $\xi$ is a wave number associated with the optical curves for the dissimilar system and determined using the algorithm outlined in 
the Supplementary Material \ref{sec_crack_disp}. As above, there exists a finite collection of the points $\xi$ and the cardinality of this collection depends on the crack speed. These points define waves possessing a group velocity $v_g>v$, and hence will propagate ahead of the crack during the failure process.
The preceding condition is equivalent to imposing
\begin{eqnarray*}
&&\left[(\Lambda+1)(\lambda_2(k, {\rm i}kv)\Big|_{k=\xi}-1)-\frac{\mu}{\alpha_2}\right]\frac{{\rm d}\lambda_1}{{\rm d} k}(k, {\rm i }kv)\Big|_{k=\xi}\\
&&\qquad \qquad +\left[(\Lambda+1)(\lambda_1(k, {\rm i}kv)\Big|_{k=\xi}-1)-\frac{\mu}{\alpha_1}\right]\frac{{\rm d}\lambda_2}{{\rm d} k}(k, {\rm i }kv)\Big|_{k=\xi}\ne 0
\end{eqnarray*}
where $\Lambda$ is given in (\ref{con1}).
 In this case, the function $L^+(k)$ has simple zeros at $\xi$ and
\begin{equation}\label{Lpsimp}
L^+(k)\sim \text{Const} \;(0-\text{i}(k-\xi))\quad  \text{ for }\quad  k\to \xi\;.
\end{equation}
This asymptote is connected with the propagation of transmitted waves with constant amplitude along the interface of the dissimilar medium during the failure process. This phenomenon does not exist in fracture in  homogeneous lattices, as studied by \cite{LSbook}. Indeed, in those cases the derived kernel function $L$ has a square root type behaviour near all  singular and zero points, that leads to waves radiated by the crack with  amplitudes having a square root type decay.

\section{Solution of the Wiener-Hopf equation and the lattice dynamics}\label{solWH}
\subsection{Solution of the problem corresponding to a remote load}\label{solWHp1}
We begin with obtaining the functions $\Psi^\pm$  that
are the solutions of (\ref{eq:WienerHopf_1}). Note that (\ref{eq:WienerHopf_1}) has a trivial solution. To have a non-trivial solution we have to impose a load on the system. As mentioned in Section \ref{sec2}, we consider the load prescribed at infinity that leads to the solution behaving like the crack tip field encountered in the analogous continuum problem, involving an interfacial crack propagating between two dissimilar elastic media. This leads to the equation 
\begin{equation*}
L^+(k)\Psi^+(k)+\frac{1}{L^-(k)}\Psi^-(k)=\frac{C}{0-{\rm i}k}+\frac{C}{0+{\rm i}k},
\end{equation*}
where $C$ is an unknown constant to be determined (see \citet{LSbook}). One can check using the asymptotes of $L^{\pm}(k)$, given in \eqref{eq:Asymptotics_L+-_inf} and \eqref{eq:Asymptototics_L+-_zero}  that the above right hand-side corresponds 
an energy flux from infinity that drives the crack propagation. 
As the above right-hand side is a sum of a function analytic for $\text{Im}(k)>0$ and one analytic for $\text{Im}(k)<0$, the solutions $\Psi^\pm$ of the Wiener-Hopf equation are readily obtained in the form
\begin{equation}
\Psi^+(k)=\frac{C}{0-{\rm i}k}\frac{1}{L^+(k)}\quad\text{ and }\quad \Psi^-(k)=\frac{C}{0+{\rm i}k}L^-(k)\;.
\label{eq:SolutioFourier_Psi}
\end{equation}
{In addition, owing to (\ref{eq:WienerHopf_relation}) we then have
\begin{equation}
\Phi(k)=-\frac{CM(k)}{L^+(k)}\frac{1}{0-{\rm i}k}\; .
\label{eq:WienerHopf_relationA}
\end{equation}}

\vspace{0.1in}{\bf Behaviour of the solution  near the crack tip and determination of $C$.\rm}
Next we establish the asymptotes of these solutions for large wave numbers.
From the asymptotes \eqref{eq:Asymptotics_L+-_inf} and the solutions \eqref{eq:SolutioFourier_Psi} it follows that at infinity, one has
\begin{equation}
\Psi^{\pm}(k)\sim C\left(\pm\frac{{\rm i}}{k}\pm \frac{Q}{k^2}\right),\quad k\to\infty\;,
\label{eq:Asymptotics_Lattice_Fourier_1}
\end{equation}
where this limit corresponds to the location of the crack tip in the lattice.

These asymptotes, 
 through the inverse Fourier transform, show that the  asymptotic behaviour of solution $\psi(\eta)$ at zero is as follows
\begin{equation}\label{psiasymp}
\psi(\eta)=C(1-Q\eta)+O(\eta^2),\quad \eta\to0\;,
\end{equation}
and this demonstrates, as expected, the stresses at the crack tip in the lattice are bounded. Here, the constant  $C$ can then be determined from the  fracture criterion \eqref{eq:FractureCondition_psi} and (\ref{psiasymp}). We obtain that
\begin{equation}\label{Ceq}
C=\epsilon_c\;.
\end{equation}
{ Hence, (\ref{eq:SolutioFourier_Psi}) and (\ref{eq:WienerHopf_relationA})  can be updated to
\begin{equation}
\Psi^+(k)=\frac{\varepsilon_c}{0-{\rm i}k}\frac{1}{L^+(k)}\;, \quad \Psi^-(k)=\frac{\varepsilon_c}{0+{\rm i}k}L^-(k) \quad \text{ and } \quad
\Phi(k)=-\frac{\varepsilon_cM(k)}{L^+(k)}\frac{1}{0-{\rm i}k}\; .
\label{eq:WienerHopf_relationAB}
\end{equation}}

\vspace{0.1in}{\bf Behaviour of the solution in the continuum limit of the lattice problem.\rm}
If the wavenumber is sufficiently small, using  \eqref{eq:Asymptototics_L+-_zero},  \eqref{eq:SolutioFourier_Psi} and (\ref{Ceq}),  we can assert that
\begin{equation}
\Psi^+(k)\sim \frac{\epsilon_c}{R\Theta}\frac{1}{\sqrt{0-{\rm i}k}}\;,\quad k\to0\;,
\label{eq:Asymptotics_Lattice_Fourier_2}
\end{equation}
and
\begin{equation}
\Psi^-(k)\sim\frac{\epsilon_c \Theta}{R}\frac{1}{(0+{\rm i}k)^{3/2}}\;,\quad k\to0\;.
\label{eq:Asymptotics_Lattice_Fourier_3}
\end{equation}
These limits allow one to deduce the behaviour of the solution far away from the crack tip and they coincide with the behaviour of the solution in the continuous limit of the problem considered. 
Thus,   by means of {contour integration and the residue theorem} applied to  \eqref{eq:Asymptotics_Lattice_Fourier_2} and \eqref{eq:Asymptotics_Lattice_Fourier_3}, we conclude that
\begin{equation}\label{asymppsi1}
\psi(\eta)\sim\frac{\epsilon_c}{R{\Theta}}\frac{1}{\sqrt{\pi\eta}}\;,\quad \text{ for }\eta\to\infty\;,
\end{equation}
and
\begin{equation}\label{asymppsi2}
\psi(\eta)\sim2\frac{\epsilon_c{\Theta}}{R}\sqrt{\frac{-\eta}{\pi}}\;, \quad \text{ for } \eta\to-\infty\;.
\end{equation}
These asymptotes show that for both behind and ahead of the crack front in the continuous problem, obtained at  infinity in the lattice,  the stresses share the same asymptotic behaviour expected in the Mode III fracture of a continuous material.

{As shown in the Supplementary Material \ref{SM4}, it follows from (\ref{eq:WienerHopf_relationA}) that $\phi(\eta)$ (defined in (\ref{psiphi})) has the asymptote}
\begin{equation}\label{asympphi1}
\phi(\eta)=\frac{\varepsilon_c\Xi_0}{\sqrt{\pi \eta}}+O\Big(\frac{1}{\eta}\Big)\;,\quad \text{ for }\eta\to\infty\;,
\end{equation}
and owing to (\ref{eq:Asymptotics_Lattice_Fourier_3})
\begin{equation}\label{asympphi2}
\phi(\eta)\sim 2\frac{K\epsilon_c{\Theta}}{R}\sqrt{\frac{-\eta}{\pi}}+O(1)\;, \quad \text{ for } \eta\to-\infty\;.
\end{equation}
where 
\begin{equation}\label{KXi0}
K=
\frac{\Upsilon}{\Theta^2}\qquad \text{ and }\qquad \Xi_0=\frac{1}{2}\left(\mu\left\{\frac{1}{\alpha_1}-\frac{1}{\alpha_2}\right\}+\left\{2-\mu\left[\frac{1}{\alpha_1}+\frac{1}{\alpha_2}\right]\right\}K\right)\frac{1}{R{\Theta}}\;.
\end{equation}
Interestingly, $K$ is analogous to Dundur's parameter for the continuous dissimilar material with a moving Mode III crack. Note also that, as expected, it does not depend on the stiffness of the interface parameter.
Consulting (\ref{psiphi}) and  (\ref{asymppsi1})--(\ref{asympphi2}), the asymptotes for $u_1$ ans $w_1$ far away from the crack tip then follow as
\[u_1(\eta)\sim \frac{\epsilon_c}{2} \left(\Xi_0+\frac{1}{R\Theta} \right)\frac{1}{\sqrt{\pi\eta}}\quad\text{ and }\quad w_1(\eta)\sim \frac{\epsilon_c}{2} \left(\Xi_0 -\frac{1}{R\Theta}\right)\frac{1}{\sqrt{\pi\eta}}\quad \text{ for }\eta\to \infty\;, \]
and 
{\[u_1(\eta)\sim \frac{\epsilon_c\Theta(K+1)}{R} \sqrt{\frac{-\eta}{\pi}}\quad\text{ and }\quad w_1(\eta)\sim \frac{\epsilon_c\Theta(K-1)}{R}\sqrt{\frac{-\eta}{\pi}}\quad \text{ for }\eta\to -\infty\;.\]}

\vspace{0.1in}{\bf Micro-oscillations in the lattice.\rm} Finally, we comment on the waves radiated in the lattice during the failure process. These waves, which are radiated behind the crack tip as this propagates, are associated with the singular branch points of the solution $\Psi^-$ in (\ref{eq:SolutioFourier_Psi}). They are defined. by the points $k=\pm p_{2j+1}$, $1\le j \le f$ and $k=\pm q_{2j+1}$, $1\le j \le g$, which give rise to singular branch points in $L^-$ {(see (\ref{rLm1}) and (\ref{rLm2}))}.
Similar to \citet{LSbook} one can show that these points describe  waves, released during the propagation of the interfacial crack, that have slowly decreasing amplitudes of the order $O((-\eta)^{-1/2})$ for ${\eta \to -\infty}$. 

On the other hand, waves radiated ahead of the crack tip are obtained by a straight-forward calculation of  complex residues at the simple poles of $\Psi^+(k)$. The simple poles occur at wave numbers defined by the optical curves  (see (\ref{Lpsimp})). The result implies that the crack can excite an interfacial wave that propagates ahead of the crack tip and that is exponentially localised to the interface.

\subsection{Dynamic features accompanying the fracture process}\label{DFLattice} {The lattice displacements are expressed in terms of inverse Fourier transform:
\begin{equation}
u_{m+1}(\eta)=\frac{1}{2\pi}\int_{-\infty}^{\infty}[\lambda_1(k, 0+{\rm i} kv)]^{m}U_1(k)e^{-{\rm i}k\eta}\,dk,\quad m\ge 0,
\label{eq:SolutionLatticeInverseFourier_Layersa}
\end{equation}
and
\begin{equation}
{w_{-m+1}(\eta)}=\frac{1}{2\pi}\int_{-\infty}^{\infty}[\lambda_2(k, 0+{\rm i} kv)]^{-m}W_1(k)e^{-{\rm i}k\eta}\,dk,\quad m\le  0\;,
\label{eq:SolutionLatticeInverseFourier_Layers}
\end{equation}
which utilises the recursive relations (\ref{eq:LambdaIntroduction}).
The information provided in the Supplementary Material \ref{sec_crack_disp}, concerning non-zero wave numbers that are linked to waves released in the lattice when the crack propagates with a given speed, can be used to determine the  form of the waves radiated.  Here, we will only concentrate on the main features of the solution and the associated details are in the {Supplementary Material (see  \ref{SM5} and \ref{SM6})}.

Note that $|\lambda_j|\le 1$, $j=1,2,$ and the functions $U_1$ and $W_1$ can be obtained from $\Phi$ and $\Psi$, identified in the last section, with the relations 
 \[U_1(k)=\frac{\Psi(k)+\Phi(k)}{2}\;, \qquad W_1(k)=\frac{\Psi(k)-\Phi(k)}{2}\;,\]
and (\ref{eq:WienerHopf_relationAB}).
 In addition, they can also be split into ``+" and ``$-$" functions with known behaviour at infinity, near zero and in the vicinity of singular points (see Section \ref{Disp_prop} where the analysis at these points is carried out). In the following, we only work with (\ref{eq:SolutionLatticeInverseFourier_Layersa}) corresponding to the displacements in the upper lattice as a similar analysis can be carried out for (\ref{eq:SolutionLatticeInverseFourier_Layers}).

First let's represent the  formula (\ref{eq:SolutionLatticeInverseFourier_Layersa}) as 
\begin{equation}
u_{m+1}(\eta)=\frac{1}{\pi}\text{Re}\Big\{\int_{0}^{\infty}|\lambda_1|^{m}e^{\text{i}m\text{Arg}(\lambda_1)}U_1(k)e^{-{\rm i}k\eta}\,dk\Big\},\quad m>0.
\label{eq:SolutionLatticeInverseFourier_Layersa1}
\end{equation}
 This may then be written as
\begin{eqnarray}
u_{m+1}(\eta)&=&\frac{1}{\pi}\text{Re}\Big\{ \int_0^{p_1}|\lambda_1|^{m}U_1(k)e^{-{\rm i}k\eta}\,dk+(-1)^m\int_{p^*_{2f^*+1}}^\infty |\lambda_1|^{m}U_1(k)e^{-{\rm i}k\eta}\,dk\nonumber\\
&&+ \int_{\Sigma_1}U_1(k)e^{{\rm i}(m\text{Arg}(\lambda_1)-k\eta)}\,dk+\int_{\Sigma^*} 
|\lambda_1|^{m}U_1(k)e^{-{\rm i}k\eta}\,dk\Big\}\label{formulaUint}
\end{eqnarray}
where
\begin{equation}\label{SSIG1}
\Sigma_1:=\Big(\bigcup_{j=0}^{f-1}{(p_{2j+1} , p_{2j+2})}\Big)\cup(p_{2f+1}, p_1^*)\cup  \Big(\bigcup_{j=1}^{f^*}{(p_{2j}^* , p_{2j+1}^*)}\Big)
\end{equation}
with $p^*_{2j+1}$, $1\le j \le f^*$ being the zeros of the function $\Omega(k, 0+{\rm i}kv)+1$ and 
 $\Sigma_*:=\mathbb{R}_+\backslash( [0, p_1] \cup \Sigma_1 \cup [p^*_{2f^*+1}, \infty))$, with $\mathbb{R}_+:=\{ x> 0, x\in \mathbb{R}\}$,
 (see the Supplementary Material \ref{SM6} for further details). 

Next we estimate the behaviour of $u_m$ along the ray $\eta=am$, where $-\infty<a<\infty$.
If the function $U_1$ did not possess any irregular points and decayed at infinity, then each of  the four  integrals above would decay, which can be proved by using appropriate asymptotic methods such as steepest descent and the stationary phase method when $d=\sqrt{m^2+\eta^2} \to \infty$. However, in our case, we have singular points in the function $U_1$ that admit the following type of singularity $k^{-3/2}$ and $k^{-1/2}$  in the vicinity of $k=0$ and $({k- p_{2j+1}})^{-1/2}$ near $k= p_{2j+1}$, $0\le j \le f$. In addition,  $U_1$ may have simple poles  for wave numbers associated with the intersection points between the line $\omega =kv$ and the optical branches on the dispersion diagram (see Section \ref{Disp_prop}).

Most of those integrals can be estimated by $O(1/\sqrt{d})$ or $O(1/d)$ when $d$ goes to infinity. 
The leading asymptotic term when $\eta$ and $m\to \infty$ is defined by the singular point of the above integrands at $k=0$ and the solution behaves as
\begin{equation}
u_m(\eta)\sim \epsilon_c\sqrt{\frac{d}{2\pi}}\frac{\Theta(K+1)}{R}\left(\sqrt{1+\Big({\frac{v_1^2-v^2}{2\alpha_1}}
-1\Big)\sin^2{\theta}}-\cos{\theta}\right)^{1/2},\quad d\to\infty, \quad m\ge1\;,
\label{eq:Asymptotic_ma}
\end{equation}
for the upper half-plane and in the case of the lower half-plane we have 
\begin{equation}
w_{-m}(\eta)\sim \epsilon_c\sqrt{\frac{d}{2\pi}}\frac{\Theta(K-1)}{R}\left(\sqrt{1+\Big({\frac{v_2^2-v^2}{2\beta\alpha_2}}-1\Big)\sin^2{\theta}}-\cos{\theta}\right)^{1/2},\quad d\to \infty, \quad m\le 1,
\label{eq:Asymptotic_m}
\end{equation}
where $\cos(\theta)=\eta/\sqrt{\eta^2+m^2}$ and $\sin(\theta)=m/\sqrt{\eta^2+m^2}$. The asymptotes (\ref{eq:Asymptotic_ma}) and (\ref{eq:Asymptotic_m}) are derived in the Supplementary Material \ref{SM5}.

Note that this asymptotic behaviour at infinity fully corresponds to the asymptotic behaviour of the solution near the crack tip in the analogous continuum model for an interfacial crack moving with a speed $v$ between two anisotropic materials (see for example the work of  \citet{MishurisetalLiv}). 

The essential terms in the asymptotics of $u_m$ and $w_{-m}$ at infinity will decay like $O(m^{-1/3})$ only along the rays 
\begin{equation}\label{wave_rays}
\eta=(-1)^j(m-1)H^{\prime}_j(k^*), \quad j =1,2\;, \qquad \text{ with } \qquad H_j(k)=2\arctan\Big(\frac{\sqrt{1-\Omega_j(k, 0+{\rm i}kv)}}{
\sqrt{1+\Omega_j(k, 0+{\rm i}kv)}}\Big)\;,
\end{equation}
and this is discussed in the Supplementary Material \ref{SM6}.
In addition, there may exist additional phenomena related to solution concerning the transmission of waves {with constant amplitude along the intact interface, that are exponentially localised along in the directions perpendicular to the interface}. For brevity, we refer the reader to the Supplementary Material \ref{SM7} where this case is analysed.
All other terms in the asymptotics would decay as $O(1/\sqrt{m})$  or $O(1/{m})$ or faster. {A number of important features in those asymptotics, also include dissipated waves radiated behind  the crack tip (see \citet{LSbook}).}
}
As the main focus of the section \ref{sec3}, we present numerical results illustrating all of these analytical findings.

\section{Strain energy release rates and the energy release rate ratio}\label{secERR}

\citet{LSbook} derived  energetic relations  that compare (i) the energy release rate $G_0$ for a crack propagating through a lattice and (ii) the corresponding quantity $G$ arising  from the problem  of a crack propagating in the  analogous continuous medium.  This ratio takes the form:
\begin{equation}
\frac{G_0}{G}=R^2.
\label{eq:ERR_R}
\end{equation}
The analysis of this energy release rate ratio was also carried out for a triangular cell lattice by \citet{Fineberg} and  \citet{MG}, for a chain with anisotropic properties (see \citet{GM2}) and non-local interactions (see \citet{GM3}). 
We note that the expression in (\ref{eq:ERR_R}) is also valid for the Mode III crack propagating in dissimilar media and this is derived in the Supplementary Material \ref{SM8}.

In addition to the local strain energy release rate $G_0$ associated with lattice,  one can define analogous quantities for the strain energy associated with the horizontal and vertical links on the upper and lower lattices.  We set
\begin{equation}\label{strainsexz1}
2(\varepsilon_{xz}(\eta, m))^{{2}}=
\frac{1}{2}\left(u_m(\eta+1)-u_m(\eta)\right)^{{2}}
\;,
\end{equation}
and 
\begin{equation}
2{\gamma} (\varepsilon_{xz}(\eta, m))^{{2}}=
\frac{\gamma}{2}
\left(w_m(\eta+1)-w_m(\eta)\right)^{{2}}
\;,\label{strainsexz2}
\end{equation}
where the left-hand sides define the strain energy release rates corresponding to the breakage horizontal links in the upper and lower lattices.
Then, we may also introduce the quantity describing the normalised strain energy in vertical links of the upper lattice as
\begin{equation}
2\alpha_1(\varepsilon_{yz}(m, \eta))^2=\frac{\alpha_1}{2} (u_{m+1}(\eta)-u_m(\eta))^2\label{strainseyz}
\end{equation}
and for the lower lattice half-plane we will refer to the same formula with $u_j$ replaced by $w_j$, $j=m, m+1$ and $\alpha_1$ replaced by $\alpha_2$. 
As discussed above, we do not analyse the case of the failure in the horizontal and vertical links in both the upper and lower lattices, but will consider the strains $\varepsilon_{xz}$ and $\varepsilon_{yz}$  when studying the lattice deformations below. Thus we focus only on the fracture along the interface and we  define the following classes of solutions:
\begin{itemize}
\item We call a regime of crack propagation at a certain speed {\it admissible} if both conditions of  \eqref{eq:FractureCondition_psi} are satisfied by the solution.
\item When the second condition of  \eqref{eq:FractureCondition_psi}  is not satisfied  by the solution, but the first condition is valid,  such a regime will be classed as a {\it not admissible}.\footnote{The solution always exists, however, the admissibility of this solution relates to the applied  fracture criterion as shown by  \citet{GGM}. Therefore, a solution deemed not admissible for the  fracture criterion we consider may be admissible for other types of fracture criteria.}
\end{itemize}

\section{Numerical illustrations of the dynamic response of the dissimilar lattice undergoing failure}\label{sec3}
Here we present several examples that demonstrate how the heterogeneity of the structured medium can affect the dynamic behaviour and associated physical quantities of the system.  We note that all the computations presented here are based on the results in (\ref{eq:Factorisation}) and Sections \ref{DFLattice} and \ref{secERR}.
Note that in the case of the uniform anisotropic lattice (i.e. $c_1=c_2$, $d_1=d_2$, $m_1=m_2$ and $c\ne c_1$) the analysis of the behaviour of the solution was {carried out by \citet{MMS}}.
In particular, it was shown that in considering a highly anisotropic system, one can ensure the existence of  admissible regimes for the   steady state movement of the crack at low speeds. Clearly, this feature would also appear in the considered problem and for this reason we do not analyse this effect here. Consequently, in this section we consider lattices that are isotropic and joined by an interface. We will also concentrate on the analysis of effects associated with the dissimilarity between lattices and the role of the interface in the dynamics of the lattice undergoing steady state failure.

\subsection{Identical lattices joined by a structured interface}\label{sec7.1}
First we begin with the case when both lattices are the same ($\beta=\gamma=\alpha_1=\alpha_2=1$) and  are connected by an interface, which may have different material properties from the ambient medium.  It is clear in this case, and as will be demonstrated below, the lattice deformation is symmetrical with respect  the line of the crack.

In Figure \ref{fig:ERR_same_speeds1}, we consider $\mu=1/5, 1$ and $ 5$ and demonstrate the behaviour of the ratio $G_0/G$ (based on (\ref{eq:ERR_R})) in the subsonic speed regime. Here the second case corresponds to the isotropic lattice  with a propagating crack and the associated results are well-known (see \citet{LSbook}). This case is presented in Figure \ref{fig:ERR_same_speeds1} for the sake of comparison. 

\begin{figure}[htbp]
\center{\includegraphics[width=0.55\linewidth]{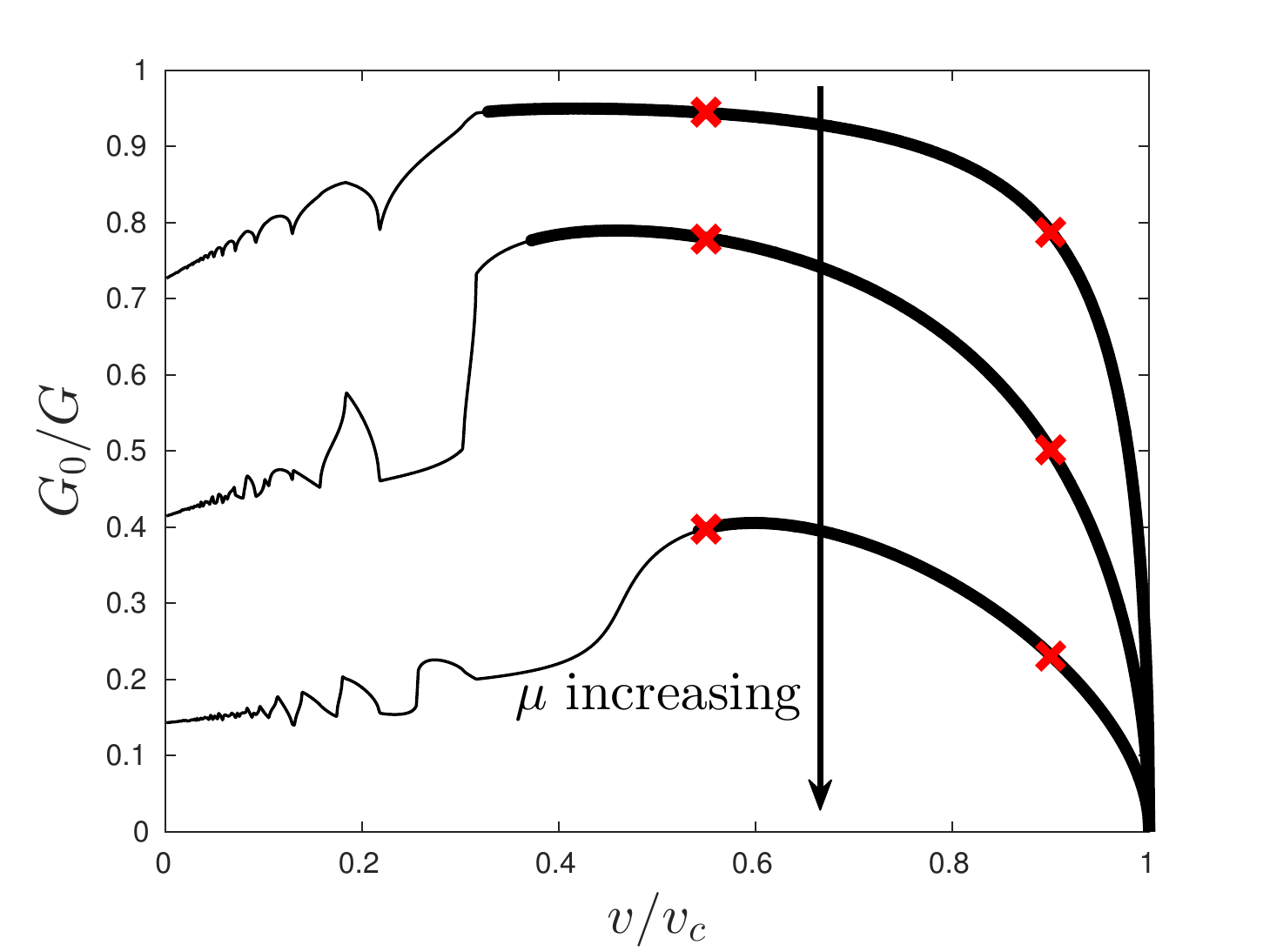} }
\caption[ ]{Dependency of energy release rate ratio $G_0/G$ on ratio $v/v_c$ based on (\ref{eq:ERR_R}).
The computations are performed for the parameters $\beta=1$, $\gamma=1$, $\alpha_1=\alpha_2=1$ and $\mu=1/5, 1,$ and 5.
 In each case, $v_1=v_2$, i.e. the Rayleigh wave speed in both the upper and lower half-planes are the same.
In addition, we indicate the speeds where admissible regimes are realised  in accordance with (\ref{eq:FractureCondition_psi}). Here, regimes which are not admissible correspond to normal lines.} 
\label{fig:ERR_same_speeds1}
\end{figure}


Figure~\ref{fig:ERR_same_speeds1} shows that, as expected, the ratio $G_0/G<1$ is non-monotonic within some neighborhood of   $v=0$, which corresponds to the static problem of a  lattice with a stationary crack.  For the case $\mu=1/5$, an example of such an interval is approximately $0< v<0.3v_c$. Within this region, the derivative of $G_0/G$ with respect to the crack speed can  be undefined at given values of $v$. Additionally, this speed interval exhibiting the non-monotonicity of $G_0/G$ increases as the stiffness of the interfacial bonds increase. Moreover, the ratio $G_0/G$  for fixed $v/v_c$ is monotonically decreasing for increasing $\mu$. 

In each case presented in Figure \ref{fig:ERR_same_speeds1}, beyond a particular value of $v/v_c$,  the ratio $G_0/G$ monotonically decreases to zero as $v$ approaches $v_c$. As an example, we refer to speed interval $0.58<v/v_c<1$ for $\mu=5$. This region is often associated with the admissibility of the solution under the criteria (\ref{eq:FractureCondition_psi}) considered.  For high crack velocities, as shown in the examples below (for instance, see Figure \ref{fig:Plots_1}), the lattice undergoes significant deformations along the crack faces. Thus if one considers additionally the strength of the horizontal links together with (\ref{eq:FractureCondition_psi}),  the crack may develop localised damage along the crack faces that could correspond to roughening of the crack surfaces with increase of the crack speed (see \citet{Fineberg,  MG, GM}).  If instead of the lattice one considers the finite-width strip loaded on the external  boundaries, most of the physical effects related to the fracture appear ahead of the crack and those effects also include branching (see \citet{Marder2}).  

%
%
%

As shown in Figure  \ref{fig:ERR_same_speeds1}, not all speeds $v$ yield admissible regimes for steady state crack propagation. In  Figure \ref{fig:ERR_same_speeds1}, admissible regimes are marked along the curves $G_0/G$ with thick lines, where conditions \eqref{eq:FractureCondition_psi}  are satisfied. The regimes that are not admissible appear in Figure  \ref{fig:ERR_same_speeds1} as normal lines, where the criterion  $(\ref{eq:FractureCondition_psi})_2$  is not obeyed by the solution.


\begin{figure}[htbp]
(a) \minipage{0.3\textwidth}
\center{\includegraphics[width=\linewidth] {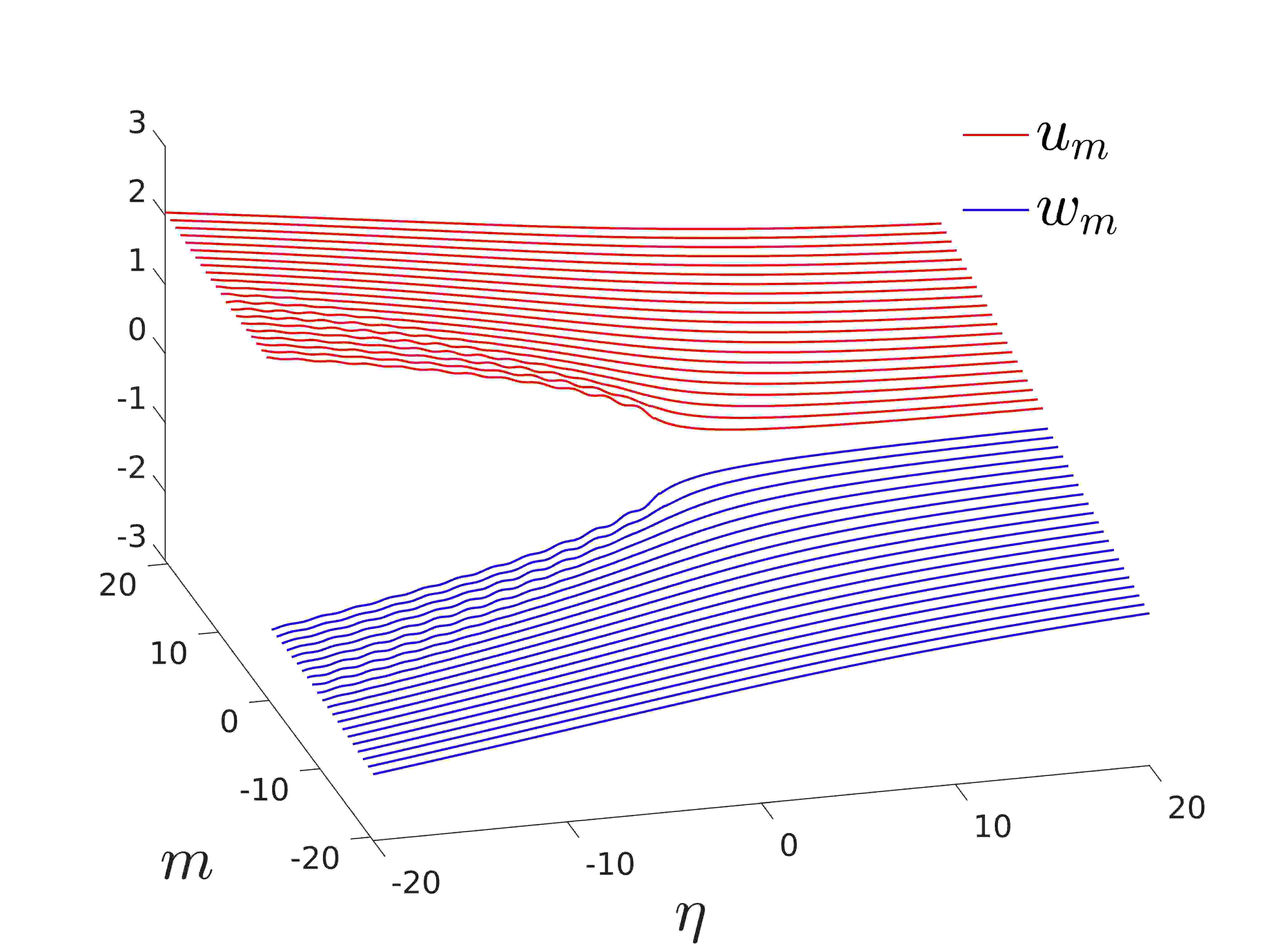}}
\endminipage
\hfill
(b) \minipage{0.3\textwidth}
\center{\includegraphics[width=\linewidth] {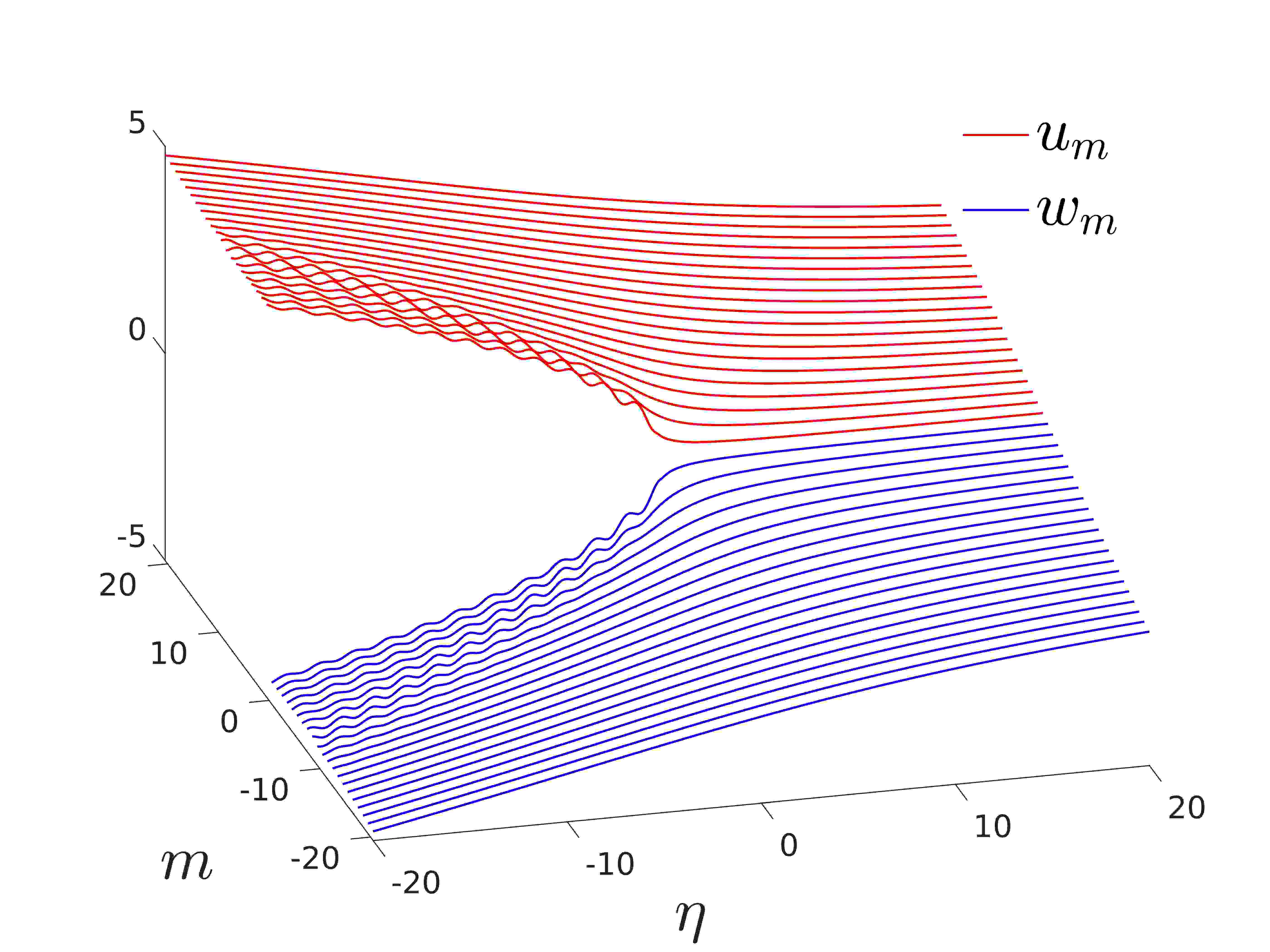}}
\endminipage
\hfill
(c) \minipage{0.3\textwidth}
\center{\includegraphics[width=\linewidth] {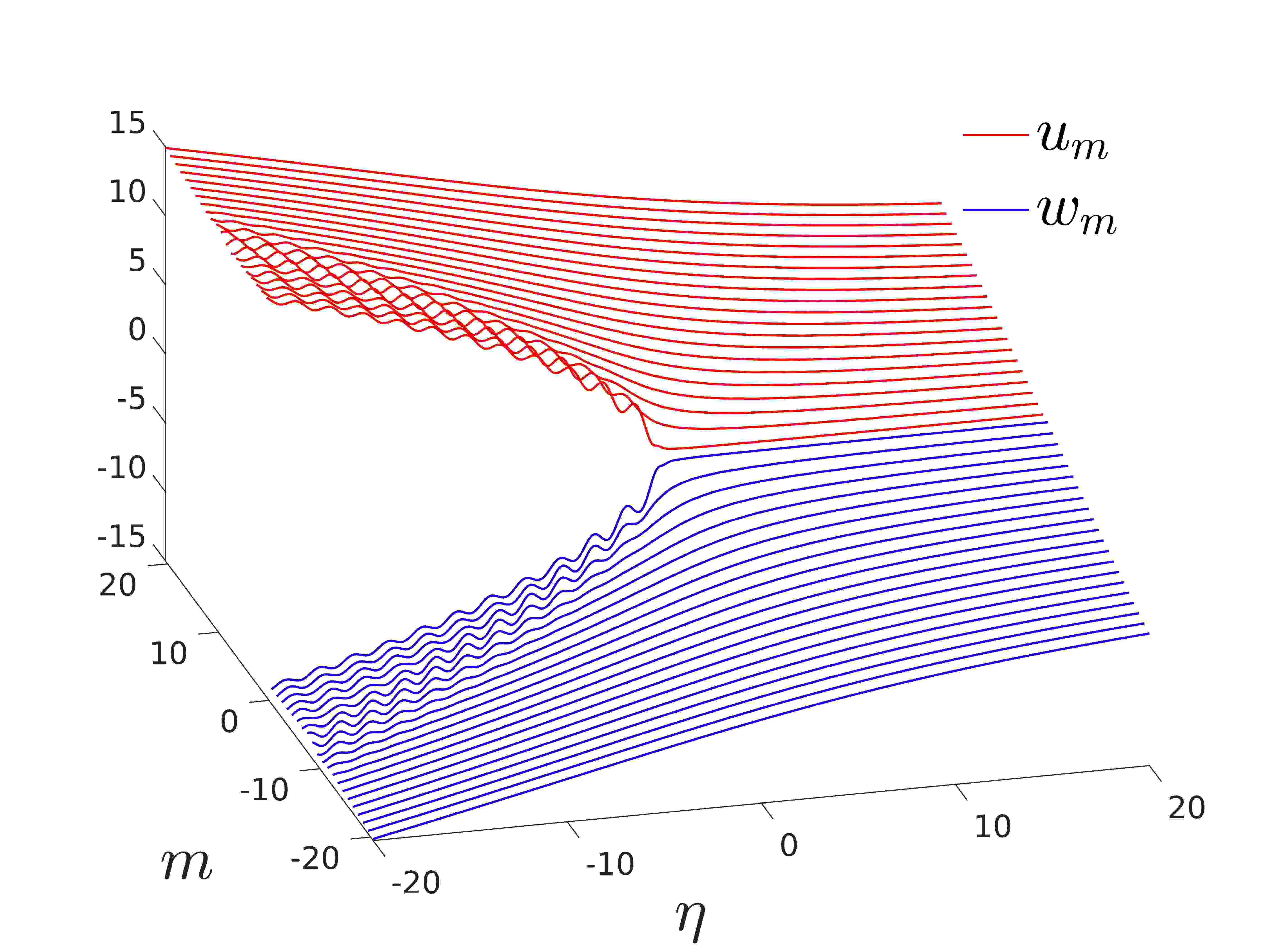}}
\endminipage 
\\
(d) \minipage{0.3\textwidth}
\center{\includegraphics[width=\linewidth] {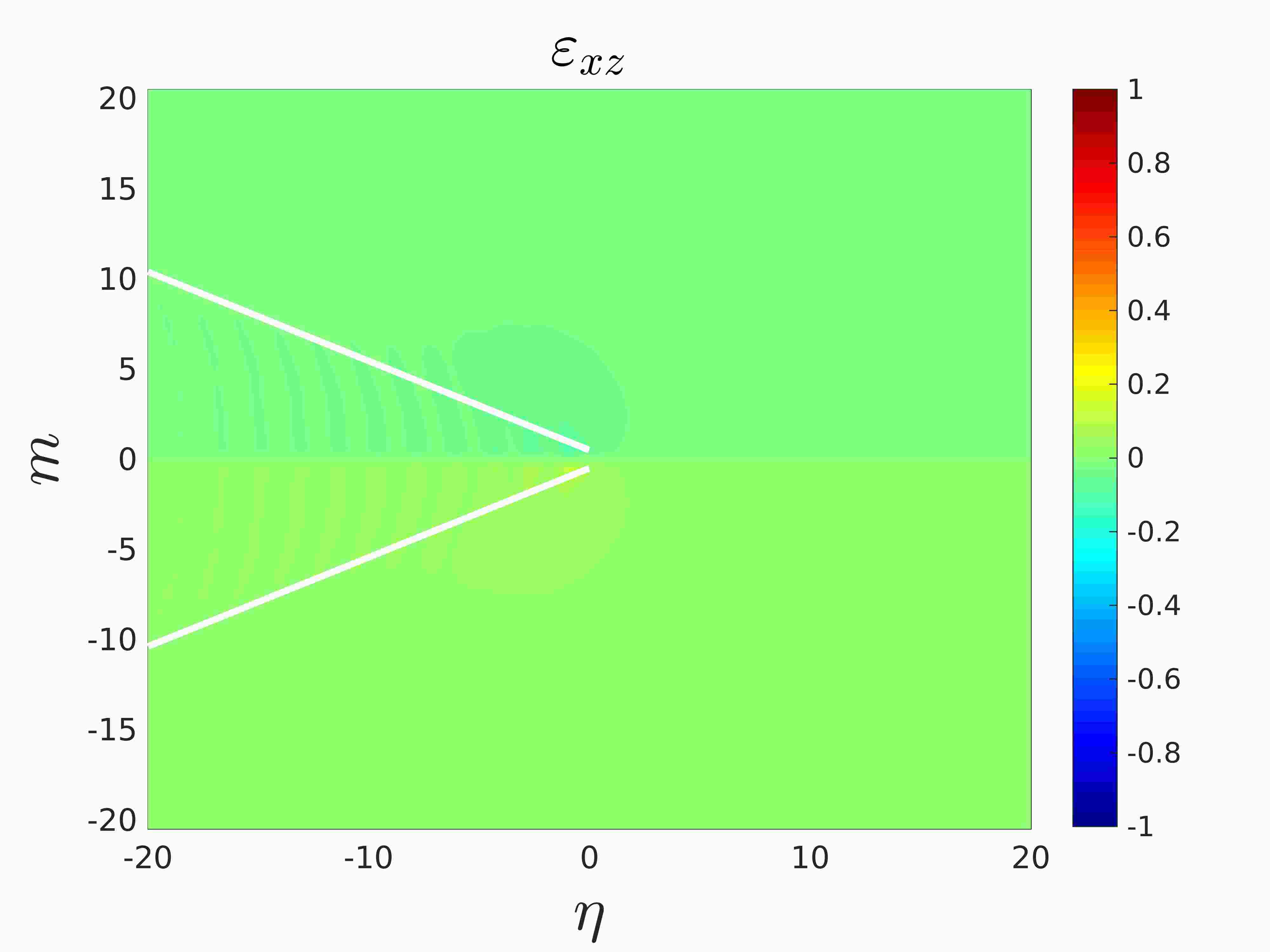}}
\endminipage
\hfill
(e) \minipage{0.3\textwidth}
\center{\includegraphics[width=\linewidth] {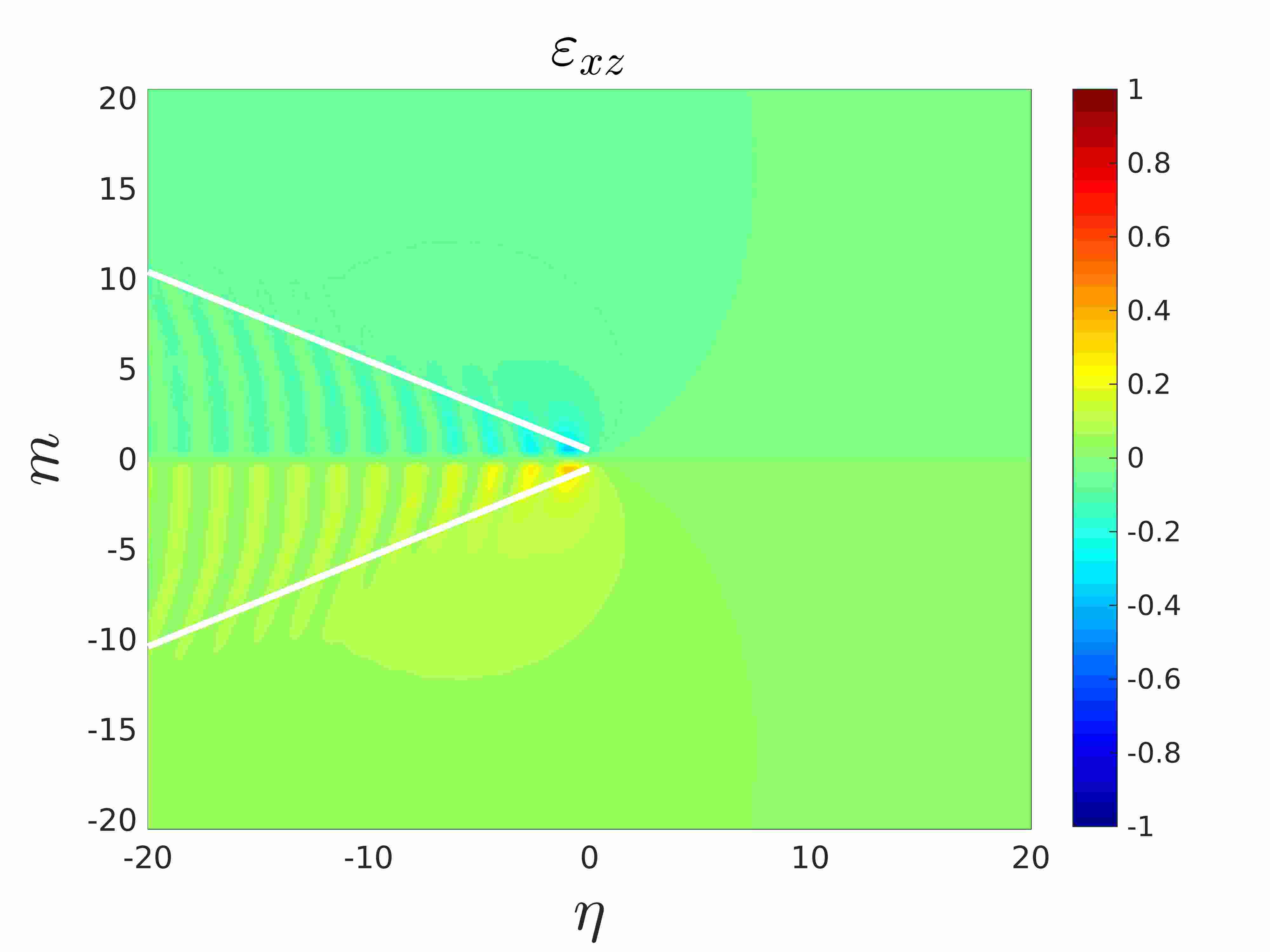}}
\endminipage
\hfill
(f) \minipage{0.3\textwidth}
\center{\includegraphics[width=\linewidth] {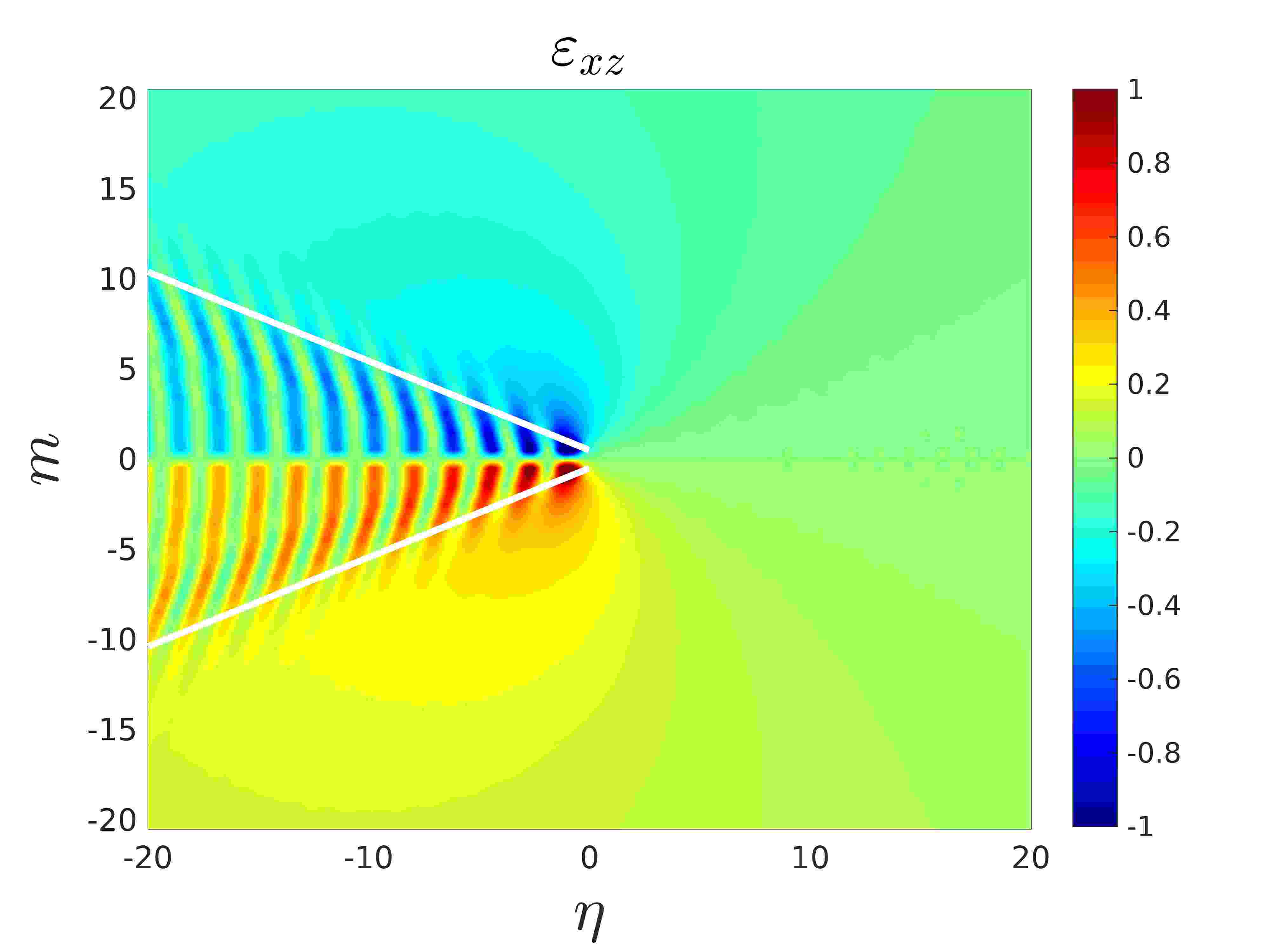}}
\endminipage
\\
(g) \minipage{0.3\textwidth}
\center{\includegraphics[width=\linewidth] {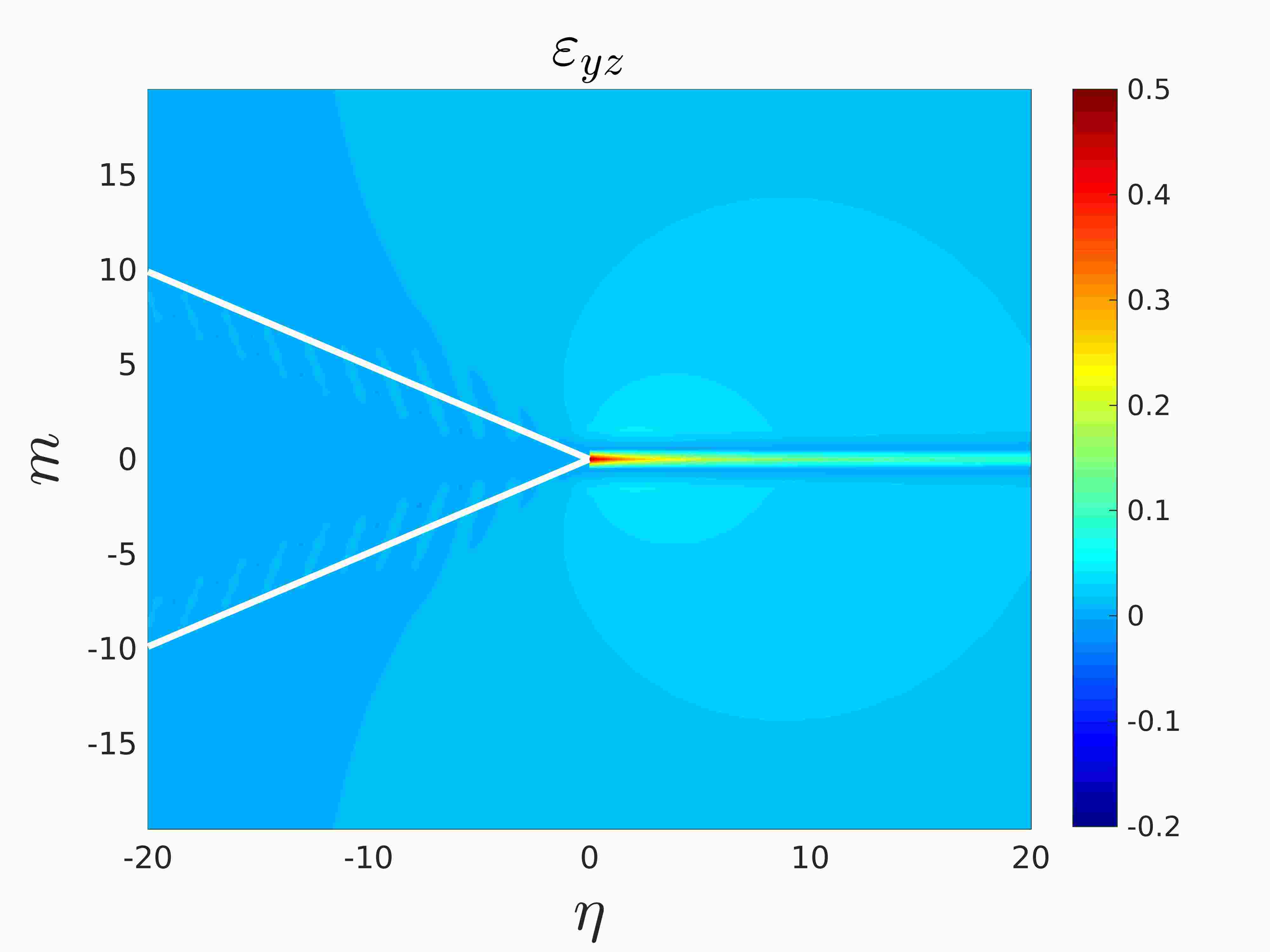}}
\endminipage
\hfill
(h) \minipage{0.3\textwidth}
\center{\includegraphics[width=\linewidth] {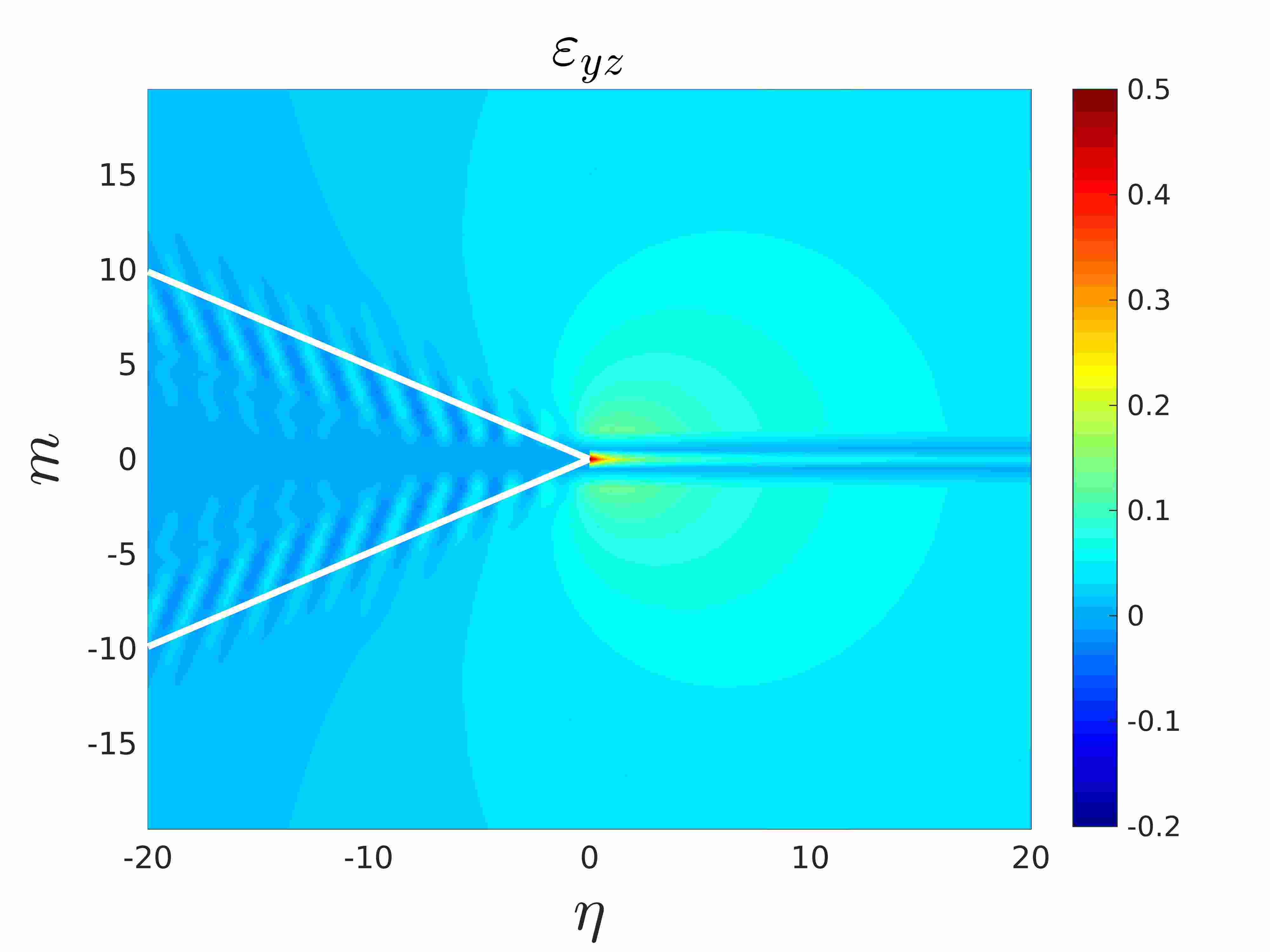}}
\endminipage
\hfill
(i) \minipage{0.3\textwidth}
\center{\includegraphics[width=\linewidth] {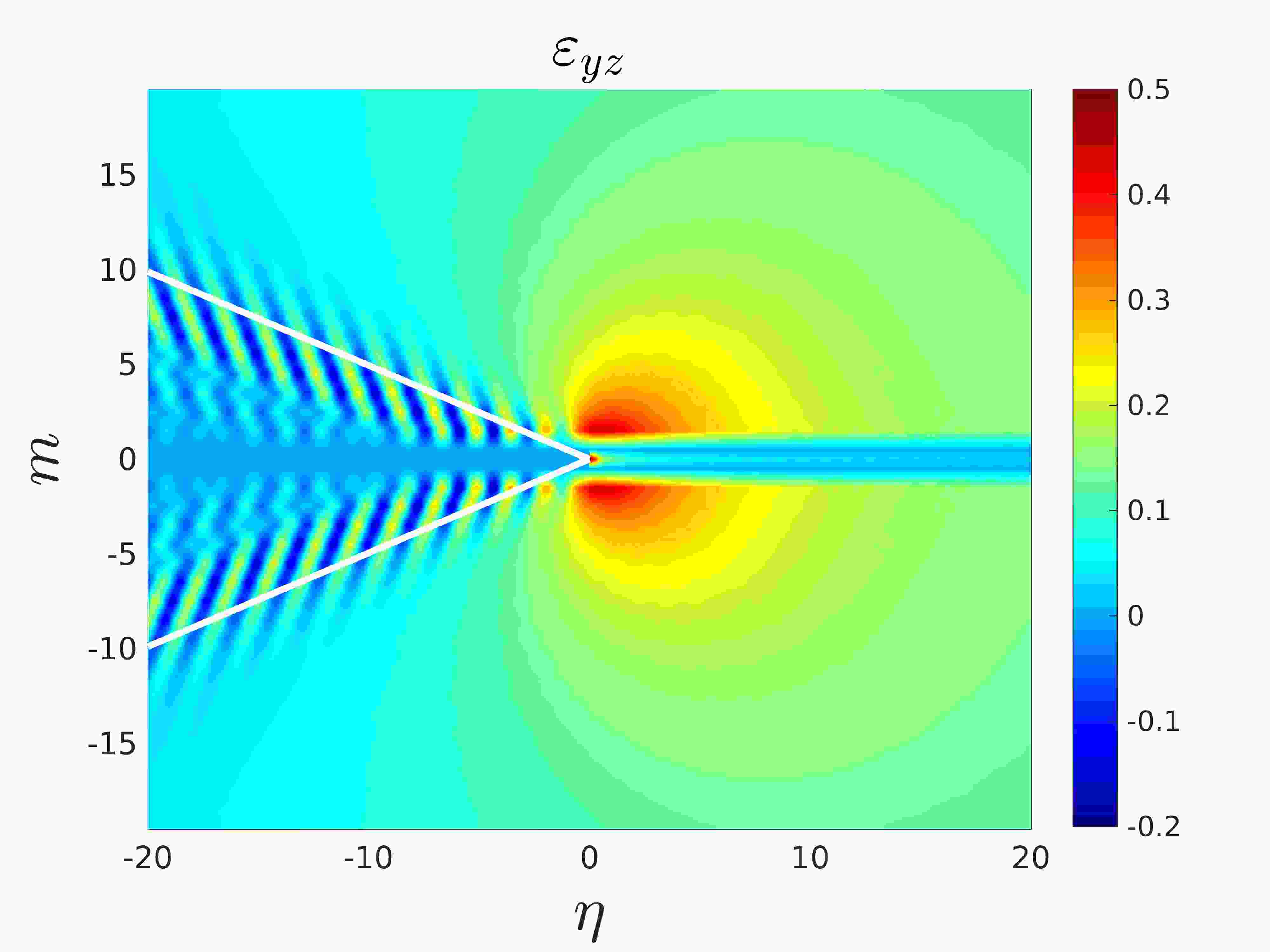}}
\endminipage
\caption[ ]{Displacements and strains in the lattice with parameters $\alpha_1=\alpha_2=\beta=\gamma=1$ and $v/v_c=0.55$ computed using (\ref{eq:SolutionLatticeInverseFourier_Layersa}), (\ref{eq:SolutionLatticeInverseFourier_Layers}).
The panels (a)-(c) show the  displacement field of the lattice with a crack. The panels (d)--(f) and (g)--(i) illustrate the associated strain fields $\varepsilon_{xz}$ and $\varepsilon_{yz}$, respectively. 
Here, the results correspond to  $\mu=1/5$, $\mu=1$ and $\mu=5$ shown in the panels contained in the first, second and third columns, respectively, of the figure. The wave radiation rays are shown in (d)--(i) by white lines and are based on (\ref{wave_rays}).}
\label{fig:Plots_1a}
\end{figure}

The crosses along each curve in Figure \ref{fig:ERR_same_speeds1} also correspond to  admissible regimes of crack propagation in the lattice and the associated dynamic behaviour of the structure  is considered in Figures \ref{fig:Plots_1a} and \ref{fig:Plots_1} for these points.
In particular, Figures \ref{fig:Plots_1a} and \ref{fig:Plots_1}  provide information about the lattice behaviour at  intermediate crack speeds ($v/v_c=0.55$) and at high crack speeds ($v/v_c=0.9$), respectively. 

The first row of panels in Figure \ref{fig:Plots_1a} show the lattice displacements in the vicinity of the crack tip. Since we assume the strength of the interface does not depend on $\mu$, as we increase the interface stiffness we have to increase the crack opening. These figures also demonstrate that softer interfaces lead to 
a decrease in the amount of energy radiated by the crack tip as it propagates. This is further supported by Figure \ref{fig:ERR_same_speeds1}, where $G_0/G$ for $\mu=1/5$ takes its largest values  from the cases presented. Consequently, the  deformations behind crack tip for softer interfaces are less pronounced, as evidenced when comparing Figure \ref{fig:Plots_1a}(a) with  (c).

 The second row of panels in Figure \ref{fig:Plots_1a} show the strains $\varepsilon_{xz}$ based on (\ref{strainsexz1}) and (\ref{strainsexz2}). They give an indication of the behaviour of the lattice behind the crack tip. Located there are several ripples in the strains $\varepsilon_{xz}$  that are confined within the region defined by the wave radiation rays, represented by white lines in Figures \ref{fig:Plots_1a}(d)--(i). Note, as shown in Section \ref{DFLattice} and the {Supplementary Material \ref{SM6}}, the wave radiation rays are independent of the interfacial bond stiffness  $\mu$, as they are linked to the rate of decay of the field in the ambient lattices that is governed by $\lambda_j$, $j=1,2$. As a result of the structural  symmetry about the crack line, these radiation rays are also symmetric about this line.
 Increasing $\mu$ causes an increase in the intensity of the deformations along the ripples in Figures \ref{fig:Plots_1a}(d)--(f) in the displacements behind the crack tip (compare Figure \ref{fig:Plots_1a}(d) with (f)).  As we will show later, higher speeds will decrease the number of ripples behind the crack tip (e.g. see Figure \ref{fig:Plots_1}). In particular, in all examples shown here, the strains $\varepsilon_{xz}$ in the lattice are small for soft interfaces.

 The third row  of Figure \ref{fig:Plots_1a} illustrates  the behaviour of the strains $\varepsilon_{yz}$ based on (\ref{strainseyz}). They show the strains are symmetric with respect to the crack line. Naturally, the 
 strain concentration is near the crack tip. For lattices with soft interfaces, the most prominent strains $\varepsilon_{yz}$ are found in the vicinity and ahead of the crack tip (see Figure \ref{fig:Plots_1a}(g)). The bulk lattice strains are usually smaller in comparison, again confirming that propagating cracks within soft interfaces are less likely to promote the radiation of lattice vibrations.

In addition, Figures \ref{fig:Plots_1a}(h) and (i) show the emergence of a region of localised deformation in the vicinity of the crack tip. The intensity of the deformation there  increases as the interface stiffness becomes larger for a fixed speed. It is clear that larger $\mu$ causes this deformation to be comparable with that at the crack tip. This effect could potentially lead to a  solution of the problem which is not admissible for this given speed, as discussed above. 
The strains shown in  Figures \ref{fig:Plots_1a}(h) and (i) also  illustrate clearly the preferential directions for the lattice wave radiation, along which the vibration amplitude decays the slowest, and these directions coincide with the wave radiation segments predicted by (\ref{wave_rays}).
 Further, the intensity of the waves along the radiation rays increase if the stiffness of the interface is also increased.

We note that Figure \ref{fig:Plots_1a} shows the displacement field has a square root type growth like $O(\sqrt{d})$, $d=\sqrt{m^2+\eta^2}$, far away from the crack tip, but both strains tend to zero in the far-field of the lattice as $O(1/\sqrt{d})$ for evident reasons. This behaviour in the remote regions of the lattice is consistent with the field in the vicinity of a crack tip propagating within the continuous bimaterial. On the other hand, along the wave radiation rays in the lattice, as discussed in Section \ref{DFLattice} and the Supplementary Material, the wave amplitudes decay like $O(1/d^{1/3})$.

\begin{figure}[htbp]
(a) \minipage{0.3\textwidth}
\center{\includegraphics[width=\linewidth] {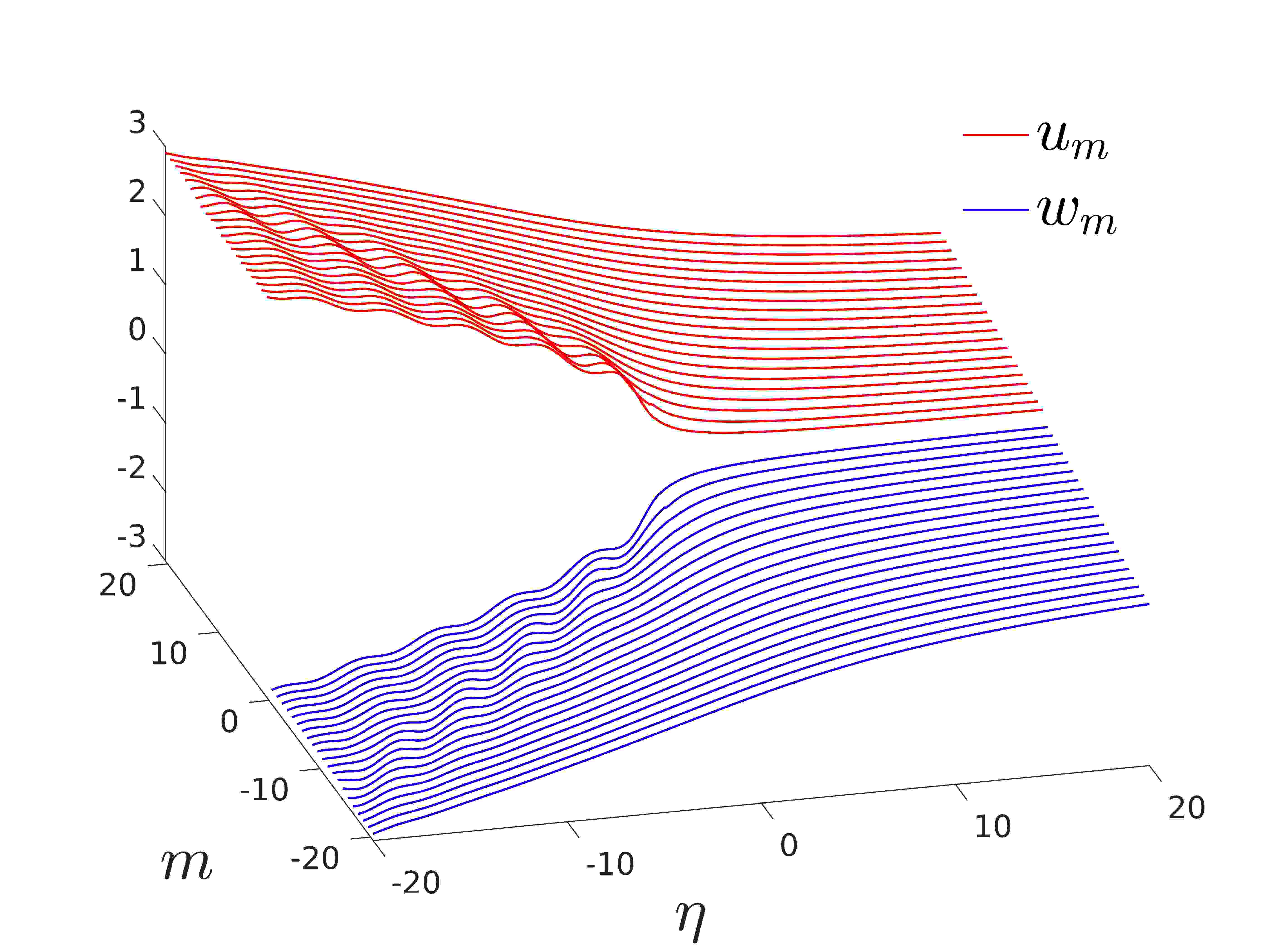}}
\endminipage
\hfill
(b) \minipage{0.3\textwidth}
\center{\includegraphics[width=\linewidth] {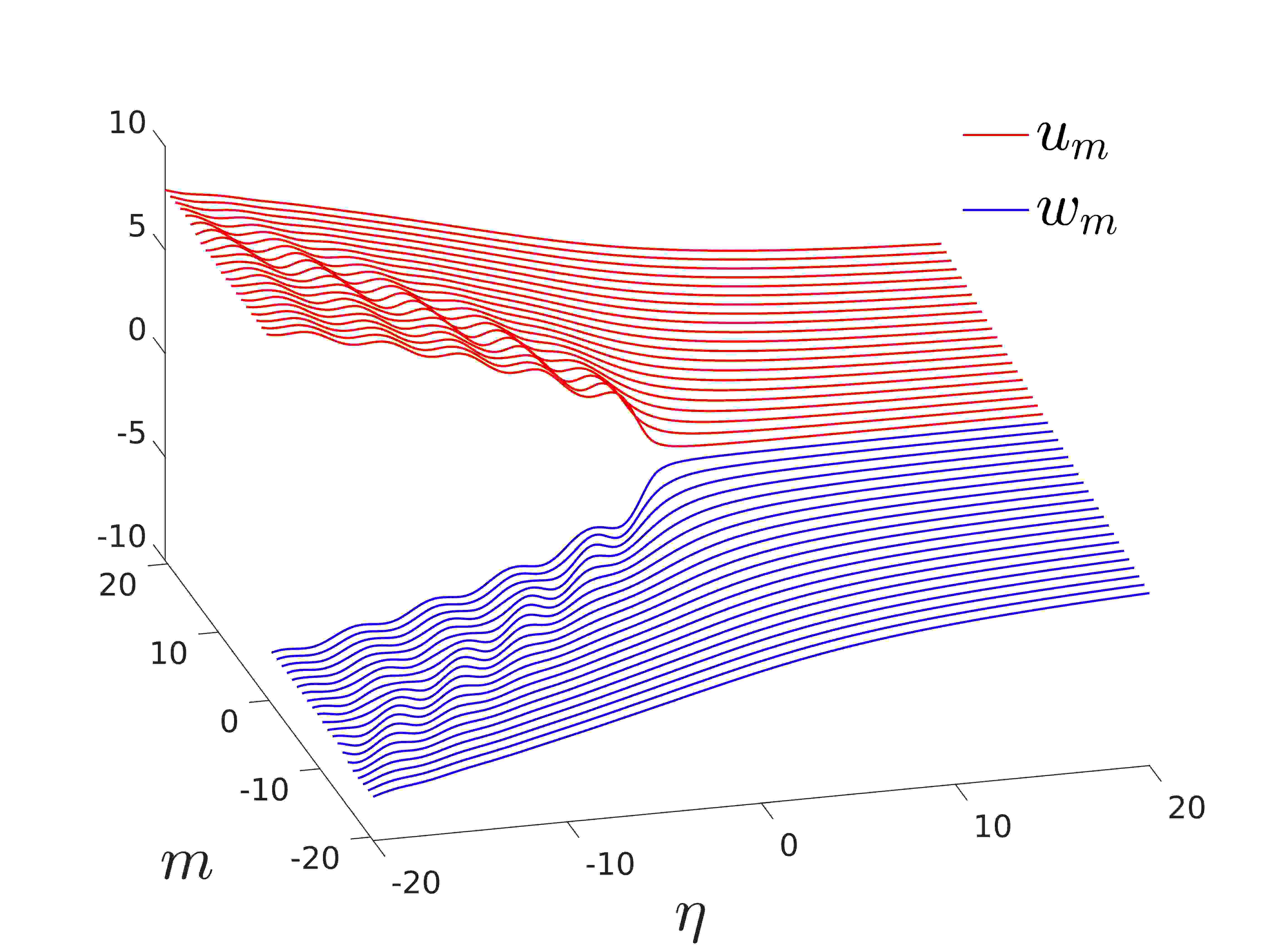}}
\endminipage
\hfill
(c) \minipage{0.3\textwidth}
\center{\includegraphics[width=\linewidth] {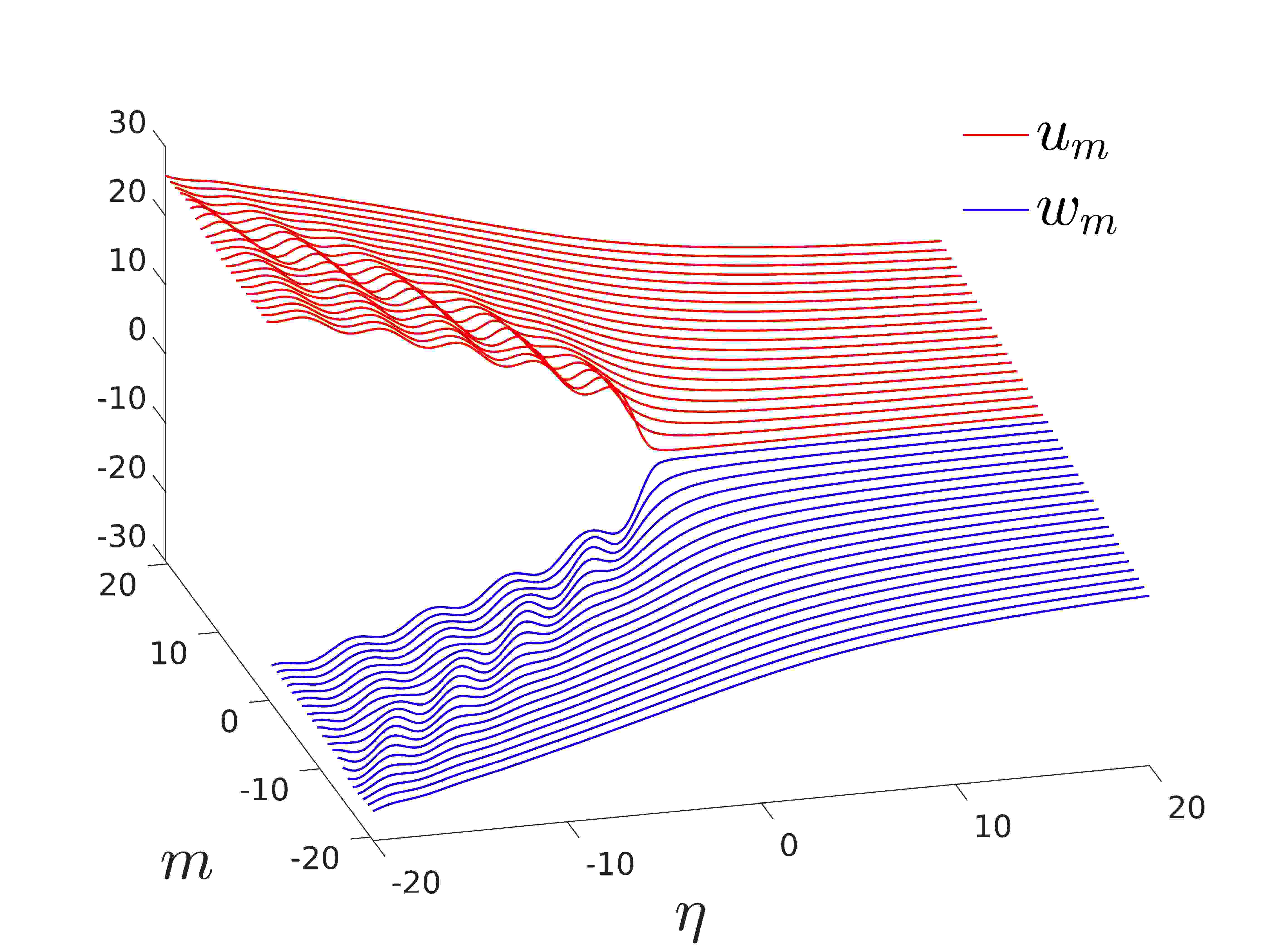}}
\endminipage 
\\
(d) \minipage{0.3\textwidth}
\center{\includegraphics[width=\linewidth] {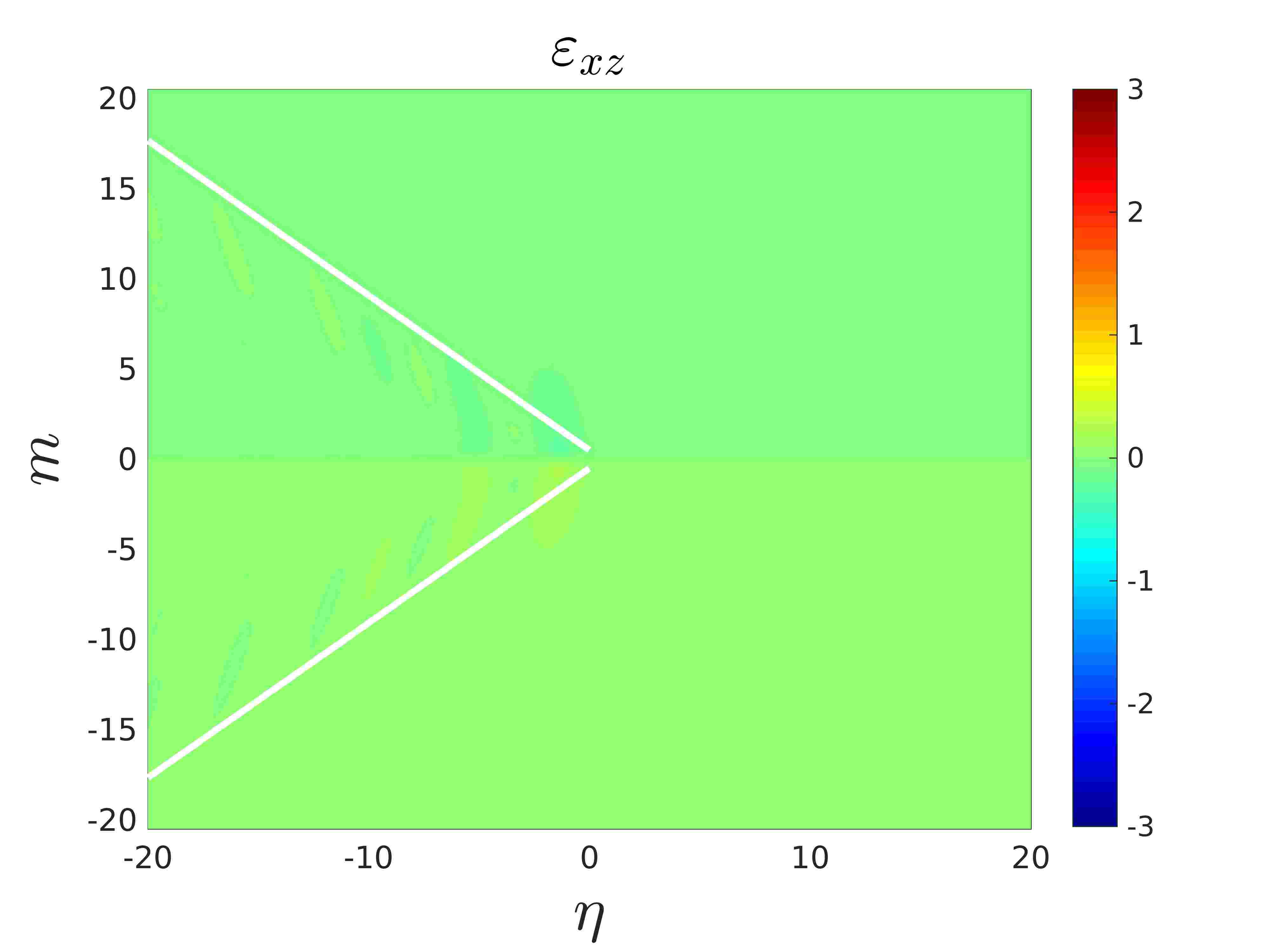}}
\endminipage
\hfill
(e) \minipage{0.3\textwidth}
\center{\includegraphics[width=\linewidth] {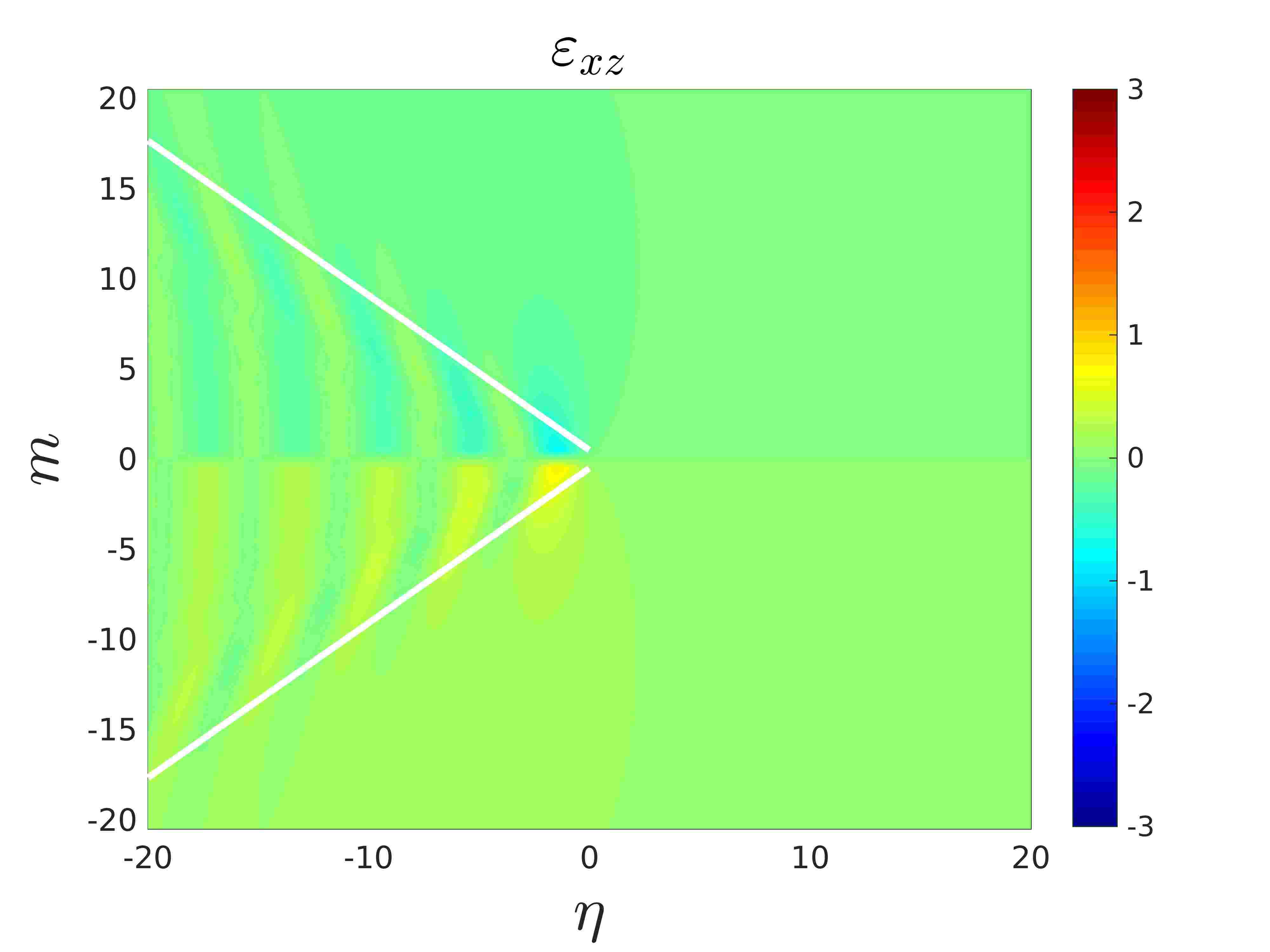}}
\endminipage
\hfill
(f) \minipage{0.3\textwidth}
\center{\includegraphics[width=\linewidth] {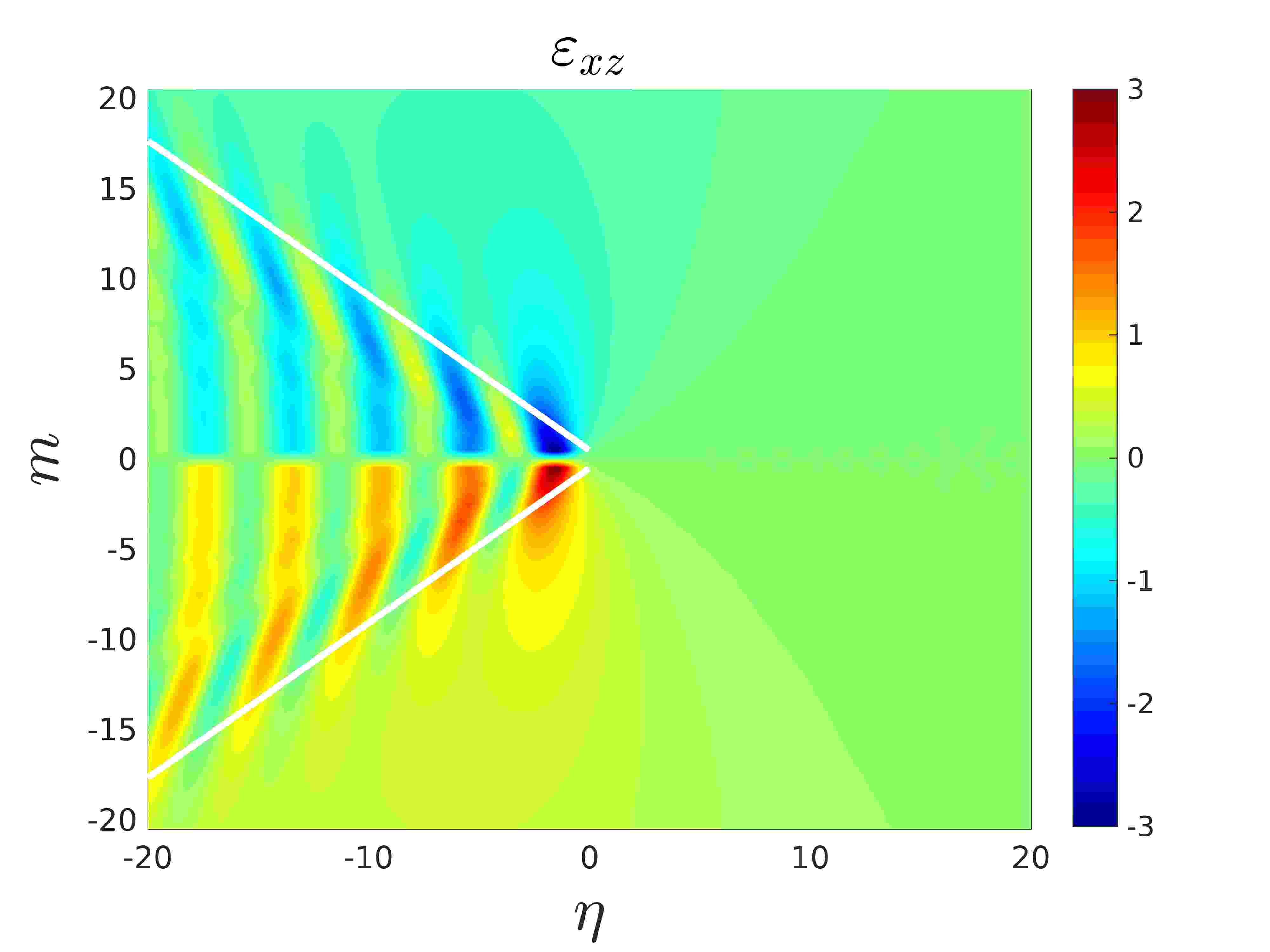}}
\endminipage
\\
(g) \minipage{0.3\textwidth}
\center{\includegraphics[width=\linewidth] {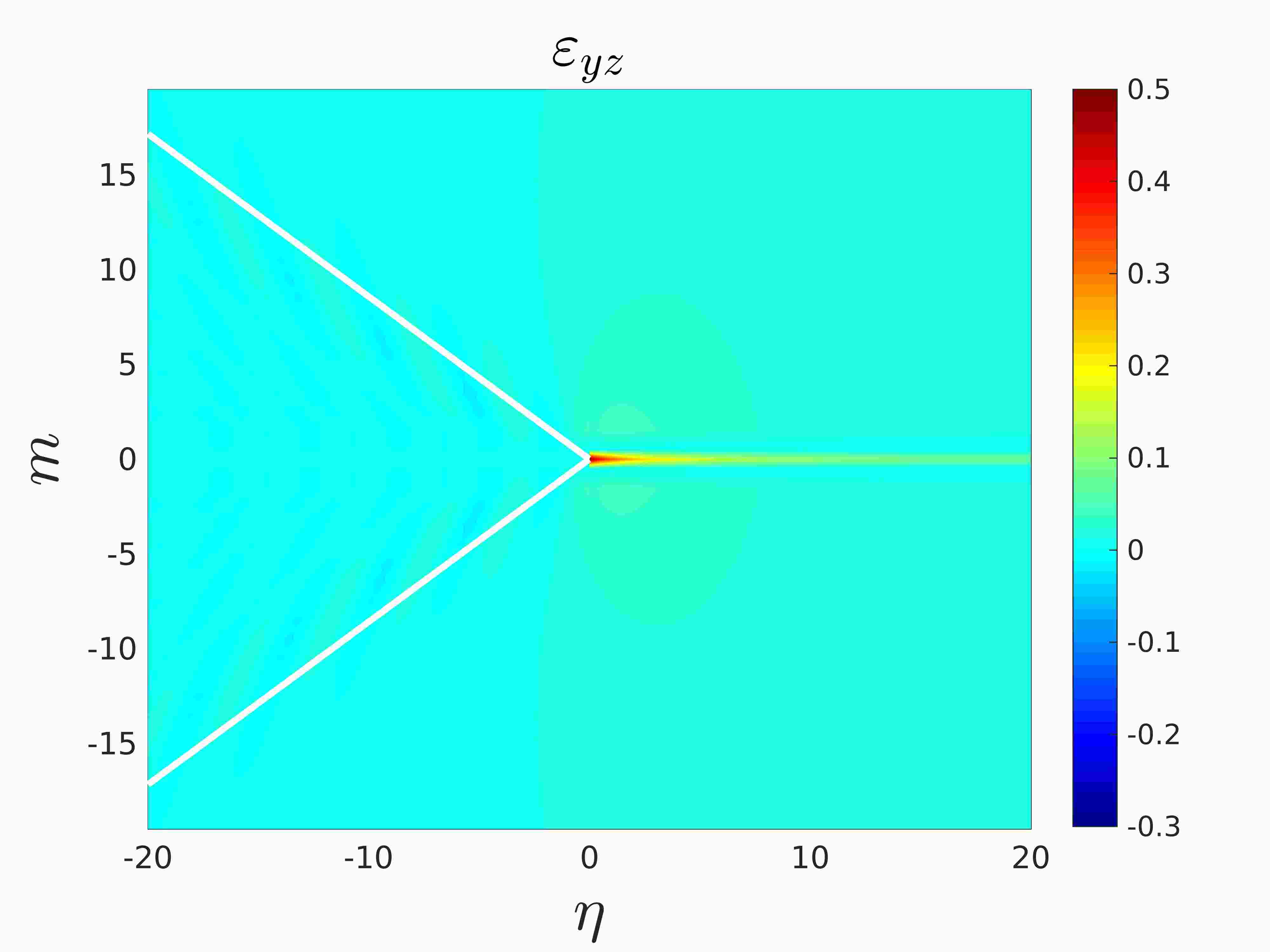}}
\endminipage
\hfill
(h) \minipage{0.3\textwidth}
\center{\includegraphics[width=\linewidth] {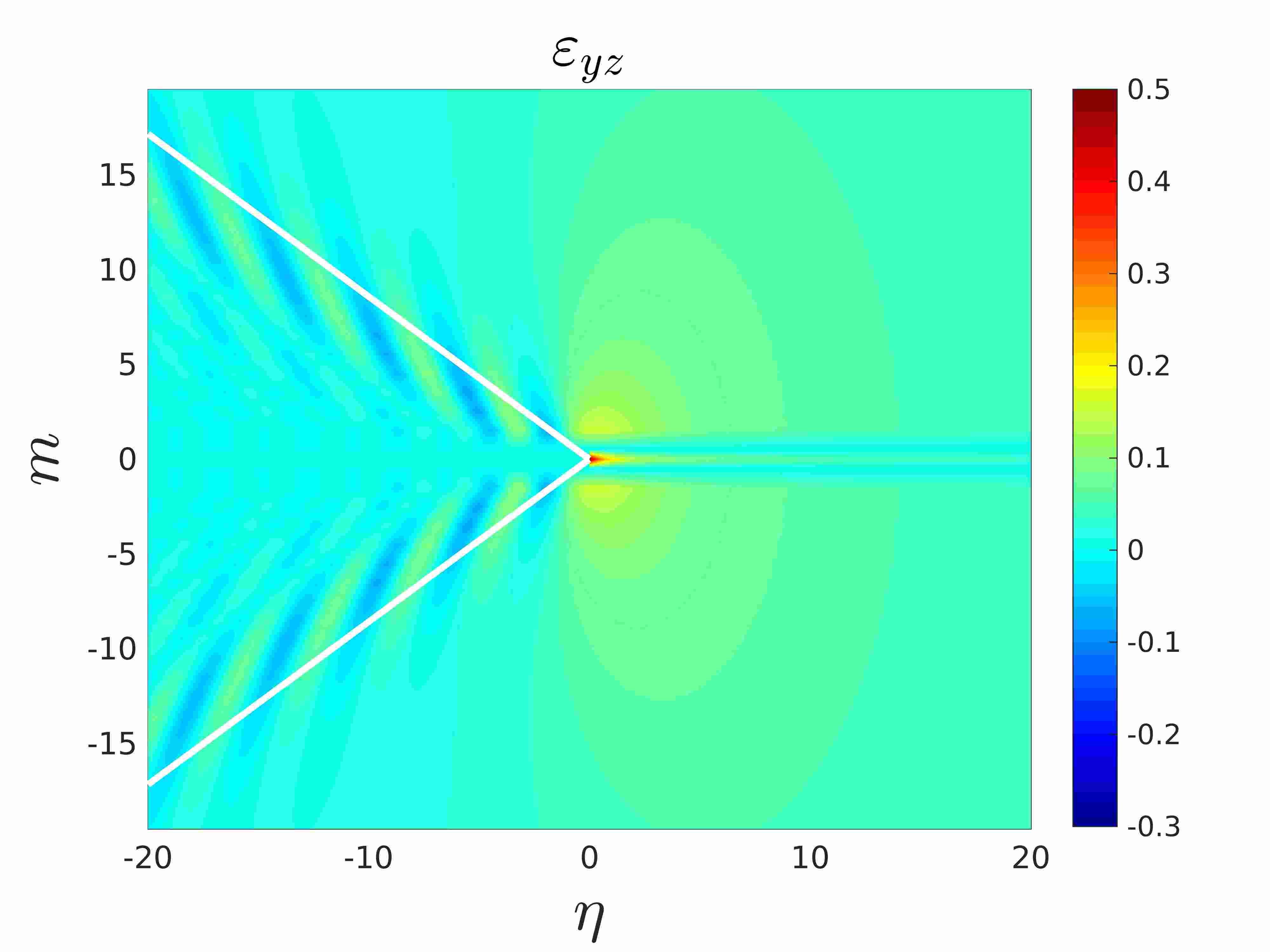}}
\endminipage
\hfill
(i) \minipage{0.3\textwidth}
\center{\includegraphics[width=\linewidth] {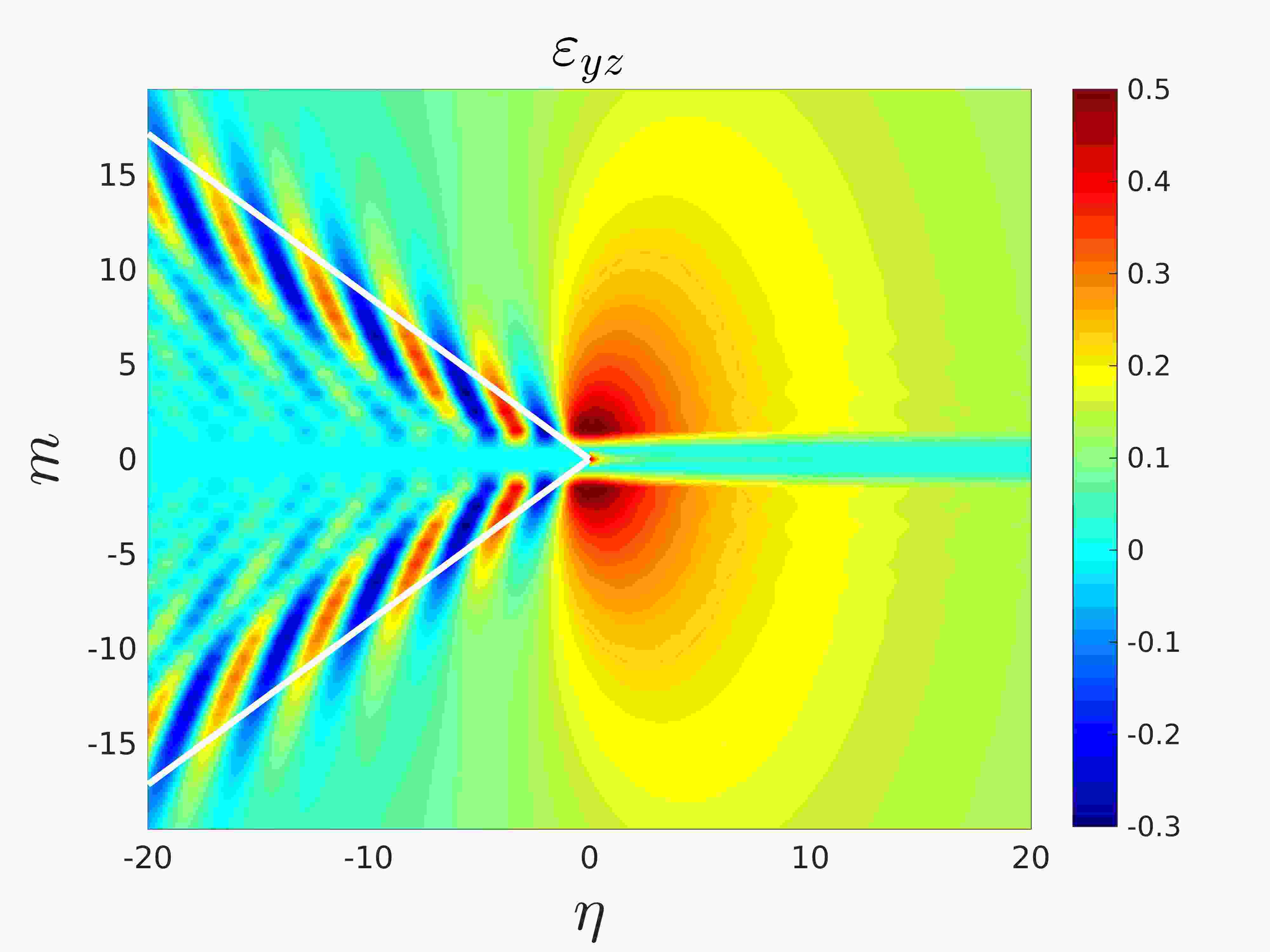}}
\endminipage
\caption[ ]{Displacements and strains in the lattice with parameters $\alpha_1=\alpha_2=\beta=\gamma=1$ and $v/v_c=0.9$ computed using (\ref{eq:SolutionLatticeInverseFourier_Layersa}) and (\ref{eq:SolutionLatticeInverseFourier_Layers}).
The panels (a)-(c) show the  displacement field of the lattice with a crack. The panels (d)--(f) and (g)--(i) illustrate the associated strain fields $\varepsilon_{xz}$ and $\varepsilon_{yz}$, respectively. 
Here, the results correspond to  $\mu=1/5$, $\mu=1$ and $\mu=5$ shown in the panels contained in the first, second and third columns, respectively, of the figure. The wave radiation rays are shown in (d)--(i) by white lines and are based on (\ref{wave_rays}). }
\label{fig:Plots_1}
\end{figure}

Next, we discuss the effect  of increasing the fracture speed $v$. Figure \ref{fig:Plots_1} shows the behaviour of the lattice with an interfacial crack  for the case of $v/v_c=0.9$. For a given $\mu$, comparing the top rows of Figures \ref{fig:Plots_1a} and \ref{fig:Plots_1}, it is clear  that increase of the failure speed leads to an increase in the crack opening (e.g. see Figure \ref{fig:Plots_1a}(a) with Figure \ref{fig:Plots_1}(a)). In addition, the oscillations in the wake of the tip of the faster crack are now more pronounced and have a longer wavelength than those observed in Figure \ref{fig:Plots_1a}. These effects are localised to the region defined by the wave radiation rays shown in the second and third rows of Figure \ref{fig:Plots_1}, corresponding to the results for the strain fields $\varepsilon_{xz}$ and $\varepsilon_{yz}$, respectively, obtained using (\ref{strainsexz1}), (\ref{strainsexz2}) and (\ref{strainseyz}). Note these strain fields indicate that the deformations behind the crack tip are now less localised about the crack line in comparison with the results of Figure \ref{fig:Plots_1a}. 

The increase in the crack speed has also led to an increase in the area bounded by the radiation rays, within which a highly oscillatory behaviour of the lattice displacements can occur. In particular,  comparing Figures \ref{fig:Plots_1a}(h) and (i) with Figures \ref{fig:Plots_1}(h) and (i),  we see that for high interfacial bond stiffness the vibrations distributed along the radiation rays have a greater intensity in the case of $v/v_c=0.9$. It is also apparent that the likelihood of a crack path instability or roughening of the crack faces is larger for high interfacial bond stiffness and crack speeds. This is evidenced by the fact that there exists a substantial region local to the crack tip in Figure \ref{fig:Plots_1}(i), where the the lattice deformations are very large in comparison with those in the ambient lattice.

\subsection{Dissimilar lattice system  exhibiting the same bulk wave speeds}\label{sec7.2}

Now we consider the dissimilar lattice system composed of two isotropic lattices, with the same wave speeds that may  be directly connected or  connected by an interface. Specifically, we will consider the parameters 
$\beta=1/5$, $\gamma=\alpha_2=5$ and $\alpha_1=1$ and  the interfacial bond stiffnesses  $\mu=1/5, 1$ and 5. Here, $\mu=1$ and 5 represent the perfect join of the two lattices (i.e. a perfect interface).

\begin{figure}[htbp]
\center{\includegraphics[width=0.55\linewidth]{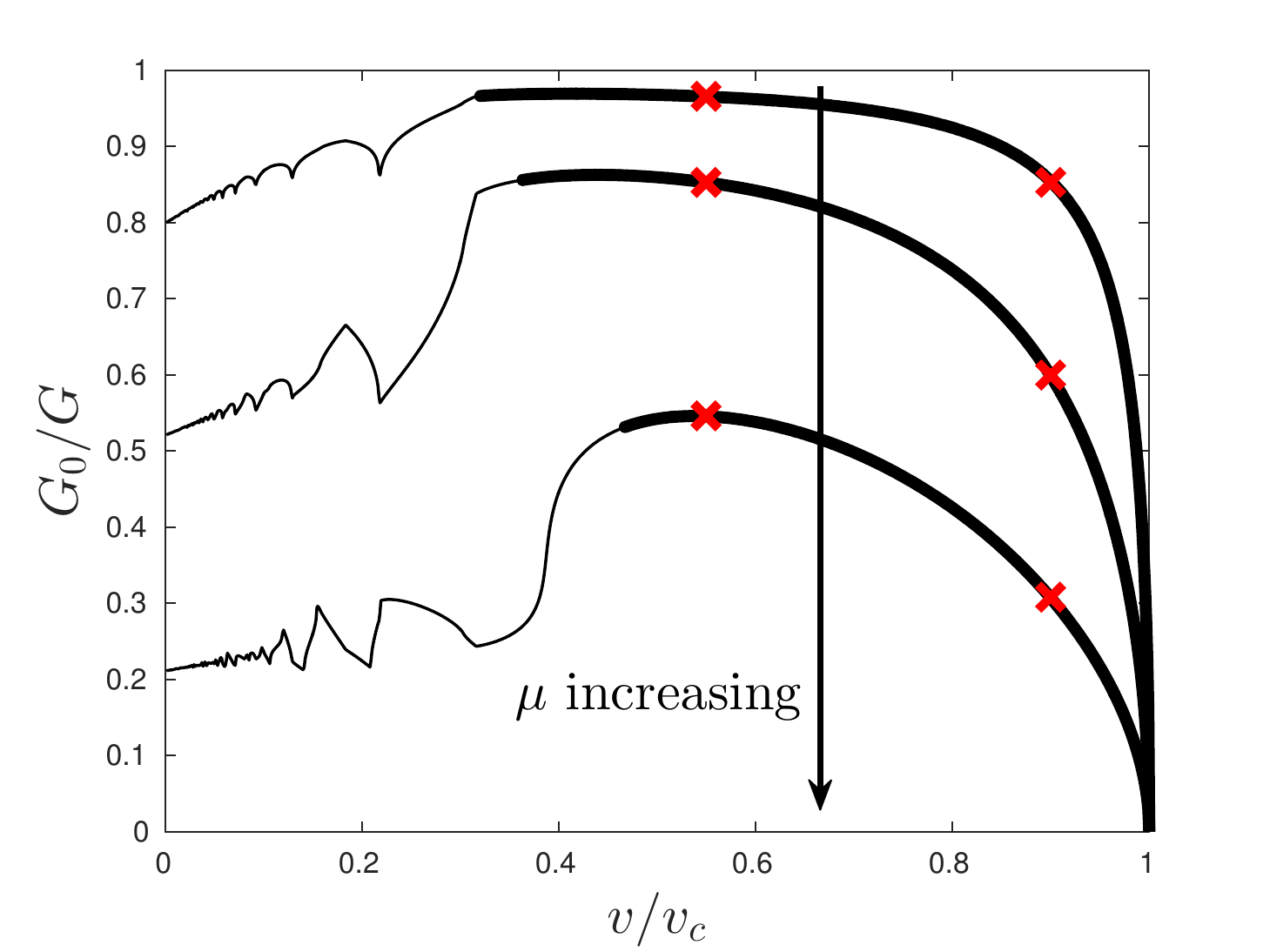} }
\caption[ ]{Dependency of energy release rate ratio $G_0/G$ on ratio $v/v_c$ based in (\ref{eq:ERR_R}).
The computations correspond to the choice of parameters $\beta=1/5$, $\gamma=\alpha_2=5$ and $\alpha_1=1$  
and $\mu=1/5, 1,$ and 5.
In addition, we indicate the speeds where admissible regimes are realised in accordance with (\ref{eq:FractureCondition_psi}). Here  regimes which are not admissible correspond to normal lines. 
}
\label{fig:ERR_same_speeds2}
\end{figure}

Figure \ref{fig:ERR_same_speeds2} shows the ratio $G_0/G$ as a function of the normalised crack speed $v/v_c$.  The same behaviour of the ratio $G_0/G$ observed in Figure \ref{fig:ERR_same_speeds1} is present also in Figure \ref{fig:ERR_same_speeds2}.
 Note that once again the regions representing admissible crack propagation regimes appear as continuous intervals located beyond  a sufficiently high crack speed. In consulting Figures \ref{fig:ERR_same_speeds1} and \ref{fig:ERR_same_speeds2}, we see that the joining of two contrasting lattices, in both stiffness and density, reduces the  energy distributed to wave radiation processes when a crack propagates between these media.  This can be seen by comparing the curves in Figures \ref{fig:ERR_same_speeds1} and \ref{fig:ERR_same_speeds2} for the same $\mu$. 
 
 The crosses along each curve in Figure \ref{fig:ERR_same_speeds2} represent scenarios for crack propagation within the dissimilar structure at two different crack speeds. These cases are further investigated in Figures \ref{fig:Plots_5} and \ref{fig:Plots_6}. There, once more the wave radiation rays are presented and are again symmetric about the crack path, despite the fact there is a clear dissimilarity between the properties of the bulk lattices in each case.

\begin{figure}[htbp]
(a) \minipage{0.3\textwidth}
\center{\includegraphics[width=\linewidth] {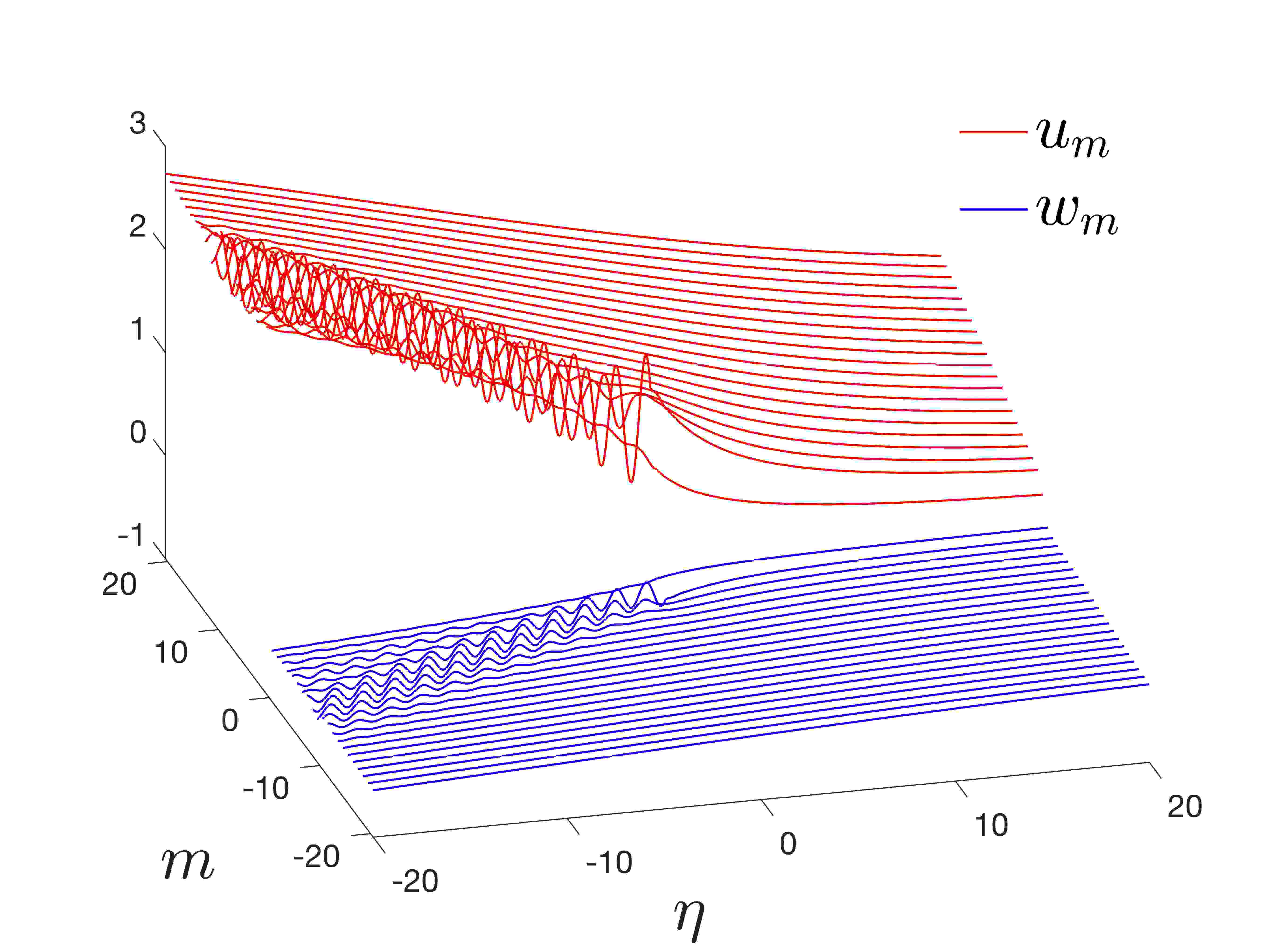}}
\endminipage
\hfill
(b) \minipage{0.3\textwidth}
\center{\includegraphics[width=\linewidth] {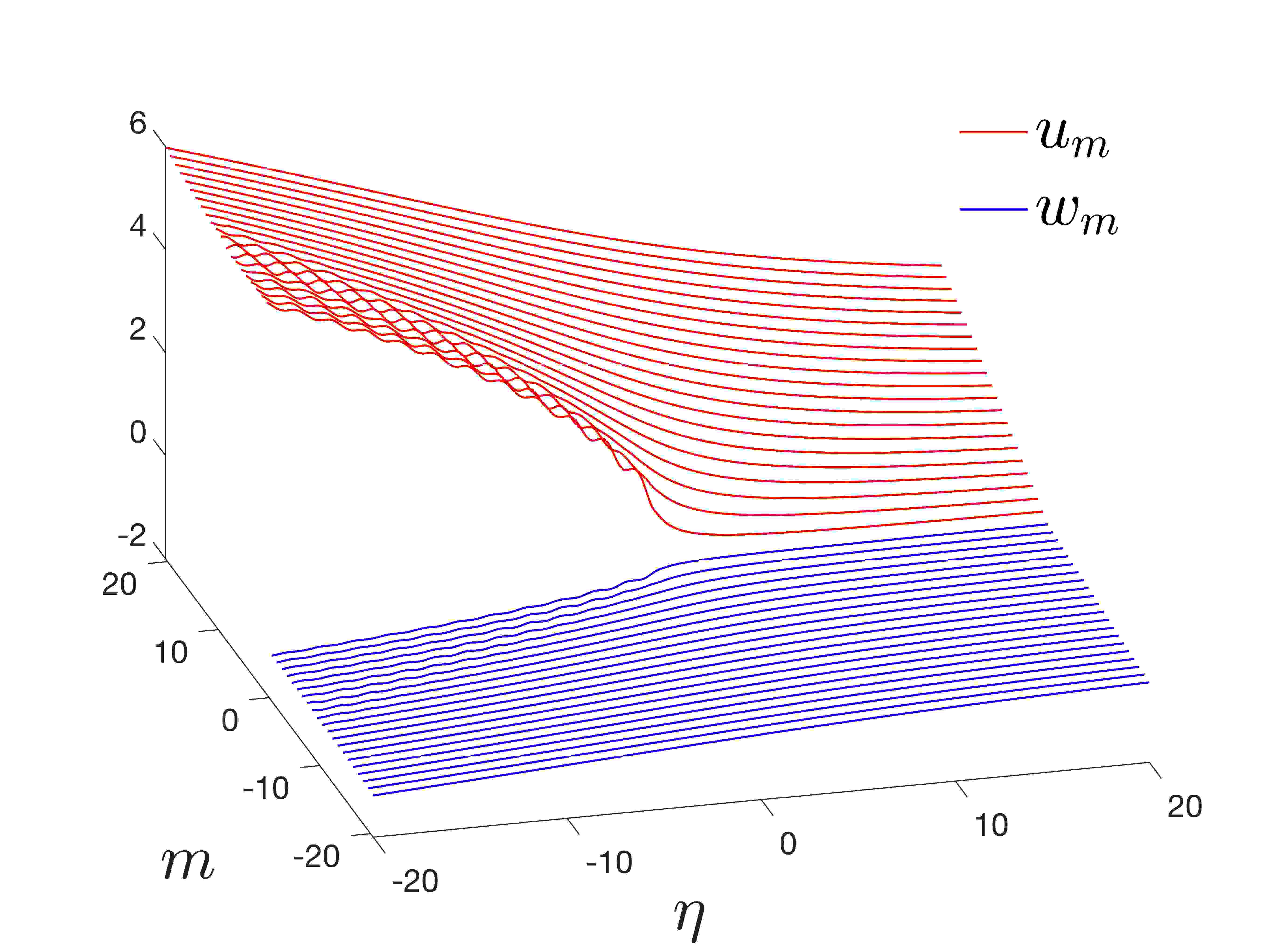}}
\endminipage
\hfill
(c) \minipage{0.3\textwidth}
\center{\includegraphics[width=\linewidth] {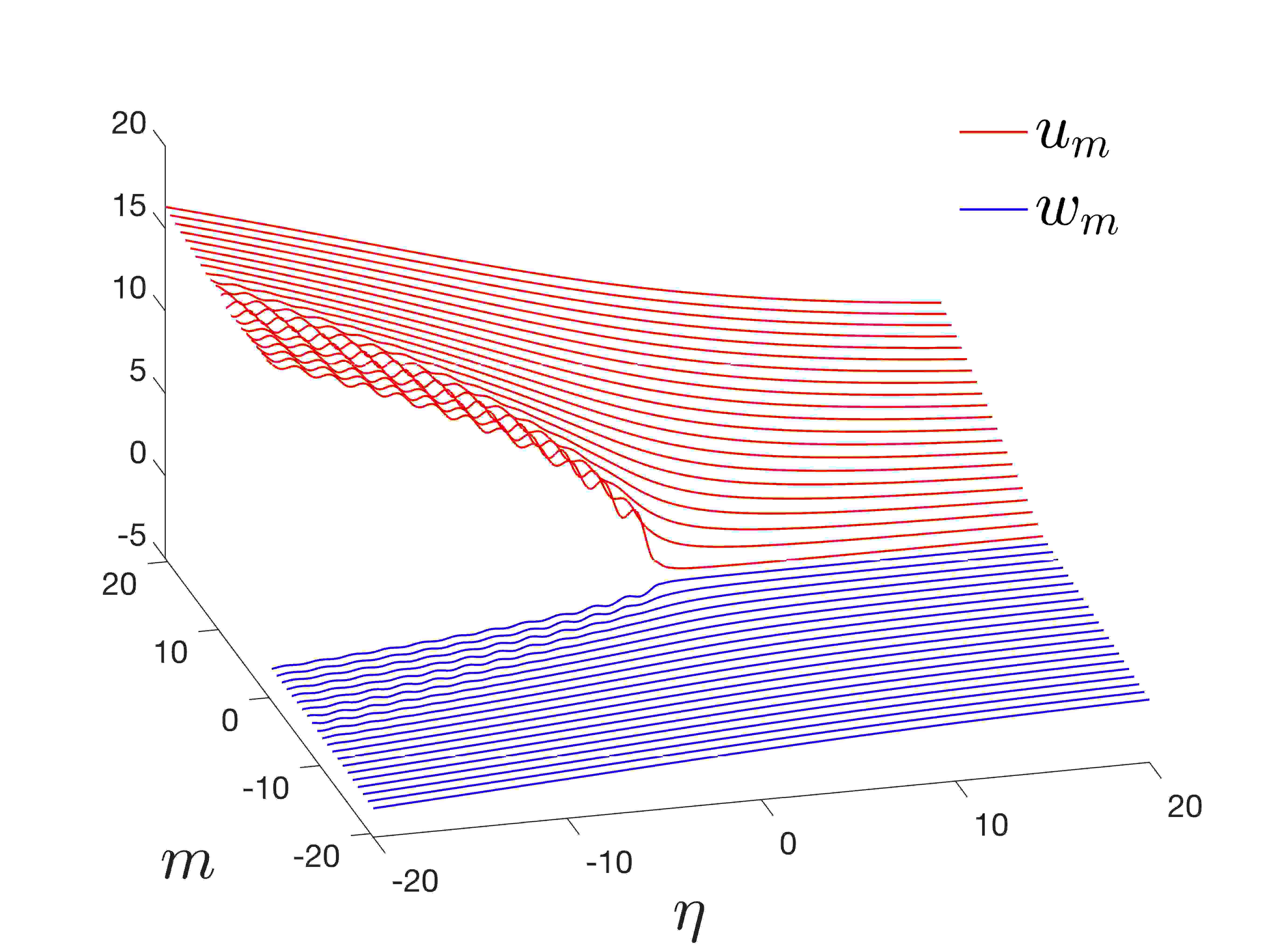}}
\endminipage 
\\
(d) \minipage{0.3\textwidth}
\center{\includegraphics[width=\linewidth] {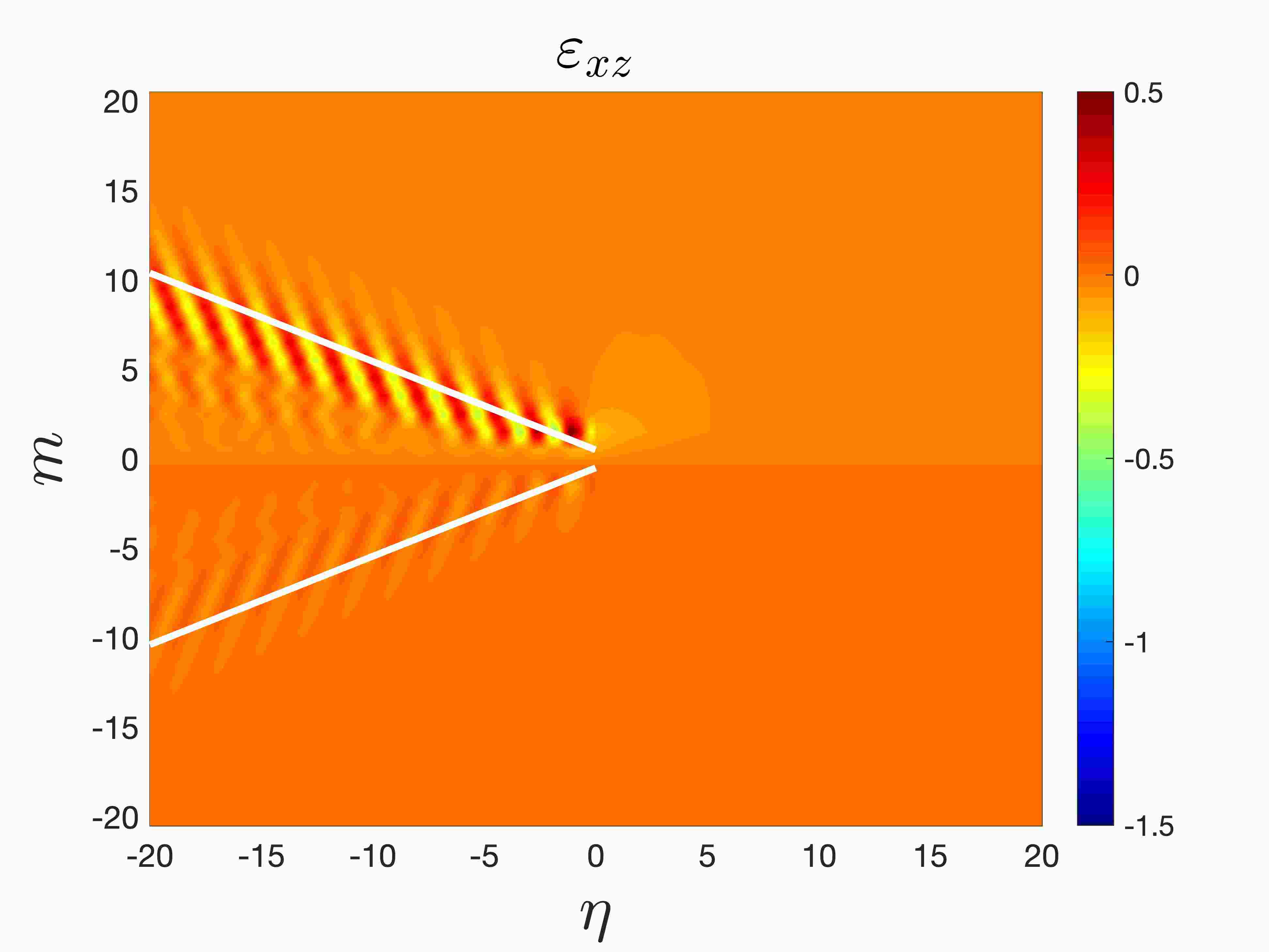}}
\endminipage
\hfill
(e) \minipage{0.3\textwidth}
\center{\includegraphics[width=\linewidth] {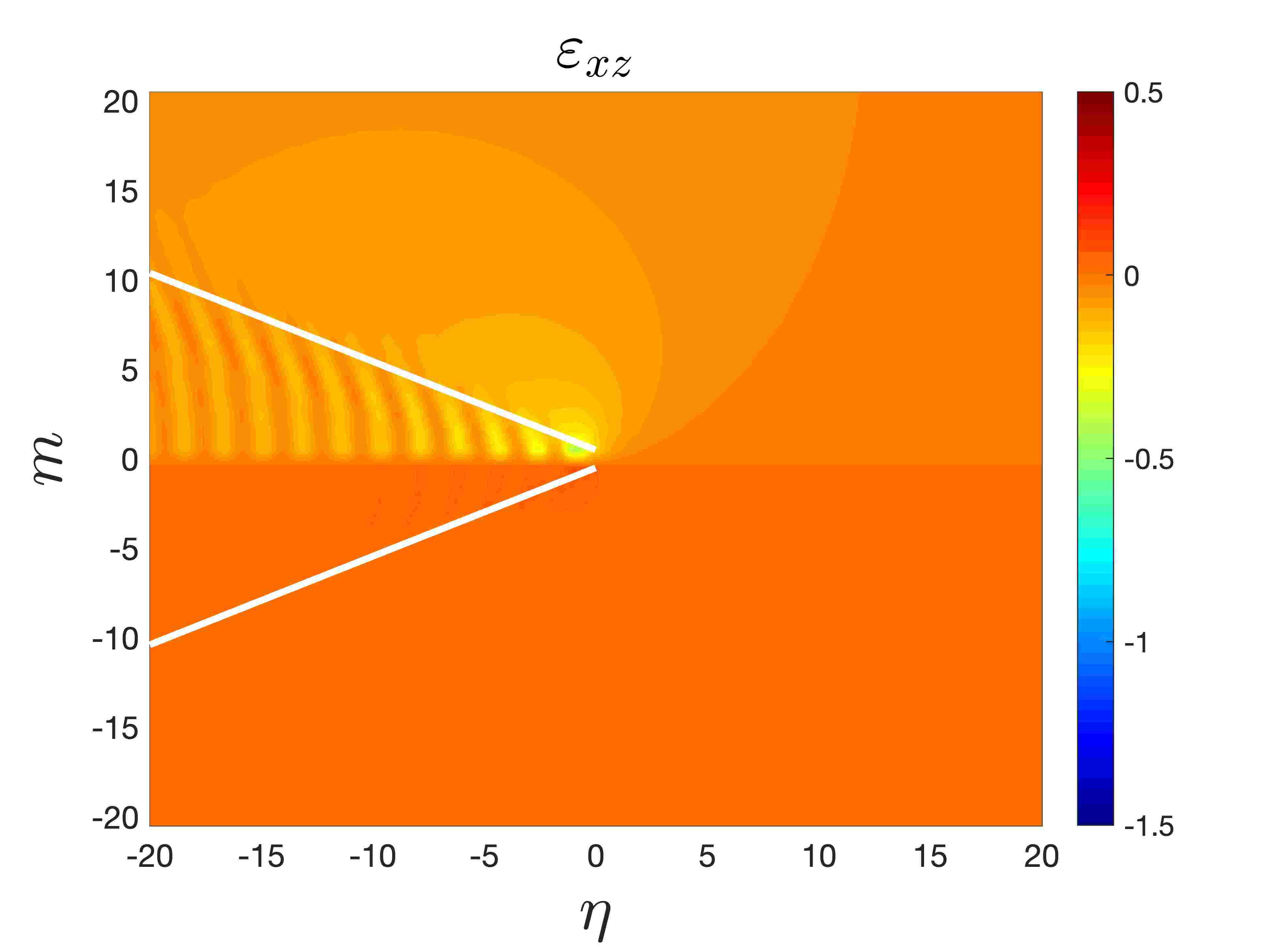}}
\endminipage
\hfill
(f) \minipage{0.3\textwidth}
\center{\includegraphics[width=\linewidth] {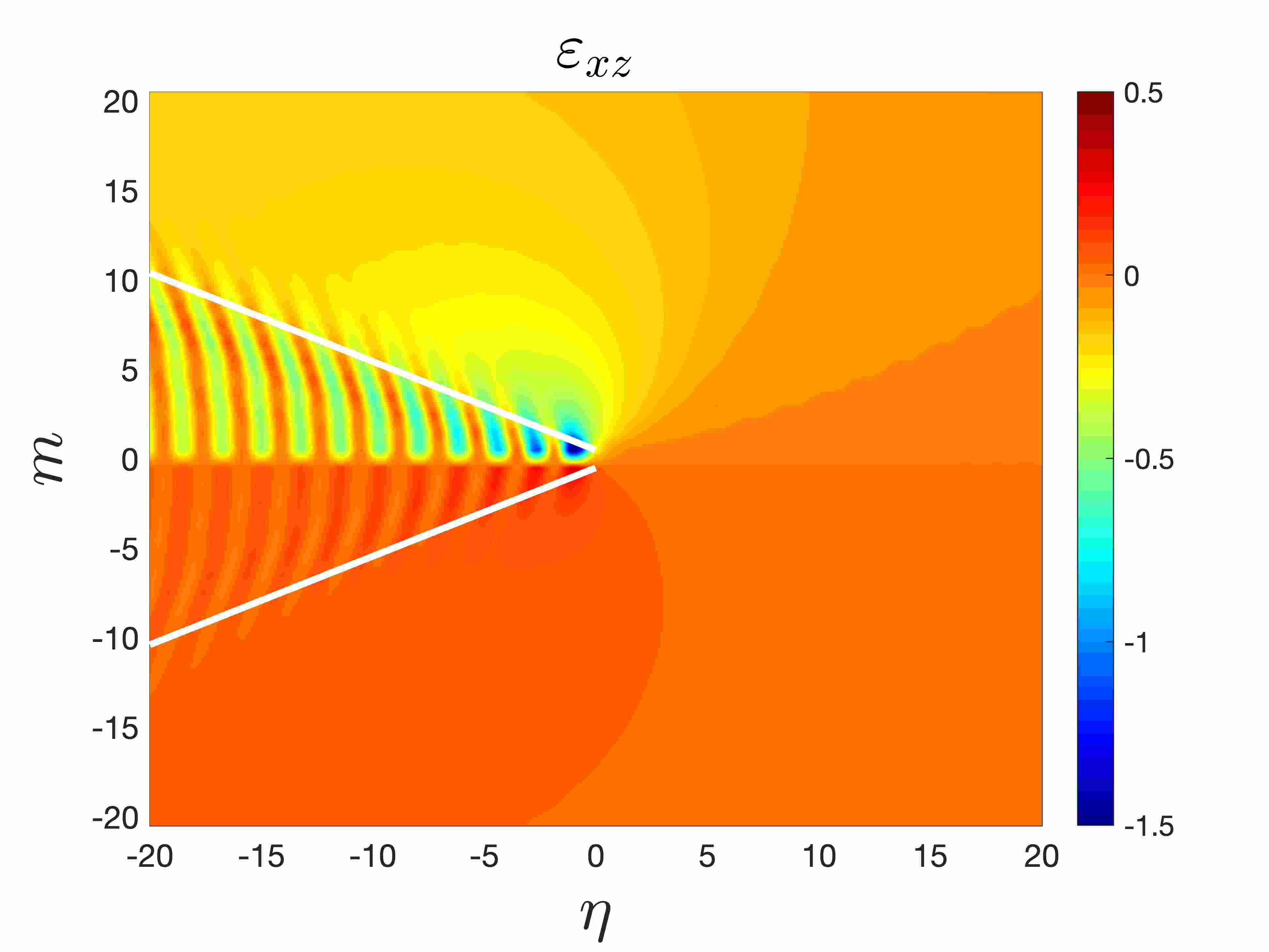}}
\endminipage
\\
(g) \minipage{0.3\textwidth}
\center{\includegraphics[width=\linewidth] {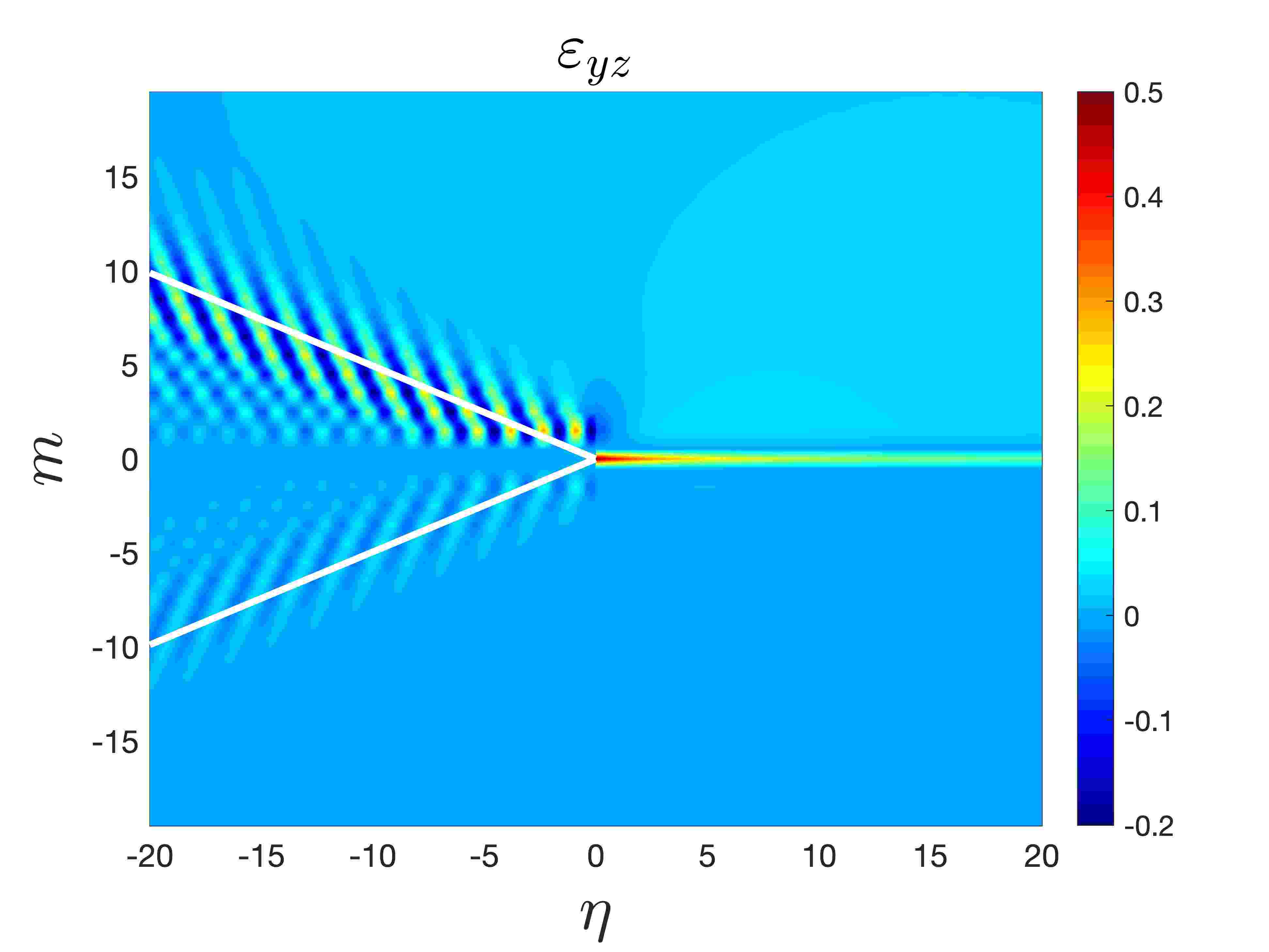}}
\endminipage
\hfill
(h) \minipage{0.3\textwidth}
\center{\includegraphics[width=\linewidth] {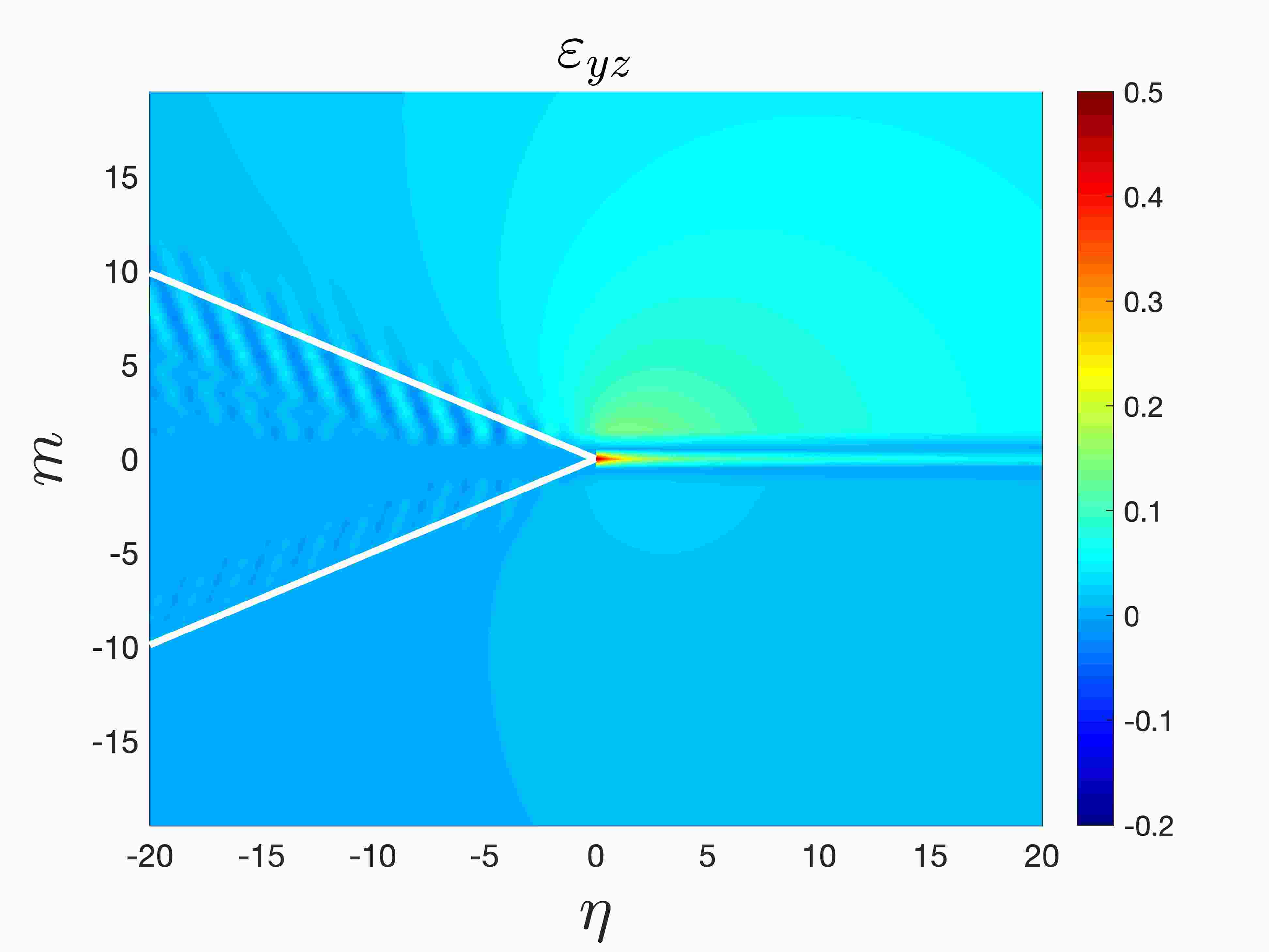}}
\endminipage
\hfill
(i) \minipage{0.3\textwidth}
\center{\includegraphics[width=\linewidth] {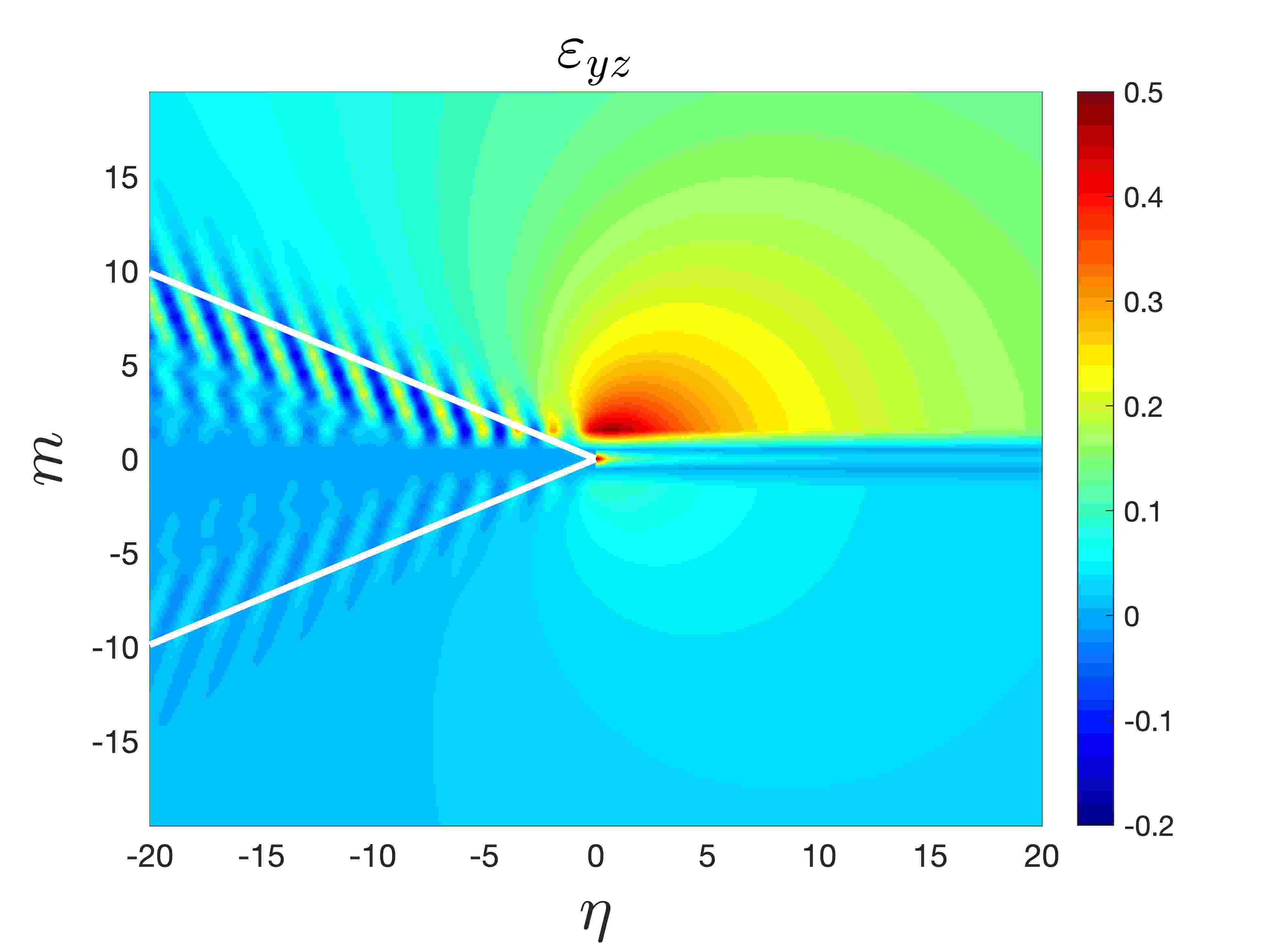}}
\endminipage
\caption[ ]{Displacements and strains in the lattice with parameters $\beta=1/5$, $\gamma=\alpha_2=5$, $\alpha_1=1$ and  $v/v_c=0.55$ computed using (\ref{eq:SolutionLatticeInverseFourier_Layersa}) and (\ref{eq:SolutionLatticeInverseFourier_Layers}).
The panels (a)-(c) show the  displacement field of the lattice with a crack. The panels (d)--(f) and (g)--(i) illustrate the associated strain fields $\varepsilon_{xz}$ and $\varepsilon_{yz}$, respectively, computed with  (\ref{strainsexz1})--(\ref{strainseyz}). 
Here, the results correspond to  $\mu=1/5$, $\mu=1$ and $\mu=5$ shown in the panels contained in the first, second and third columns, respectively, of the figure. The wave radiation rays are shown in (d)--(i) by white lines and are based on (\ref{wave_rays}).}
\label{fig:Plots_5}
\end{figure}

\begin{figure}[htbp]
(a) \minipage{0.3\textwidth}
\center{\includegraphics[width=\linewidth] {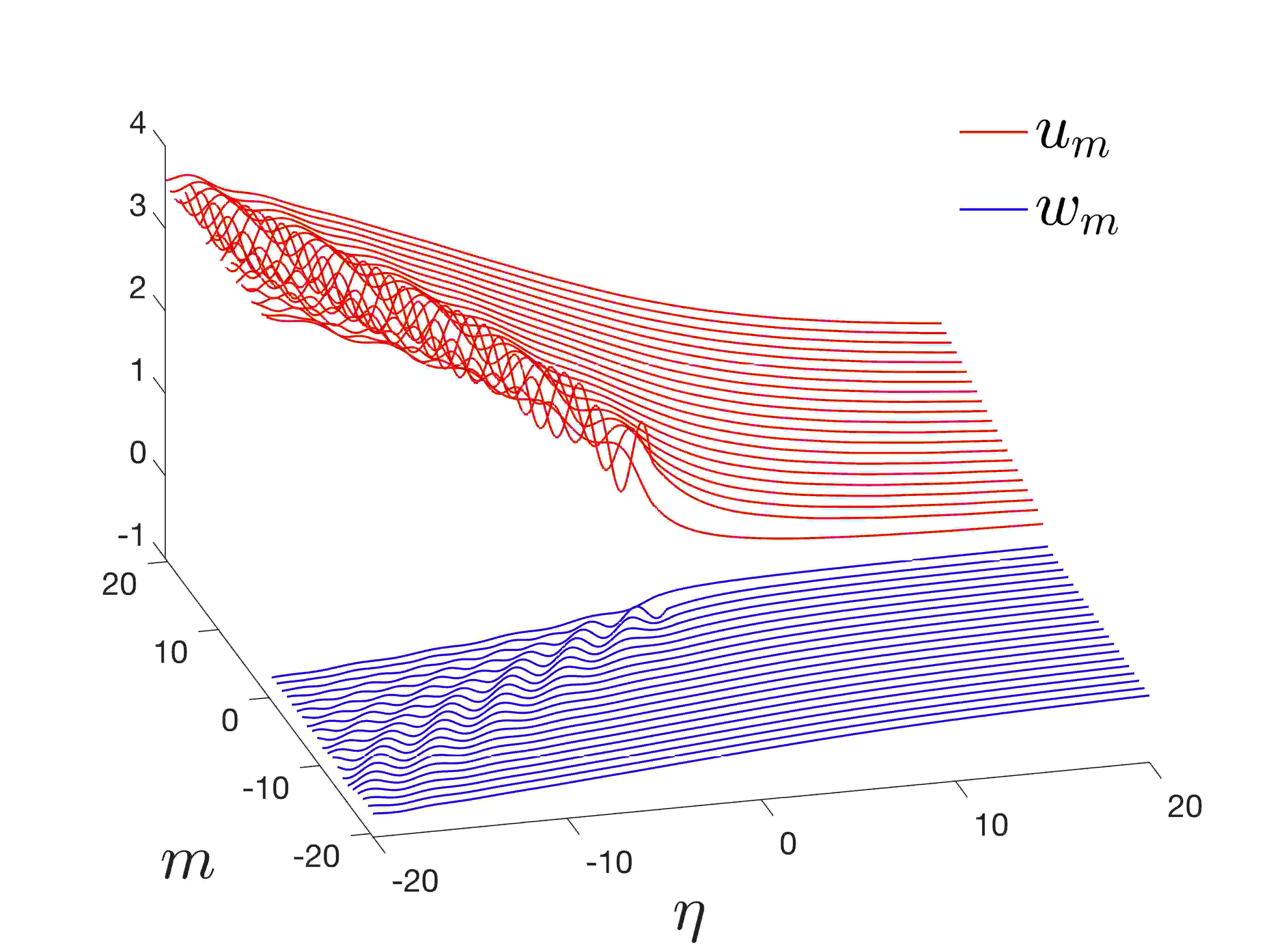}}
\endminipage
\hfill
(b) \minipage{0.3\textwidth}
\center{\includegraphics[width=\linewidth] {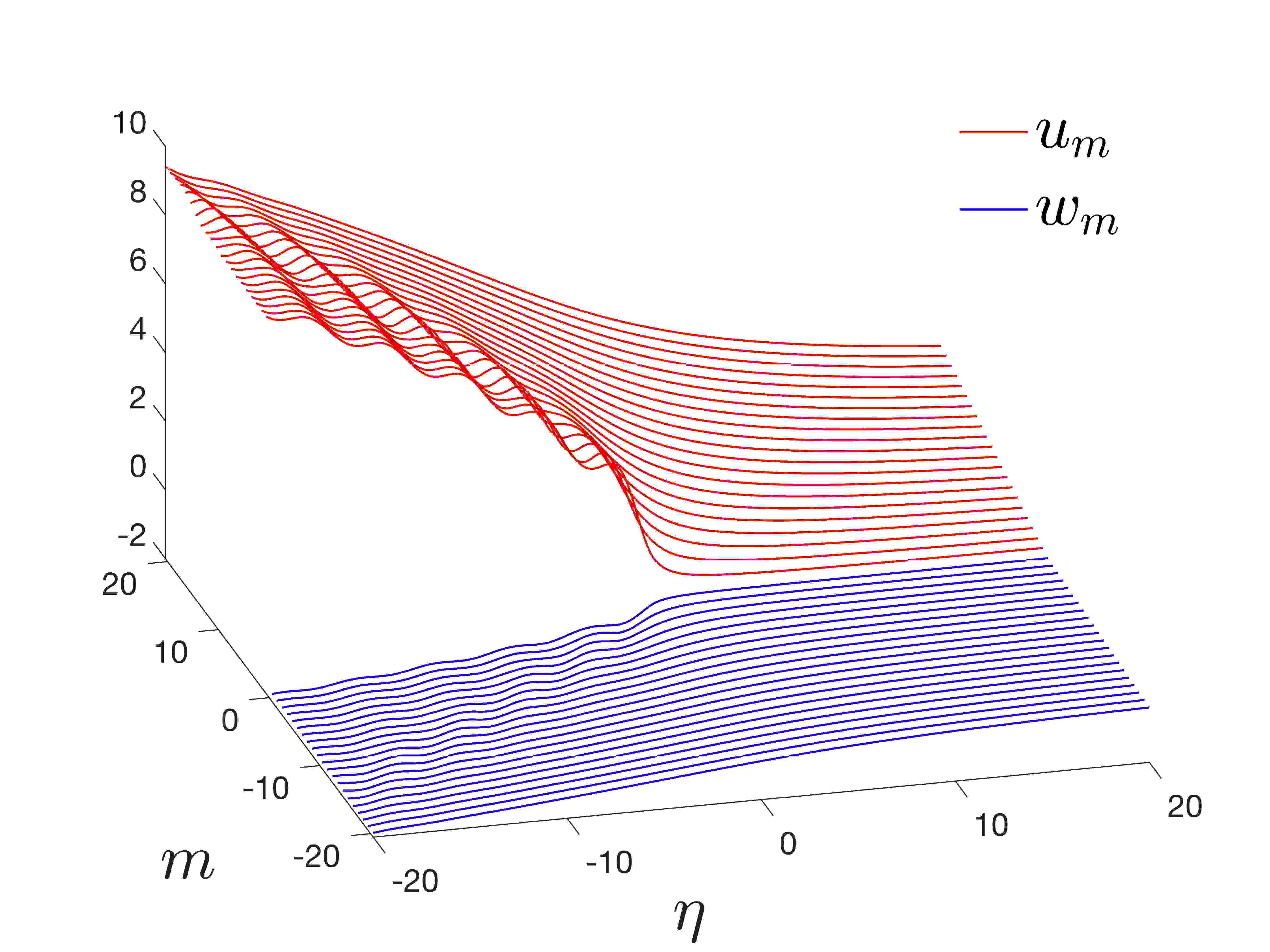}}
\endminipage
\hfill
(c) \minipage{0.3\textwidth}
\center{\includegraphics[width=\linewidth] {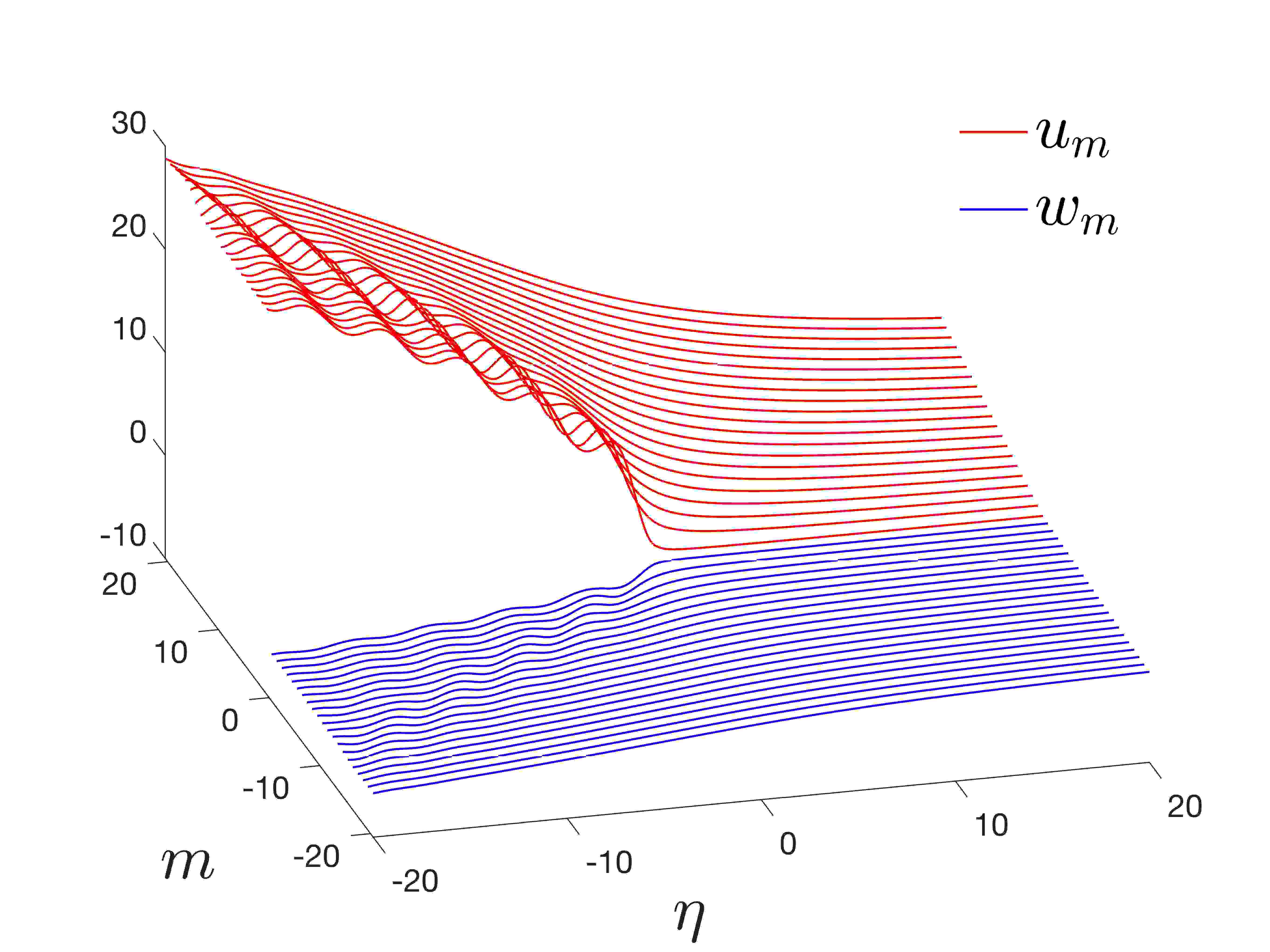}}
\endminipage 
\\
(d) \minipage{0.3\textwidth}
\center{\includegraphics[width=\linewidth] {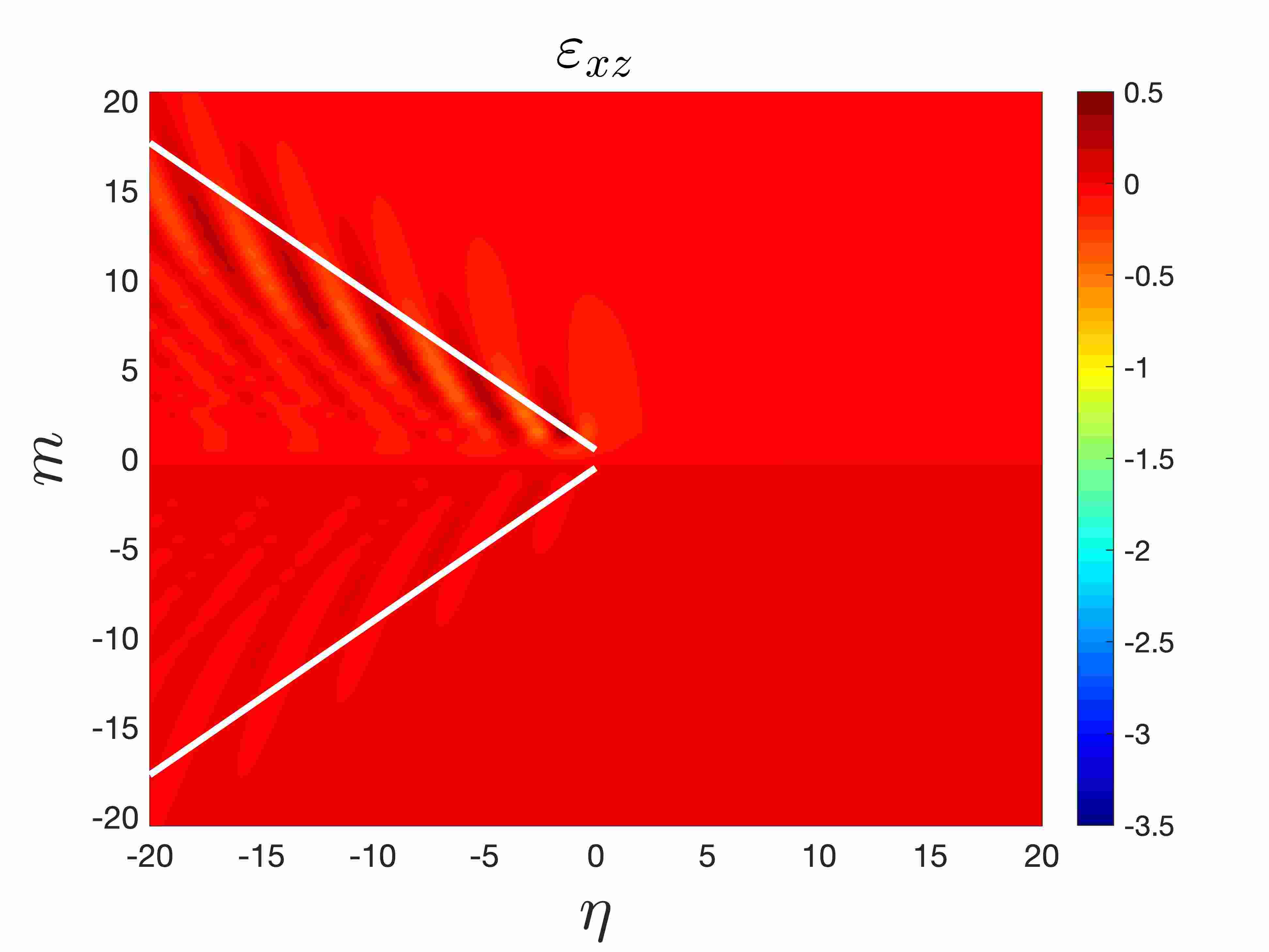}}
\endminipage
\hfill
(e) \minipage{0.3\textwidth}
\center{\includegraphics[width=\linewidth] {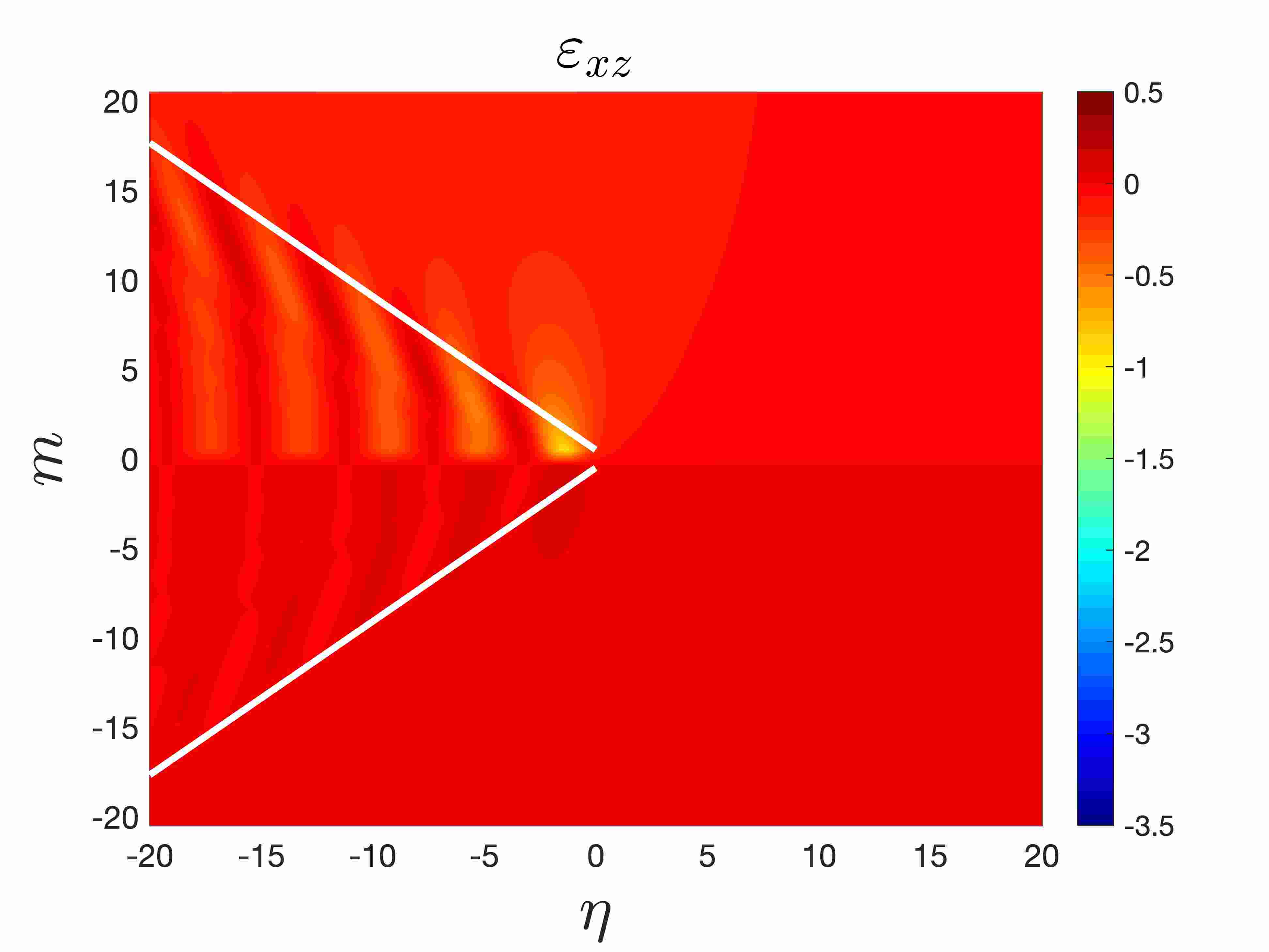}}
\endminipage
\hfill
(f) \minipage{0.3\textwidth}
\center{\includegraphics[width=\linewidth] {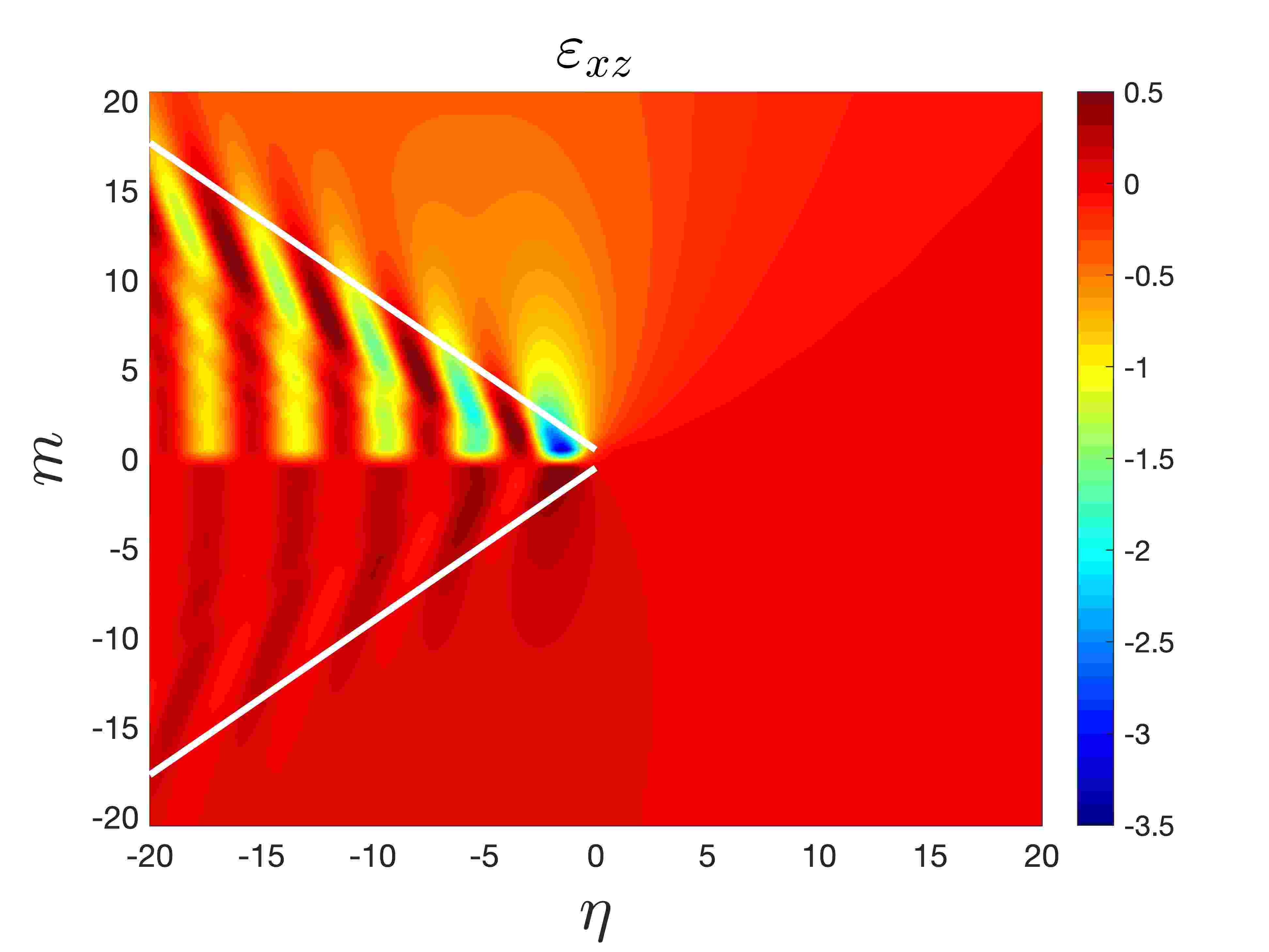}}
\endminipage
\\
(g) \minipage{0.3\textwidth}
\center{\includegraphics[width=\linewidth] {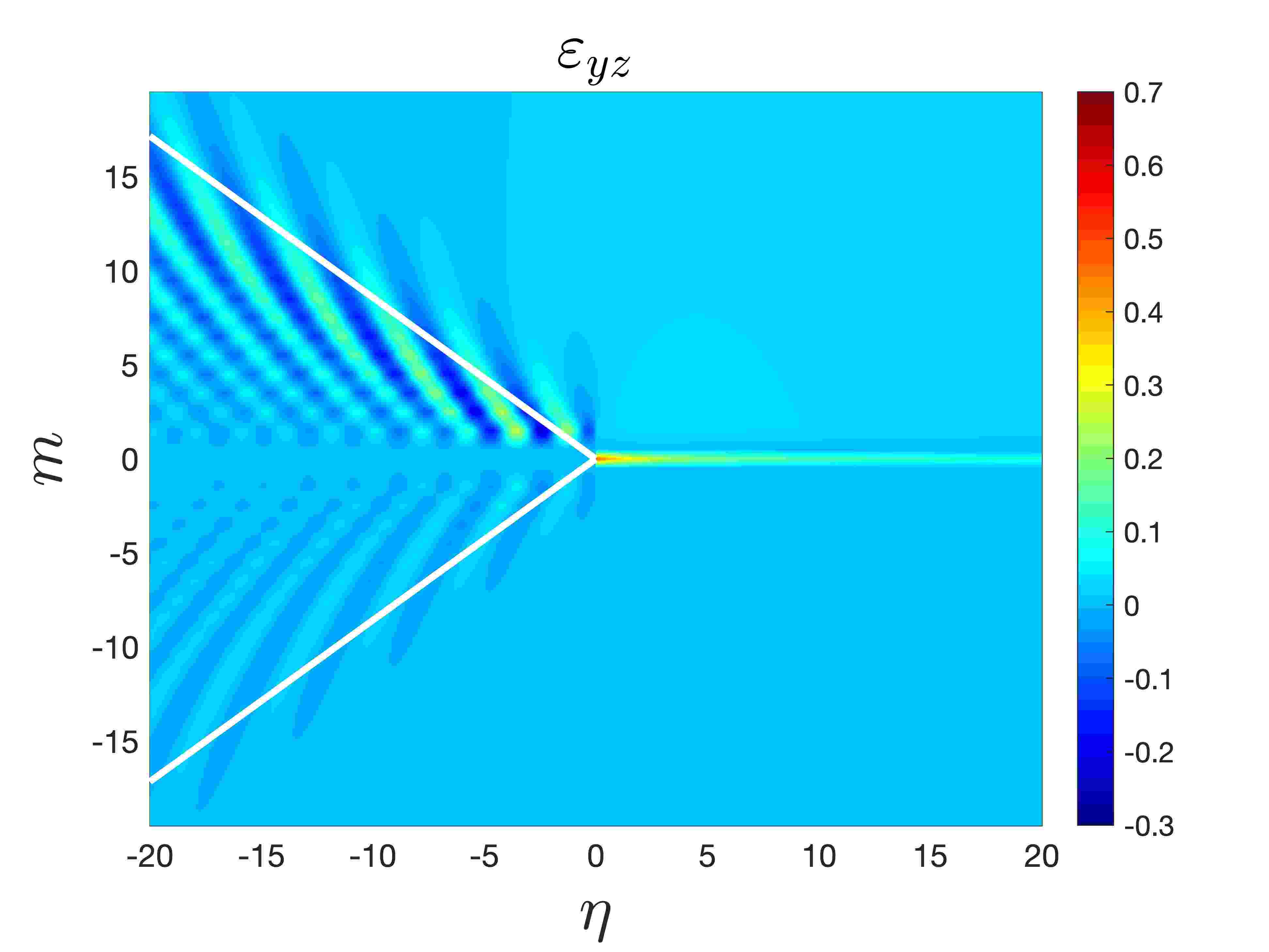}}
\endminipage
\hfill
(h) \minipage{0.3\textwidth}
\center{\includegraphics[width=\linewidth] {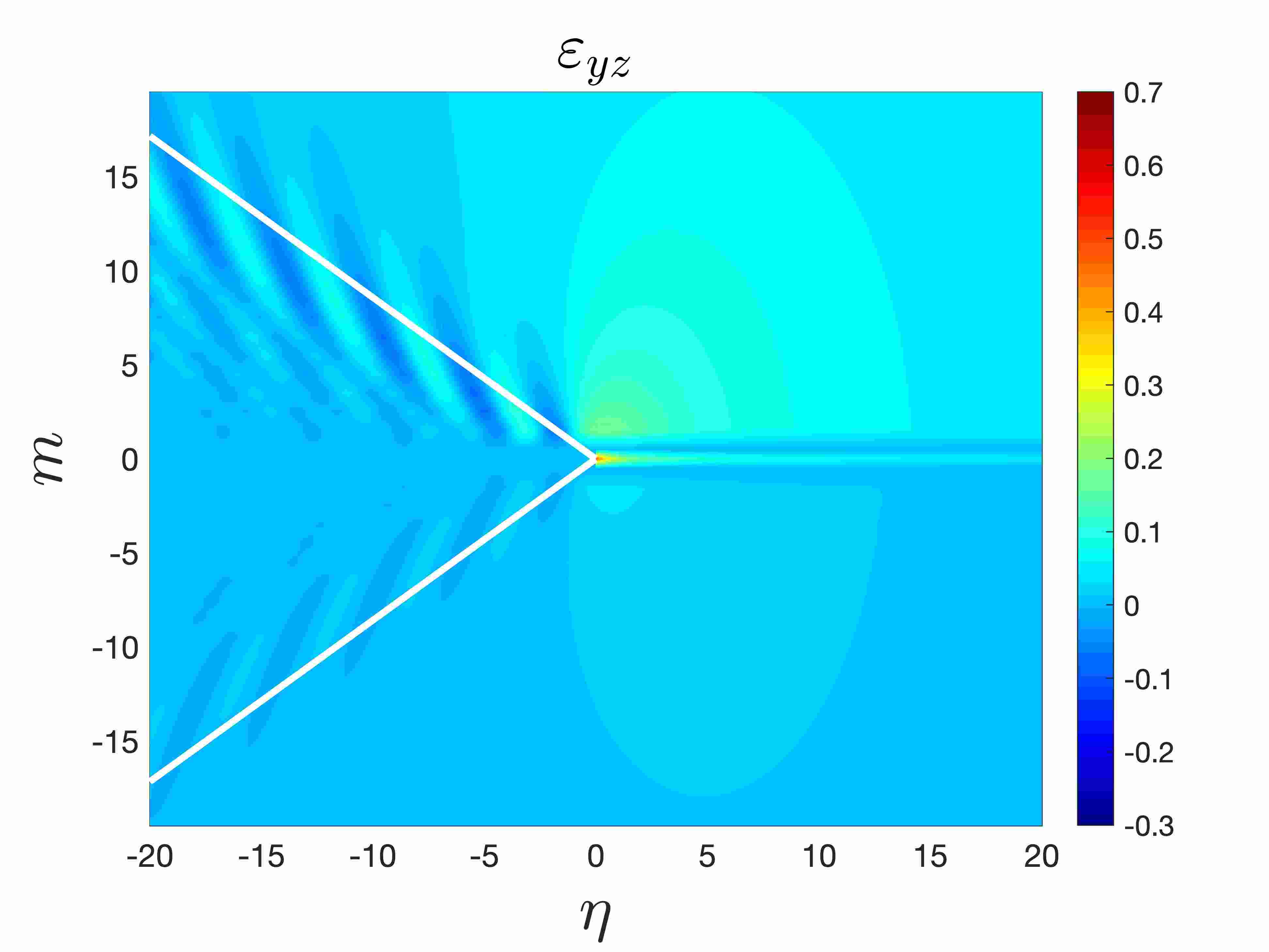}}
\endminipage
\hfill
(i) \minipage{0.3\textwidth}
\center{\includegraphics[width=\linewidth] {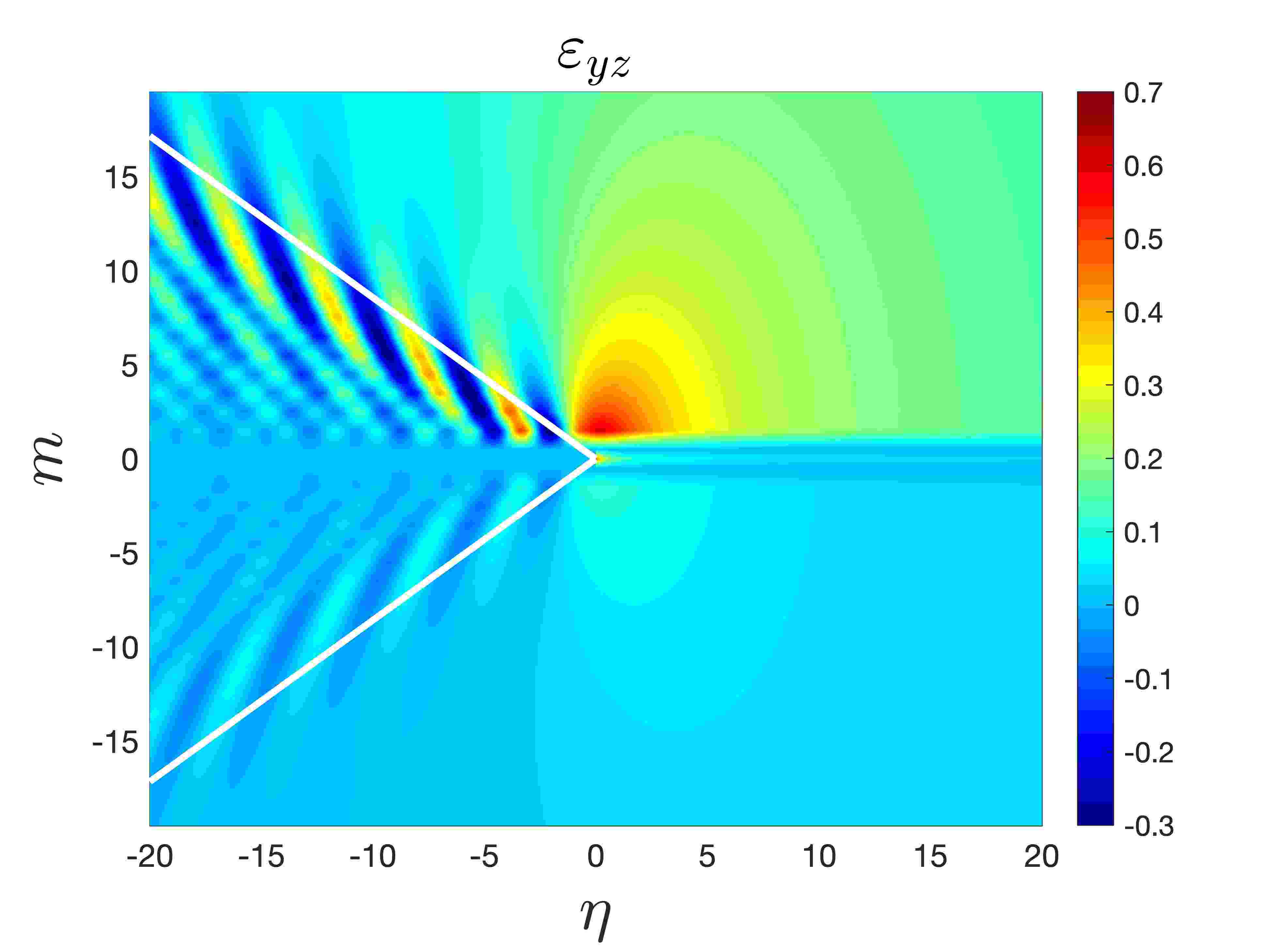}}
\endminipage
\caption[ ]{Displacements and strains in the lattice with parameters $\beta=1/5$, $\gamma=\alpha_2=5$, $\alpha_1=1$ and  $v/v_c=0.9$ computed using (\ref{eq:SolutionLatticeInverseFourier_Layersa}) and  (\ref{eq:SolutionLatticeInverseFourier_Layers}).
The panels (a)--(c) show the  displacement field of the lattice with a crack. The panels (d)--(f) and (g)--(i) illustrate the associated strain fields $\varepsilon_{xz}$ and $\varepsilon_{yz}$, respectively, computed with  (\ref{strainsexz1})--(\ref{strainseyz}). 
Here, the results correspond to  $\mu=1/5$, $\mu=1$ and $\mu=5$ shown in the panels contained in the first, second and third columns, respectively, of the figure. The wave radiation rays are shown in (d)--(i) by white lines and are based on (\ref{wave_rays}).}
\label{fig:Plots_6}
\end{figure}

 
Figures \ref{fig:Plots_5}(a)--(c) show the lattice displacements for $v/v_c=0.55$. There is a substantial difference in the displacements between the upper and lower  lattices,  as expected, that results from their dissimilarity. Here, the lighter and softer upper lattice deforms more than the heavier, stiffer lower lattice during the failure process. There is also evidence that soft interfaces lead to large deformations, induced by vibration, behind the crack predicted by the solution. Such an example is shown in  Figure \ref{fig:Plots_5}(a) and this solution is admissible.

The difference in the behaviour of the upper and lower lattices becomes more apparent when one considers the distribution of strains $\varepsilon_{xz}$, as shown in Figures \ref{fig:Plots_5}(d)--(f). There, as in Section \ref{sec7.1}, vibrations appearing in the wake of the crack tip in the upper lattice are concentrated inside the  region enclosed by the crack and the wave radiation ray. The increase in the interface stiffness cause prominent oscillations in the strains that are almost perpendicular to  crack faces in the vicinity of $m=0$ far behind the crack tip.
These local effects may be attributed 
the behaviour of the acoustic modes outlined in Section \ref{Disp_prop}, where the lattice with free boundary is considered and waves travelling parallel to the free boundary were identified. In the failure problem considered, those effects are coupled with the those concerning waves propagating along the internal boundary of the dissimilar lattice. Thus, in the problem of the propagating crack, 
travelling waves must decay to zero in the direction perpendicular to the lattice as $O(1/m^{1/2})$.

Figures \ref{fig:Plots_5}(g)--(i) show the the strains $\varepsilon_{yz}$, where the effects brought by the far-field waves propagating parallel to the crack in the lower lattice are suppressed. However, these strains reveal the preferential directions taken by the vibrations propagating along the wave radiation rays and whose intensity increases with increase of the interface stiffness.

Note that  Figures \ref{fig:Plots_5}(g)--(i) show that due to the dissimilarity of the lattices and their responses, the strain $\varepsilon_{yz}$  is also concentrated at the crack tip and ahead of the  crack. 
For soft interfaces this effect is the most prominent as the bulk lattice effects are quite small in comparison. As the stiffness of the interface is increased, effects such as micro-structural vibrations and localised deformations within in the bulk lattices become more noticeable. In particular Figure \ref{fig:Plots_5}(i) shows the emergence of a region local to  the crack tip possessing comparable deformations.  {In fact, our definition of admissibility allows for such a phenomenon.  Indeed, if one checks the admissibility conditions (\ref{eq:FractureCondition_psi}) along the interface, then they will be satisfied. If simultaneously, the toughness of this upper lattice is much higher than the interface toughness then the fracture will occur along the interface only.
However, if the toughness of the upper lattice is comparable to the interface such a scenario may lead to the deflection of the crack into the softer lattice. }

In Figure \ref{fig:Plots_6}, we consider the dissimilar lattice behaviour when a crack propagates with a higher speed. As in Sections \ref{sec7.1}, increasing to the crack speed leads to more significant lattice deformations and a wider crack opening, as evidenced by lattice displacements Figures \ref{fig:Plots_6}(a)--(c). Similarly to  Section \ref{sec7.1}, the increase of the crack speeds leads once again to wave radiation rays, symmetric about the crack path and separated by a larger angle than in Figures \ref{fig:Plots_5}. In comparison with that case (see Figures \ref{fig:Plots_5}(e), (f) and \ref{fig:Plots_6}(e), (f)),  this results in a weaker localisation of the field in the upper lattice. The effect is also observed for the strains $\varepsilon_{yz}$ in Figures \ref{fig:Plots_6}(h) and (i). For the lower lattice,  we again encounter vibrations with a wavefront perpendicular to the crack and  that decay slowly in the same direction. In addition, the vibrations  induced  by the  crack propagating at high speed have a longer wavelength than those identified in Figure \ref{fig:Plots_5}.

\subsection{Essentially different dissimilar lattice system with an interface}\label{sec7.3}

Next, we investigate the case when the upper and lower lattices possess different wave speeds and are either perfectly connected or connected through an imperfect  interface. 
\begin{figure}[htbp]
\center{\includegraphics[width=0.55\linewidth]{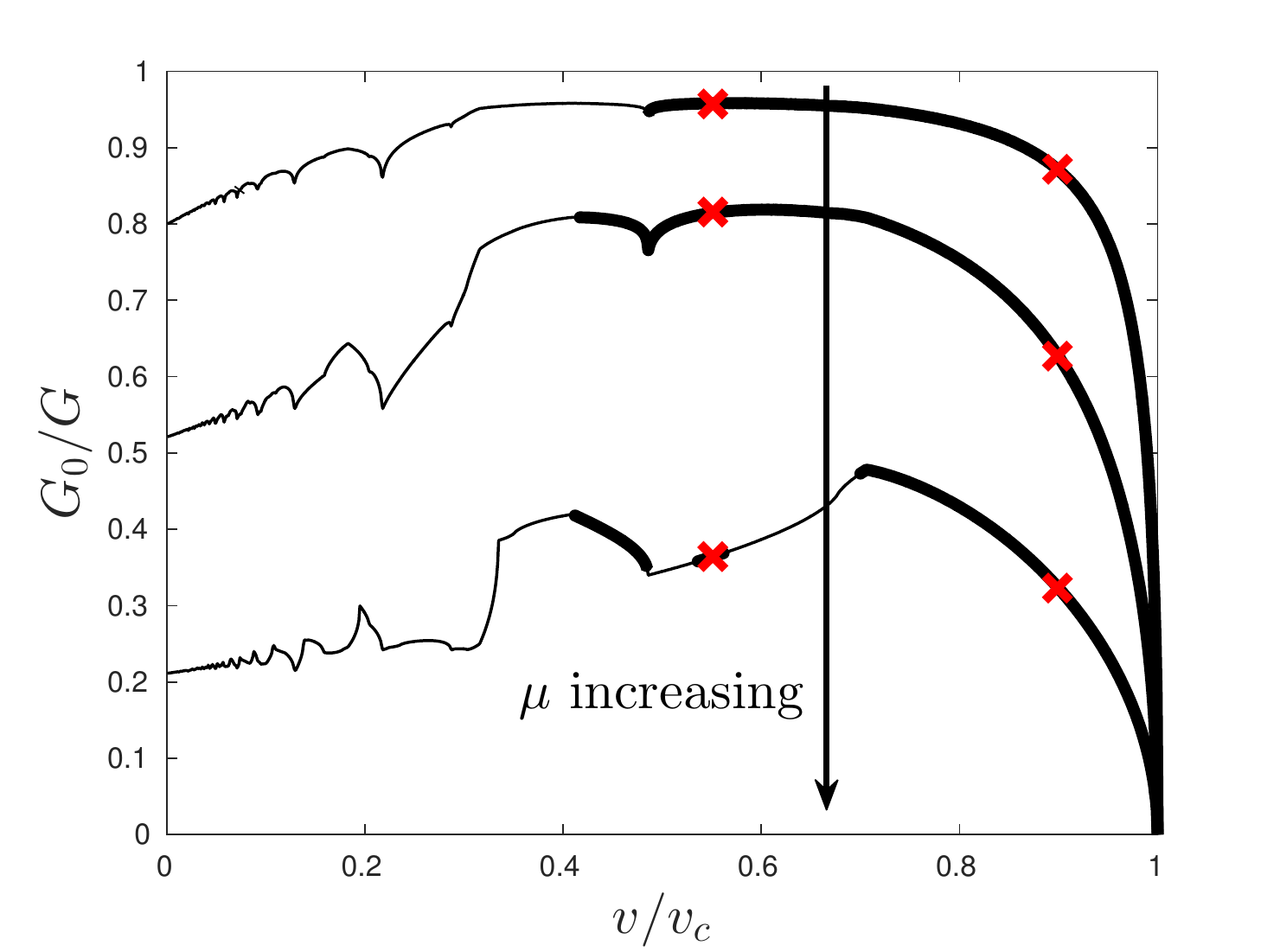} }
\caption[ ]{Dependency of energy release rate ratio $G_0/G$ on ratio $v/v_c$ based in (\ref{eq:ERR_R}).
The computations performed for  $\beta=1=\alpha_1=1$, $\gamma=\alpha_2=5$ and $\mu=1/5, 1,$ and 5.
In addition, we indicate the speeds where admissible regimes are realised in accordance with (\ref{eq:FractureCondition_psi}). Here,  regimes that are not admissible correspond to normal lines.}
\label{fig:ERR_same_speeds2aa}
\end{figure}

The computations are now performed for a lattice system characterised by $\beta=\alpha_1=1$ and $\gamma=\alpha_2=5$. We again choose the interfacial bond stiffness to be $\mu=1/5$, 1 and 5. Here, the cases $\mu=1$ and $\mu=5$ represent the  perfect joining of a soft and stiff lattice at $m=1$ and $m=-1$, respectively, (see Figure \ref{fig:Infinite Lattice}).

We first consider the behaviour of ratio $G_0/G$ for these structures in Figure \ref{fig:ERR_same_speeds2aa}. As a function of $v$, this ratio possesses similar properties as  those described in the previous sections. Here for $\mu \le 1$, by comparison with Figures \ref{fig:ERR_same_speeds1} and \ref{fig:ERR_same_speeds2}, it appears  the lattice dissimilarity leads to a reduction in the size of the speed intervals for admissible propagation regimes satisfying (\ref{eq:FractureCondition_psi}).  
In addition, for the high interfacial bond stiffness the resulting admissible speed regimes are represented by a collection of disjoint intervals, as evidenced by the case $\mu=5$ in Figure \ref{fig:ERR_same_speeds2aa}. The overall size of the union of these intervals  is again less than the size of the continuous interval describing the admissible regimes in Figures \ref{fig:ERR_same_speeds1} and \ref{fig:ERR_same_speeds2} for $\mu=5$. As shown in Figure \ref{fig:ERR_same_speeds2aa} for $\mu=5$, these disjoint intervals may also coincide with the non-smooth behaviour of $G_0/G$. Once more,  each curve in Figure \ref{fig:ERR_same_speeds2aa} is supplied with crosses at two particular speeds corresponding to admissible regimes for failure that are further investigated in Figures \ref{fig:Plots_3} and \ref{fig:Plots_4}.
\begin{figure}[htbp]
(a) \minipage{0.3\textwidth}
\center{\includegraphics[width=\linewidth] {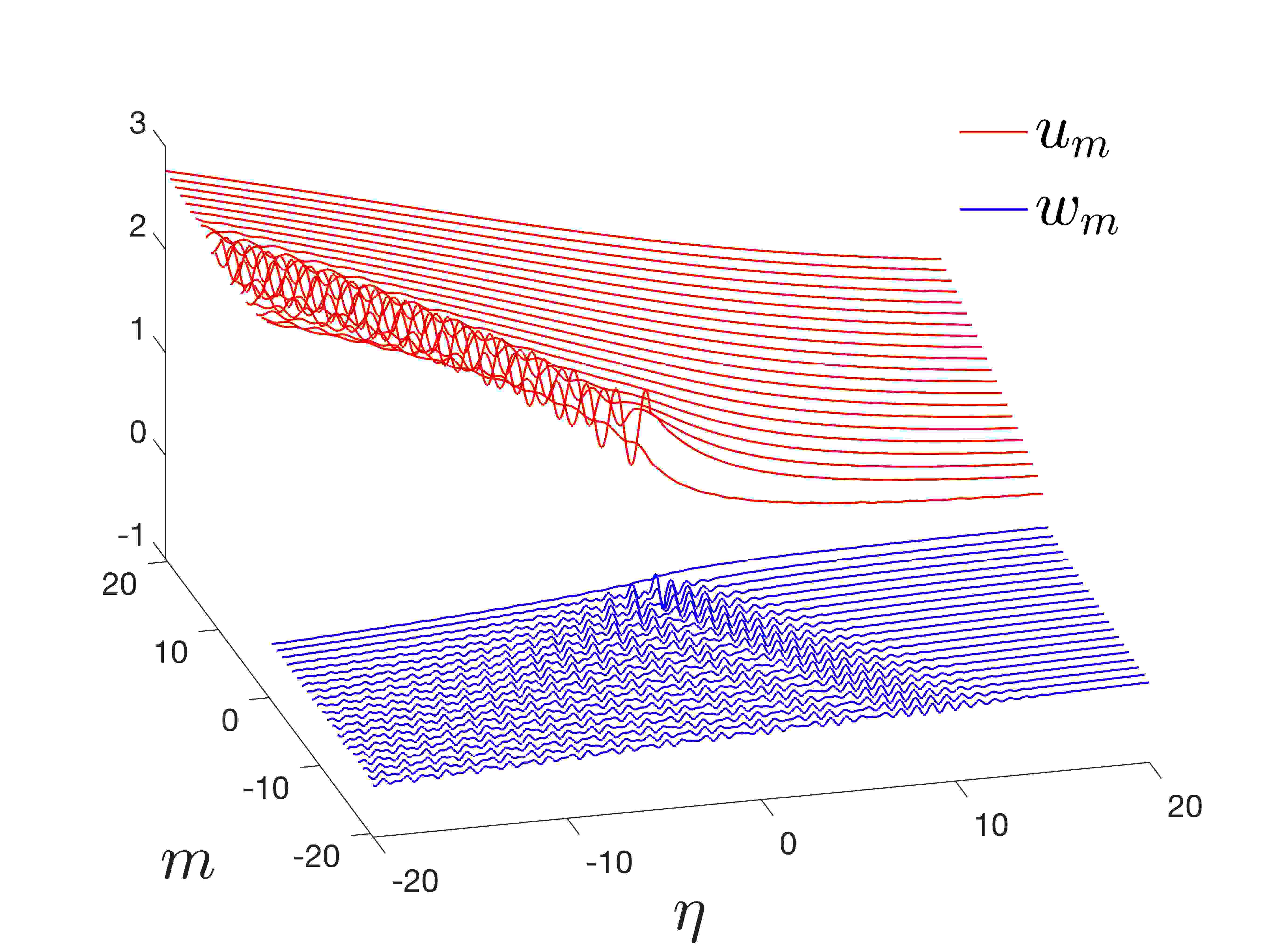}}
\endminipage
\hfill
(b) \minipage{0.3\textwidth}
\center{\includegraphics[width=\linewidth] {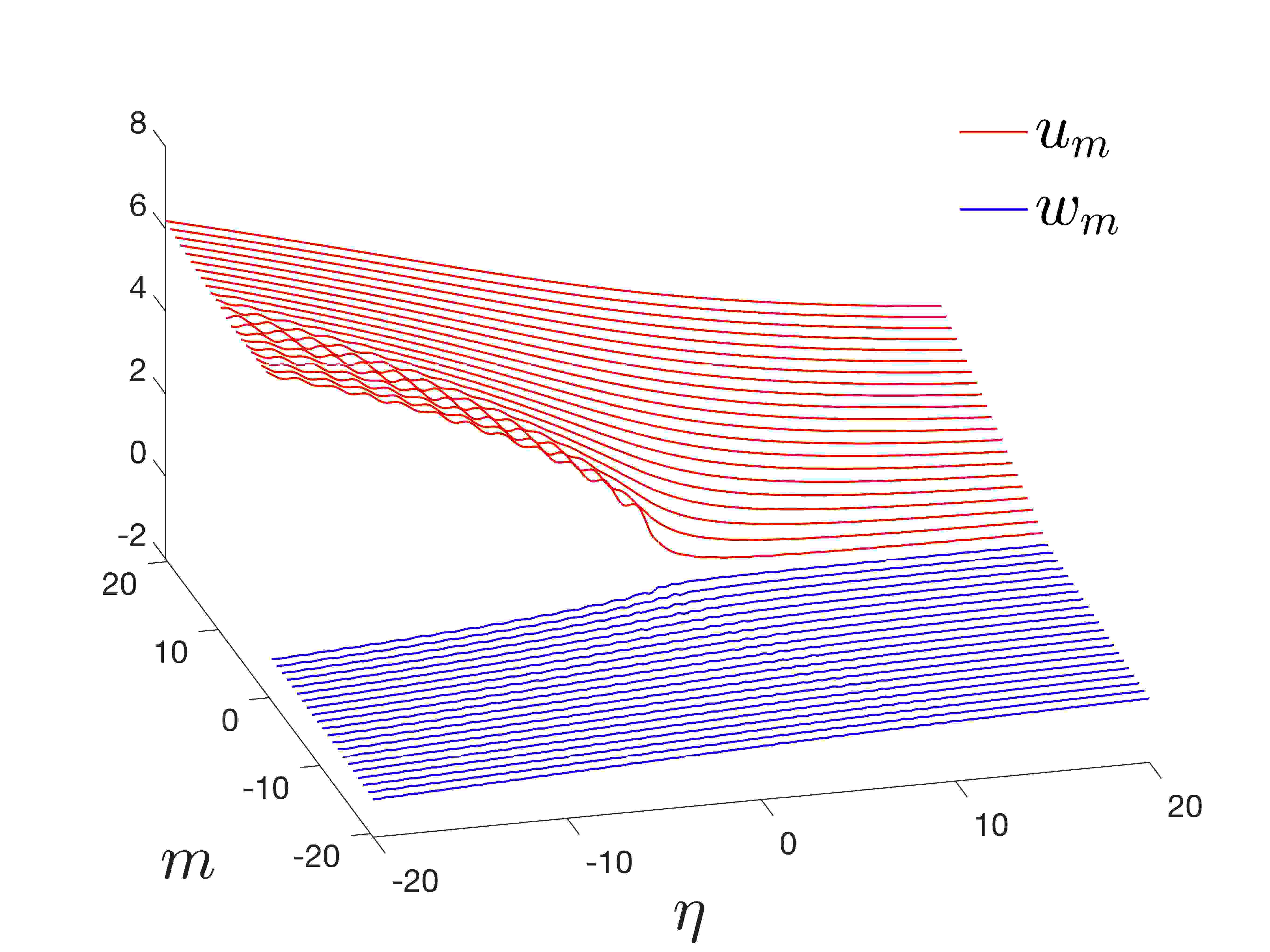}}
\endminipage
\hfill
(c) \minipage{0.3\textwidth}
\center{\includegraphics[width=\linewidth] {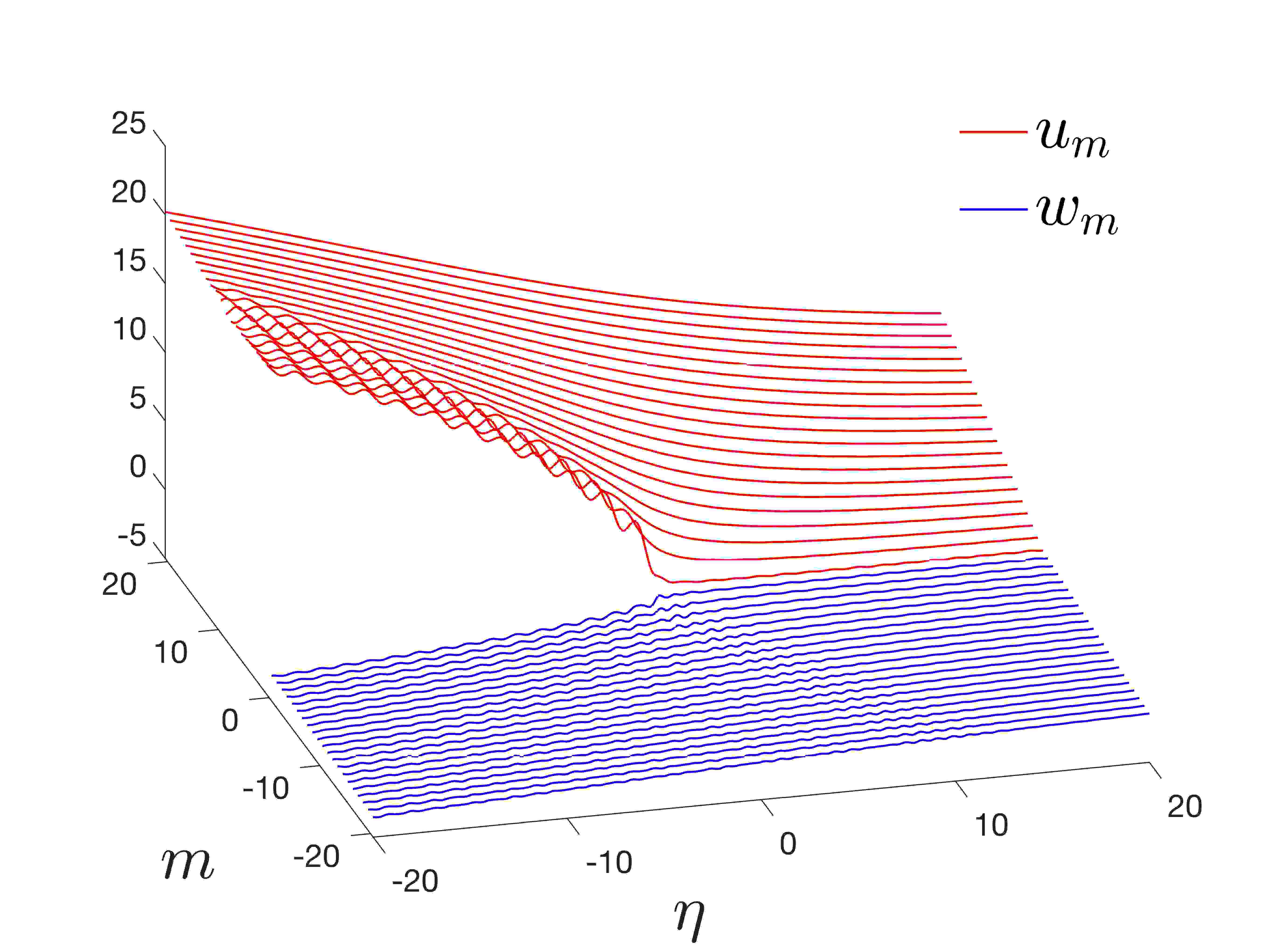}}
\endminipage 
\\
(d) \minipage{0.3\textwidth}
\center{\includegraphics[width=\linewidth] {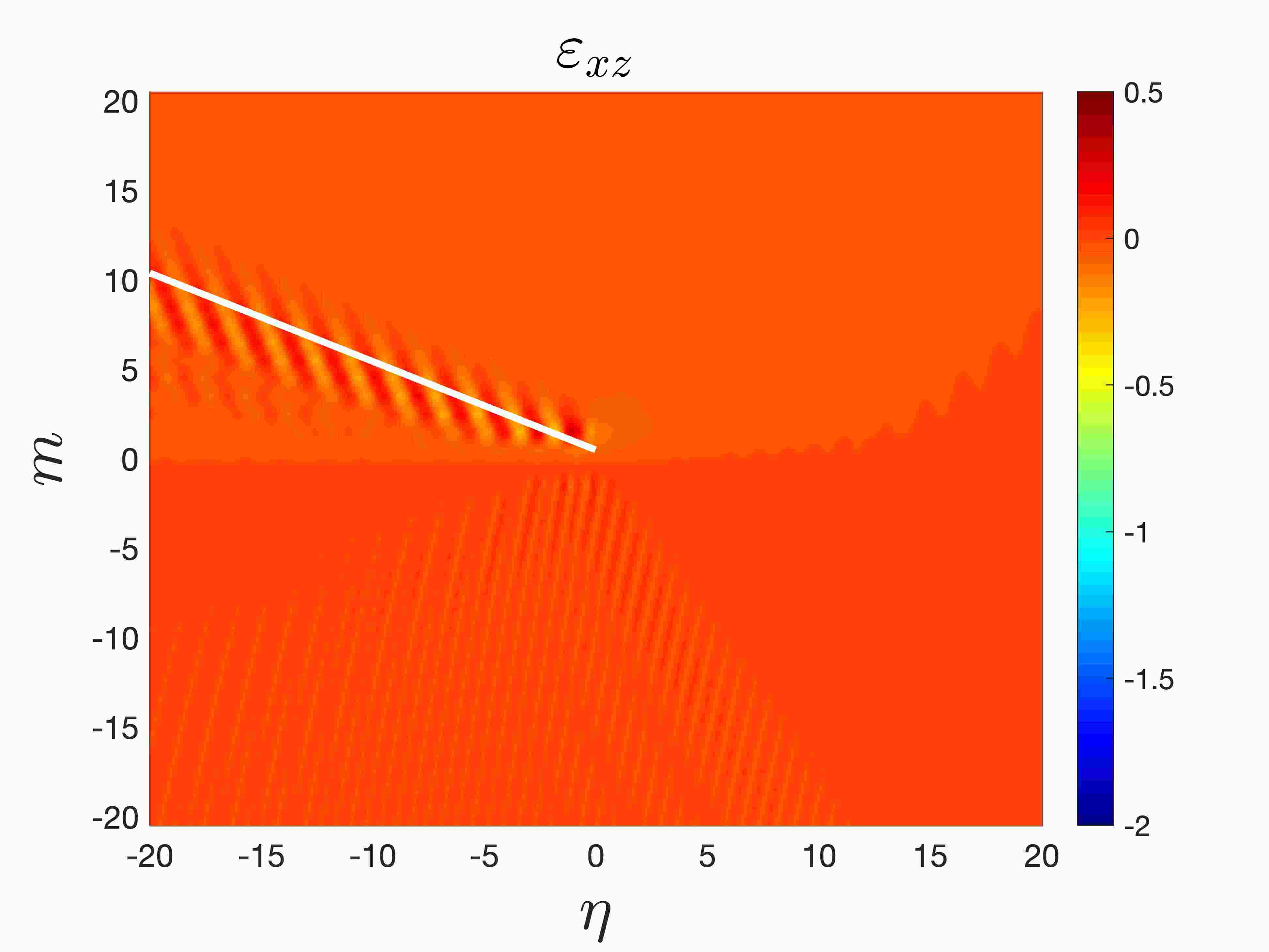}}
\endminipage
\hfill
(e) \minipage{0.3\textwidth}
\center{\includegraphics[width=\linewidth] {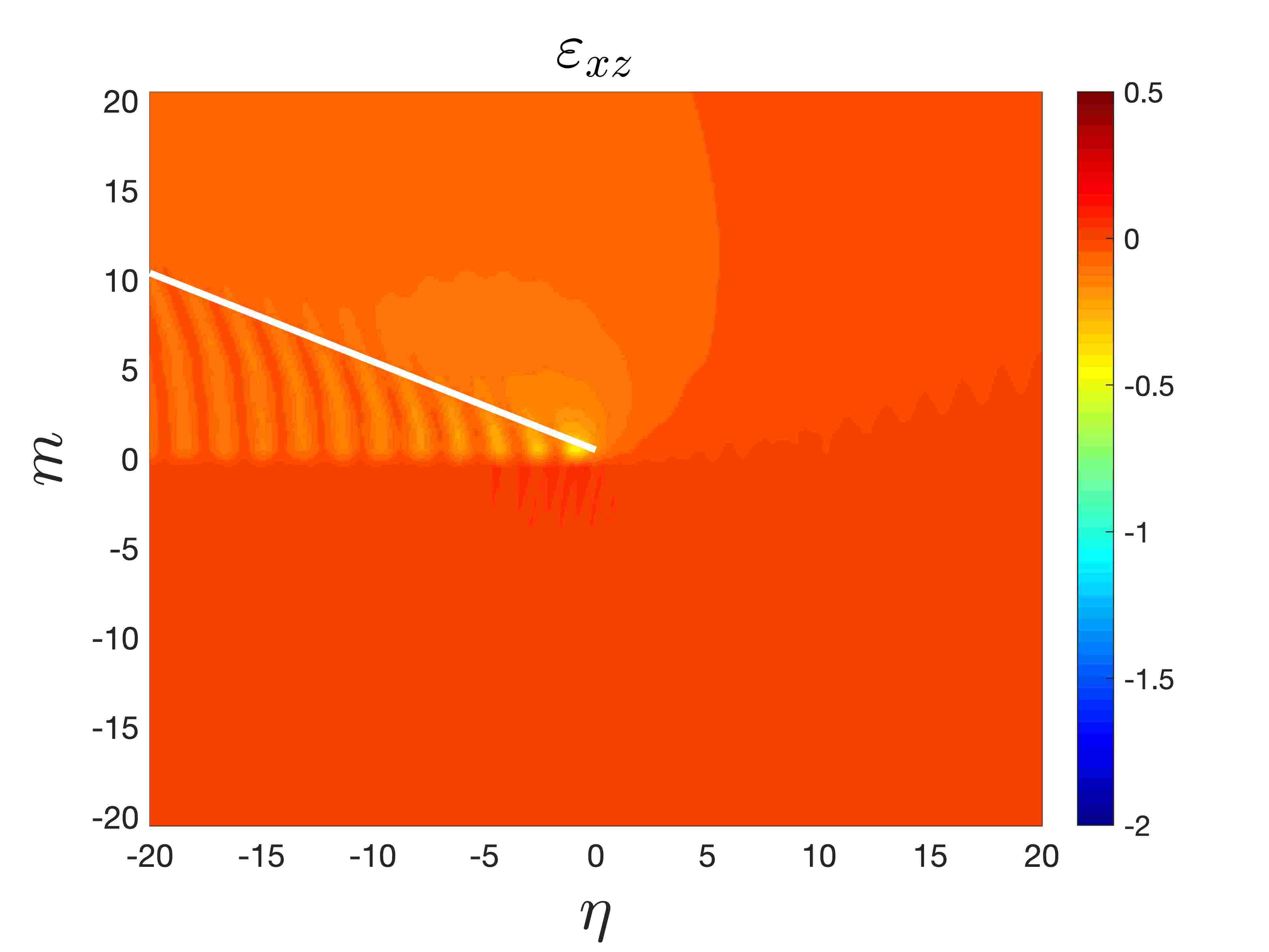}}
\endminipage
\hfill
(f) \minipage{0.3\textwidth}
\center{\includegraphics[width=\linewidth] {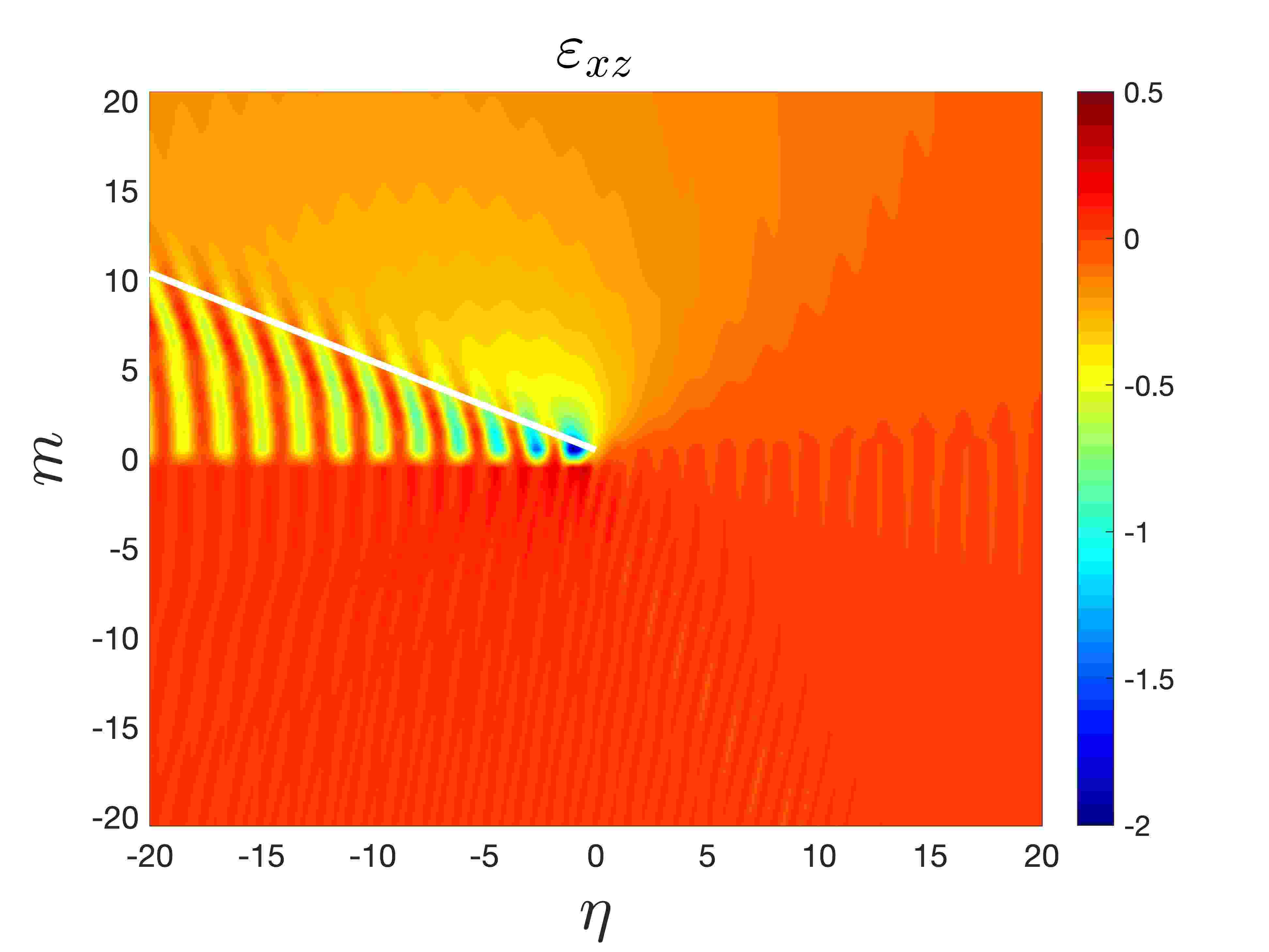}}
\endminipage
\\
(g) \minipage{0.3\textwidth}
\center{\includegraphics[width=\linewidth] {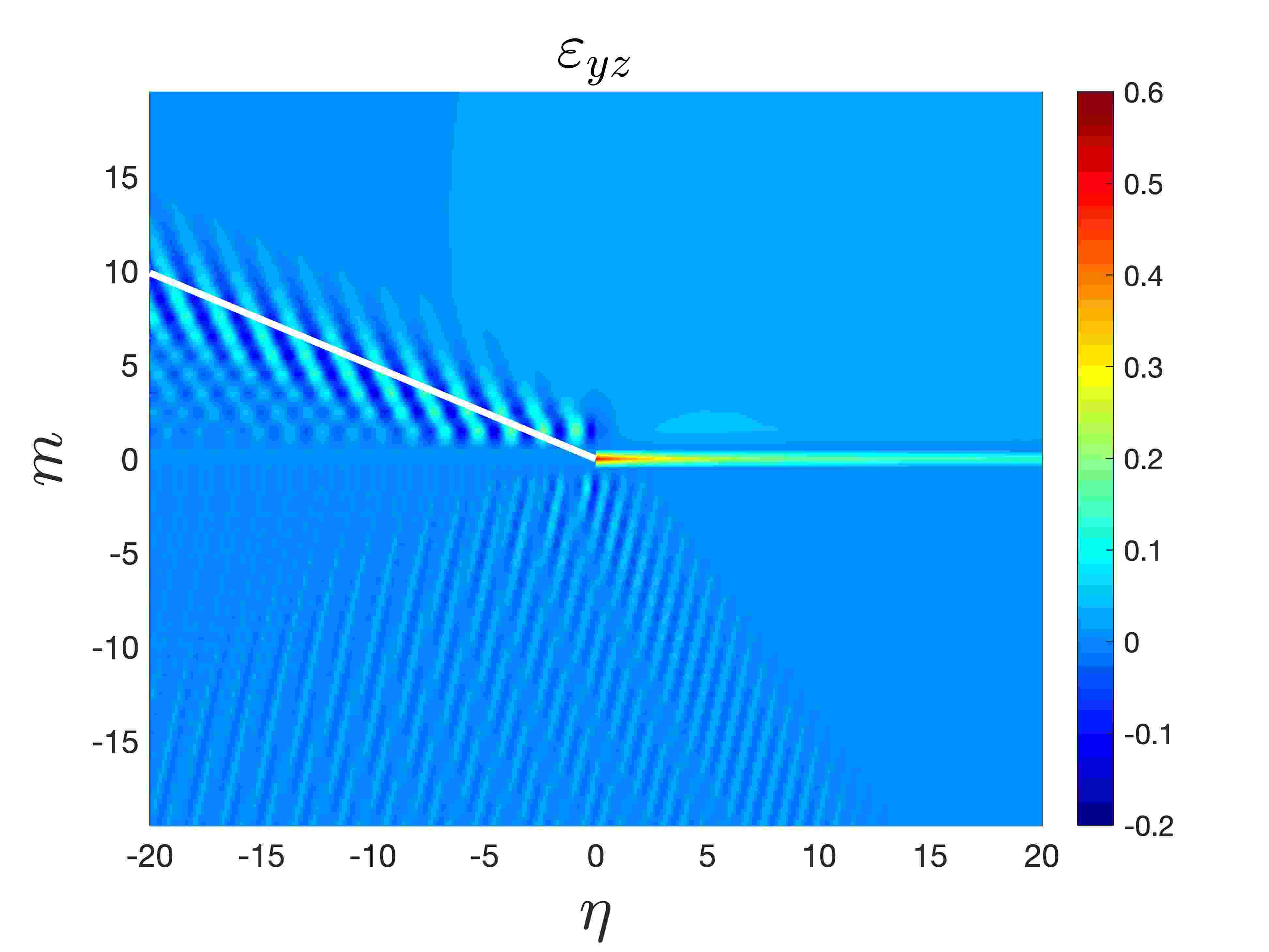}}
\endminipage
\hfill
(h) \minipage{0.3\textwidth}
\center{\includegraphics[width=\linewidth] {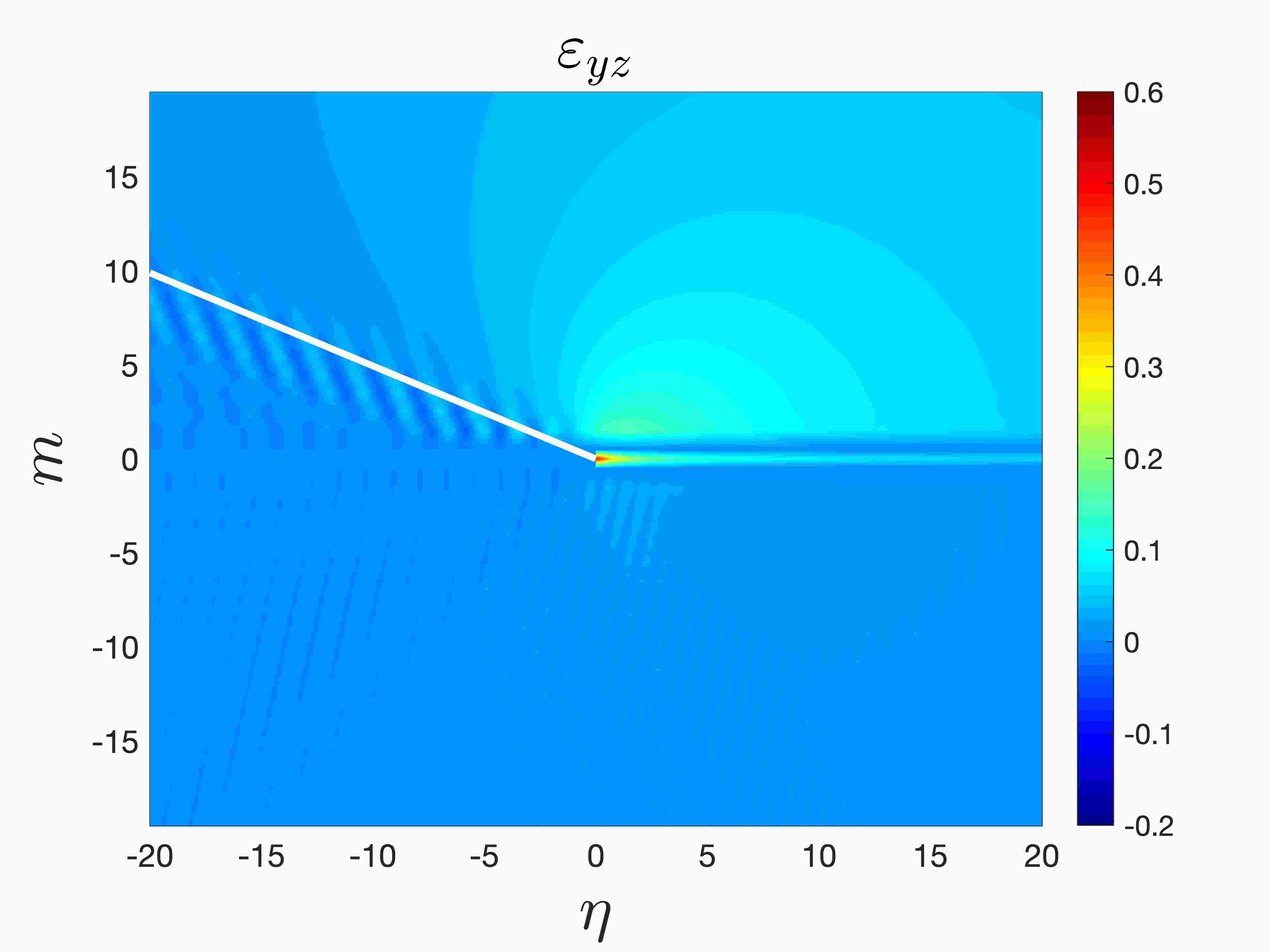}}
\endminipage
\hfill
(i) \minipage{0.3\textwidth}
\center{\includegraphics[width=\linewidth] {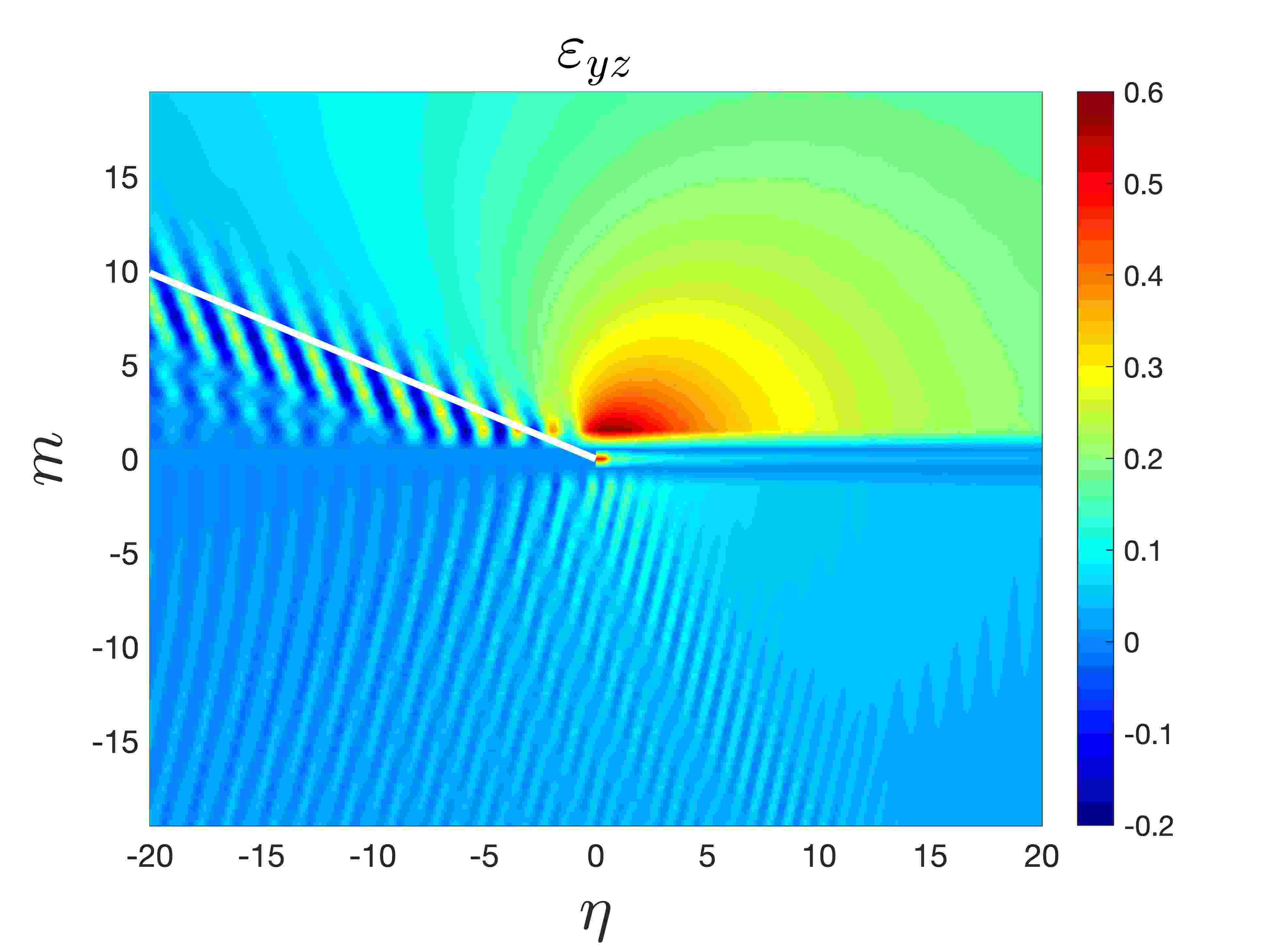}}
\endminipage

\caption[ ]{Displacements and strains in the lattice with parameters $\beta=1=\alpha_1=1$, $\gamma=\alpha_2=5$ and $v/v_c=0.55$ computed using (\ref{eq:SolutionLatticeInverseFourier_Layersa}) and (\ref{eq:SolutionLatticeInverseFourier_Layers}).
The panels (a)--(c) show the  displacement field of the lattice with a crack. The panels (d)--(f) and (g)--(i) illustrate the associated strain fields $\varepsilon_{xz}$ and $\varepsilon_{yz}$, respectively, computed with  (\ref{strainsexz1})--(\ref{strainseyz}). 
Here, the results correspond to  $\mu=1/5$, $\mu=1$ and $\mu=5$ shown in the panels contained in the first, second and third columns, respectively, of the figure. The wave radiation rays are shown in (d)--(i) by white lines and are based on (\ref{wave_rays}).  }
\label{fig:Plots_3}
\end{figure}

\begin{figure}[htbp]
(a) \minipage{0.3\textwidth}
\center{\includegraphics[width=\linewidth] {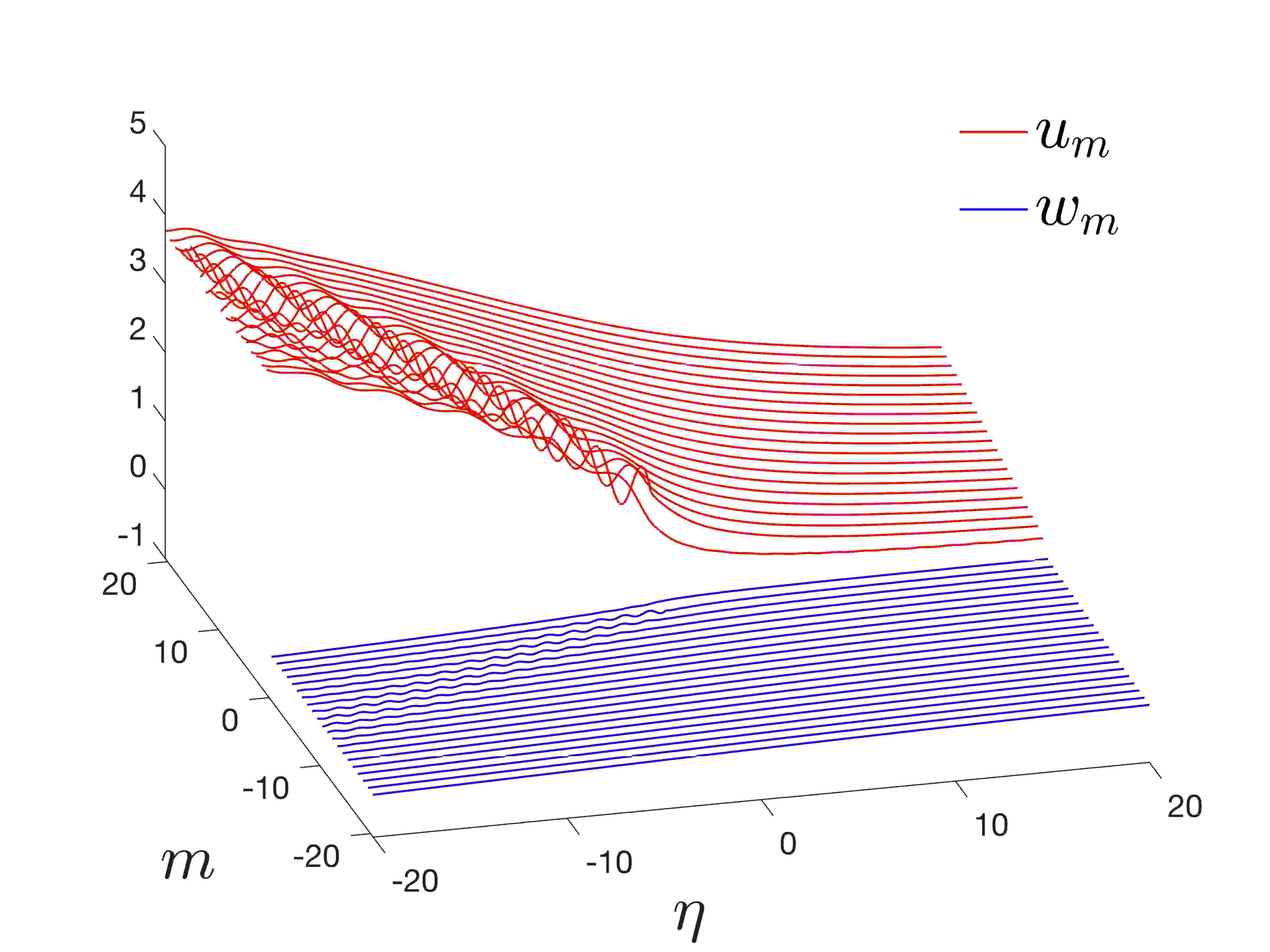}}
\endminipage
\hfill
(b) \minipage{0.3\textwidth}
\center{\includegraphics[width=\linewidth] {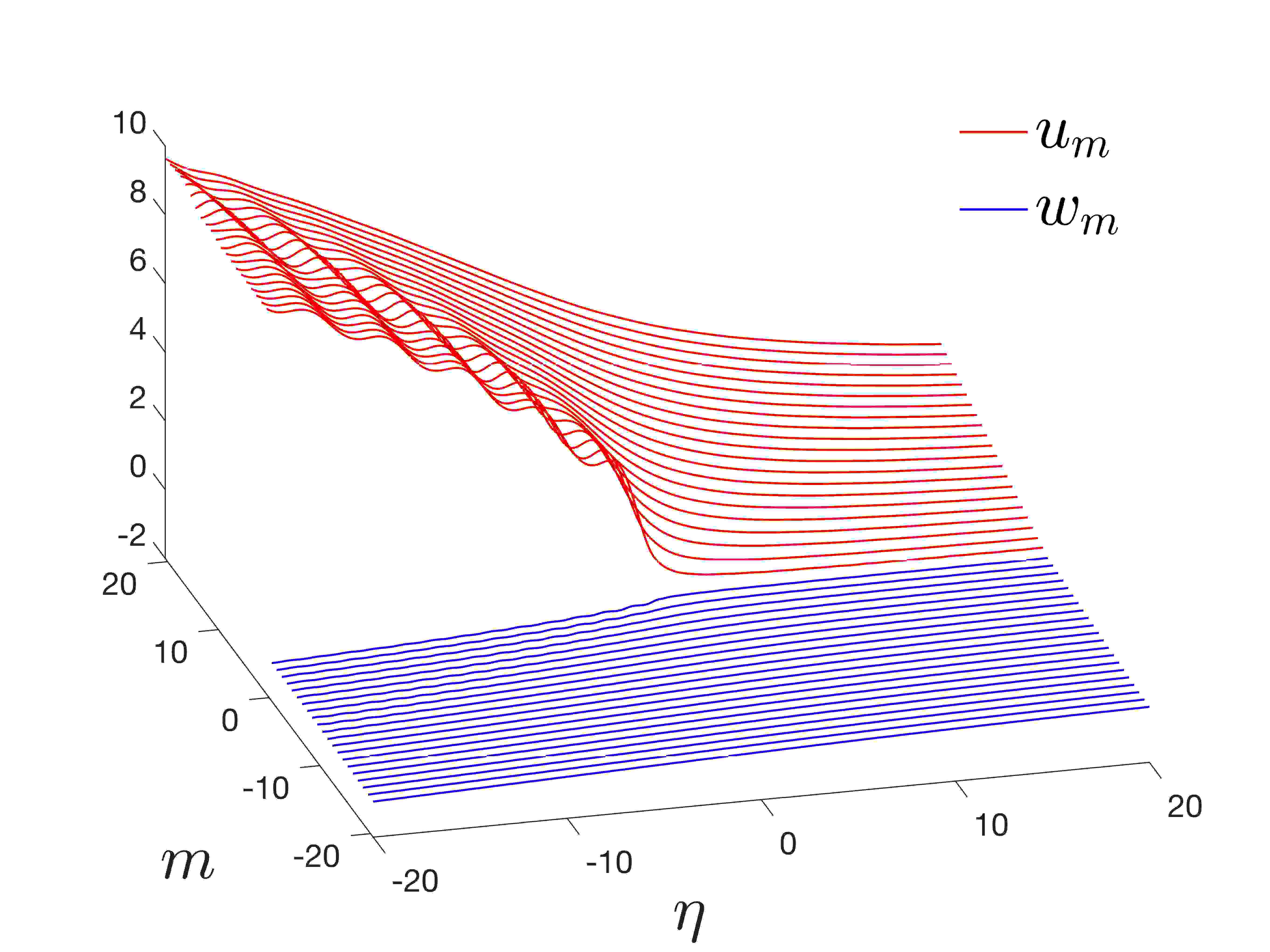}}
\endminipage
\hfill
(c) \minipage{0.3\textwidth}
\center{\includegraphics[width=\linewidth] {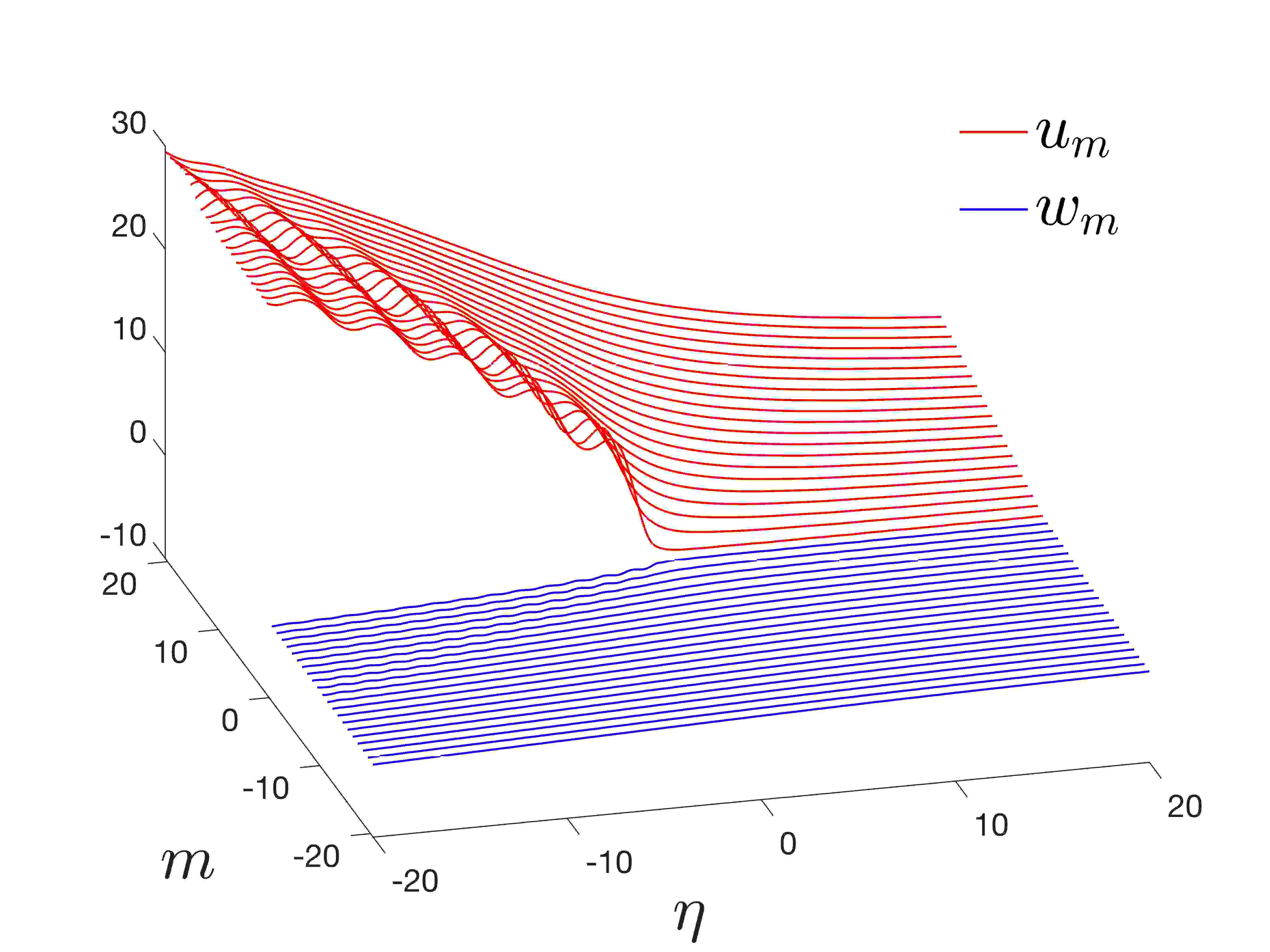}}
\endminipage 
\\
(d) \minipage{0.3\textwidth}
\center{\includegraphics[width=\linewidth] {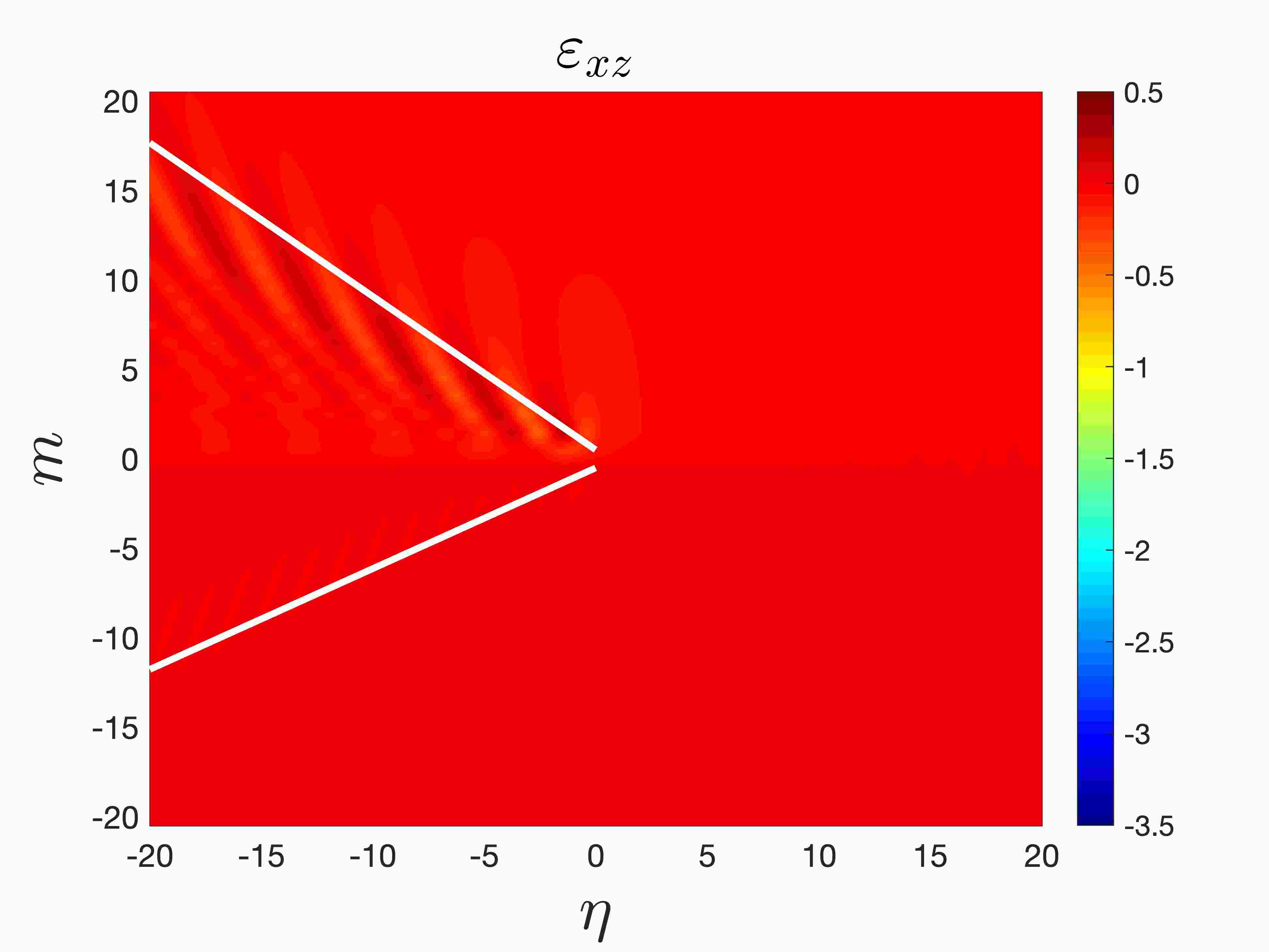}}
\endminipage
\hfill
(e) \minipage{0.3\textwidth}
\center{\includegraphics[width=\linewidth] {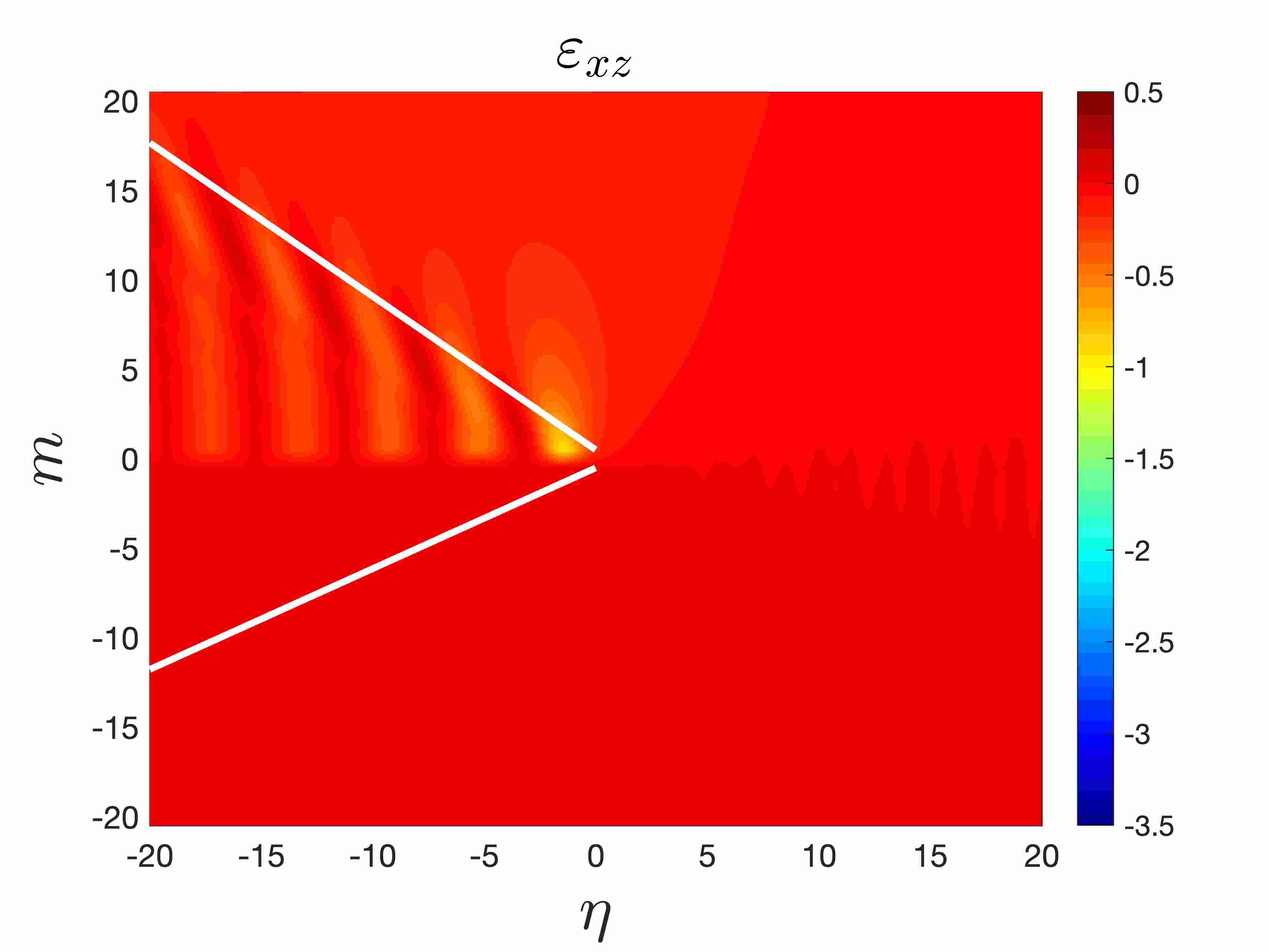}}
\endminipage
\hfill
(f) \minipage{0.3\textwidth}
\center{\includegraphics[width=\linewidth] {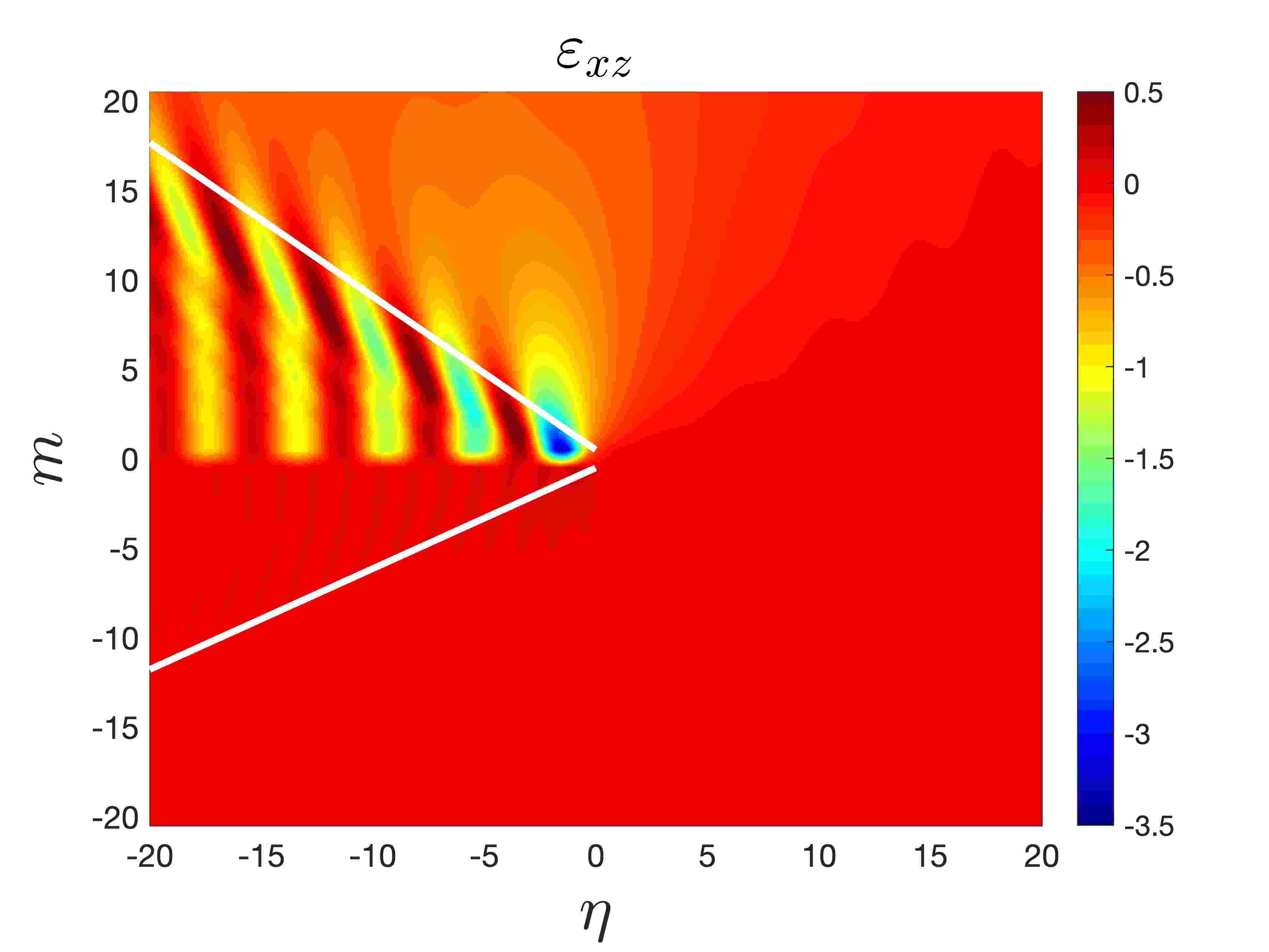}}
\endminipage
\\
(g) \minipage{0.3\textwidth}
\center{\includegraphics[width=\linewidth] {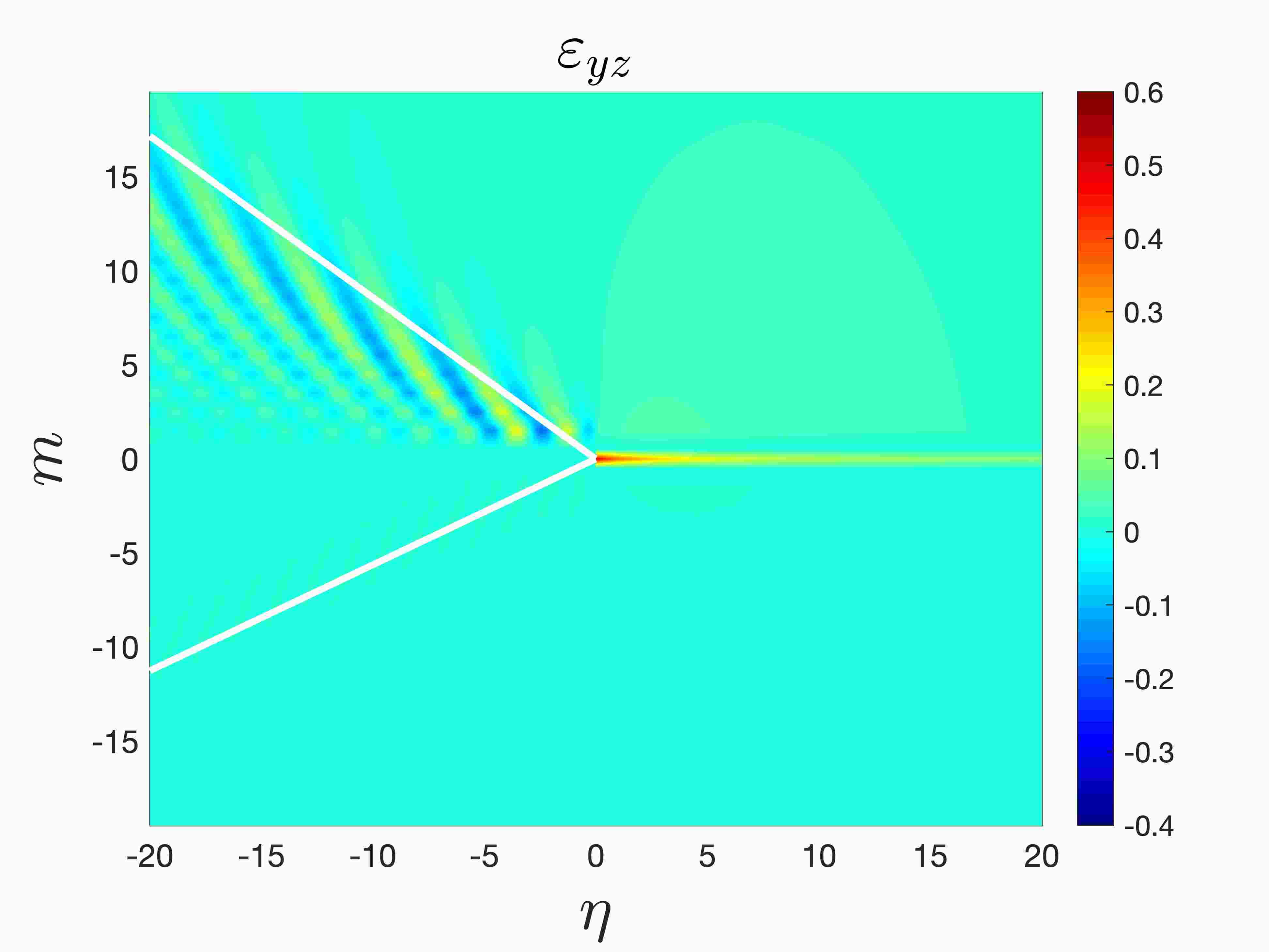}}
\endminipage
\hfill
(h) \minipage{0.3\textwidth}
\center{\includegraphics[width=\linewidth] {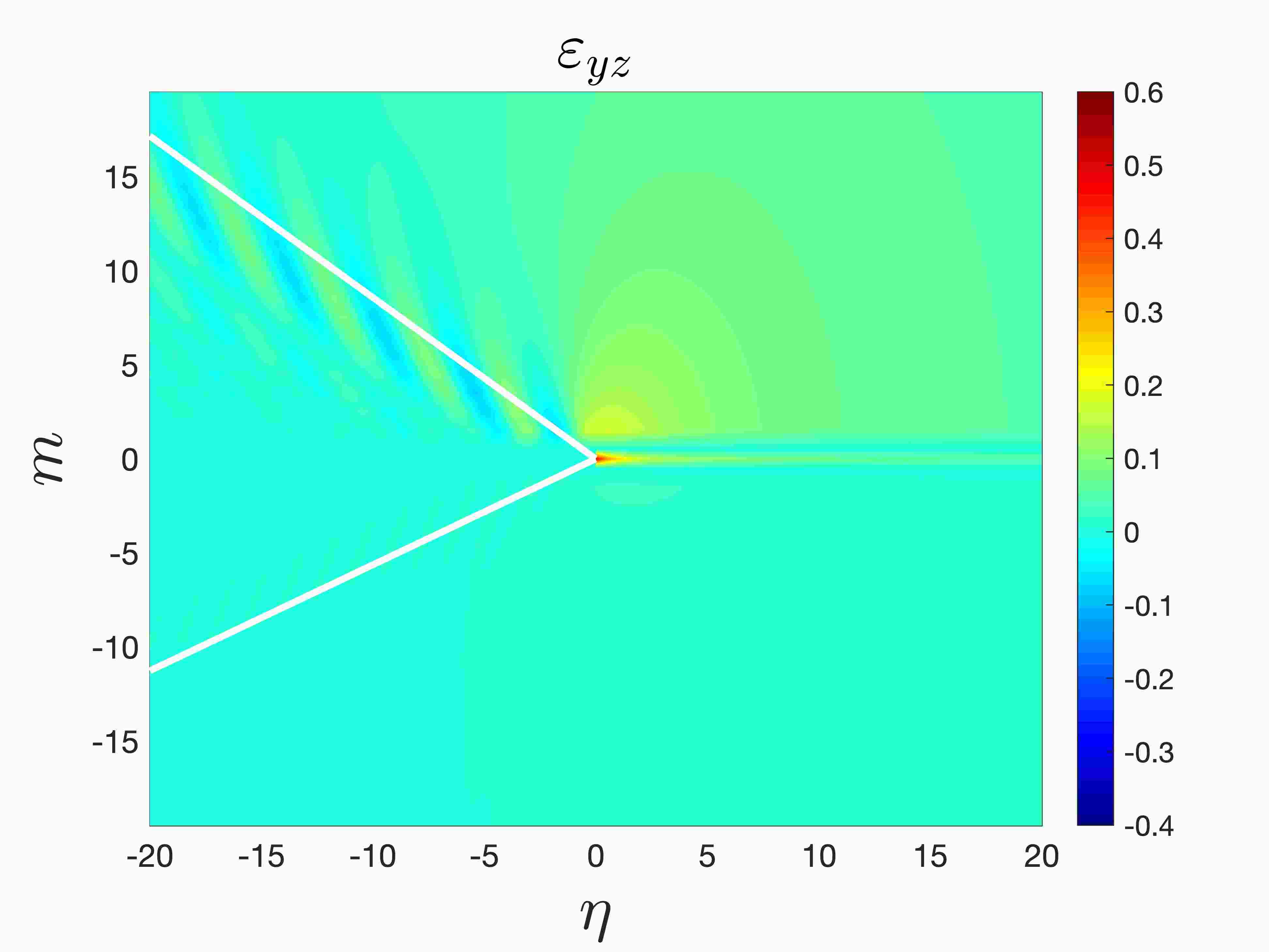}}
\endminipage
\hfill
(i) \minipage{0.3\textwidth}
\center{\includegraphics[width=\linewidth] {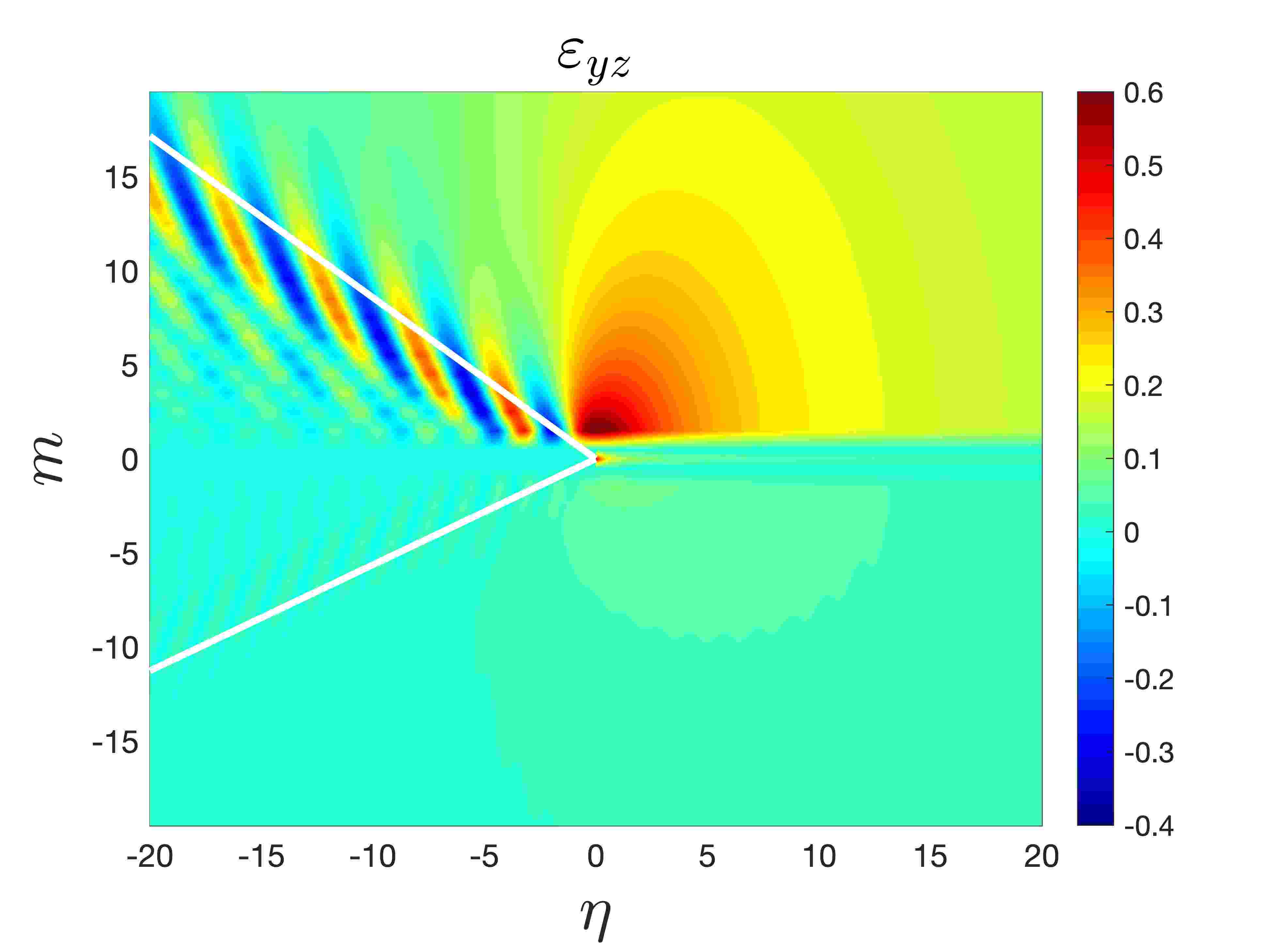}}
\endminipage
\caption[ ]{Displacements and strains in the lattice with parameters $\beta=1=\alpha_1=1$, $\gamma=\alpha_2=5$ and $v/v_c=0.9$ computed using (\ref{eq:SolutionLatticeInverseFourier_Layersa}) and (\ref{eq:SolutionLatticeInverseFourier_Layers}).
The panels (a)--(c) show the  displacement field of the lattice with a crack. The panels (d)--(f) and (g)--(i) illustrate the associated strain fields $\varepsilon_{xz}$ and $\varepsilon_{yz}$, respectively, based on (\ref{strainsexz1})--(\ref{strainseyz}). 
Here, the results correspond to  $\mu=1/5$, $\mu=1$ and $\mu=5$ shown in the panels contained in the first, second and third columns, respectively, of the figure. The wave radiation rays are shown in (d)--(i) by white lines and are based on (\ref{wave_rays}).  }
\label{fig:Plots_4}
\end{figure}

Figures \ref{fig:Plots_3}(a)--(c) show the resulting displacement plots for the dissimilar lattices for the lowest speed considered. These  clearly demonstrate that the softer lattice undergoes larger deformations than the stiffer lattice and again the deformation is no longer symmetrical about the crack path. This is also reflected in the results concerning the  strain fields $\varepsilon_{xz}$ and $\varepsilon_{yz}$  shown in  Figures \ref{fig:Plots_3}(d)--(f) and (g)--(i), respectively. 

These figures show the upper lattice admits a ray along which waves are radiated into the bulk, whereas the lower lattice does not possess such a ray.  Indeed, it corresponds to the theoretical findings of Section \ref{solWH}, where functions $H_j(k)$, $j=1,2,$ are identical only in the  case when the upper and lower lattice share the same wave speed. 

In addition, Figures \ref{fig:Plots_3}(d)--(f) and (g)--(i) illustrate the  disparity in the micro-structural behaviour of the  upper and lower lattice. 
 For instance, Figures \ref{fig:Plots_3}(e), (f), (h) and (i) show that if the interface is not soft then the upper lattice admits visible  wave radiation patterns. In the lower lattice, these effects are much weaker as this medium possesses both a higher density and stiffness than the upper lattice.
 Vibrations in the upper lattice appear to be confined between the crack and the wave radiation ray as  shown in these figures. On the other hand, in the lower lattice the micro-structural deformations  are less localised.
 As before, for softer interfaces the wave radiation effects are less visible
  and there the macro-scale deformations linked to the propagating crack are the dominant features (see  Figures \ref{fig:Plots_3}(d) and (g)).

Figures \ref{fig:Plots_3}(g)--(i) also reveal a prominent strain distribution $\varepsilon_{yz}$  ahead of the crack tip, due to the difference in the displacements in upper and lower lattices. In addition, these figures illustrate that as the interfacial bond stiffness is increased the magnitude of the vibrations distributed along the wave radiation ray in the upper lattice also becomes more significant. Moreover, a concentrated region of high deformation can appear in the softer lattice local to the crack tip as in Section \ref{sec7.2}.

Comparing  Figures \ref{fig:Plots_4}(a)--(c)  with Figures \ref{fig:Plots_3}(a)--(c),we see that when the crack speed is increased the deformations of lattice can increase.  Figure \ref{fig:Plots_4}(a)--(c) provide the displacement fields for the medium and, again for the dissimilar structures considered,  the lower lattice does not undergo any significant deformations  in comparison with the upper lattice. 

Figures \ref{fig:Plots_4}(d)--(i) illustrate that for a faster crack,  the lattice can admit wave radiation rays in both the upper and lower bulk lattices that are not symmetric about the crack line. Figures \ref{fig:Plots_4}(d)--(f) demonstrate that for higher speeds, micro-structural oscillations in the strains $\varepsilon_{xz}$ of the lower lattice ahead of the crack tip can be encountered. These oscillations become more prominent with increase of the interfacial bond stiffness. In particular, Figure \ref{fig:Plots_4}(f) shows there exist visible ripples in the lattice strains $\varepsilon_{xz}$ localised to the region bounded by the wave radiation rays for $\mu=5$, corresponding to the perfect join of the lattices at $m=1$. Again, having a softer interface  suppresses this effect in the lower lattice (see Figures \ref{fig:Plots_4}(d) and (e)).

Figures \ref{fig:Plots_4}(g)--(h) show the strains $\varepsilon_{yz}$ in the upper lattice possess a similar distribution to those presented in Figures \ref{fig:Plots_3}(g)--(h). The main difference here being that the increase in the crack speed causes the wave radiation effects and the deformations in crack tip neighborhood to be more localised.
 For the stiffer  lattice the distribution of these strains appears to be almost uniform, with increase in the interfacial bond stiffness bringing barely visible effects in terms of wave radiation (see Figure \ref{fig:Plots_4}(i)).

{
\subsection{Failure mode inducing an interfacial wave}
In the previous examples we focused on failure modes where the wave radiation, attributed to the propagation of failure in the lattice, occurred behind the crack tip.
It follows from the analysis of Section \ref{secdisp_description}, that one can  find interfacial wave modes that can be initiated ahead of the crack tip as it propagates (see also Supplementary Material \ref{sec_crack_disp}). Here we consider an example showing an admissible failure mode of this type for a lattice described by the parameters $\beta=10$, $\gamma=2$,  $\alpha_1=1/30$, $\alpha_2=1/40$ and $\mu=1$.  In this case, the lattice properties are fully dissimilar and they represent a structure involving an interface joining two different anisotropic lattices having sufficiently softer transverse links in comparison with the interfacial bonds. Additionally, the mass density of the upper lattice is much greater than that of the lower lattice.

Figure \ref{fig:Plots_1trdisp} shows the dispersion diagram for this case, where in addition to the acoustic curves, both optical curves describing waves in the dissimilar medium are present. 
\begin{figure}[h!]
\centering
\includegraphics[width=0.5\textwidth]{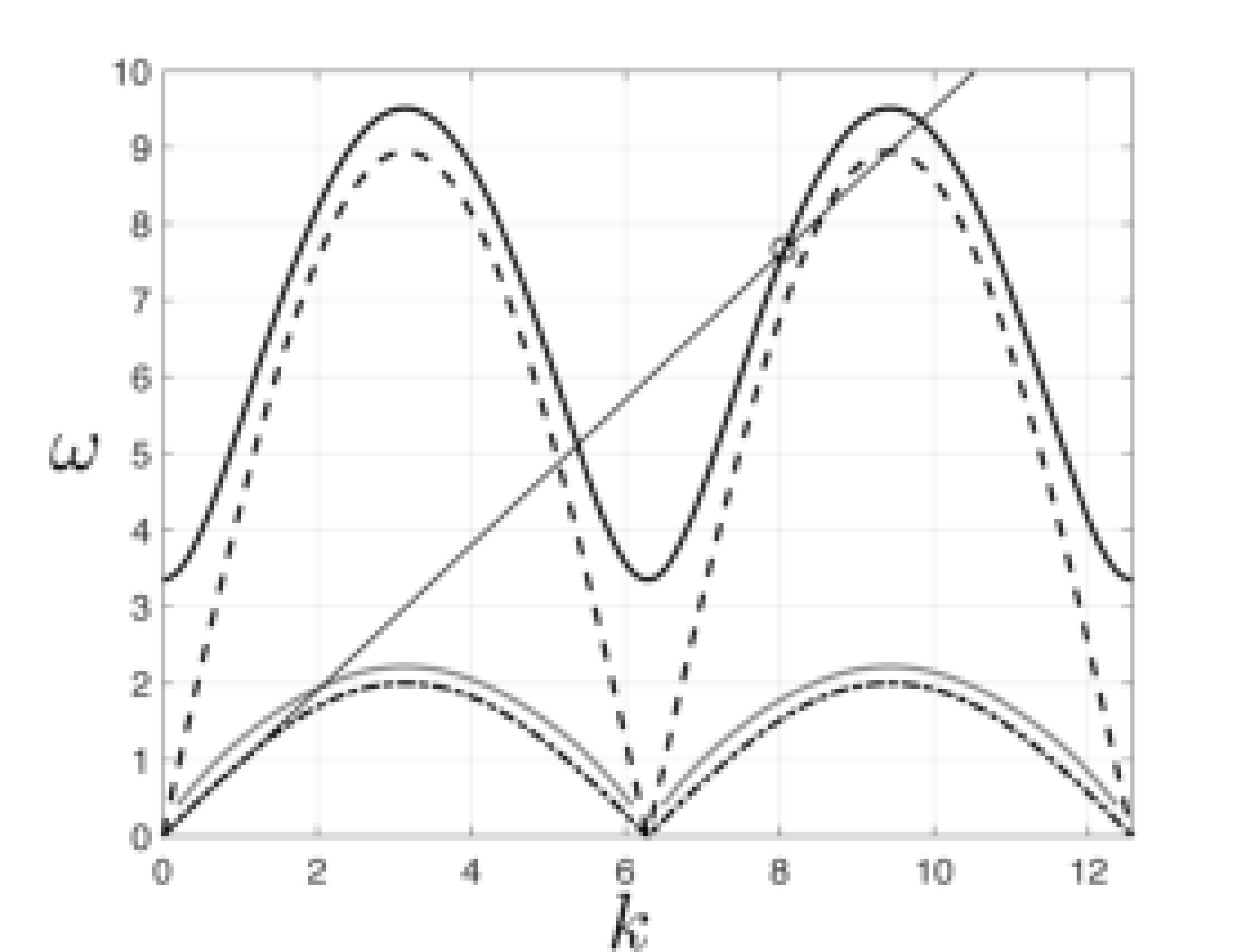}
\caption{Dispersion diagram for the dissimilar lattice with $\beta=10$, $\gamma=2$,  $\alpha_1=1/30$, $\alpha_2=1/40$ and $\mu=1$. The ray $\omega=kv$, with $v/v_c=0.95$ is also shown. Acoustic branches $\omega_1$ and $\omega_2$ are represented by the  dot-dashed and dashed curves, respectively (see (\ref{eq:DispersionRelations})). The optical curves $\omega_1^{\text{op}}$ and  $\omega_2^{\text{op}}$ are provided by the respective black and grey curves. This diagram indicates the possibility to propagate a crack within the lattice at the speed indicated, whilst generating an interfacial wave mode with constant amplitude that is transmitted ahead of the crack tip. This wave mode is approximately defined by  $(k, \omega)=(8.06, 7.66)$, shown here by the circle representing the intersection point of the highest optical branch with  the line $\omega=kv$.}
\label{fig:Plots_1trdisp}
\end{figure}

Figure \ref{fig:Plots_1tr} illustrates the behaviour of this lattice as it undergoes steady state failure  with a speed below the shear wave speed of the lower lattice half-plane. In Figure \ref{fig:Plots_1tr}(a) the function $\psi(\eta)$ defined in (\ref{psiphi}) is shown. This function indicates the square root growth of the solution behind the crack tip ($\eta=0$) that is super-positioned with waves that radiate behind the tip as the crack advances. Here the failure criterion is satisfied at $\eta=0$ and for $\eta>0$, the solution does not possess displacements greater than the critical threshold for failure (indicated by unity in the vertical axis in Figures \ref{fig:Plots_1tr}(a) by the dash-dot line). Figure \ref{fig:Plots_1tr}(b) shows the magnification of the plot in Figure \ref{fig:Plots_1tr}(a), where it is evident that ahead of the crack tip, a wave of constant amplitude is transmitted as the fault propagates. We note the oscillations along the interface induced by this transmitted wave are attributed to the movement of the nodes along the rows $m=\pm 1$ as shown in Figure \ref{fig:Plots_1tr}(c), which also shows the upper lattice particles along $m=1$ ahead of the crack tip have a significantly smaller displacement amplitude than those along $m=-1$.

As predicted in Section \ref{Disp_prop} and the Supplementary Material \ref{SM7}, the transmitted wave is exponentially localised about $m=-1$ and $m=1$. Far ahead of the crack tip in the lattice, one can expect the transmitted mode to behave as in the case of the free vibration problem for the dissimilar lattice, studied in Section \ref{Disp_prop}. This mode exhibits a strong localisation about the interface is shown in Figure \ref{fig:Plots_1tr}(d) and corresponds to the frequency and wavenumber defined by intersection point of the line $\omega=kv$ with the highest optical curve in Figure \ref{fig:Plots_1trdisp}.

The transverse strains in the  lattice are shown in Figure \ref{fig:Plots_1tr}(e). They reveal that in addition to the transmitted wave ahead of the crack tip, the upper lattice undergoes high deformations both in a region just ahead of the crack tip and behind it, where the lattice is subjected to waves radiated by the progression of the fault. In comparison with the upper lattice behaviour, during the failure process the spatial variation of the  deformation in the lower lattice is not as significant. These effects are also present in the plot of the horizontal strains in the lattice, illustrated in Figure \ref{fig:Plots_1tr}(f). In both Figure \ref{fig:Plots_1tr}(e) and (f), rays corresponding to the directions with the most significant wave radiation are shown. In particular, these examples show the lighter lattice may possess several directions where wave radiation is prominent, whereas in the upper lattice only a single ray is found.

The case considered here was investigated for a high crack speed. As illustrated by the examples in this  section, high crack speeds lead to possible scenarios where crack deviation or roughening of the crack surfaces can occur if the integrity of the upper and lower lattices is considered in addition to the integrity of the interfacial bonds. In this case, the solution presented in Figure \ref{fig:Plots_1tr} would be not admissible. We note the occurrence of transmitted modes is most significant when considering low crack velocities,   but as we have shown in the above examples, at these velocities the solution constructed is typically  not admissible. This may result from the fact that even if the transmitted wave has finite amplitude that does not induce critical strains ahead of the crack, its combination with the effects local to the crack tip may yield a significant crack path displacement rendering the solution not admissible. Hence, this makes the  effects of the transmitted failure mode illustrated here  difficult to capture without inducing non-steady fracture mechanisms.}
\begin{figure}[h!]
\centering
\includegraphics[width=0.85\textwidth]{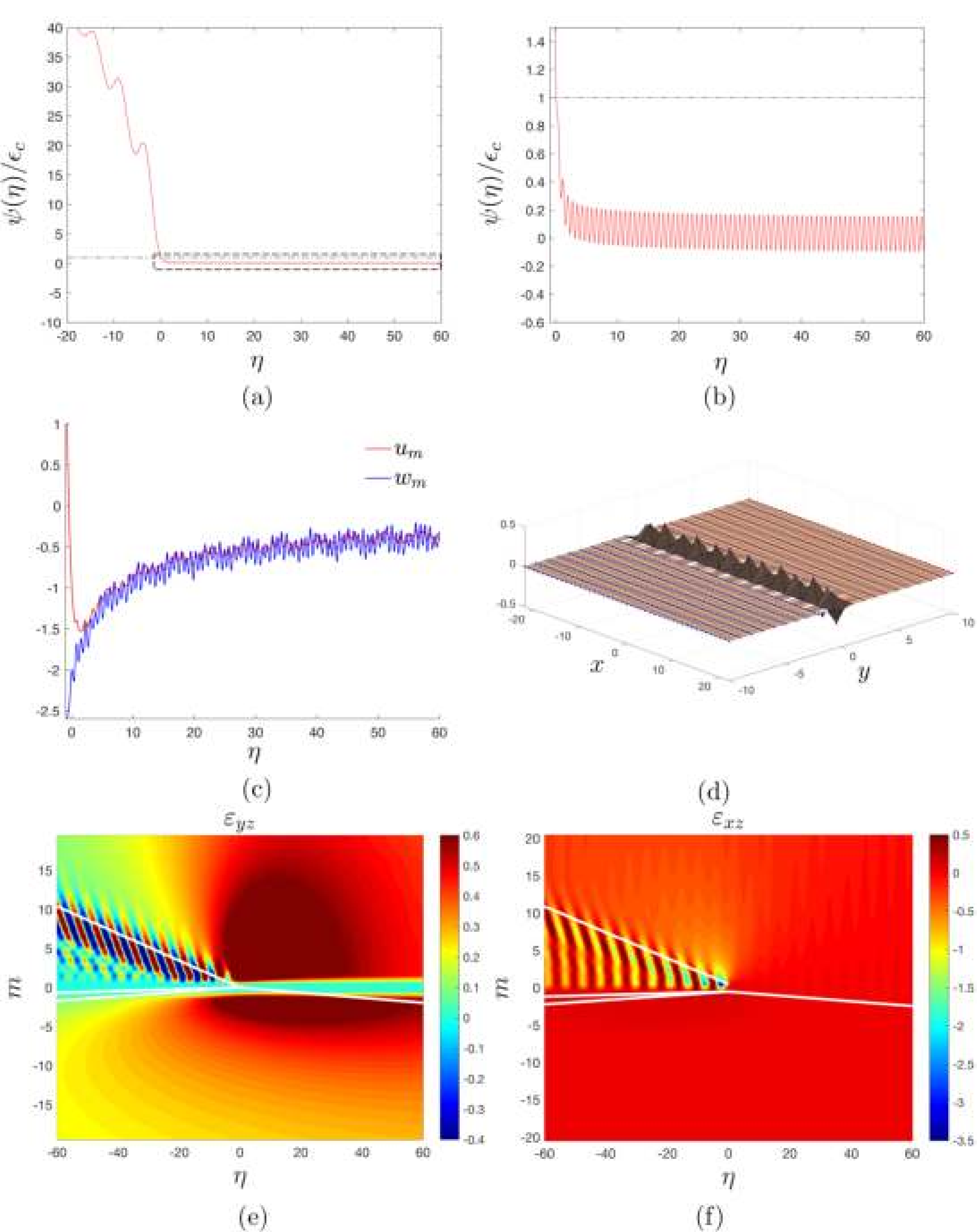}
\caption{The dissimilar lattice undergoing steady-state failure, whilst supporting interfacial wave modes. (a) The quantity $\psi(\eta)/\epsilon_c$ as a function of $\eta$ (see (\ref{psiphi})). (b) Magnification of the dashed box in (a). (c) Displacements $u_1(\eta)$ and $w_1(\eta)$ along $m=1$ and $m=-1$, respectively. (c) Eigenmode of the dissimilar lattice corresponding to the intersection point highlighted in Figure \ref{fig:Plots_1trdisp}. Red (blue) dots correspond to the nodes in the upper (lower) lattice. (e) Transverse strains $\varepsilon_{zy}$ and (f) horizontal strains $\varepsilon_{zx}$ in the lattice. The computations are performed for $v/v_c=0.95$, $\beta=10$, $\gamma=2$,  $\alpha_1=1/30$, $\alpha_2=1/40$ and $\mu=1$.}
\label{fig:Plots_1tr}
\end{figure}

}



\section{Conclusions}\label{conclusions}
We have presented the analysis  of a discrete dissimilar  lattice formed by two contrasting structured media joined by interfacial links that undergo steady fracture. In this problem, the fracture is assumed to be caused by the remote load that allows the solution to recreate the behaviour encountered near the tip of an interfacial crack propagating between two elastic solids, but we note that the formulation developed here allows for general classes of loads to be considered that also embed oscillations.

The problem has been solved via a reduction to a Wiener-Hopf equation through the Fourier transform by taking advantage of the assumption the crack propagates with a uniform speed. The detailed analysis of this functional equation and its solution has been carried out.  This has enabled  the full characterisation, through  a dispersion and asymptotic analysis, of the dynamic processes that can occur during the fracture process. This includes vibrations that excite the lattice microstructure and the global deformation associated  with the behaviour of the solution of the continuous problem for a crack propagating in a continuous bi-material. 

The study of the wave radiation process due to the propagation of the crack has also been presented and  it was shown that there exist crack propagation regimes where vibrations are transmitted and reflected or only reflected from the crack tip. In particular, the energy release ratio, comparing the lattice energy release rate with the global energy release rate, has been used to identify the amount of energy distributed to wave radiation processes. This was also accompanied by the analysis of the influence of the dissimilar lattice parameters on this quantity. Additionally, the solution of the problem has been implemented to determine admissible regimes for steady crack propagation and where the crack will propagate non-uniformly. Several examples have also shown how the admissibility of the steady fracture regimes are affected by the properties of the dissimilar lattice.

Applications of the considered problem include civil engineering, where the control of vibrations of dissimilar frame-like structures is of importance, and advanced material manufacturing, where the customisation of the material properties of structured medium can lead to failure resistant materials for specific configurations.

\vspace{0.1in}{\bf Acknowledgements:} {A.P. would like to acknowledge financial support from the
European Union's Seventh Framework Programme FP7/2007-2013/ under REA grant
agreement number PCIG13-GA-2013-618375-MeMic. 
G.M. acknowledges financial support from the ERC Advanced Grant ``Instabilities and nonlocal multiscale modelling of materials": ERC-2013-ADG-340561-INSTABILITIES.
G.M, also thanks  M. Kachanov for his illuminating comments during G.M.'s visit to Nizhny Novgorod Technical
University  supported by the project no. 14.Z50.31.0036 from the Ministry of Education and Science of
the Russian Federation. He is also thankful to Royal Society for the Wolfson Research
Merit Award. M.J.N.  gratefully acknowledges the support of the EU H2020 grant MSCA-IF-2016-747334-CAT-FFLAP.
G.M and M.N. also thank the Simons Foundation for their financial support.
The authors would like to thank the Isaac Newton Institute for
Mathematical Sciences, Cambridge, for the support and hospitality during the
programme ``\emph{Bringing pure and applied analysis together via the
Wiener-Hopf technique, its generalisations and applications}", where some
work on this paper was undertaken with the financial support through EPSRC
grant no EP/R014604/1.}

\setcounter{equation}{0} \renewcommand{\theequation}{A.\arabic{equation}}
\section*{Appendix: Parameters governing the appearance of interfacial waves}

\renewcommand{\thesubsection}{A.\arabic{subsection}}

Here, in accordance with the description of the dispersion features in Section \ref{secdisp_description}, we detail when the optical curves $\omega_j^{\text{op}}$, $j=1,2,$ can exist. We assume the parameters characterising the upper and lower lattice half-planes are given apriori and determine intervals for the crack bond stiffness for when  the optical curves appear.
To enable  this,
we define the following four critical values for the crack bond stiffness $\mu$:
\begin{eqnarray}
&&\mu_1=\frac{2\alpha_1\alpha_2[1+\alpha_1-\beta\gamma]^{1/2}}{(\alpha_1+\alpha_2)[1+\alpha_1-\beta\gamma]^{1/2}-\alpha_1[1+\alpha_1-\beta(\gamma+\alpha_2)]^{1/2}}\;, \\[3mm]
&&\mu_2=\frac{2\alpha_1^{3/2}\alpha_2}{\alpha_1^{3/2}+\alpha_2\alpha^{1/2}_1-\alpha_1[\alpha_1-\alpha_2\beta]^{1/2}}\;, \\[3mm]
&&\mu_3=\frac{2\alpha_1\alpha_2\beta^{1/2}\alpha_2^{1/2}}{(\alpha_1+\alpha_2)\beta^{1/2}\alpha_2^{1/2}-\alpha_2[\beta \alpha_2-\alpha_1]^{1/2}}\;
\end{eqnarray}
and 
\begin{equation}
\mu_4=\frac{2\alpha_1\alpha_2[\beta(\gamma+\alpha_2)-1]^{1/2}}{(\alpha_2+\alpha_1)[\beta(\alpha_2+\gamma)-1]^{1/2}-\alpha_2[\beta(\gamma+\alpha_2)-\alpha_1-1]^{1/2}}\;.
\end{equation}
We recall that the presence of the optical branches implies there exists frequency ranges where the interfacial waves propagate along the line inhomogeneity in the lattice and are exponentially localised about this defect. In Tables \ref{table1} and \ref{table2} we report the results concerning the intervals of $\mu$ where the dissimilar lattice will support high frequency interfacial waves based on the choices of  $\gamma\le \alpha_2/\alpha_1$ or $\gamma>\alpha_2/\alpha_1$, respectively, and the parameters $\alpha_j$, $j=1,2$ and $\beta$ (see the first columns of these tables for bounds on these parameters). These waves are connected with the optical curve  $\omega_1^{\text{op}}$ that can be either continuous or piecewise defined and for these scenarios  the intervals for $\mu$  are shown in the first and second columns, respectively, of Tables \ref{table1} and \ref{table2}.

The low frequency interfacial waves are associated with the optical branch  $\omega^{\text{op}}_2$, which is always piecewise defined. The conditions for the appearance of this curve are shown in Table \ref{table3}.
 
\begin{table}[htbp]
\begin{center}
\renewcommand\arraystretch{2}
\begin{tabular}{|c| c| c |}
\hline
 Upper and lower lattice parameters &  Continuous  $\omega^{\text{op}}_1$ & Piecewise defined $\omega^{\text{op}}_1$ \\ 
 \hline
 \hline
$\frac{\alpha_2}{\alpha_1}(1-\alpha_1)<\gamma$, $\frac{1}{\gamma+\alpha_2}\le \beta\le \frac{\alpha_1}{\alpha_2}$&  $\mu>\max\{\mu_1, \mu_2\}$ &  $\min\{\mu_1, \mu_2\} <\mu<\max\{\mu_1, \mu_2\}$  \\
\hline
 $\beta\le \min\Big\{\frac{\alpha_1}{\alpha_2}, \frac{1}{\gamma+\alpha_2}\Big\}$&  $\mu>\max\{\mu_1, \mu_2\}$ &  $\min\{\mu_1, \mu_2\} <\mu<\max\{\mu_1, \mu_2\}$  \\
 \hline
  $\max \Big\{\frac{\alpha_1}{\alpha_2} , \frac{1}{\gamma+\alpha_2}\Big\} <\beta\le \frac{1+\alpha_1}{\gamma+\alpha_2}$ & $\mu>\max\{\mu_1, \mu_3\}$ & $\frac{2\alpha_1\alpha_2}{\alpha_1+\alpha_2} <\mu<\max\{\mu_1, \mu_3\}$ \\
 \hline
 $\frac{1+\alpha_1}{\gamma+\alpha_2}< \beta
 $ & $\mu>\max\{\mu_3 , \mu_4\}$ &$\min\{\mu_3 , \mu_4\} <\mu <\max\{\mu_3 , \mu_4\}$ \\
 \hline
\end{tabular}
\end{center}
\caption{Admissible intervals of the crack bond stiffness $\mu$ for the existence of the continuous or piecewise defined optical curve $\omega_1^{\text{op}}$,  based on the parameters for $\gamma\le \alpha_2/\alpha_1$, $\alpha_j$, $j=1,2$, and $\beta$.}
\label{table1}
 \end{table}
 
 \begin{table}[htbp]
\begin{center}
\renewcommand\arraystretch{2}
\begin{tabular}{|c| c| c |}
\hline
 Upper and Lower lattice parameters &  Continuous  $\omega^{\text{op}}_1$ & Piecewise defined $\omega^{\text{op}}_1$ 
  \\ 
 \hline
 \hline
$\beta\le \frac{1+\alpha_1}{\gamma+\alpha_2}$&  $\mu>\max\{\mu_1, \mu_2\}$ &  $\min\{\mu_1, \mu_2\} <\mu<\max\{\mu_1, \mu_2\}$ 
\\
\hline
  $ \frac{1+\alpha_1}{\gamma+\alpha_2} <\beta\le \min\Big\{\frac{\alpha_1}{\alpha_2} ,\frac{1+\alpha_1}{\gamma}\Big\}$ & $\mu>\max\{\mu_2, \mu_4\}$ & $\frac{2\alpha_1\alpha_2}{\alpha_1+\alpha_2} <\mu<\max\{\mu_2, \mu_4\}$
  \\
 \hline
$\frac{\alpha_2(1+\alpha_1)}{\alpha_1}<\gamma$,  $ \frac{1+\alpha_1}{\gamma}
<\beta\le \frac{\alpha_1}{\alpha_2}$ & $\mu>\max\{\mu_2 , \mu_4\}$ &$\frac{2\alpha_1\alpha_2}{\alpha_1+\alpha_2} <\mu<\max\{\mu_2, \mu_4\}$ 
\\
\hline
$\frac{\alpha_1}{\alpha_2}<\beta$&  $\mu>\max\{\mu_3, \mu_4\}$ &  $\min\{\mu_3, \mu_4\} <\mu<\max\{\mu_3, \mu_4\}$ 
\\
 \hline

\end{tabular}
\end{center}
\caption{Admissible intervals of the crack bond stiffness $\mu$ for the existence of the continuous or piecewise defined optical curve $\omega_1^{\text{op}}$ based on the parameters $\gamma>\alpha_2/\alpha_1$,  $\alpha_j$, $j=1,2$, and $\beta$. }
\label{table2}
 \end{table}

 \begin{table}[htbp]
\begin{center}
\renewcommand\arraystretch{2}
\begin{tabular}{|c| c| }
\hline
 Upper and Lower lattice parameters &    $\omega^{\text{op}}_2$ 
  \\ 
 \hline
 \hline
$\beta<\min\Big\{ \frac{1}{\gamma+\alpha_2},  \frac{\alpha_1+1-(\alpha_2+1)^2}{(\alpha_1+2\alpha_2)(\gamma+\alpha_2)}\Big\}$&  $\mu>\mu_4$ \\
\hline
  $\beta>\max\Big\{ \frac{1+\alpha_1}{\gamma},  \frac{(2\alpha_1+\alpha_2)(1+\alpha_1)}{(2\alpha_1+\alpha_2)\gamma-\alpha^2_1}\Big\}$ & $\mu>\mu_1$ 
\\
 \hline

\end{tabular}
\end{center}
\caption{Admissible intervals of the crack bond stiffness $\mu$ for the existence of the optical curve $\omega_2^{\text{op}}$ based on the parameters $\gamma$,  $\alpha_j$, $j=1,2$, and $\beta$. }
\label{table3}
 \end{table}

\setcounter{equation}{0} \renewcommand{\theequation}{SM.\arabic{equation}}
\setcounter{figure}{0} \renewcommand{\thefigure}{SM.\arabic{figure}}

\newpage\section*{Supplementary Material}
\setcounter{page}{1}
\setcounter{figure}{1}
\renewcommand{\thesubsection}{SM.\arabic{subsection}}
\subsection{Derivation of the transformed equations and the matrix Wiener-Hopf problem}\label{SM1}
By introducing the continuous variable $\eta$  and (\ref{form1}), the  dimensionless equations of motion \eqref{eqLattice1a}--\eqref{eqLattice4a} are updated to
\begin{eqnarray}
v^2u_m^{\prime\prime}(\eta)=u_m(\eta+1)+u_m(\eta-1)-2u_m(\eta)
+\alpha_1 (u_{m+1}(\eta)+u_{m-1}(\eta)-2u_m(\eta)) \label{eq:LatticeProblemEta_1}
\end{eqnarray}
for $m>1$,
\begin{eqnarray}
 v^2w_{-m}^{\prime\prime}(\eta)=\beta\big[\gamma (w_{-m}(\eta+1)+w_{-m}(\eta-1)-2w_{-m}(\eta))
+\alpha_2 (w_{-m+1}(\eta)+w_{-m-1}(\eta)-2w_{-m}(\eta))\big]\;,\nonumber \\\label{eq:LatticeProblemEta_2}
\end{eqnarray}
for $m<-1$, whereas along the lines containing the propagating defect we have
\begin{eqnarray}
v^2u_1^{\prime\prime}(\eta)=u_1(\eta+1)+u_1(\eta-1)-2u_1(\eta)-\mu(u_1(\eta)-w_1(\eta))H(\eta)
+\alpha_1(u_2(\eta)-u_1(\eta))\;,
\label{eq:LatticeProblemEta_1a}
\end{eqnarray}
and
\begin{eqnarray}
 v^2w_1^{\prime\prime }(\eta)=\beta\big[\gamma(w_1(\eta+1)+w_1(\eta-1)-2w_1(\eta))+\mu (u_1(\eta)-w_1(\eta))H(\eta)
+\alpha_2(w_2(\eta)-w_1(\eta))\big]\;.\label{eq:LatticeProblemEta_2a}
\end{eqnarray}
Here, the prime in the above equations represents  differentiation with respect to $\eta$.

Let us first consider  \eqref{eq:LatticeProblemEta_1} and \eqref{eq:LatticeProblemEta_2}. The application of Fourier transform with respect to $\eta$ (see (\ref{FTeta})) and use of the following formulae 
\[\mathcal{F} [\{u_m^{\prime\prime}(\eta), w_{-m}^{\prime\prime}(\eta)\}]=(0+{\rm i}k)^2 \{U_m(k), W_{-m}(k)\}\]
and 
\[ \mathcal{F} \left[\{u_m(\eta\pm 1), w_{-m}(\eta\pm 1)\}\right]=e^{\mp {\rm i} k}\{U_m(k), W_{-m}(k)\} \;,\]
leads to 
\begin{eqnarray}
(0+{\rm i}kv)^2U_m(k)=2(\cos(k)-1)U_m(k)+\alpha_1 (U_{m+1}(k)+U_{m-1}(k)-2U_m(k)) \label{eq:LatticeProblemEta_1tr}
\end{eqnarray}
for $m>1$ and
\begin{eqnarray}
(0+{\rm i}kv)^2W_{-m}(k)=\beta\big[\gamma (2(\cos(k)-1)W_{-m}(k)+\alpha_2 (W_{-m+1}(k)+W_{-m-1}(k)-2W_{-m}(k))\big]\;,\label{eq:LatticeProblemEta_2tr}
\end{eqnarray}
for $m<-1$.
The preceding equations  can be then rearranged to give (\ref{eq:FourierInsideLattice}).

Taking advantage from the fact that the coefficients in the system of linear equations (\ref{eq:FourierInsideLattice})  do not depend on the value of $m$ and $n$, 
we seek the solutions in the form (\ref{eq:LambdaIntroduction})
where the exponents $\lambda_{1,2}$ are independent of $m$. For a physically acceptable solution we also impose (\ref{eq:Lambda_condition}).

Substituting \eqref{eq:LambdaIntroduction} into \eqref{eq:FourierInsideLattice}, leads to  a quadratic equation to determine the functions $\lambda_{1,2}$:
\begin{equation}
\lambda_j^2(k, 0+{\rm i}kv)-2\Omega_j(k, 0+{\rm i}kv)\lambda_j(k, 0+{\rm i}kv)+1=0\;,\quad j=1,2\;,
\label{eq:LambdaDerivation}
\end{equation}
with $\Omega_j(k, s)$ defined in (\ref{eqOM}). Note that from the preceding equation, the second relation in (\ref{eqOM}) follows immediately.
The equations in (\ref{eq:LambdaDerivation}) deliver two solutions for $\lambda_{j}$, $j=1,2,$ and  we choose the branch to satisfy the requirement (\ref{eq:Lambda_condition}). Thus, we have
\begin{eqnarray*}
\lambda_j(k, 0+{\rm i}kv)&=&\Omega_j(k, 0+{\rm i}kv)-\text{sgn}(\Omega_j(k, 0+{\rm i}kv)) \sqrt{[\Omega_j(k, 0+{\rm i}kv)]^2-1}\;, \text{ for } k \in \mathbb{R}
\end{eqnarray*}
and if $k \in \mathbb{C}$ we may write $\lambda_j$ as in (\ref{eq:Lambda_final}).

Next we apply the Fourier transform defined in (\ref{FTeta}) to (\ref{eq:LatticeProblemEta_1a}) and (\ref{eq:LatticeProblemEta_2a}). We obtain
\begin{eqnarray}
(0+{\rm i}kv)^2U_1(k)=2(1-\cos(k))U_1(k)-\mu(U^+_1(k)-W^+_1(k))
+\alpha_1(u_2(\eta)-u_1(\eta))\;,
\label{eq:LatticeProblemEta_1atr}
\end{eqnarray}
and
\begin{eqnarray}
(0+{\rm i}k v)^2W_1(k)=\beta\big[2\gamma(\cos(k)-1)W_1(k)+\mu (U^+_1(k)-W^+_1(k))
+\alpha_2(W_2(k)-W_1(k))\big]\;.\label{eq:LatticeProblemEta_2atr}
\end{eqnarray}
We then use the solution for the lattice half-plane problems defined by (\ref{eq:LambdaIntroduction}). Equations (\ref{eq:LatticeFourierTransform}) then follow from this by using (\ref{psieq2}), which represents the Fourier transform of the function corresponding to the crack opening for $\eta<0$ and elongation of the interfacial bonds for $\eta \ge 0$.
Moreover, using (\ref{eq:LambdaIntroduction})  and invoking the additive splits (\ref{addsplitUW}), in (\ref{eq:LatticeProblemEta_1atr}) and (\ref{eq:LatticeProblemEta_2atr}) we can gather terms corresponding to ``$+$" and ``$-$" to retrieve  (\ref{matform1}) and (\ref{eqSj}).
\subsection{Eigenmodes for lattice problems connected with the failure of the dissimiliar structure}\label{SM2}
Here we describe the method for constructing the eigenmodes for the problems linked to the propagation waves in the lattice half-planes with free boundaries and in the dissimilar structured medium, on which the computations in Section \ref{Disp_prop} are based.

\vspace{0.1in}{\emph{Eigenmodes of the semi-infinite lattices with free boundaries.}} The branches  $\omega_1(k)$ and $\omega_2(k)$ in (\ref{eq:DispersionRelations}) 
are connected with the vibration modes of the separated  semi-infinite upper or lower lattices, respectively.
Along the branch $\omega_1$ ($\omega_2$) the associated eigenmode in the upper (lower) lattice half-plane has constant amplitude in the direction perpendicular to boundary and oscillates in the direction parallel to the  boundary. If $\omega>0$, both half-planes produce a different  dynamic response provided $\gamma \beta\ne  1$. For a given frequency $\omega$, the eigenmodes can be determined from (\ref{eqLattice1a}) and (\ref{eqLattice4a}), assuming that 
\begin{equation}\label{eigmode1}
 u_{m, n}(t)=f_m e^{{\rm i}(\omega t - kn)}\;, \qquad w_{-m, n}(t)=g_{-m} e^{{\rm i}(\omega t - kn)}
 \end{equation}
where  the amplitudes  $f_m$ and $g_{-m}$ are sought in the form
\begin{equation}\label{eigmode2}
f_m=[\lambda_1(k, {\rm i}\omega)]^{m-1} f_1\;,\quad \text{ for }m>1 \qquad \text{ and } \qquad g_{-m}=[\lambda_2(k, {\rm i}\omega)]^{-m-1} g_{1},  \qquad \text{ for }m<-1 \;.
\end{equation}
These assumptions lead to the equations
\begin{equation}\label{eigmode3}
S_1(k, {\rm i}\omega)f_1=0 \qquad \text{ and }\qquad S_2(k, {\rm i}\omega)g_1=0\;,
\end{equation}
which are satisfied by $\omega=\omega_1(k)$ and $\omega=\omega_2(k)$, respectively. Note in this case, $f_1$ and $g_1$ are independent.

\vspace{0.1in}{\emph{Eigenmodes 
 along the high frequency optical curves.} }For the dissimilar lattice with an interface, the eigenmodes can be  computed using the governing equations  (\ref{eqLattice1a})--(\ref{eqLattice4a}) (assuming $n>n^*$), with  (\ref{eigmode1}) and (\ref{eigmode2}). They lead to the following homogeneous system that couples the amplitudes $f_1$ and $g_1$:
\begin{equation}\label{eigmode4}
\begin{bmatrix}
{\alpha_1}\mathcal{S}_1(k, {\rm i}\omega )+{\mu}& -{\mu}\\[3mm]
-{\beta\mu}&\beta\big[{\alpha_2}\mathcal{S}_2(k, {\rm i}\omega)+{\mu}\big]
\end{bmatrix}\begin{bmatrix}
f_1\\[3mm]
g_1
\end{bmatrix} =\begin{bmatrix}
0\\[3mm]
0
\end{bmatrix}\;,
\end{equation}
where the coefficient matrix is degenerate along the curves $\omega_j^{\text{op}}$, $j=1,2$, discussed in Section \ref{Disp_prop}.

\subsection{Dynamic features of the dissimilar lattice undergoing steady-state fracture}\label{sec_crack_disp}
To determine the behaviour of the dissimilar lattice undergoing fracture, we use the dispersion properties of the two lattices considered in Section \ref{Disp_prop}. The associated dispersion curves  define waves incident on and  radiated from the crack front during the steady failure process.
The waves are identified by considering the line $\omega=kv$ on the dispersion diagram as in the work of \citet{LSbook}, with $v$ being the crack speed satisfying (\ref{speedbound}). 

\vspace{0.1in}{\bf {Incident and radiated waves} propagating behind the crack front. \rm}In the subsonic regime, the kernel function $L(k, {\rm i}kv)$ has a {singular point} at zero associated with the intersection of the acoustic branches $\omega_j$, $j=1,2,$ with $\omega=kv$ at the origin on the dispersion diagram. It corresponds to the continuum  or long wavelength limit of the lattice.
In fact, as discussed in the main text, this singular point is connected with the application of a remote load that leads to a solution corresponding to the field local to the crack tip of a dynamic interfacial defect propagating through a dissimilar continuous elastic medium. 

Additionally, owing to the intersections of the line $\omega=kv$ with $\omega_1(k)$, the function $L(k, {\rm i}kv)$ has one, three or more pairs of {singular points} at $k= \pm p_1, \pm p_2$, $\dots$, $\pm p_{2 f+1}$, where $f\ge 0$, $f \in \mathbb{Z}$, is dependent on the crack speed. As an example, we refer to  Figure \ref{crack_disp_1}, where it can be seen for the lines $\omega=kv$ with $v=0.08$, 0.15 and 0.4 we have $f=3, 1$ and 0, respectively.

Similarly, analysing the intersections of $\omega=kv$ with $\omega_2$ on the dispersion diagram, $L(k, {\rm i}kv)$ can have additional {singular points} at $k= \pm q_1, \pm q_2$, $\dots$, $\pm q_{2 g+1}$, $g\ge 0$, $g \in \mathbb{Z}$. Once more, $g$ is dependent on the crack speed. The example presented in Figure \ref{crack_disp_1} shows for $v=0.08$, 0.15 and 0.4 that $g=2, 1, 0$, respectively.

\begin{figure}[htbp]
\begin{center}
{\includegraphics[width=0.5\textwidth]{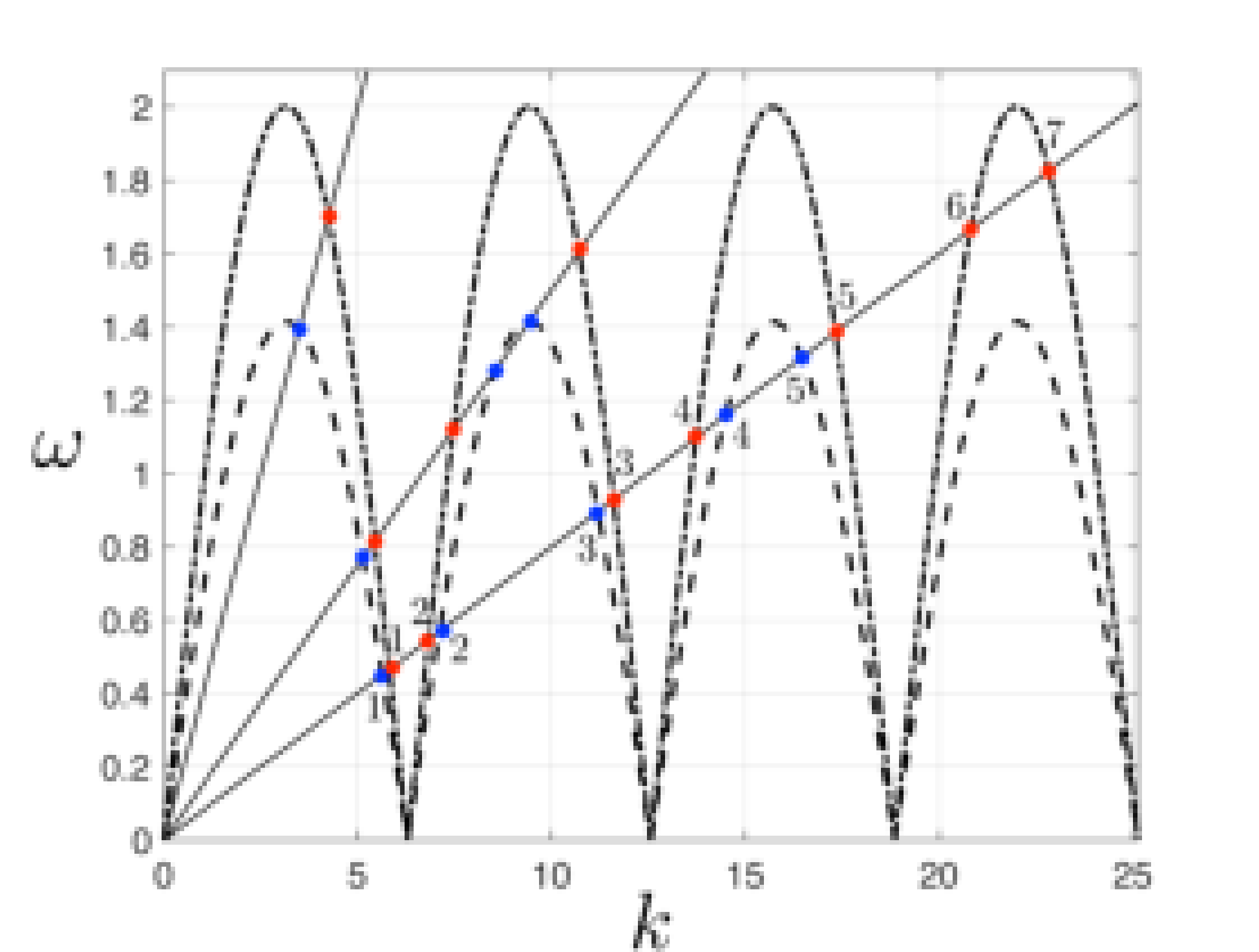}
 }
\caption{Dispersion diagram for the case when (\ref{con1}) is not satisfied. The lines $\omega=kv$ are shown for $v=0.4, 0.15$ and $0.08$. Intersections of this line with the curve for  $\omega_1(k)$ (dash-dot curve) and $\omega_2(k)$ (dashed curve) are marked with red and blue dots, respectively. For  $v=0.08$, we supply each red (blue) dot with an integer $j$ and the coordinate of this dot is $(p_j, \omega_1(p_j))$ $((q_j, \omega_2(q_j))$). Here, $\beta \gamma=1/2$ and (\ref{eq:DispersionRelations}) has been used.}
\label{crack_disp_1}
\end{center}
\end{figure}

We also compare the crack speed $v$ (slope of $\omega=kv$) with the group velocity $v_g=\frac{{\rm d}\omega}{{\rm d} k}$ at each intersection point and note that  for intersection points with the wave numbers:
\begin{enumerate}
\item $k=\pm p_1, \pm p_3, \dots, \pm p_{2 f+1}$    and      $k=\pm q_1, \pm q_3, \dots, \pm q_{2 g+1}$, with $f,g \ge 0$, $f, g \in \mathbb{Z}$, the group velocity $v_g<v$. These wave numbers are connected with waves located at $n<n^*$ {that appear as the  crack tip advances and are radiated behind it}.
\item $k=\pm p_2, \pm p_4, \dots, \pm p_{2 f}$    and      $k=\pm q_2, \pm q_4, \dots, \pm q_{2 g}$, with $f,g \ge 1$, $f, g \in \mathbb{Z}$, the group velocity $v_g>v$. These wave numbers define waves incident on the crack front that propagate in the region $n<n^*$. In this article, we do not consider  loading that produces such a dynamic response and these wave numbers can be neglected.
\end{enumerate}

In particular the singular wave numbers of $L(k, {\rm i}kv)$ identified here define branch points of  this function.



\vspace{0.1in}{\bf Transmitted waves ahead of the crack front.\rm} Now we describe the waves that can appear ahead of the crack front during the fracture process. These waves are connected with the intersection points of the line $\omega=kv$ with the optical branches $\omega_j^{\text{op}}$, $j=1,2$. 
As mentioned in the previous section, these curves  may not exist and their appearance is related to a specific choice of the lattice material parameters. The analysis is separated into the following scenarios:
\begin{figure}[htbp]
\begin{center}
{\includegraphics[width=1\textwidth]{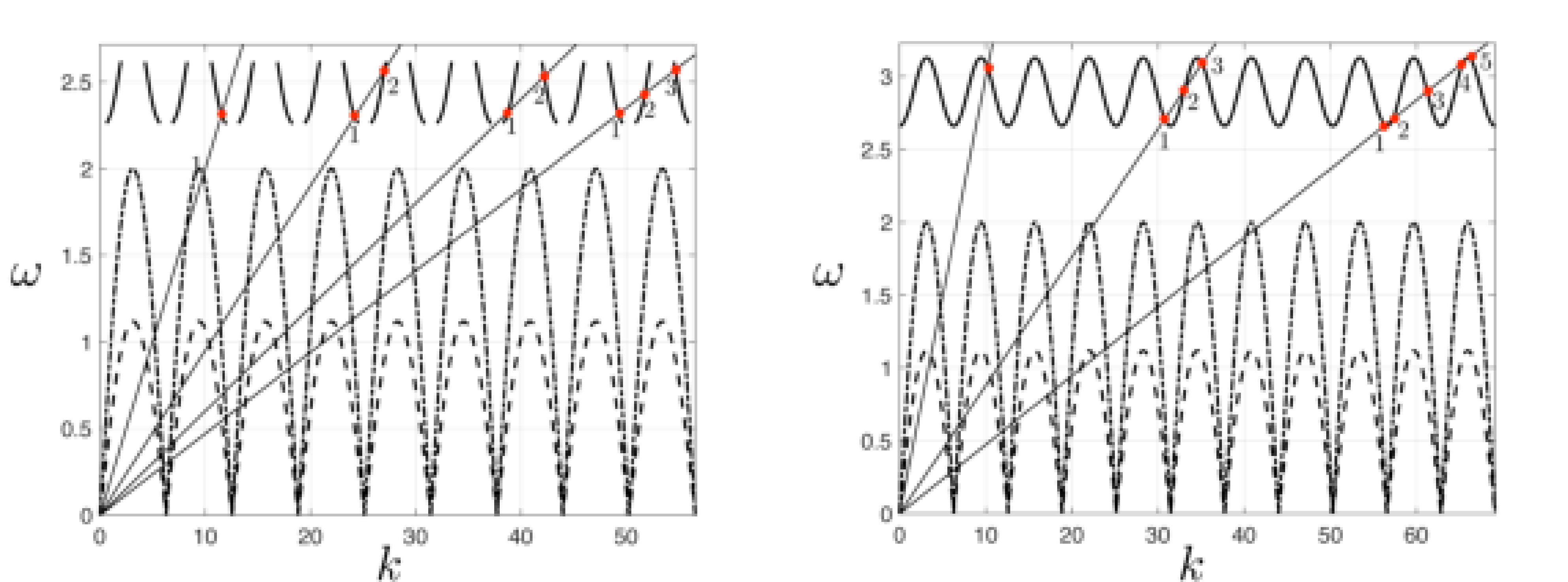}
 }
 (a)~~~~~~~~~~~~~~~~~~~~~~~~~~~~~~~~~~~~~~~~~~~~~~~~~~~~~~~~~~~~~~~~~~~~(b)
\caption{Some representative dispersion diagrams for the case when (\ref{con1})  is satisfied. We show the acoustic branches $\omega_1$ (dash-dot curve) and  $\omega_2$ (dashed curve)
in (\ref{eq:DispersionRelations}) and the optical branch $\omega^{\text{op}}_1$ (black solid curve)  as the solution of (\ref{eqzeros}). 
The diagrams are also  supplied with the lines $\omega=kv$. These illustrations  are presented  for $\beta=5/4$, $\gamma=1/4$, $\alpha_1=\alpha_2= 1$ and (a) $\mu=5/4$ and (b) $\mu=5/2$. The slope of the lines in (a) are $v=0.047, 0.06$,  $0.095$ and $0.2$, whereas in (b) $v=0.0473, 0.088$ and 0.3.  The red dots in both diagrams represent the intersections of the lines $\omega=kv$ with  $\omega_1^{\text{op}}$. 
 We also supply each red  dot with an integer $j$ and the coordinate of this dot is $(h_j, \omega_1^{\text{op}}(h_j))$. 
}
\label{crack_disp_2}
\end{center}
\end{figure}

\begin{enumerate}[$\bullet$]
\item \underline{Case 1}: If (\ref{con1}) does not hold, then the  optical branches $\omega_j^{\text{op}}$, $j=1,2,$ do not exist. In such a case, only evanescent waves can propagate ahead of the crack front. These are associated with the zero points of $L(k, {\rm i}kv)$, located in the lower half of the complex plane defined by $k$ {having non-negligible imaginary parts}.  In this scenario the dispersion diagram is as reported in Figure \ref{crack_disp_1}.

\item \underline{Case 2}: When the high frequency  optical branch 
$\omega_1^{\text{op}}$  exists and this is either a piecewise defined curve or a continuous curve. Let the pairs of wave numbers associated with the intersection points of  $\omega^{\text{op}}_1$ with the line $\omega=kv$ be $k=\pm h_1, \pm h_2$, $\dots$, $\pm h_N$, ${N}\in \mathbb{Z}$, $N\ge 1$. They represent zero points of $L(k, {\rm i}kv)$.
Here, if $\omega_1^{\text{op}}$ is piecewise defined, the number of pairs $N$ can be even or odd depending on $v$. Figure \ref{crack_disp_2}(a) shows several examples for this configuration, where $N=3$ and $2,$ for $v=0.047$ and $v=0.06$ and  $0.095$, respectively.
In the situation when $\omega_1^{\text{op}}$ is continuous, $N$ is always odd as illustrated in Figure \ref{crack_disp_2}(b).

\item \underline{Case 3}: Additionally to Case 2, if  one of the conditions (\ref{con2}) holds,  then the low frequency  optical branch 
$\omega_2^{\text{op}}$ may also appear (and this depends on the value of $\mu$ not present in (\ref{con2})).  This curve  is always  piecewise defined as discussed in the main text. Let the pairs of wave numbers associated with the intersection points of  $\omega^{\text{op}}_2$ with the line $\omega=kv$ be $k=\pm r_1, \pm r_2$, $\dots$, $\pm r_P$, ${P}\in \mathbb{Z}$, $P\ge 1$. As a result of $\omega^{\text{op}}_2$ being always piecewise  defined, the integer $P$ can be either even or odd depending on the crack speed. Figure \ref{crack_disp_3} demonstrates this, where for $v=0.027$ and 0.05  the number of pairs of zero points of $L(k, {\rm i}kv)$ are  $P=3$ and 2, respectively.
\end{enumerate}
\begin{figure}[htbp]
\begin{center}
{\includegraphics[width=0.6\textwidth]{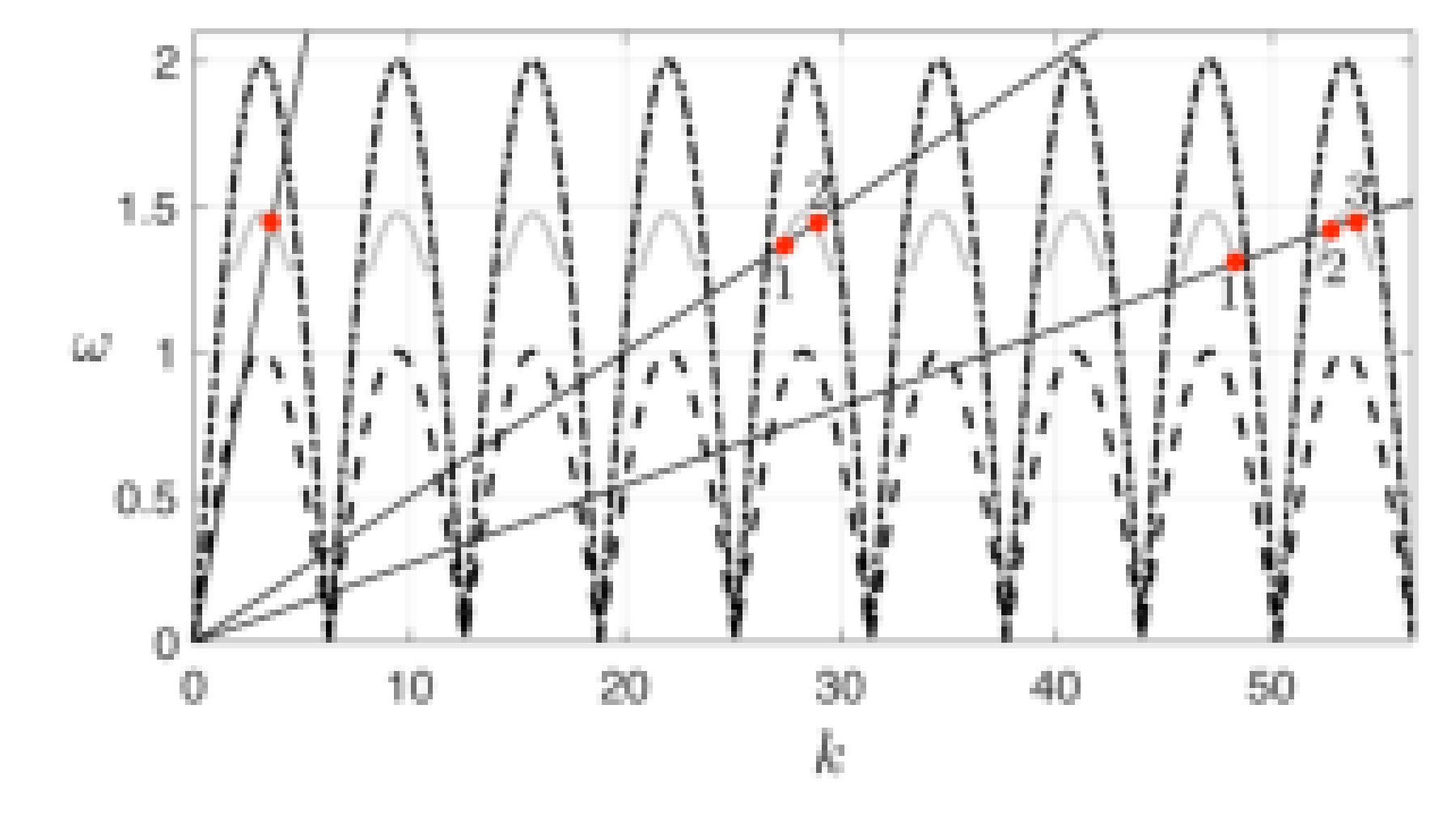}
 }
\caption{{We show the acoustic branches $\omega_1$ (dash-dot curve) and  $\omega_2$ (dashed curve)
in (\ref{eq:DispersionRelations}) and the optical branch $\omega^{\text{op}}_2$ (grey solid curve)  that is a solution of (\ref{eqzeros}). 
In addition,  lines corresponding to $\omega=kv$ are shown. The dispersion diagram is for a lattice with $\beta=\gamma=\alpha_1=\alpha_2=1/2$, $\mu=6$ and the slope of the lines are $v=0.027, 0.05$ and 0.4. The red dots indicate the intersections of the lines $\omega=kv$ with  $\omega_2^{\text{op}}$. For  $v=0.027$ and $0.05$, the intersection points are accompanied by integers $j$ and the coordinates of the points are $(r_j, \omega_2^{\text{op}}(r_j))$. }
}
\label{crack_disp_3}
\end{center}
\end{figure}

We can again compare the slope of the line $\omega=kv$ with the group velocity of the optical branches at the intersection points identified in Cases 2 and 3. Then for the intersection points with the wave numbers:

\begin{enumerate}
\item 

\begin{equation}\label{htr}
k=\left\{\begin{array}{ll}
\pm h_2, \pm h_4, \dots, \pm h_{2\lfloor \frac{N}{2}\rfloor}\;, & \quad \text{ if } N>1 \text{ and } v_g(h_1)<v\;,\\
 \pm h_1, \pm h_3, \dots, \pm h_{2\lceil \frac{N}{2}\rceil-1}\;, & \quad \text{ if } N>1 \text{ and } v_g(h_1)>v
 \end{array}\right.
 \end{equation}
 and 
\begin{equation}\label{rtr}
k=\left\{\begin{array}{ll}
\pm r_2, \pm r_4, \dots, \pm r_{2\lfloor \frac{P}{2}\rfloor}\;, &\quad  \text{ if $P>1$ and $v_g(r_1)<v$,} \\
\pm r_1, \pm r_3, \dots, \pm r_{2\lceil \frac{P}{2}\rceil-1}\;, &\quad  \text{ if $P>1$ and $v_g(r_1)>v$,} 
\end{array}\right.
\end{equation}
the group velocity $v_g>v$ (see Figures \ref{crack_disp_2} and \ref{crack_disp_3}). Here $\lfloor \cdot \rfloor$ and $\lceil\cdot \rceil$ denote the floor and ceil functions, respectively. These wave numbers are connected with simple poles of the function $L(k, {\rm i}kv)$ and lead to waves transmitted  ahead of the crack front.

\item  
\[k=\left\{\begin{array}{ll}
\pm h_1, \pm h_3, \dots, \pm h_{2\lceil \frac{N}{2}\rceil-1} & \quad \text{ if } N>1 \text{ and } v_g(h_1)<v\;,\\
 \pm h_2, \pm h_4, \dots, \pm h_{2\lfloor \frac{N}{2}\rfloor} & \quad \text{ if } N>1 \text{ and } v_g(h_1)>v
 \end{array}\right. \]
and 
\[k=\left\{\begin{array}{ll}
\pm r_1, \pm r_3, \dots, \pm r_{2\lceil \frac{P}{2}\rceil-1} &\quad  \text{ if $P>1$ and $v_g(r_1)<v$,} \\
\pm r_2, \pm r_4, \dots, \pm r_{2\lfloor \frac{P}{2}\rfloor} &\quad  \text{ if $P>1$ and $v_g(r_1)>v$,} 
\end{array}\right.\]
the group velocity $v_g<v$ (see Figures \ref{crack_disp_2} and \ref{crack_disp_3}). These points are linked to waves created by  a remote oscillating load  ahead of the crack front. This problem {will be considered somewhere else and hence these wave numbers} can be neglected when characterising the dynamics of the lattice.
\end{enumerate}


The above information concerning the group velocity at the intersection points is used in Section \ref{asympL} onwards in the tracing the form of the waves in the lattice. Indeed, it allows us to determine the location of the wave numbers in the complex plane defined by $k$, prior to taking the limit (\ref{eq:Limit_s}).  This concludes the analysis of dispersive properties for the system undergoing failure.

\subsection{Derivation of asymptotes for $\phi(\eta)$}\label{SM4}
Here we derive the asymptotic behaviour of the function $\phi(\eta)$ (see (\ref{psiphi})) in the vicinity of the crack tip.
We consult its Fourier transform $\Phi(k)$, which owing to  (\ref{eq:WienerHopf_relationA})
takes the equivalent  form
\begin{equation}\label{Psik_rep_LM}
\Phi(k)=\epsilon_c\left[\frac{K}{0+{\rm i}k}+\left(K-\frac{M(k)}{L(k)}\right)\frac{1}{0-{\rm i}k}\right]L^-(k)\;,
\end{equation}
where (\ref{eq:Asymptotics_L_zero}) and (\ref{eq:Asymptotics_M_zero}) show that for small wave numbers 
\begin{equation}\label{MoL}
\frac{M(k)}{L(k)}\sim K+\hat{K}\sqrt{(0+{\rm i}k)(0-{\rm i}k)},
\end{equation}
with $K$ is defined in (\ref{KXi0}) and $\bar{K}$ is another constant that we do not determine here.
with $K$ being the constant in (\ref{KXi0}).
Using (\ref{eq:Asymptototics_L+-_zero}) we can also obtain the asymptote 
\begin{equation}\label{asympratio1}
\left(K-\frac{M(k)}{L(k)}\right)\frac{L^-(k)}{0-{\rm i}k}\sim\frac{\Xi_0}{\sqrt{0-{\rm i}k}},\quad k\to0\;,
\end{equation}
where $\Xi_0$ is defined in (\ref{KXi0}).
Note that (\ref{eq:Asymptotics_L_inf}), (\ref{eq:Asymptotics_M_inf}) and (\ref{eq:Asymptotics_L+-_inf}) provide the following asymptote for large wave numbers:
\begin{equation*}
\left(K-\frac{M(k)}{L(k)}\right)\frac{L^-(k)}{0-{\rm i}k}= \frac{K}{0-\text{i}k}\left(1+\frac{Q}{0+{\rm i}k}\right)+O\Big(\frac{1}{k^3}\Big),\quad k\to\infty\;,
\end{equation*}
{where $Q$ is the constant defined in (\ref{eq:Asymptotics_L+-_inf}).} 

Let $\Xi^+(k)$ denote a ``+" function behaving at infinity and near zero in the following manner
\begin{equation}\label{eqxiP1}
\Xi^+ (k)= \frac{\Xi_0}{\sqrt{0-{\rm i}k}}+O(1)\;,\qquad  k\to 0\;,\end{equation}
and 
\begin{equation}\label{eqxiP2}
\Xi^+(k)=\frac{K}{0-\text{i}k}+O\Big(\frac{1}{k^2}\Big)\;,\qquad k \to \infty\;.
\end{equation}
Further, using 
 (\ref{eq:SolutioFourier_Psi}), with (\ref{Ceq}), let $\Phi(k)$  take the form
\begin{equation}
\Phi(k)=K\Psi^-(k)+\epsilon_c\left(K-\frac{M(k)}{L(k)}\right)\frac{L^-(k)}{0-{\rm i}k}\;.
\label{eq:Function_Phi}
\end{equation}
Now from this we have
\begin{equation}\label{PpPm}
\Phi^+(k)+\Phi^-(k)-K\Psi^-(k)-\varepsilon_c \Xi^+(k)=\varepsilon_c \left[ \left(K-\frac{M(k)}{L(k)}\right)\frac{L^-(k)}{0-{\rm i}k}-\Xi^+(k)\right]=\varepsilon_c [\mathfrak{G}^+(k)+\mathfrak{G}^-(k)]\;.
\end{equation}
Here, the last term brackets  is a bounded function near zero and is $o(1/k)$ as $k\to \pm \infty$. Additionally,  the Cauchy split has been employed  to represent this term as the sum of a ``+"  and a ``$-$" function. As a result, we can obtain $\Phi^{\pm}$ as
\[\Phi^+=\varepsilon_c \Xi^+(k)+\varepsilon_c \mathfrak{G}^+(k)\;,
\quad\text{ and }\quad
\Phi^-=K\Psi^-(k)+\varepsilon_c \mathfrak{G}^-(k)\;.\]
Thus, the asymptote of the function $\phi$, using (\ref{eq:Asymptotics_Lattice_Fourier_1}), (\ref{eq:Asymptotics_Lattice_Fourier_3}), (\ref{eqxiP1}) and (\ref{eqxiP2}) near the crack tip is
\begin{equation}\label{phiasymp}
\phi(\eta)=\varepsilon_c(K+\mathfrak{G}_\infty)+O(\eta),\quad \eta\to0\;,
\end{equation}
with 
\[\mathfrak{G}_{\infty}=\frac{1}{2\pi}\int_{-\infty}^\infty \left[\left(K-\frac{M(\xi)}{L(\xi)}\right)\frac{L^-(\xi)}{0-{\rm i}\xi}-\Xi^+(\xi) \right]d\xi\]
and far away ahead of the crack tip we have (\ref{asympphi1}), owing to (\ref{asympratio1}) and (\ref{eqxiP1}), whereas in the far-field behind the crack tip we receive and (\ref{asympphi2}) using (\ref{eq:Asymptotics_Lattice_Fourier_3}).

\subsection{Derivation of the far-field behaviour in the dissimilar lattice undergoing fracture}\label{SM5}
To estimate the asymptotics of $u_m(\eta)$ and $w_m(\eta)$ when $\eta,m\to\infty$ we need to consider their Fourier transforms $U_m(k)$ and $W_m(k)$, given in (\ref{eq:LambdaIntroduction}), when $k\to0$. First, we consider the solutions (\ref{eq:WienerHopf_relationAB}). The function $\Phi(k)$ can be written as in (\ref{eq:Function_Phi}).
 Owing to (\ref{MoL}), (\ref{eq:Asymptototics_L+-_zero}) and (\ref{eq:Asymptotics_L_zero}) 
it can be shown (\ref{eq:Asymptotics_Lattice_Fourier_3}) holds.

 It follows from (\ref{eq:Asymptotics_Lattice_Fourier_3}),  (\ref{asympratio1}) and  (\ref{eq:Function_Phi})  that
\begin{equation}
\Phi(k)\sim \epsilon_c\frac{\Theta}{R}\frac{K}{(0+{\rm i}k)^{3/2}}\quad \text{ for } k\to 0\;.
\end{equation}

Using  (\ref{psiphi}),  (\ref{FTPsiPhi}),  (\ref{psieq2}) and  (\ref{eq:Asymptotics_Lattice_Fourier_3}), the Fourier transforms $U_1(k)$ and $W_1(k)$ behave as follows when $k\to 0$:
\begin{equation}
U_1(k)\sim\epsilon_c\frac{\Theta}{R}\frac{K+1}{2}\frac{1}{(0+{\rm i}k)^{3/2}},\quad \text{ and }
W_1(k)\sim\epsilon_c\frac{\Theta}{R}\frac{K-1}{2}\frac{1}{(0+{\rm i}k)^{3/2}}
\end{equation}
The displacements of the remaining layers are found through (\ref{eq:LambdaIntroduction}).
The factors $\lambda_{1,2}(k)$ contained there possess the following asymptotic behaviour:
\begin{equation}
\lambda_{1,2}(k)\sim 1-a_{1,2}\sqrt{(0+{\rm i}k)(0-{\rm i}k)}\sim e^{a_{1,2}\sqrt{(0+{\rm i}k)(0-{\rm i}k)}},\quad k\to0,
\end{equation}
where constants $a_{1,2}>0$ are
\begin{equation}
a_j=\sqrt{\frac{v_j^2-v^2}{2\beta^{j-1}\alpha_j}},\quad j=1,2.
\end{equation}
This gives the following leading asymptotic terms of $U_m(k)$ and $W_{-m}(k)$ when $k\to 0$:
\begin{equation}\label{Um0asymp}
U_m\sim \epsilon_c\frac{\Theta}{R}\frac{K+1}{2}\frac{1}{(0+{\rm i}k)^{3/2}}e^{a_1(m-1)\sqrt{(0+{\rm i}k)(0-{\rm i}k)}}
\end{equation}
and 
\begin{equation}
W_{-m}\sim \epsilon_c\frac{\Theta}{R}\frac{K-1}{2}\frac{1}{(0+{\rm i}k)^{3/2}}e^{a_2(-m-1)\sqrt{(0+{\rm i}k)(0-{\rm i}k)}}\;.
\end{equation}
To obtain the far-field asymptotes, we need to calculate the inverse Fourier transforms of these leading terms. 
We concentrate on $U_m$ as the derivation involving $W_m$ is essentially the same up to constants. 
We need to calculate:
\begin{equation}\label{intSM1A}
\frac{1}{2\pi}\int_{-\infty}^{\infty}\frac{1}{(0+{\rm i}k)^{3/2}}e^{a_1(m-1)\sqrt{(0+{\rm i}k)(0-{\rm i}k)}}e^{-{\rm i}k\eta}\,dk\;.
\end{equation}
First, we can take $\eta=\alpha(m-1)$ where $\alpha$ is some constant and we have
\begin{equation}\label{intSM1}
\frac{1}{2\pi}\int_{-\infty}^{\infty}\frac{1}{(0+{\rm i}k)^{3/2}}e^{(a_1\sqrt{(0+{\rm i}k)(0-{\rm i}k)}-{\rm i}\alpha k)(m-1)}\,dk\;.
\end{equation}
There are two possible cases: $\alpha>0$ and $\alpha<0$. Moreover, the integrand has branch cuts resulting from the terms $\sqrt{0\mp {\rm i}k}$ contained between the points $\mp 0{\rm i}$ to $\mp {\rm i}\infty$, respectively. When $\alpha>0$ we make a contour defined for $\text{Im }k\le 0$ which runs along the real line, a semi-circle with a sufficiently large radius and the branch cut  from $-0{\rm i}$ to $-{\rm i}\infty$. Allowing the radius of the semi-circle to tend to  infinity and taking $k=-{\rm i}z$ we reduce the integration from the real line to the integration along the branch cuts with respect to $z>0$ due to Jordan's lemma of the complex analysis. We have $\sqrt{0-{\rm i}k}=-{\rm i}z$ on the right side of the branch cut for which $k=z\exp{(-{\rm i}\pi/2)}$ whereas $\sqrt{0-{\rm i}k}={\rm i}z$ on its left side where $k=z\exp{(3i\pi/2)}$. Therefore, we get:
\begin{eqnarray}
&&\frac{1}{2\pi}\int_{-\infty}^{\infty}\frac{1}{(0+{\rm i}k)^{3/2}}e^{(a_1\sqrt{(0+{\rm i}k)(0-{\rm i}k)}-{\rm i}\alpha k)(m-1)}\,dk\nonumber \\
&=&\frac{-i}{2\pi}\left[\int_{\infty}^{0}\frac{e^{-{\rm i}a_1(m-1)z}}{z\sqrt{z}}e^{-\alpha (m-1)z}\,dz+\int_{0}^{\infty}\frac{e^{{\rm i}a_1(m-1)z}}{z\sqrt{z}}e^{-\alpha (m-1)z}\,dz\right]\;.
\end{eqnarray}
As the last two terms on the right-hand side are complex conjugate to each other, this is equal to 
\[\frac{1}{\pi}\int_{0}^{\infty}\frac{\sin{(a_1(m-1)z)}}{z\sqrt{z}}e^{-\alpha(m-1)z}\,dz=\frac{\sqrt{a_1(m-1)}}{\pi}\int_{0}^{\infty}\frac{\sin{z}}{z\sqrt{z}}e^{-(\alpha/a_1) z}\,dz\]
where a change of variable was made in moving to the last equality.
After computing the  standard integral in the right-hand side we have this is equal to 
\[\sqrt{\frac{2}{\pi}(m-1)}\frac{a_1}{\sqrt{\alpha+\sqrt{\alpha^2+a_1^2}}}\]
Thus, we have shown that 
\begin{eqnarray}
&&\frac{1}{2\pi}\int_{-\infty}^{\infty}\frac{1}{(0+{\rm i}k)^{3/2}}e^{(a_1\sqrt{(0+{\rm i}k)(0-{\rm i}k)}-{\rm i}\alpha k)(m-1)}\,dk\nonumber \\
&=&a_1\sqrt{\frac{2}{\pi}}\frac{(m-1)}{\sqrt{\eta+\sqrt{\eta^2+a_1^2(m-1)^2}}}
=\sqrt{\frac{2}{\pi}}\sqrt{\sqrt{\eta^2+a_1^2(m-1)^2}-\eta},\quad \eta>0\;.\label{SM34}
\end{eqnarray}
For $\alpha<0$, we have to handle the singular branch point of the kernel in (\ref{intSM1}) at $k={\rm i}0$. We first represent the original integral in the following way:
\begin{equation}
\begin{gathered}
\frac{1}{2\pi}\int_{-\infty}^{\infty}\frac{1}{(0+{\rm i}k)^{3/2}}e^{(a_1\sqrt{(0+{\rm i}k)(0-{\rm i}k)}-i\alpha k)(m-1)}\,dk\\
=\frac{1}{2\pi}\int_{-\infty}^{\infty}\frac{1}{(0+{\rm i}k)^{3/2}}\left(e^{a_1(m-1)\sqrt{(0+{\rm i}k)(0-{\rm i}k)}}-1\right)e^{-i\alpha k(m-1)}\,dk+\frac{1}{2\pi}\int_{-\infty}^{\infty}\frac{1}{(0+{\rm i}k)^{3/2}}e^{-{\rm i}\alpha k(m-1)}\,dk
\end{gathered}
\label{eq:Integral_neg_alpha}
\end{equation}
We construct a similar contour to as before but only in the upper half-plane $\text{Im }k>0$ with the branch cut of $\sqrt{0+{\rm i}k}$ running from $0{\rm i}$ to ${\rm i}\infty$. With the same arguments the integration can be calculated along the branch cuts through the change of variables $k={\rm i}z,\,z>0$. On the right side of the cut with $k=z\exp{({\rm i}\pi/2)}$ we have $\sqrt{0+{\rm i}k}={\rm i}z$, whereas on the left side of the cut when $k=z\exp{(-3{\rm i}\pi/2)}$ we have $\sqrt{0+{\rm i}k}=-{\rm i}z$. After integration by parts, the second integral in \eqref{eq:Integral_neg_alpha} gives:
\begin{equation}
\frac{1}{2\pi}\int_{-\infty}^{\infty}\frac{1}{(0+{\rm i}k)^{3/2}}e^{-{\rm i}\alpha k(m-1)}\,dk=\frac{1}{\pi}\int_{-\infty}^{\infty}\frac{-\alpha(m-1)}{\sqrt{0+{\rm i}k}}e^{-{\rm i}\alpha k(m-1)}\,dk\;.
\label{eq:integral_1}
\end{equation}
Next, repeating similar steps to as before, the above described  contour integration in ${\text{Im } }k>0$ allows us to write:
\begin{eqnarray}
\frac{1}{\pi}\int_{-\infty}^{\infty}\frac{-\alpha(m-1)}{\sqrt{0+{\rm i}k}}e^{-{\rm i}\alpha k(m-1)}\,dk&=&-\frac{-\alpha(m-1)}{\pi}\left(\int_{\infty}^0\frac{1}{\sqrt{z}}e^{\alpha(m-1)z}\,dz-\int_0^{\infty}\frac{1}{\sqrt{z}}e^{\alpha(m-1)z}\,dz\right)\nonumber \\
&=&\frac{-\alpha(m-1)}{\pi}\frac{2\sqrt{\pi}}{\sqrt{-\alpha(m-1)}}=\frac{2}{\sqrt{\pi}}\sqrt{-\eta},\quad \eta<0.\label{SM36}
\end{eqnarray}
Next we concentrate on the evaluation of the first integral in \eqref{eq:Integral_neg_alpha}. Owing to a change variable this becomes:
\begin{eqnarray}
&& \frac{1}{2\pi}\int_{-\infty}^{\infty}\frac{1}{(0+ik)^{3/2}}\left(e^{a_1(m-1)\sqrt{(0+ik)(0-ik)}}-1\right)e^{-i\alpha k(m-1)}\,dk\nonumber \\
&=&-\frac{1}{2\pi}\left[\int_{\infty}^{0}\frac{(e^{ia_1(m-1)z}-1)}{-z\sqrt{z}}e^{\alpha (m-1)z}\,dz+\int_{0}^{\infty}\frac{(e^{-ia_1(m-1)z}-1)}{z\sqrt{z}}e^{\alpha (m-1)z}\,dz\right]\label{SM37}
\end{eqnarray}
where again as the terms in the above right-hand side are complex conjugate to each other, we combine them to obtain the integral
\[\frac{1}{\pi}\int_{0}^{\infty}\frac{1-\cos{(a_1(m-1)z)}}{z\sqrt{z}}e^{\alpha (m-1)z}\,dz\;.\]
Introducing next a change of variable then leads to 
\[\frac{\sqrt{a_1(m-1)}}{\pi}\int_0^{\infty}\frac{1-\cos{z}}{z\sqrt{z}}e^{(\alpha/a_1)z}\,dz\]
that can be evaluated to give
\begin{eqnarray}
&&\frac{1}{2\pi}\int_{-\infty}^{\infty}\frac{1}{(0+{\rm i}k)^{3/2}}\left(e^{a_1(m-1)\sqrt{(0+{\rm i}k)(0-{\rm i}k)}}-1\right)e^{-{\rm i}\alpha k(m-1)}\,dk\nonumber \\
&=& \frac{2}{\sqrt{\pi}}\left(-\sqrt{-\alpha(m-1)}+\left(\frac{\alpha^2}{a_1^2}+1\right)^{1/4}\cos{\left(\frac{\text{acot}(-\alpha/a_1)}{2}\right)}\right)\nonumber \\
&=&\frac{2}{\sqrt{\pi}}\left(-\sqrt{-\eta}+\left(\frac{\alpha^2}{a_1^2}+1\right)^{1/4}\cos{\left(\frac{\text{acot}(-\alpha/a_1)}{2}\right)}\right),\quad \eta,\alpha<0, \label{SM38}
\end{eqnarray}
where $\text{acot } x$ is the inverse cotangent function. We recall the following properties:
\begin{equation}
\text{acot}\left(\frac{1}{x}\right)=\text{atan}(x),\quad \text{atan}(x)=2\text{atan}\left(\frac{x}{1+\sqrt{1+x^2}}\right),\quad \cos{(\text{atan}(x))}=\frac{1}{\sqrt{1+x^2}}
\end{equation}
and by combining this with \eqref{SM38}  and \eqref{eq:Integral_neg_alpha}--\eqref{SM36}, 
we obtain:
\begin{equation}
\begin{gathered}
\frac{1}{2\pi}\int_{-\infty}^{\infty}\frac{1}{(0+{\rm i}k)^{3/2}}e^{(a_1\sqrt{(0+{\rm i}k)(0-{\rm i}k)}-{\rm i}\alpha k)(m-1)}\,dk=a_1\sqrt{\frac{2}{\pi}}\frac{(m-1)}{\sqrt{-\eta+\sqrt{\eta^2+a_1^2(m-1)^2}}}\\
=\sqrt{\frac{2}{\pi}}\sqrt{\sqrt{\eta^2+a_1^2(m-1)^2}+\eta},\quad \eta<0\;,
\end{gathered}
\end{equation}
which is the same result as for $\alpha>0$.
This and the result (\ref{SM34}) with (\ref{intSM1A}) and (\ref{intSM1}) show that the asymptotic behaviour for large $|\eta|$ or $|m|$ of $u_m$ is
\begin{equation}\label{Um0asymp}
u_m(\eta) \sim \epsilon_c\frac{\Theta}{R}\frac{K+1}{\sqrt{2\pi}}\sqrt{\sqrt{\eta^2+a_1^2(m-1)^2}-\eta}\;.
\end{equation} 
If we then make the substitution $\eta=\rho\cos{\theta}$ and $m=\rho\sin{\theta}$, we can obtain (\ref{eq:Asymptotic_ma}) for  $\rho\to\infty$.
In a similar way to as discussed here, one can show that for large $|\eta|$ or $|m|$ that 
\begin{equation}\label{Wm0asymp}
w_{-m}(\eta) \sim \epsilon_c\frac{\Theta}{R}\frac{K-1}{\sqrt{2\pi}}\sqrt{\sqrt{\eta^2+a_1^2(m-1)^2}-\eta}\;.
\end{equation} 
from which (\ref{eq:Asymptotic_m}) follows.

\subsection{Wave radiation segments}\label{SM6}

Let $k=p^*_{j}$, $1\le j \le 2f^*+1$ denote the zeros of the function $\Omega_1(k, {\rm i}kv)+1$ and $k=q^*_{j}$, $1\le j \le 2g^*+1$ be the zeros of $\Omega_2(k, {\rm i}kv)+1$. 
Here $f^*$ and $g^*$ are dependent on the crack speed. 
The wave numbers $k=p^*_{j}$, $1\le j \le 2f^*+1$ correspond to intersection points of the line $\omega=kv$ with
\begin{equation}
\omega_1^*(k)=\sqrt{\omega_1(k)^2+4\alpha_1}\;,
\end{equation}
on the dispersion diagram.
On the other hand, $k=q^*_{j}$, $1\le j \le 2g^*+1$ represent the wave numbers where the expression $\omega=kv$ and 
\begin{equation}
\omega_2^*(k)=\sqrt{\omega_2(k)^2+4\beta \alpha_2}\;,
\end{equation}
are equal.

Let $\mathcal{K}_1=(p_{2f+1}, p_1^*)$ and $\mathcal{K}_2=(q_{2g+1}, q_1^*)$. 
We define the sets $\Sigma_1$ as in (\ref{SSIG1}) and 
\[ \Sigma_2:=\Big(\bigcup_{j=0}^{f-1}{(q_{2j+1} , q_{2j+2})}\Big)\cup \mathcal{K}_2 \cup  \Big(\bigcup_{j=1}^{f^*}{(q_{2j}^* , q_{2j+1}^*)}\Big)\;. \]
Here, $\mathcal{K}_j$, $j=1,2,$ are radiation or $\mathcal{K}$-segments (see \citet{Slepyan_wave_rad}), that define  principal directions of the lattice along which waves can be radiated.

\begin{figure}[htbp]
\begin{center}
{\includegraphics[width=1\textwidth]{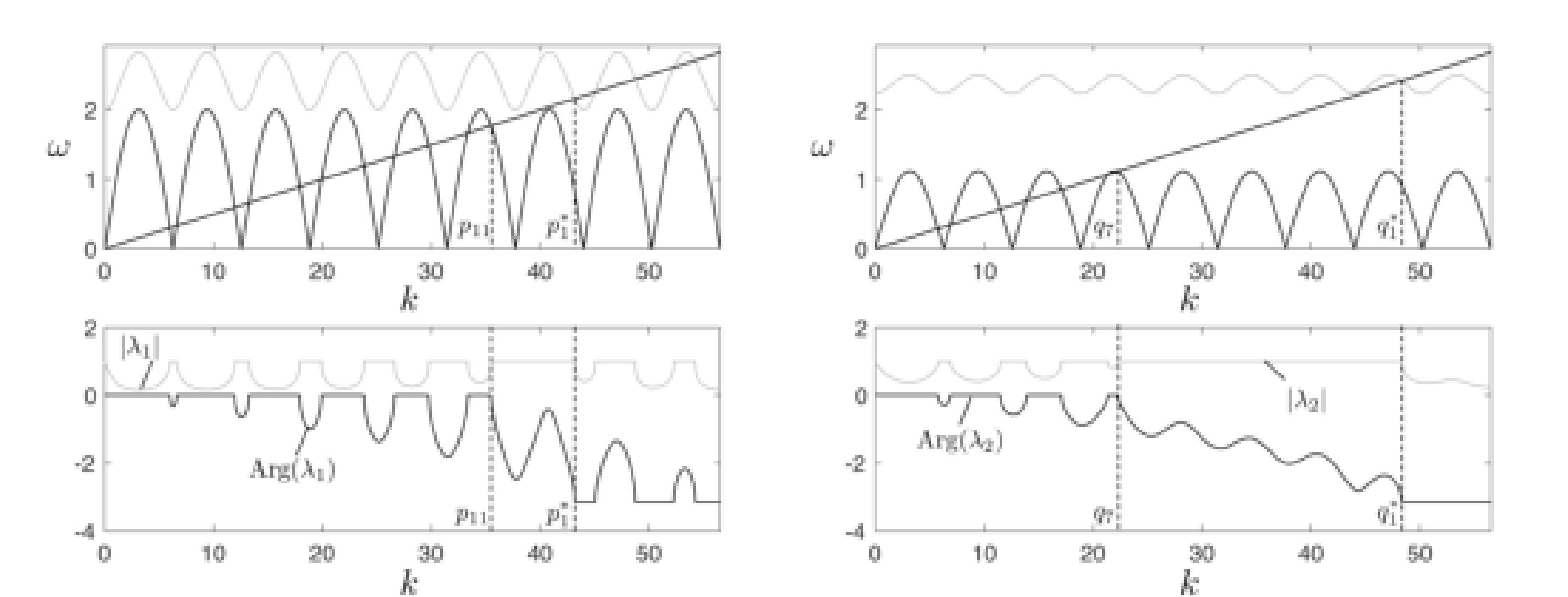}
 }~~~~(a)~~~~~~~~~~~~~~~~~~~~~~~~~~~~~~~~~~~~~~~~~~~~~~~~~~~~~~~~~~~~~~~~~~(b)
\caption{Here we show in the top panels of (a)  $\omega_1(k)$ and $\omega_1^*(k)$ and (b)   $\omega_2(k)$ and $\omega_2^*(k)$ as functions of the wavenumber. In these diagrams the line $\omega=kv$ is also supplied, where $v=0.05$. In correspondence with the top panels of (a) and (b), in the bottom panels we present the behaviours of $|\lambda_j|$ and  $\text{Arg}(\lambda_j)$, $j=1,2,$ as functions of the wave number. The diagrams are presented for the case $\beta=5/4$, $\gamma=1/4$ and $\alpha_1=\alpha_2=1$.
 }
\label{wave_rad}
\end{center}
\end{figure}

Inside $\Sigma_j$ the argument of $\lambda_j$ is negative. Note for $k \in {\cal K}_j$, $\lambda_j$ in (\ref{eq:Lambda_final}) is complex and can be written as
\begin{equation}\label{lam_complex}
\lambda_j(k, 0+{\rm i} k v)=e^{-{\rm i}H_j(k)}\;, \qquad H_j(k)=2\arctan\Big(\frac{\sqrt{1-\Omega_j(k, 0+{\rm i}kv)}}{
\sqrt{1+\Omega_j(k, 0+{\rm i}kv)}}\Big)\;, \quad j=1,2\;.
\end{equation}
Outside of the sets $\Sigma_j$,  we have $|\lambda_j|< 1$, $j=1,2$.
We use the above facts  to determine the far-field behaviour of the lattice displacements  through the inverse Fourier transform.

The behaviour of $\lambda_j$, $j=1,2,$  is demonstrated for a representative example in Figure \ref{wave_rad}. The top panels of Figure \ref{wave_rad}  show the curves of $\omega_j(k)$ and $\omega^{*}_j(k)$, $j=1,2$, along with the line $\omega=kv$ for a given crack speed $v$. 
In the bottom panel of Figure \ref{wave_rad}(a), the behaviour of $|\lambda_1|$ and $\text{Arg}(\lambda_1)$ is shown. The argument of $\lambda_1$ is non-uniform for $k\in \Sigma_1$, where $|\lambda_1|=1$, and $\text{Arg}(\lambda_1)$ has several stationary points. Furthermore, inside ${{\cal K}_1}$ $\text{Arg}(\lambda_1)=-H_1(k)$. For $k \notin \overline{\Sigma_1}$,  with the over line indicating the closure of this set, $\lambda_1$ takes its values on the interval $(-1, 1)$. Similar comments are also valid for $\lambda_2$, where for $k \in {\cal K}_2$ we have  $\text{Arg}(\lambda_2)=-H_2(k)$.

Additionally, in the  example of  Figure \ref{wave_rad}(a),  $f=5$ and the interval   ${\cal K}_1=(p_{11}, p_1^*)$ is indicated.
For Figure \ref{wave_rad}(b), $g=3$ and the interval ${\cal K}_2=(q_7, q_1^*)$ is also shown. In these intervals, the bottom panels of Figure \ref{wave_rad} show the arguments of $\lambda_j$, $j=1,2,$ can oscillate. In particular, 
Figure \ref{wave_rad}(a) shows $\text{Arg}(\lambda_1)$ has an inflection point, whereas in Figure \ref{wave_rad}(b), the function  $\text{Arg}(\lambda_2)$ has several inflection points. 

\subsubsection*{Wave radiation in the lattice bulk}

Here, we discuss how the information about $\mathcal{K}$-segments  is used to 
determine the propagation direction of slowly decreasing waves in the lattice. As shown above, there may exist several stationary points of the function $\text{Arg}(\lambda_j)$  inside the set ${\Sigma}_j$, $j=1,2$. Moreover, there can exist inflection points of $H_j(k)$ in the radiation segment  $\mathcal{K}_j$, $j=1,2$. In association with the inflection points of $H_j$, $j=1,2$, we define the sets
\[Q^+_j:=\{k: H^{\prime\prime}_j(k)=0 \text{ and } H^{\prime}_j(k)>0\}\;,\]
and 
\[Q^-_j:=\{k: H^{\prime\prime}_j(k)=0 \text{ and } H^{\prime}_j(k)<0\}\;,\] 
for $j=1,2$, (see (\ref{lam_complex})). 
According to Section \ref{DFLattice}, the third integral in the right-hand side of (\ref{formulaUint}) can be used to find the rays in the lattice where  wave radiation created by the propagating crack is the most significant. This term leads to the following  contribution to the lattice solution $u_{m+1}$, $m \ge 0$:
\begin{equation}\label{u_wave_rad}
u^{(W)}_{m+1}(\eta)=\frac{1}{\pi} \text{Re}\Big\{\int_{\Sigma_1}U_1(k)e^{-{\rm i}\mathcal{P}(k)}dk\Big\}\;,
\end{equation}
where
\[\mathcal{P}(k)=
- \text{Arg}(\lambda_1)m +k\eta  
\]
Note for $k\in {\cal K}_1$, using (\ref{lam_complex}) we have  $\mathcal{P}(k)=H_1(k)m +k\eta$. The sets 
$Q_1^{\pm}\subset {\cal K}_1$  then  define preferential directions relative  propagating crack tip, where wave radiation amplitude is most significant in the limit $m\to \infty$.
Indeed, we use $k^*\in Q^{\pm}_1$ to define lines in the upper and lower lattice half-planes in the form
\begin{equation}\label{eta_star}
\zeta_1(k^*)=-mH^{\prime}_1(k^*)\;.
\end{equation}
along which the phase $\mathcal{P}(k)$ in (\ref{u_wave_rad}) has a stationary point, i.e. $\mathcal{P}^{\prime}(k^*)=0$. As the cardinality of $Q^{\pm}_1$ may exceed one, several lines of this type may exist. If $k^*\in {Q}^+_1$,  then (\ref{eta_star}) represents a line emanating from the crack tip located in the upper  half space at $\eta<0$.
For $k\in Q^-_1$, these lines are located at $\eta>0$.

Then, employing a stationary phase argument as adopted by  \citet{Slepyan_wave_rad}, one determines an asymptote for the function $u_{m+1}^{(W)}$ as $m\to \infty$ that reveals along the lines (\ref{eta_star}) the term $u_{m+1}^{(W)}$ has the order $O(m^{-1/3})$. A  similar approach also shows that such rays exist in the lower lattice defined by $\zeta_2(k^*)=mH^{\prime}_2(k^*)$, where now $k^*\in Q_2^\pm$. As noted by  \citet{Slepyan_wave_rad}, this phenomenon is an effect of the discrete medium and is not observed in the continuum model of  dynamic Mode III crack propagation, as wave radiation due to the crack movement does not occur there.

\subsection{The total transmitted wave field} \label{SM7}
Waves propagating ahead of the crack tip are defined by the wave numbers  in (\ref{htr}) and (\ref{rtr}).
The group velocity of these waves $v_g>v$ and these points are also associated with simple poles in the lower half of the complex plane defined by $k$ in the solution to the problem. 

Indeed, before to taking the limit (\ref{eq:Limit_s}), these simple poles are located at $k=\pm \xi-{\rm i}0$, where $\xi$ is a wavenumber in (\ref{htr}) and (\ref{rtr}). One can also show that the exponents $\lambda_j$ have the following asymptotes at such points:
\begin{equation}\label{lam_asymp_opt}
\lambda_j(k, 0+{\rm i}kv)\sim\lambda_j(\xi, {\rm i}\xi v)+ \frac{{\rm d} \lambda_j}{{\rm d} k}(k, {\rm i}kv)\Big|_{k=\xi}(k-\xi+{\rm i}0)\;, \quad k=\pm \xi-{\rm i}0\;, \quad j=1,2\;.
\end{equation}
{Here, the first term on the right is a real number with modulus less than unity and for $j=1,2,$ these numbers satisfy (\ref{eqzeros}).
In addition,  the coefficient  of the linear term on the right-hand side is generally non-zero. By referring to Figures \ref{disp2_cont}--\ref{disp2opt}, we see this coefficient may only be zero only at  $k=\pi (2z+1)$, $z \in \mathbb{Z}$, whereas according to Supplementary Material \ref{sec_crack_disp} the wave numbers in (\ref{htr}) and (\ref{rtr}) are located between these points (see Figure \ref{crack_disp_2} and \ref{crack_disp_3}).
 
This  information is important in obtaining the form of the interfacial wave modes that propagate along the interface ahead of the crack.}
Through the inverse Fourier transform in (\ref{eq:SolutionLatticeInverseFourier_Layersa}),  the simple poles at $k=\xi-{\rm i}0$ and $k=-\xi-{\rm i}0$ lead to residues complex conjugate to each other that combine to give a real wave.   

Thus, for $\eta>0$, 
by applying a contour integration and the residue theorem,  one can show that the transmitted wave fields $u^{(t)}_m(\eta)$ and $w^{(t)}_m(\eta)$ arising due to the propagation of the crack are 
\begin{eqnarray*}
u^{(t)}_m(\eta)&=&\chi_{\mathbb{R}^-}(v_g(h_1)-v)\sum_{j=1}^{\lfloor N/2\rfloor}
\mathcal{U}^{( m)}(\eta, h_{2j})
+\chi_{\mathbb{R}^+}(v_g(h_1)-v)\sum_{j=1}^{\lceil N/2\rceil}\mathcal{U}^{(m)}(\eta, h_{2j-1})
\\&&+\chi_{\mathbb{R}^+}(v_g(r_1)-v)\sum_{j=1}^{\lceil P/2\rceil}
\mathcal{U}^{(m)}(\eta, r_{2j-1})
+\chi_{\mathbb{R}^-}(v_g(r_1)-v)\sum_{j=1}^{\lfloor P/2\rfloor}\mathcal{U}^{(m)}(\eta, r_{2j})
\end{eqnarray*}
for $m\ge 1$ and
\begin{eqnarray*}
w_m^{(t)}(\eta)&=&\chi_{\mathbb{R}^-}(v_g(h_1)-v)\sum_{j=1}^{\lfloor N/2\rfloor}\mathcal{W}^{(m)}(\eta, h_{2j})
 +\chi_{\mathbb{R}^+}(v_g(h_1)-v)\sum_{j=1}^{\lceil N/2\rceil}\mathcal{W}^{(m)}(\eta, h_{2j-1})
\\
&&+\chi_{\mathbb{R}^+}(v_g(r_1)-v)\sum_{j=1}^{\lceil P/2\rceil}
\mathcal{W}^{(m)}(\eta, r_{2j-1})
+\chi_{\mathbb{R}^-}(v_g(r_1)-v)\sum_{j=1}^{\lfloor P/2\rfloor} \mathcal{W}^{(m)}(\eta, r_{2j})
\end{eqnarray*}
for $m\le -1$, where 
\[\mathcal{U}^{(m)}(\eta, \xi)=[\lambda_1(\xi, {\rm i} \xi v)]^{m-1} | \Upsilon^{(+)}(\xi)|\cos(\xi \eta-\tau^{(+)}(\xi))\;,\]
and 
\[\mathcal{W}^{(m)}(\eta, \xi)=[\lambda_2(\xi, {\rm i} \xi v)]^{-m-1} | \Upsilon^{(-)}(\xi)|\cos(\xi \eta-\tau^{(-)}(\xi))\;.\]
In addition,
\[\Upsilon^{(\pm )}(z)=\lim_{k\to z} (0-{\rm i}(k-z))[\Psi(k)\pm \Phi(k)]\quad \text{ and } \quad \tau^{(\pm)}(z)=\text{arg}(\Upsilon^{(\pm )}(z))\;.\]
Here, we define $\sum_{j=1}^0=0$ and $\chi_{\mathbb{R}^{\pm}}(z)$ is the characteristic function of the set $\mathbb{R}^{\pm}=\{x: x\gtrless 0\}$. 

Note that for $m=\pm1$ unlike the waves radiated behind the crack tip having slowly decaying amplitudes (see \citet{LSbook}),
here the transmitted waves have a constant amplitude. Therefore, in these regimes the interface will support waves running ahead of the crack front. Moreover, here the fields are majorised by powers of the exponents $\lambda_j$, $j=1,2,$ indicating that the transmitted field is localised about the defect.





\subsection{The strain energy release rate ratio for the interfacial crack propagating through dissimilar media}\label{SM8}
Here $G$, for the interfacial crack in a continuous bimaterial formed from two dissimilar anisotropic half spaces, can be computed according to \citet{Suo_err} with
\[G=\lim_{\tau\to 0}\frac{1}{2\tau}\int_0^\tau \boldsymbol{\sigma}(x)\cdot [\mathbf{u}](x-\tau)dx\;, \]
where  $\boldsymbol{\sigma}$ represents stresses ahead of the crack tip and $[\mathbf{u}]$ is the displacement jump computed behind the crack tip.
As in the monograph of   \citet{LSbook}, for the continuous bimaterial this may also take the form
\begin{equation}\label{eqG}
G=\frac{1}{2} \lim_{k\to \infty} k^2 \boldsymbol{\sigma}^{F}(\text{i} k) [\mathbf{u}]^{F}(-\text{i} k)\;.
\end{equation}
In addition, the equivalent quantity to (\ref{eqG}) can be obtained from the dissimilar lattice with the formula
\begin{equation}\label{eqGlat}
G=\frac{\mu}{2} \lim_{k\to 0} k^2 \Psi_+(\text{i} k) \Psi_-(-\text{i} k)\;.
\end{equation} 

It then follows from this, (\ref{eq:Asymptotics_Lattice_Fourier_2}) and (\ref{eq:Asymptotics_Lattice_Fourier_3}) that 
\begin{equation}\label{GD}
G=\frac{1}{2}\mu \varepsilon^2_c R^{-2}\;,
\end{equation}
where $R$ is given in (\ref{eq:Asymptototics_L+-_zero}). We  also note that if the crack advances a single lattice unit through the breakage of a crack front bond, the non-dimensional strain energy release rate is 
\begin{equation}
\label{G01}
 G_{0}=
 \frac{\mu}{2} (\psi(0))^{{2}}=
 \frac{\mu}{2} \epsilon_c^{{2}}\;.
 \end{equation}
Taking the ratio of (\ref{G01}) to (\ref{GD}) then gives (\ref{eq:ERR_R}).

\end{document}